%% file: orbreview-rpp.tex
\documentclass[rmp,aps,onecolumn,superscriptaddress]{revtex4-1}
\usepackage{amsmath,amssymb,mathrsfs}
\usepackage{graphicx}
\usepackage{paralist} 
\usepackage{esint}
\usepackage{dcolumn}
\usepackage{color}
\usepackage{bbold}

\usepackage[unicode=true, bookmarks=true,bookmarksnumbered=false,bookmarksopen=false,breaklinks=false,pdfborder={0 0 1},backref=false,colorlinks=true] {hyperref}
\hypersetup{ colorlinks=false,linkbordercolor={1 0 .5},urlbordercolor={0 1 0}, citebordercolor={0 1 0}, pdfstartview={FitH}, unicode=true}

\def\be{\begin{equation}}
\def\ee{\end{equation}}
\def\nn{\nonumber}

\def\tbf{\textbf}

\def\bea{\begin{eqnarray}}
\def\eea{\end{eqnarray}}

\newcommand{\figfolder}{figures}

\begin{document}


\title{Physics of higher orbital bands in optical lattices: a review} 
\author{Xiaopeng Li} 
\affiliation{Condensed Matter Theory Center and Joint Quantum Institute, Department of Physics, 
University of Maryland, College Park, MD 20742-4111, USA}

\author{W. Vincent Liu} 
\email{wvliu@pitt.edu}
\affiliation{Department of Physics and Astronomy, University of Pittsburgh, Pittsburgh, Pennsylvania 15260, USA}
\affiliation{Wilczek Quantum Center, Zhejiang University of Technology, Hangzhou 310023, China}

\date{\today}

\begin{abstract}
  The orbital degree of freedom plays a fundamental role in understanding the
  unconventional properties in solid state materials.  Experimental progress in
  quantum atomic gases has demonstrated that high orbitals in optical lattices
  can be used to construct quantum emulators of exotic models beyond natural
  crystals, where novel many-body states such as complex Bose-Einstein
  condensates and topological semimetals emerge.  A brief introduction of
  orbital degrees of freedom in optical lattices is given and a summary of
  exotic orbital models and resulting many-body phases is provided. Experimental
  consequences of the novel phases are also discussed.
\end{abstract}

\begin{titlepage}
\maketitle
\thispagestyle{empty}
\end{titlepage}

\tableofcontents  

\thispagestyle{empty}

\newpage  



\input{1introduction}

\input{2bandstructure} 
\input{3phases}

\input{4ExpProbes}

\input{5Discussion}

\section*{Acknowledgement} The authors are grateful to Andreas Hemmerich, Sankar Das Sarma, Ivan H. Deutsch, Philipp Hauke, Chiu Man Ho, Randy Hulet, Hsiang-Hsuan Hung, Maciej Lewenstein, Chungwei Lin, Bo Liu, Joel Moore, Arun Paramekanti, Vladimir Stojanovic, Kai Sun, Biao Wu, Congjun Wu, Hongwei Xiong, Yong Xu, Zhixu Zhang, Zhenyu Zhou, and Erhai Zhao for their close collaboration and important contributions reviewed in this paper. 
This work is supported by ARO (W911NF-11-1-0230), AFOSR (FA9550-16-1-0006), the Charles E. Kaufman Foundation, and The Pittsburgh Foundation (W. V. L.) and by LPS-MPO-CMTC, JQI-NSF-PFC and ARO-Atomtronics-MURI (X. L.). Part of the work reviewed in this paper is the outcome of Overseas Collaborative Program of NSF of China No. 11429402 sponsored by Peking University, which is deeply acknowledged. We  want to thank  the International Center for Quantum Materials at Peking University and Wilczek Quantum Center at Zhejiang University of Technology, where the manuscript is completed, for the hospitality.  

\appendix

\input{6Appendix}

\bibliography{references}
\bibliographystyle{apsrmp4-1}

\end{document}

%% file: 1introduction.tex
\section{Introduction}

Optical lattices play a central role in studying strongly interacting many-body physics with ultracold atoms~\cite{2008_Bloch_Dalibard_RMP,2012_Lewenstein_Sanpera_book,2015_Dutta_Gajda_RPP}. Because of their unprecedented controllability, atomic gases confined in optical lattices {enable quantum simulation} of various lattice Hamiltonians, e.g., Bose and Fermi Hubbard models, where different aspects have been intensively investigated. With single-species of bosons, e.g., $^{87}$Rb, a quantum Mott-to-superfluid transition has been observed. Multi-component lattice models have been reached with atomic internal degrees of freedom. The SU(2)  spinfull Fermi-Hubbard simulator has been carried out by using hyperfine states of $^{6}$Li or $^{40}$K. One theme along this direction  is to emulate complex correlated phenomena of strongly interacting electrons. Such multi-component quantum simulators with atomic internal degrees of freedom have been very successful in simulating  Hamiltonians with high symmetries. 

For electrons, one important ingredient besides spin is the orbital degrees of freedom, which 
arises in a variety of condensed matter systems~\cite{2000_Tokura_Nagaosa_Science}. 
In solid state materials, 
orbitals originate from electron clouds surrounding the ions in the crystal. With tunnelings, 
these orbitals form Bloch bands. Orbitals are Wannier states corresponding to different bands. 
Degenerate orbitals (or bands) could emerge in presence of point group symmetries, but the symmetry for orbitals is much lower than for spins. Understanding such orbitals degree of freedom is crucial to obtain a simple and yet powerful model that captures the essence of many complicated materials, such as transition metal oxides, pnictides, etc.  A task of this kind however remains outstanding, much due to the complexity of multiple types of degrees of freedoms coupled together, including orbital, charge, spin, and crystal field.  The intricate coupling makes it an expensive challenge, both analytically and numerically,  to understand orbital physics alone first and to attempt to compare with any electronic solid state materials in experiments.

 Given one important application of optical lattices is to simulate complex phenomena of electrons, it is rather essential to find ways to emulate electron orbitals with atoms. 
Actually with optical lattices, the ionic crystal trapping electrons is replaced by an artificial crystal of light, created by standing waves of laser beams. The Wannier orbitals in the lattice naturally mimic properties of that in ionic crystals. Due to the intrinsic spatial nature, orbital degree of freedom in both of these ionic and light crystals respect space point group symmetries rather than internal continuous group symmetries, which defines its uniqueness. Such symmetry properties of orbitals make them  fundamentally difficult to be simulated with internal atomic degrees of freedom such as hyperfine spins. On this regard, the orbital states of an atom in an optical lattice provide a natural avenue to emulating the electronic orbital related physics.


Exploration of orbital physics in optical lattices is certainly not restricted to quantum simulations of electrons in solids.  For example, orbital bosons are able to bring to the study of quantum matter some really novel concepts that have no prior analogue in systems of (fermionic) electrons.  Moreover,  bosons (e.g.,  $^{87}$Rb atoms)  are more widely used in optical lattice experiments. In the first experimental demonstration of many-body orbital physics, bosons were loaded into the $p$-bands of an optical lattice, for which earlier theoretical studies had predicted interesting phenomena such as time-reversal symmetry breaking and spontaneous angular momentum order~\cite{2005_Isacsson_Girvin_PRA,2006_Liu_Wu_PRA,2006_Kuklov_PRL}. 

Strong interactions which are achievable in optical lattice experiments also lead to interesting orbital physics. Firstly, with strongly repulsive bosons loaded into higher orbital bands, they would form a Mott state with orbital degree of freedom. Orbital ordering in such a Mott state is drastically different from spin ordering in Mott states. For Mott states formed by spinor bosons (assuming no spin-orbital coupling), the super-exchange Hamiltonian typically has high symmetries. The orbital super-exchange Hamiltonian is generally more complicated and at the same time promises richer physics. 
Secondly, for strongly interacting atoms in a lattice (e.g., lattice bosons in the Mott regime, or a Feshbach resonant Fermi gas~\cite{2010_Chin_Grimm_RMP} in a lattice), even without deliberately loading atoms into higher orbital bands, population of those bands is unavoidable due to interaction effects. This is because local interactions would mix  all different orbitals. Recent works~\cite{2011_Zhou_Porto_PRAR,2012_Soltan_Luhmann_NatPhys} have shown that the interaction-induced high-band population could give rise to significant physical effects, such as  condensation of boson pairs, and exotic symmetry breaking orders. It is therefore essential rather than an option to account for orbital physics in modeling strong interaction effects in optical lattices.

Research of fermions in higher orbitals adds a remarkably distinct venue.  Theoretical studies have also found  quantum phases with angular momentum ordering that spontaneously break time-reversal symmetry. For fermions, this symmetry breaking leads to even more dramatic effects than the bosonic counterpart. Considering the angular momentum order and mixing of orbitals with opposite parities (like $s$ and $p$, or $p$ and $d$ orbitals), the fermionic atoms experience effective gauge fields, which then gives rise to topological phenomena, like quantum Hall, topological insulator, or certain topologically protected gapless phases. This route of engineering topological matter offers one way different from the Raman-induced synthetic gauge fields~\cite{2011_Dalibard_Gerbier_RMP} or the artificial spin-orbit couplings~\cite{2015_Zhai_RPP,2013_Galitski_Nat}. It has fundamentally distinct properties and is advantageous in certain aspects.  For example, it does not involve complications of Raman couplings, and the resultant topological phases would have longer lifetime due to less heating effects. The finite temperature behaviors of the spontaneously generated gauge fields are also different from the the Raman-induced case.


In this review, we start by describing basics of modeling orbitals in optical lattices.  Then by using particular examples, we present a selection of many-body aspects of orbital physics that we find most interesting and novel, as sketched above. Along with developing theoretical concepts and models pedagogically, we review the recent experimental developments and the current status in this field, and outline several future directions.

%% file: 2bandstructure.tex
\section{High orbitals and band structures in optical lattices}

Previous studies in optical lattices largely focused on atoms trapped in 
the lowest band and the resultant single-band Hubbard model, where 
correlated effects of bosons, e.g., the Mott-superfluid transition, have been 
intensively investigated~\cite{2008_Bloch_Dalibard_RMP,2012_Lewenstein_Sanpera_book,2015_Dutta_Gajda_RPP}.  
In this section we present the procedure to construct tight binding models involving high orbital degrees of 
freedom, which is one essential step to study correlation effects in interacting atoms in lattices. 
To demonstrate the validity condition of tight binding models, we also show the exact results 
from plane-wave expansion for the tunneling amplitudes, band structures and Wannier functions of higher bands. 
A two dimensional square lattice 
is assumed in this section. 

\subsection{Harmonic approximation and tight binding models} 

In the tight binding regime, an optical lattice can be treated as 
individual harmonic oscillators, 
which are coupled by quantum tunnelings. On each harmonic oscillator centered at 
a lattice site labeled by its position $\tbf{R}$, we have discrete energy levels with orbital wavefunctions 
$\phi_{\alpha} (\tbf{x} - \tbf{R})$. Associated with the localized orbital wavefunctions, 
we can define the lattice operators $b_{\alpha} (\tbf{R})$. To do this,
it has to be enforced that the orbital wavefunctions are orthonormal. 
The simple eigen wavefunctions of harmonic oscillators do not satisfy 
orthonormal condition, for the reason that there are overlaps between 
orbital wavefunctions on neighboring sites. 

The procedure to construct the orthogonal basis from the 
localized harmonic oscillator wavefunctions is the following. 
We start with the harmonic oscillator wavefunctions 
$\phi_{\alpha} (\textbf{x} - \tbf{R})$ localized on site $\textbf{R}$. 
These wavefunctions are already approximately orthogonal, i.e., 
$$
\int d \textbf{x}  \phi_{\alpha} ^* (\textbf{x} - \tbf{R})  
    \phi_{{\alpha}'} (\textbf{x} - \tbf{R}') = 
    \delta_{{\alpha} {\alpha}'} \delta_{\tbf{R} \tbf{R}'} + \epsilon_{{\alpha} \tbf{R}, {\alpha}'\tbf{R}'},
$$ 
where $\epsilon_{{\alpha} \tbf{R}, {\alpha}'\tbf{R}'}$ are small numbers. By definition we know that 
$[\epsilon]$ is a traceless Hermitian matrix. 
Then we improve this basis by introducing 
\bea 
\tilde{\phi}_{\alpha} (\textbf{x}) 
  = \phi_{\alpha} (\textbf{x}) 
      - \frac{1}{2} \sum_{{\alpha}' \tbf{R}'}  \epsilon_{{\alpha}' \tbf{R}', {\alpha} \tbf{R}} \phi_{{\alpha}'} (\tbf{x} - \tbf{R}') 
\eea
After that $\phi_{\alpha}(\tbf{x}-\tbf{R})$  is renormalized as
$$ 
 \tilde{\phi}_{\alpha} (\textbf{x} ) \to \tilde{\phi}_{\alpha} (\textbf{x} ) 
 \displaystyle /
      \sqrt{ \int d \textbf{x}' |\tilde{\phi}_{\alpha}  (\textbf{x}')|^2 }. 
$$  
The improved wavefunctions satisfy a better  approximate orthogonal condition 
$$
\int d \textbf{x} \tilde{\phi}_{\alpha} ^* (\textbf{x}-\tbf{R}) 
	      \tilde{\phi}_{{\alpha}'} (\textbf{x}-\tbf{R}') 
  = \delta_{{\alpha} {\alpha}'} \delta_{\tbf{R} \tbf{R}'}  + \mathcal{O} ( \epsilon^2). 
$$
The above procedure can be iterated $N$ times to get the orthonormal basis to the 
precision of $\mathcal{O} (\epsilon^{2^N})$.

Once we have the orthonormal basis, the tunnelings  between $\tbf{R}$ and $\tbf{R}'$ 
are calculated as 
\be 
t_{{\alpha} {\alpha}'} (   {\tbf{R} -\tbf{R}'} )
= \int d \textbf{x} \tilde{\phi}_{\alpha} ^* (\textbf{x} - \tbf{R}) H (\textbf{x}) 
      \tilde{\phi}_{{\alpha}'} (\textbf{x} - \tbf{R}'), 
 \label{eq:HarmonicTunneling} 
\ee 
where $H(\textbf{x})$ is the Hamiltonian in the first quantization form 
$
H = -\frac{\hbar^2}{2m}\vec{\nabla}^2 + V(\tbf{x}). 
$ 
The lattice model Hamiltonian including tunnelings is given by 
\bea 
\hat{H} = \sum_{{\alpha} {\alpha}', \tbf{R} \tbf{R}'}
t_{{\alpha} {\alpha}'} (   {\tbf{R} -\tbf{R}'} ) b_{\alpha} ^\dag (\tbf{R}) b_{{\alpha}'} (\tbf{R}').
\eea 
 Without truncating the 
basis, the Hamiltonian is exact, from which the band structure can be calculated. 
If we only keep the lowest harmonic wave functions, this lattice Hamiltonian gives 
qualitatively correct band structures for deep lattices.

The procedure described above to construct orthogonal orbital wave functions is  one essential step if one uses harmonic approximation.  In principle, the constructed wave functions are  not the same as the maximally localized Wannier functions~
(\onlinecite{1982_Kivelson_PRB}; \onlinecite{2012_Marzari_Mostofi_RMP}; \onlinecite{2013_Uehlinger_Jotzu_PRL}; \onlinecite{2014_Ganczarek_Modugno_PRA}). The procedure to calculate such maximally localized Wannier functions is not as straightforward and is beyond the scope of this review.

\paragraph*{Multi-band Hubbard model.---}
Considering interacting bosonic atoms loaded on excited bands, 
the physics will be described by a 
 multi-band Hubbard model 
\bea 
H &=& \sum_{\tbf{R} \tbf{R}'} t ^{({\alpha})} (\tbf{R} - \tbf{R}') b_{\alpha} ^\dag (\tbf{R}) b_{\alpha} (\tbf{R}') \nn \\
&&+ \sum_{\tbf{R}} U_{{\alpha}_1 {\alpha}_2 {\alpha}_3 {\alpha}_4} 
b_{{\alpha}_1} ^\dag (\tbf{R}) b_{{\alpha}_2}^\dag (\tbf{R}) b_{{\alpha}_3} (\tbf{R}) b_{{\alpha}_4} (\tbf{R}).
\eea 
With weak interaction, the coupling constants $U_{{\alpha}_1 {\alpha}_2 {\alpha}_3 {\alpha}_4}$ can be estimated at tree-level 
as~\cite{1998_Jaksch_Bruder_PRL,2006_Liu_Wu_PRA,2011_Zhou_Porto_PRAR,2013_Li_Paramekanti_NatComm,2015_Dutta_Gajda_RPP}  
\bea 
\label{eq:2interactions} 
U_{{\alpha}_1 {\alpha}_2 {\alpha}_3 {\alpha}_4} = \frac{4\pi a_s \hbar^2}{2m} \int d^3 \tbf{x} \phi_{{\alpha}_1} ^* (\tbf{x}) \phi_{{\alpha}_2} ^* 
(\tbf{x}) \phi_{{\alpha}_3} (\tbf{x}) \phi_{{\alpha}_4} (\tbf{x}), 
\eea 
where $a_s$ is the s-wave scattering length, tunable with Feshbach Resonance techniques. 
With fermionic atoms, we have a similar Hubbard model with interactions between hyperfine states.

\subsection{Band structures} 
In terms of field operators, the Hamiltonian of particles moving in optical lattices 
is 
\bea 
\hat{H} = \int d^d  \tbf{x} \psi  ^\dag(\tbf{x}) 
	\left(-\frac{\hbar^2}{2m} \vec{\nabla}^2 + V(\tbf{x}) \right) \psi(\tbf{x}), 
\eea 
where $\psi(\tbf{x})$ is a field operator. It can be either bosonic or fermionic. 
Statistics is irrelevant here to determine single-particle band structures. 
We expand the operator $\psi(\tbf{x})$ in the momentum basis
\bea 
\psi(\tbf{x}) = \sum_{\tbf{K}} \frac{1}{\sqrt{N_s}} \sum_\tbf{k} \tilde{ a}_{\tbf{K}} (\tbf{k}) e^{i(\tbf{K} + \tbf{k})\cdot \tbf{x}}, 
\eea 
where  $\tbf{K}$ labels the reciprocal lattice vectors, $\tbf{k}$ the lattice momentum, 
and $N_s$ the number of lattice sites. 
Here and henceforth, the lattice constant is set as the length unit. 
Optical lattice potentials $V(\tbf{x})$, unlike the potentials in electronic materials, can 
typically be written as superpositions of just a few plane waves, i.e., 
$$
V(\tbf{x}) = \sum_{\tbf{K}} v(\tbf{K}) e^{i \tbf{K} \cdot\tbf{x}}.  
$$ 
For example, the potential of a square lattice created by 
laser is 
\begin{align*}
V(\tbf{x}) &= - V_0 \left[ \sin^2 (kx) + \sin^2 (ky) \right] \\ 
&= - \frac{V_0}{4} \left [ e^{2ikx} + e^{2iky} + c.c.\right] + const,  
\end{align*}
where $k$ is the wavevector of the laser beams. 
The Hamiltonian in momentum space reads as 
\bea 
\hat{H}  = \sum_{\tbf{k}} \sum_{\tbf{K}_1, \tbf{K}_2}  
	\mathcal{H}_\tbf{k} (\tbf{K}_1, \tbf{K}_2)
	  \tilde{a}_{\tbf{K}_1} ^\dag (\tbf{k}) \tilde{a}_{\tbf{K}_2} (\tbf{k}), 
\eea 
with the matrix given by  
\bea 
\mathcal{H}_\tbf{k} (\tbf{K}_1, \tbf{K}_2)
= \frac{\hbar^2 (\tbf{K}_1 + \tbf{k}) ^2}{2m} \delta_{\tbf{K}_1 \tbf{K}_2} 
+v(\tbf{K}_1 - \tbf{K}_2). 
\eea 
Diagonalizing this matrix, we get the band structure $E_n (\tbf{k})$ and 
the eigenvectors $\lambda^{(n)}_{\tbf{K}} (\tbf{k})$, with $n$ the band index. The Hamiltonian in the 
eigen-basis 
reads 
\bea 
\hat{H} = \sum_{n} \sum_{\tbf{k}}   E_n (\tbf{k}) b_n ^\dag (\tbf{k}) b_n (\tbf{k}),  
\eea  
with ${b}_n (\tbf{k}) = \sum_{\tbf{K}} \lambda ^{(n)*} _{\tbf{K}} (\tbf{k}) \tilde{a}_{\tbf{K}} (\tbf{k})$.

\begin{table}[ht]
\centering
\begin{tabular}{c c c c c c c }
 \hline \hline
 $V_0/E_R$ 	&$4 t^s _{nn}/E_R$	&$4t^s _{nnn}/E_R$		&$4t^p _{nn}/E_R$ 	&$4t^p _{nnn}/E_R$	\\
\hline 
$3$		&-0.4441	&0.0449		&2.0074		&0.3308	\\
$5$		&-0.2631	&0.0136		&1.6912		&0.2914	\\
$10$		&-0.07673	&9.1E-4		&0.9741		&0.1051	\\
$20$		&-9.965E-3	&1.2E-5		&0.2411		&5.5E-3	\\	
\hline \hline
\end{tabular}
\caption{Tunneling amplitudes in a two dimensional square lattice with potential 
$V({\bf x} ) = - V_0 \left[ \sin^2 (kx) + \sin^2 (ky)\right] $. $E_R$ is the one photon recoil energy $\frac{\hbar^2 k^2}{2m}$. 
$t^s_{nn}$ and $t^s_{nnn}$ are nearest 
neighbor and next nearest neighbor tunnelings for the lowest $s$ band. $t^p _{nn}$ and $t^p_{nnn}$ are nearest 
neighbor and next nearest neighbor tunnelings in the $x$ direction for the $p_x$ (first excited) band. 
 }
\label{table:tunnelings}
\end{table}


The Wannier basis is given by 
\bea 
{b}_n (\tbf{R}) = \frac{1}{\sqrt{N_s}} \sum_\tbf{k} {b}_n(\tbf{k}) e^{i\tbf{k}\cdot \tbf{R}}. 
\eea 
Inversely we have 
$
{b}_n(\tbf{k}) = \frac{1}{\sqrt{N_s}} \sum_\tbf{R} {b}_n (\tbf{R}) e^{-i\tbf{k} \cdot \tbf{R}}. 
$

The  Wannier wavefunctions of the Bloch bands are given by 
\bea 
w_n(\tbf{x} - \tbf{R}) = \sum_{\tbf{K}} \left[ \frac{1}{N_s} \sum_{\bf k} \lambda^{(n)} _\tbf{K} (\tbf{k}) 
	e^{i(\tbf{K} + \tbf{k}) \cdot (\tbf{x} - \tbf{R})} \right] . 
\label{eq:2wannier}
\eea 
The Hamiltonian can be rewritten in the Wannier basis as 
\bea 
\hat{H} = \sum_{\tbf{R} \tbf{R}'} t ^{(n)} (\tbf{R} - \tbf{R}') {b}_n ^\dag (\tbf{R}) {b}_n (\tbf{R}'), 
\eea 
with
\bea
t^{(n)} (\tbf{R} - \tbf{R}') =  
 \frac{1}{N_s} \sum_{\bf k} E_n (\tbf{k}) e^{i \tbf{k} \cdot (\tbf{R} - \tbf{R}') }.  
\eea
Typical values of tunnelings (tunnelings refer to tunneling matrix elements here) for $s$ and $p$ bands are listed in Table~\ref{table:tunnelings}.

\begin{figure}[htp]
\includegraphics[angle=0,width=\linewidth]{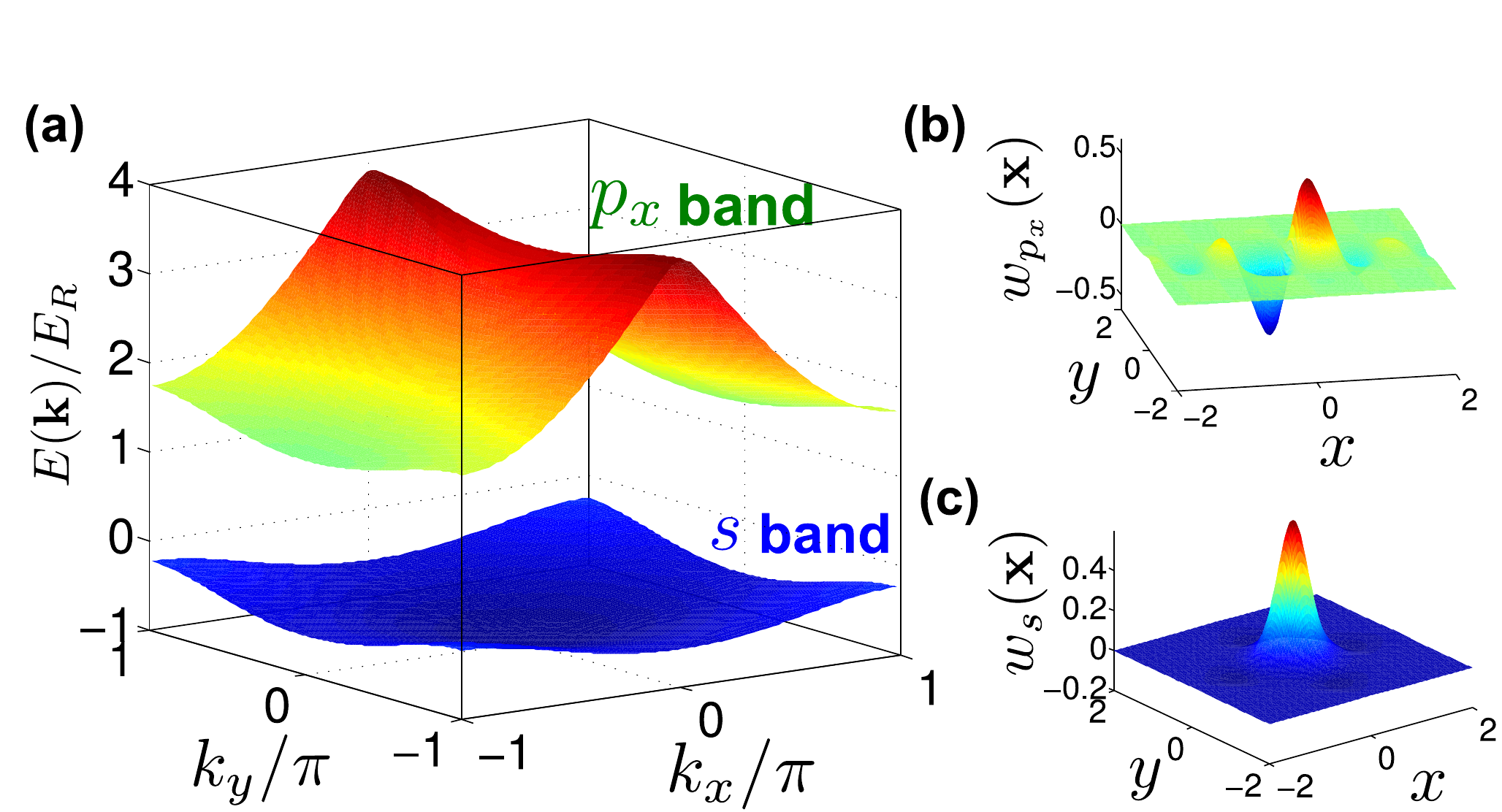}
\caption{Lowest two bands of a square lattice. The potential we choose here is 
$V(\tbf{x}) = -V_0 \left[ \sin^2 (k x) + \sin^2 (k y) \right]$, with $V_0/E_R = 4$ ($E_R$ is the one photon recoil energy). 
(a) shows band structures of $p_x$ and $s$ bands, whose Wannier functions are respectively 
shown in (b) and (c).}
\label{fig:2band_wannier}
\end{figure}

In the definition of Wannier functions (Eq.~\eqref{eq:2wannier}), there are gauge degrees of freedom 
$\lambda_\tbf{K} ^{(n)} (\tbf{k}) \to e^{i\theta _n (\tbf{k})} \lambda _\tbf{K} ^{(n)} (\tbf{k})$. One has 
to make a smooth gauge choice to get localized Wannier functions~\cite{2012_Marzari_Mostofi_RMP}. Wannier functions 
of $p$-bands of a square lattice are shown in Fig.~\ref{fig:2band_wannier}.

\subsection{Heuristics to lifetime of high orbital atoms}
\label{sec:fermigoldenlifetime}
Here the lifetime of $p$-orbital condensate in a 
{\it one dimensional} (1D) lattice 
is discussed based upon Fermi's Golden rule calculation. 
The resulting time scale is expected to apply to two dimensional (2D) 
square and three dimensional (3D) cubic lattices as well~\cite{2005_Isacsson_Girvin_PRA}.
The $p$-orbital condensate wavefunction is given as 
\bea 
|\Psi\rangle = \frac{\left[ {b}_p ^\dag (Q=\pi)  \right]^N }{\sqrt{N}!} |{\rm vac}\rangle. 
\eea 
With interactions, two particles in the $p$-band can collide and one particle would decay to the lowest $s$-band 
and the other goes to the second excited $d$-band. This process is described by the following interaction term 
$$  
 H_{\rm int}^{spd}  
 = {U/N_s} \sum_{k_1+ k_2+ k_3+ k_4 =0} 
\left\{ {b}_s^\dag ({k}_1) {b}_d ^\dag ({k}_2) 
{b}_p (k_3) {b}_p (k_4) + h.c. \right\} . 
$$ 
The final state after the collision is 
$$
| \Psi_f; k_1, k_2 \rangle = {b}_s ^\dag (k_1) {b}_d ^\dag (k_2) 
	\frac{[{b}_p ^\dag (Q) ]^{N-2}}{\sqrt{(N-2)!}} |{\rm vac}\rangle. 
$$
The transition probability from second order perturbation theory is 
$$
P(k_1 , k_2; t) \approx \frac{4 \sin^2 (\Delta \epsilon_{k_1 k_2} t/2 ) }{ \Delta \epsilon_{k_1 k_2}^2  }
|\langle \Psi_f; k_1, k_2 | H_{\rm int} ^{spd} | \Psi \rangle |^2,  
$$   
where $\Delta \epsilon_{k_1 k_2}$ is the difference of kinetic energy between $|\Psi\rangle$ and 
$|\Psi_f; k_1, k_2 \rangle $. 
The loss rate from the $p$-band is obtained as 
\begin{align*}
w_t &= \sum _{k_1 k_2}  \frac{1}{t} P(k_1 , k_2; t) |_{t\to \infty}  \\
 &\approx \sum_{k_1 k_2} \frac{2\pi}{\hbar} |\langle \Psi_f ; k_2, k_2 | H_{\rm int} ^{spd} | \Psi \rangle | ^2 
	  \delta( \Delta \epsilon_{k_1 k_2})  \\
& \approx  \frac{4\pi U^2  }{\hbar N_s } N(N-1) \left[ \frac{1}{\rho (\epsilon_s (K))} + \frac{1}{\rho (\epsilon _d (-K) )}   \right] ^{-1}, 
\end{align*}
where $\rho(\epsilon)$ is the density of states and $K$ is determined by 
\be
\epsilon_s (K) + \epsilon_d (-K) = 2 \epsilon_p (Q), 
\label{eq:2energyconservation} 
\ee
which in general has two solutions when the band gap between $s$ and $p$ matches that between $p$ and $d$. 

The loss rate per site is 
\bea 
w & \approx & \frac{4\pi (\nu U) ^2  }{\hbar}  \left[ \frac{1}{\rho (\epsilon_s (K))} + \frac{1}{\rho (\epsilon _d (-K) )}   \right] ^{-1}, 
\eea 
with $\nu$ the filling factor. 
The lifetime $1/w$ is typically short for cubic or square lattices, 
where the condition of Eq.~\eqref{eq:2energyconservation} may be satisfied. 
It was suggested that anharmonicity~\cite{2007_Muller_Folling_PRL} present in the actual 
optical lattice potential should help suppress the decay. 
Nonetheless, the lifetime can be significantly improved  
by using double-well lattice potentials to mismatch the band gaps 
as first discussed in Ref.~\cite{2008_Stojanovic_Wu_PRL} and further confirmed in the experiments~\cite{2011_Wirth_Olschlager_NatPhys}. 

%% file: 3phases.tex
\section{Many-body phases and transitions}

Orbital degrees of freedom play an important role in understanding  many
complex phases in solid state materials. For example, 
high temperature superconductivity in the
cuprates~\cite{1986_Bednorz_Muller_PBCM} 
and pnictides~\cite{2006_Kamihara_Hiramatsu_JACS}, 
chiral $p$-wave superconductivity proposed in Sr$_2$RuO$_4$~\cite{1998_Luke_Fudamoto_Nature}, 
and Ferromagnetic superconductivities in oxide heterostructures such as 
LaAlO$_3$/SrTiO$_3$~\cite{2004_Ohtomo_Hwang_Nature},  are all nucleated by strong correlation 
effects in a multi-orbital setting~\cite{2000_Tokura_Nagaosa_Science}. 
In optical lattices, recent studies have shown that the interplay of high orbitals and interaction effects give rise to unconventional many-body phenomena~\cite{2011_Lewenstein_Liu_NatPhys}.

For bosons loaded into high-orbital bands of an optical lattice, an analogue of Hund's rule coupling leads to a complex Bose-Einstein condensate with spontaneous angular momentum order~\cite{2005_Isacsson_Girvin_PRA,2006_Liu_Wu_PRA,2006_Wu_Liu_PRL,2006_Kuklov_PRL,2009_Wu_MPLB}. 
The bosonic analogue of Hund's rule basically states that repulsive contact interactions favor maximization of the local angular momentum. Different aspects of the unconventional condensate have been theoretically investigated, e.g., rotation effects~\cite{2008_Umucalar_Oktel_PRA}, manifestations of lattice geometry and trapping potential~\cite{2011_Cai_Wu_PRA,2008_Lim_Smith_PRL,2012_Pinheiro_Martikainen_PRA}, and orbital phase transitions~\cite{2011_Stasyuk_Velychko_CMP,2012_Stasyuk_Velychko_CMP,2012_Pietraszewicz_Sowinski_PRA,2013_Pinheiro_Bruun_PRL,2015_Pinheiro_Matrikainen_NJP}.
Experimentally, this complex Bose-Einstein condensate has recently been demonstrated in a checkerboard optical lattice~\cite{2011_Wirth_Olschlager_NatPhys,2013_Olschlager_Kock_NJPHYS,2015_Kock_Olschlager_PRL}. By considering strong interactions, this condensate state develops a quantum phase transition to a Mott state with very rich orbital ordering, which has been studied by mean field theories~\cite{2009_Larson_Collin_PRA,2010_Collin_Larson_PRA,2011_Li_Zhao_PRA} and also by unbiased numerical methods~\cite{2013_Hebert_Cai_PRB,2012_Li_Zhang_PRL,2013_Sowinski_Lacki_PRL}. Even without deliberately loading atoms into the higher bands, it has been shown high-band population can be stabilized by interaction effects~\cite{2011_Zhou_Porto_PRAR,2012_Soltan_Luhmann_NatPhys}.

For fermions, it has been shown that interaction effects combined with the band topology of $p$-orbitals lead to various exotic quantum phases. With $p$-orbital fermions in two dimensions, interactions cause generic instabilities towards quantum density wave orders (modulations in spin, charge or orbital density)~\cite{2006_Wu_Liu_PRL,2008_Wu_Sarma_PRB,2008_Zhao_Liu_PRL,2008_Wu_PRL2,2009_Lu_Arrigoni_PRB,2012_Zhang_Li_PRA,2012_Wu_He_PRB},  unconventional Cooper pairings~\cite{2011_Hung_Lee_PRB,2011_Cai_Wang_PRA,2011_Zhang_Hung_PRA,2010_Lee_Wu_PRA,2010_Zhang_Hung_PRA,2015_Liu_Li_arXiv}, and  novel quantum magnetism~\cite{2008_Wang_Dai_PRA,2008_Wu_Zhai_PRB,2010_Zhang_Hung_PRA2,2011_Hauke_Zhao_PRAR,2015_Zhou_Zhao_PRL} at low temperature. From quantum engineering perspectives,  the elongated spatial nature of $p$-orbitals makes them  ideal building blocks for  fascinating topological states, e.g., topological semi-metal~\cite{2011_Sun_Liu_NatPhys}, quantum Hall phases~\cite{2008_Wu_PRL,2010_Wang_Gong_PRB}, topological insulators/superconductors~\cite{2010_Liu_Liu_PRA,2013_Li_Zhao_NatComm,2014_Liu_Li_NatComm,2015_Liu_Li_arXiv2}, and even fractional states~\cite{2010_Sun_Zhao_PRL}.  

In this section, we will review a selection of quantum many-body phases of $p$-orbital bosons and fermions.

\subsection{Orbital $p+ip$ Bose-Einstein condensation}

\subsubsection{Complex $p_x+ip_y$ Bose-Einstein condensation at finite momentum}
\begin{figure}[htp]
\includegraphics[angle=0,width=\linewidth]{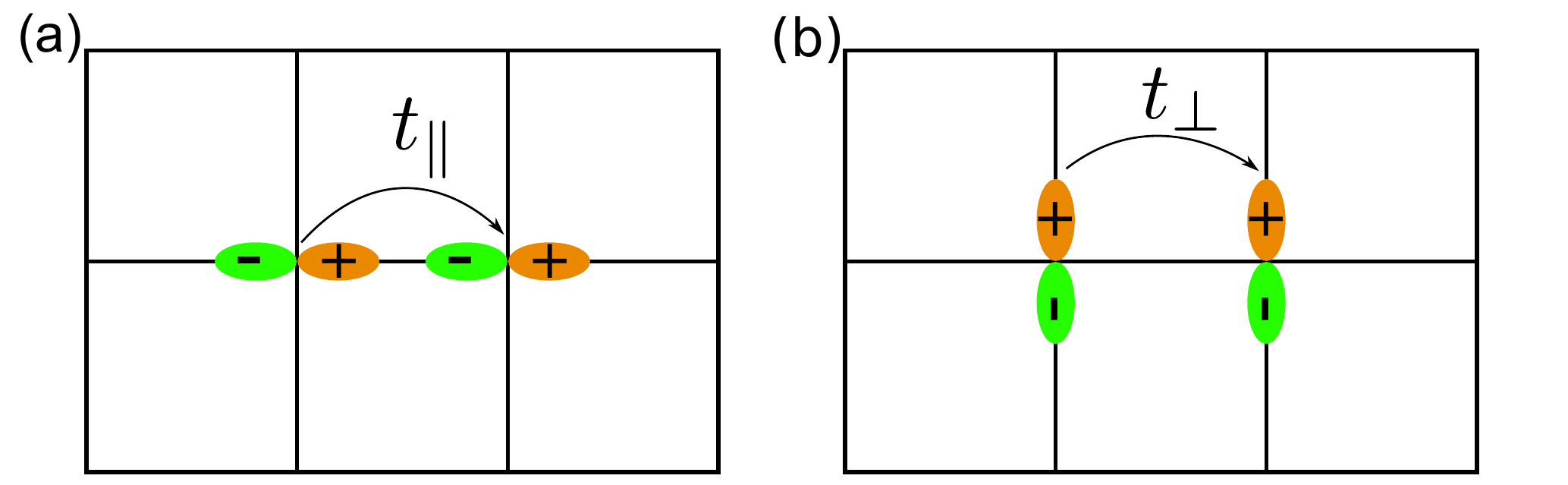}
\caption{Illustration of the tight binding model of $p$-orbital bosons on 
a square lattice~\cite{2006_Liu_Wu_PRA}. 
The longitudinal tunneling amplitude $t_\parallel$ is in general 
far greater than the transverse tunneling $t_\perp$. 
The ``$\pm$'' symbols indicate the sign 
of two lobes of $p$-orbital wave functions. }
\label{fig:3ptightbinding}
\end{figure}

For bosons loaded on $p$-orbitals of a 2D square lattice~\cite{2011_Wirth_Olschlager_NatPhys} in 
the tight-binding regime, the tunneling Hamiltonian is 
\bea 
H_{\rm tun}  &=& \sum_{\tbf{r}}\left\{  t_\parallel 
\left [ b_x ^\dag (\tbf{r})  b_x (\tbf{r} + \hat{a}_x) + 
	x \leftrightarrow y
\right] \right.\nn \\
&-&\textstyle \left. t_\perp 
\left [ b_x ^\dag (\tbf{r}) b_x (\tbf{r} + \hat{a}_y) + 
	x \leftrightarrow y
\right]
+h.c.
\right \} , 
\label{eq:3pbandHt}
\eea 
where $b_x$ and $b_y$ are bosonic annihilation operators for $p_x$ and $p_y$ orbitals, respectively (Fig.~\ref{fig:3ptightbinding}). 
After a Fourier transformation, we get the energy spectra for the $p_x$ and $p_y$ bands. The dispersion for 
the $p_x$ band is 
$$
\epsilon_x (\tbf{k}) =2 t_\parallel \cos (k_x) - 2 t_\perp \cos (k_y). 
$$ 
The dispersion for the $p_y$ band is readily obtained with a lattice rotation ($C_4$). 
There are two degenerate minima---$\tbf{Q}_x = (\pi, 0)$ and $\tbf{Q}_y = (0, \pi)$ in the $p$-bands with the degeneracy 
protected by the $C_4$ symmetry. 
The ground state manifold of non-interacting $p$-orbital bosons is spanned by 
\bea 
|N_x, N_y \rangle = \frac{\left[ {b}_x ^\dag  (\tbf{Q}_x)\right]^{N_x} 
		  \left[ {b}_y ^\dag (\tbf{Q}_y)\right]^{N_y}  }{\sqrt{N_x! N_y !}} |{\rm vac}\rangle , 
\label{eq:3groundmanifold}
\eea 
which has a large degeneracy that shall be lifted by interactions. 

The interaction terms of repulsive $p$-orbital bosons read~\cite{2005_Isacsson_Girvin_PRA,2006_Liu_Wu_PRA,2011_Li_Zhao_PRA}
\bea 
H_{\rm int} &=& \sum_{\tbf{r}} \left\{\frac{1}{2} U_1 \left[ n_x (\tbf{r}) n_x (\tbf{r})  + n_y (\tbf{r}) n_y (\tbf{r}) \right] \right. \nn\\
			      &&+ 2 U_2 n_x (\tbf{r})  n_y (\tbf{r})  \nn \\
			      &&\left. + \frac{1}{2}U_3  \left[ b_x ^\dag  (\tbf{r})b_x ^\dag  (\tbf{r})b_y (\tbf{r}) b_y (\tbf{r}) + h.c.\right] \right\}, 
\label{eq:3genericHint}  
\eea 
where the density operators $n_{\mu}= b_\mu^\dag b_\mu $. 
Approximating Wannier functions by localized harmonic wavefunctions, 
we have  
\be 
U_1 = 3 U_2 = 3 U_3 \equiv U >0, 
\label{eq:3Us}
\ee 
from which the interaction can be rewritten as 
\bea 
H_{\rm int} = \frac{U}{2} \sum_\tbf{R} \left[  n ^2 (\tbf{R})  - \frac{1}{3} L_z ^2 (\tbf{R}) \right],
\label{eq:3Hint}  
\eea 
with $n= \sum_\nu b_\nu ^\dag b_\nu$ and 
$
L_z = i b_x ^\dag b_y + h.c. 
$
We thus expect that the angular momentum order is ``universally'' favorable in $p$-orbital Bose gases. 

It is however worth emphasizing here that the angular momentum ordering does not rely on the strict equality in Eq.~\eqref{eq:3Us} or the interaction form in Eq.~\eqref{eq:3Hint}. This becomes more clear with Ginzburg-Landau or effective field theories analysis~\cite{2012_Li_Zhang_PRL,2013_Liu_Yu_PRA,2013_Li_Paramekanti_NatComm}. 
Detailed studies taking into account unharmonic corrections and trapping potentials also confirm that the angular momentum order indeed exists in the regimes accessible to optical lattice experiments~\cite{2010_Collin_Larson_PRA,2012_Pinheiro_Martikainen_PRA,2013_Sowinski_Lacki_PRL,2013_Pietraszewicz_Sowinski_PRA}.

\begin{figure}[htp]
\includegraphics[angle=0,width=\linewidth]{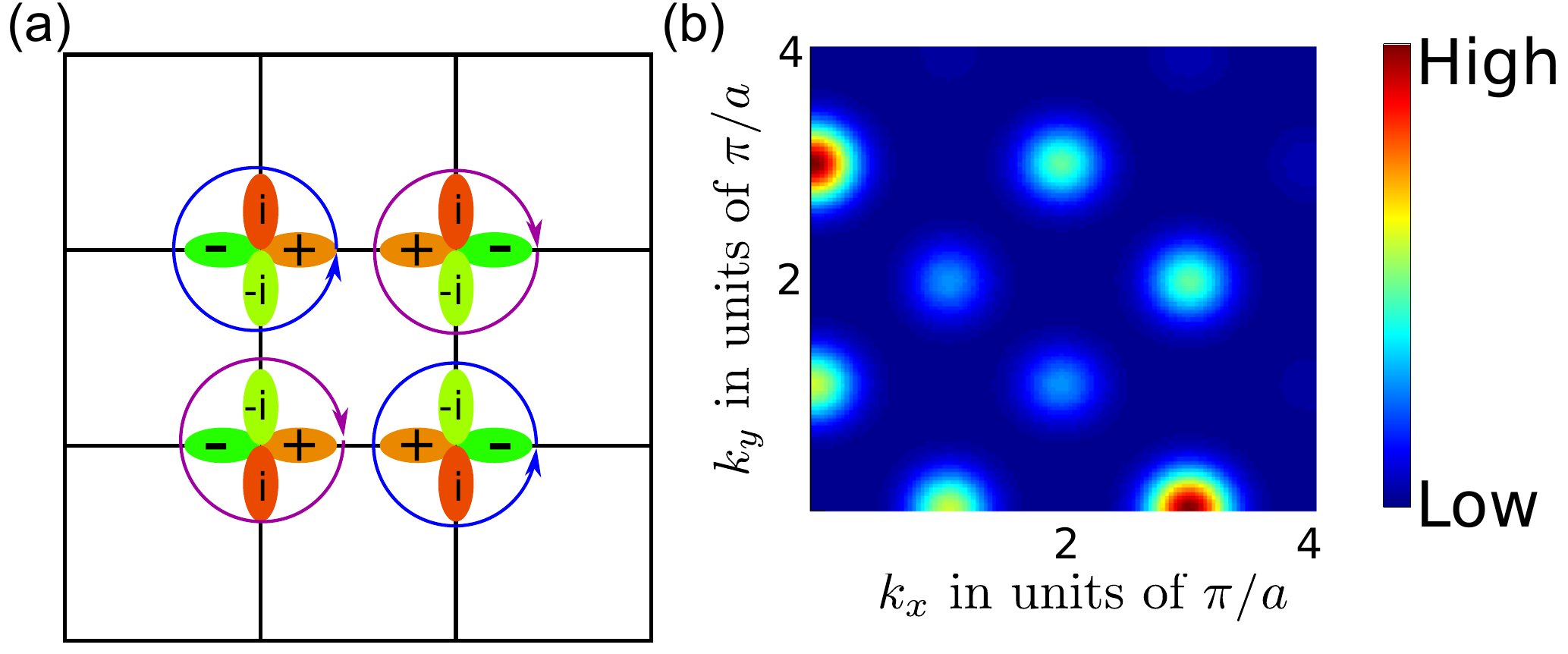}
\caption{Transverse staggered orbital current superfluid state~\cite{2006_Liu_Wu_PRA,2011_Li_Zhao_PRA}. 
(a) shows the phase configuration, from which 
one can infer the orbital current alternates from site to site. (b) shows the momentum distribution, which is confirmed 
in the experiments~\cite{2011_Wirth_Olschlager_NatPhys}.}
\label{fig:3TSOC_momdist}
\end{figure}

To capture quantum/thermal fluctuations, two slowly varying bosonic fields are introduced as 
$$
 \phi_\mu (\tbf{x}) = \sum_{\bf k} ^ \Lambda b_\mu (\tbf{Q}_\mu + \tbf{k}) e^ {i \tbf{k}\cdot \tbf{x}}, 
$$ 
where $\Lambda$ is a momentum cut off. The effective Hamiltonian of the field theory of $\phi_\mu (\tbf{x})$ is 
\bea 
H &=& \int d^2 \tbf{r} \left [ K_1 \left( \partial_x \phi_x^\dag  (\tbf{r}) \partial_x \phi_x(\tbf{r}) + x\leftrightarrow y\right) \right. \nn \\
			   && +  K_2 \left( \partial_y \phi_x ^\dag(\tbf{r})  \partial_y \phi_x (\tbf{r}) + x\leftrightarrow y \right)  \nn \\
  &&		   - \mu  \left( \phi_x ^\dag \phi_x + x\leftrightarrow y \right)   \nn \\
&& 	+ \frac{1}{2} g_1 \left( \phi_x ^\dag \phi_x \phi_x ^\dag \phi_x + x\leftrightarrow y\right) 
	+ 2 g_2  \phi_x ^\dag \phi_x \phi_y ^\dag \phi_y \nn \\
&& \left. 	+ \frac{1}{2} g_3\left( \phi_x ^\dag \phi_x ^\dag \phi_y \phi_y + h.c. \right) \right]. 
\eea 
In a superfluid state, we have $\langle \phi_\nu \rangle $ = $\varphi_\nu$. At mean field level, 
the energy of this state is 
$$
E = \frac{1}{2} g_1 \left(|\varphi_x|^4 + |\varphi_y|^4 \right) +2 g_2 |\varphi_x|^2 |\varphi_y|^2   
+ \frac{1}{2} g_3 \left(\varphi_x ^{*2}\varphi_y ^2 + c.c.\right). 
$$
From Eq.~\eqref{eq:3Us}, we have $g_1 = 3 g_2 = 3 g_3 >0$, and 
the relative phase between $p_x$ and $p_y$ is locked at $\pm \pi/2$, i.e. 
$\varphi_x = \varphi_y e^{\pm i \frac{\pi}{2}}$, where the ``$\pm$'' sign is spontaneously chosen. 
The superfluid state has a staggered angular 
momentum order $(-1)^{R_x + R_y} \langle L_z (\tbf{R}) \rangle$, which breaks time-reversal symmetry. Such 
a superfluid state is named transversely staggered orbital current (TSOC) superfluid. 
The phase configuration of this superfluid state and its momentum distribution are 
shown in Fig.~\ref{fig:3TSOC_momdist}.

An alternative way of looking at time-reversal symmetry breaking is to project 
interactions into the subspace spanned by $|N_x N_y \rangle$ in Eq.~\eqref{eq:3groundmanifold}.  
In this subspace, the interaction reads  
\bea 
\label{eq:3projectHint} 
&& \langle N_x ' N_y' | H_{\rm int} | N_x N_y\rangle \\
 &=& \frac{U_1}{2 N_s} (N_x ^2 + N_y ^2 ) \delta _{N_x N_x'} \delta _{N_y N_y'}  
  + 2 \frac{U_2}{N_s} N_x N_y \delta _{N_x N_x'} \delta_{N_y N_y'} \nn \\
 &&  + \frac{U_3}{2 N_s} \left[ 
			\sqrt{N_x' (N_x'-1) N_y (N_y-1)} \delta _{N_x' N_x+2} \delta _{N_y' N_y-2} 
			\right. \nn \\ 
&& + x\leftrightarrow y 
		      \Big]. \nn
\eea  
For the two orbital components to be miscible, we need 
\be 
 2U_2 - |U_3| < U_1, 
\ee 
as analogous to spin miscible condition in spinor condensates~\cite{2008_Pethick_Smith_book}. 
 
\begin{figure}[htp]
\includegraphics[angle=0,width=\linewidth]{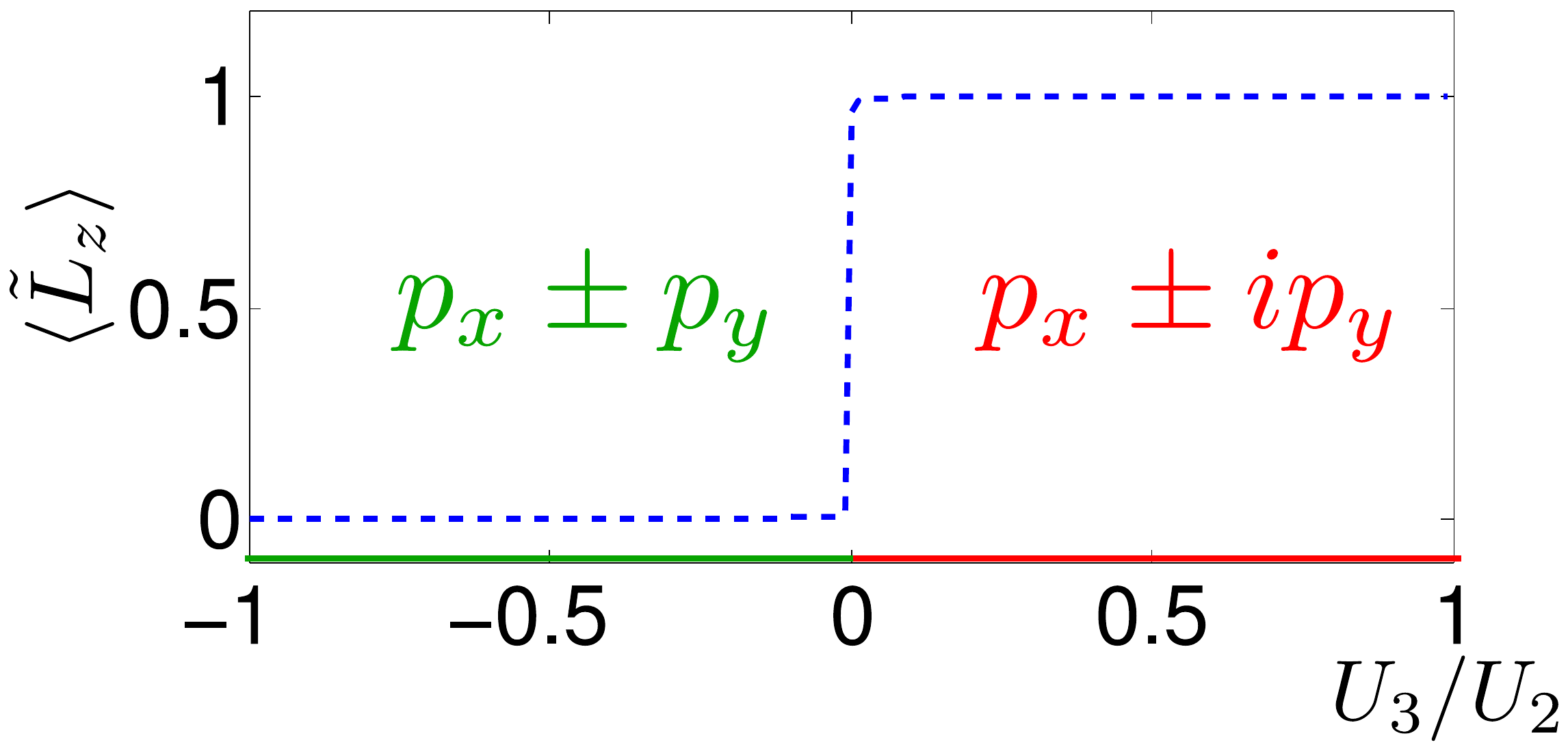}
\caption{Staggered angular momentum order. For positive $U_3$, the staggered angular momentum order is finite and 
the condensate has a staggered $p_x\pm ip_y$ (TSOC) order; while for the 
negative case, the staggered angular momentum order vanishes and the condensate has a $p_x \pm p_y$ order. 
In this plot, we assumed  $2U_2 - |U_3| < U_1$ such that the orbital mixed state has lower energy than $p_x$ or $p_y$ state.
}
\label{fig:3TSOCphasediag}
\end{figure}

The angular momentum correlation is given by 
\bea 
&& (-1)^{R_x + R_y}  \langle L_z (\tbf{R}) L_z (0) \rangle \nn \\
&=& -\frac{1}{N_s^2} 
\Big[ \left \langle \left( {b}_x ^\dag  (\tbf{Q}_x) {b}_y (\tbf{Q}_y) \right)^2 + h.c. \right \rangle \nn \\
&& - \langle n_x\rangle (\langle n_y\rangle +1) - \langle n_y \rangle (\langle n_x\rangle +1) \large \Big], 
\eea 
with $\langle \ldots \rangle$ the ground state expectation value. 
With $U_3>0$,  to minimize the energy in Eq.~\eqref{eq:3projectHint}, 
$\left \langle \left( {b}_x ^\dag  (\tbf{Q}_x) {b}_y (\tbf{Q}_y) \right)^2 + h.c. \right \rangle$ 
gets a negative value in the ground state and 
in the thermodynamical limit 
($\langle n_x \rangle \gg 1$, $\langle n_y \rangle \gg 1$), it approaches 
$ (-) 2 \langle n_x\rangle \langle n_y\rangle  $.  
The system thus has a long range correlation in angular momentum, 
i.e., $(-1)^{R_x + R_y} \langle L_z (\tbf{R}) L_z (0) \rangle \xrightarrow{|\tbf{R}| \to \infty}  const \neq 0$. 
The  corresponding Ising order parameter is a staggered angular momentum $\tilde{L}_z (\tbf{R})\equiv (-1)^{R_x + R_y} L_z (\tbf{R})$. 
When $U_3$ is negative, $\langle \left( {b}_x ^\dag  (\tbf{Q}_x) {b}_y (\tbf{Q}_y) \right)^2 + h.c. \rangle$ 
becomes positive, and the angular momentum order $\langle \tilde{L}_z \rangle $ vanishes and the system develops the other 
Ising orbital order $p_x \pm p_y$ with an order parameter $(-1)^{R_x + R_y} \langle b_x ^\dag  (\tbf{R}) b_y (\tbf{R}) + h.c. \rangle $. 
From the above analysis, the transition at $U_3=0$ is predicted to be first order~(Fig.~\ref{fig:3TSOCphasediag}), although fluctuations may 
stabilize some intermediate state and the first order transition could be replaced by a sequence of double second order transitions.

\subsubsection{Symmetry based effective field theory description}
The predicted TSOC superfluid state in the $p$-band tight binding model is also confirmed with 
effective field theory (EFT) treatment~\cite{2013_Li_Paramekanti_NatComm}, 
which infers that the TSOC superfluid does not necessarily require a deep lattice. 
In the band structure calculation for the case of lattice rotation symmetry~\cite{2011_Wirth_Olschlager_NatPhys}, 
dispersion of the relevant $p$-band $E_p (\tbf{k})$ has two degenerate minima 
at $\tbf{Q}_x = (\pi, 0)$ and $\tbf{Q}_y = (0, \pi)$, 
around which low energy modes can be excited due to quantum or thermal fluctuations.  
This leads to a two-component EFT, where the fields are
introduced as 
\bea 
\phi_\alpha (\tbf{x}) = \int ^\Lambda \frac{d^2 \tbf{q}}{(2\pi)^2}{b}   ({\bf Q}_\alpha + q) e^{i\tbf{q}\cdot \tbf{x}},  
\eea 
with $\Lambda$ a momentum cutoff and ${b}  (Q_\alpha +q)$  annihilation operators for the Bloch modes near the band minima.  
The form of EFT is determined by considering lattice rotation and reflection symmetries, under which  
the fields $\phi_\alpha$ transform as 
\bea 
\left [\begin{array}{c}
        \phi_x (x,y)\\
	\phi_y (x,y) 
       \end{array}
 \right] \to 
\left[ 	\begin{array}{c}
       	 -\phi_y (y,-x) \\ 
	  \phi_x(y, -x)
       	\end{array}
\right], 
\eea 
and 
\bea 
\left [\begin{array}{c}
        \phi_x (x,y)\\
	\phi_y (x,y) 
       \end{array}
 \right] \to 
\left[ 	\begin{array}{c}
       	 -\phi_x (-x,y) \\ 
	  \phi_y(-x, y)
       	\end{array}
\right], 
\eea 
respectively. 
The Hamiltonian density of the EFT consistent with these symmetries is 
\bea 
{\cal H}_{\rm eff} &=&  \phi_x ^\dag (\tbf{x}) \left(
      K_\parallel \frac{\partial^2}{\partial x ^2} 
      + K_\perp \frac{\partial^2}{\partial  y ^2} -\mu\right) \phi_x (\tbf{x}) + x\to y \nn \\ 
&+& \sum_{\alpha_1 \alpha_2} g_{\alpha_1 \alpha_2} 
	   |\phi_{\alpha_1} |^2|\phi_{\alpha_2}|^2 
+  g_3  \left(\phi_x ^{\dag 2} \phi_y ^2  +h.c.\right), 
\label{eq:3HpbandEFT}
\eea 
with effective couplings $K_\parallel$, $K_\perp$ and $g$'s. 
This form of EFT is solely  symmetry based, i.e., independent of microscopic details. 
For weakly interacting bosons, the coupling constants in Eq.~\eqref{eq:3HpbandEFT} can be calculated 
from microscopic models (see Appendix~\ref{sec:EFTcouplings}).  
 
In the vicinity of thermal phase transitions of the superfluid phases, classical phase fluctuations are expected to
dominate the universal physics, which allows us to ignore the subdominant density fluctuations and to replace 
$\phi_\alpha$ by $\sqrt{\rho /2} e^{i\theta_\alpha}$ with $\rho$ the total density. In terms of phases 
$\theta_\alpha$, the Hamiltonian density is rewritten  as
\bea
{\cal H}_{\rm eff} &=& (-K_\parallel  (\partial_x \theta_x )^2 -K_\perp (\partial_y \theta_x)^2 + x\leftrightarrow y) \nn \\
&& + 
  \frac{1}{2} g_3 \rho ^2 \cos \left( 2(\theta_x - \theta_y)\right).  
\eea
Bearing in mind the periodic nature of the phases $\theta_\alpha$, 
 a proper lattice regularization of this EFT leads to a coupled XY  model, 
\bea 
H^{\rm eff}_{\rm phase} \!\!\!\!&=&\!\!\!\! 
 \sum_{\tbf{r}} \left [  \left\{ 2 J_\parallel 
\cos (\Delta_x \theta_x(\tbf{r})) - 2 J_\perp 
\cos(\Delta_y \theta_x (\tbf{r}))
\right\} \right. \nn \\
&+& \left. \left\{ x \leftrightarrow y \right\} \right]
 - U \sum_{\tbf{r}} \sin^2(\theta_x(\tbf{r}) - \theta_y(\tbf{r})), 
\label{eq:3Hphase}
\eea 
where $\Delta_j \theta_\alpha(\tbf{r}) = \theta_\alpha(\tbf{r}+ {\bf a}_j)-\theta_\alpha(\tbf{r})$ with $j=x,y$. 
At zero temperature we have the TSOC superfluid where 
the phases are locked 
at $\theta_x ({\bf r})  = {r_x}\pi +  \theta_0$, $\theta_y ({\bf r}) = {r_y} \pi + \theta_0 + s \pi/2$,   
with $s=\pm$ and 
$\theta_0 \in [0, 2\pi)$ spontaneously chosen. 
At finite temperature, the coupled XY model supports two types of topological defects. The first is a vortex in 
the phase $\theta_0$, which is a point defect with logarithmic energy cost. The second is a domain wall connecting 
two Ising domains with different $s$. 
Upon heating the TSOC superfluid, vortex proliferation should drive a Kosterlitz-Thouless transition and the domain 
wall fluctuations should drive an Ising transition. Monte Carlo study finds that 
the Kosterlitz-Thouless transition temperature is lower than the Ising transition~\cite{2013_Li_Paramekanti_NatComm}.

From the effective field theory analysis, the $p$-orbital angular momentum order, or equivalently the $\pm \frac{\pi}{2}$ phase locking,  does not rely on the precise form of the interaction (Eq.~\eqref{eq:3Hint}). The requirements are $g_3>0$ and  two $p$ orbitals being miscible.

\subsubsection{Population of higher bands by interaction} 

Here, we will focus on condensation of weakly interacting bosons in a lattice potential. 
With weak interaction, the condensate is well described by Gross-Pitaevskii approach 
where the condensate wavefunction $\phi (\tbf{x})$ is obtained by minimizing an energy functional 
\bea 
E_{\rm GP}  = \int d^d \tbf{x} \phi^* (\tbf{x}) \left(-\frac{\vec{\nabla}^2}{2m} + V(\tbf{x}) - \mu\right) \phi(\tbf{x}) 
+ g|\phi(\tbf{x})|^4, \nn \\
\label{eq:3EGP}
\eea 
With infinitesimal interaction, the condensate wavefunction resembles the lowest band Bloch wavefunction with lattice 
momentum $\tbf{k} =0$, i.e., 
$\phi(\tbf{x}) \propto \sum_\tbf{R} w_0 (\tbf{x}-\tbf{R})$. 
This wavefunction preserves lattice translation and time-reversal symmetries meaning 
$\phi(\tbf{x} ) = \phi(\tbf{x} + \tbf{a})$ and $\phi = \phi^*$. 
From these preserved symmetries the generic form of the condensate wavefunction with weak interaction is 
\bea 
\phi (\tbf{x}) = \sum_n \lambda_n \sum_\tbf{R} w_n (\tbf{x} - \tbf{R}), 
\eea 
provided that there are no first order transitions. 
The coefficients $\lambda_n$ are real and the interaction induced high band condensate 
is at zero lattice momentum. 
In terms of $\lambda_n$, the energy $E_{\rm GP}$ reads as 
\bea 
&& E_{\rm GP} =\sum_n \left(E_n (\tbf{k} = 0) - \mu\right) \lambda_n ^2  
	      + U_{0000} \lambda_0^4 \nn \\
	   &&   + 4 \sum_{n\neq 0} U_{000n} \lambda_0 ^3 \lambda_n 
	      + 10 \sum_{n \neq 0, m\neq0} U_{00mn} \lambda_0^2 \lambda_m \lambda_n \nn \\
&& + {\cal O} (\lambda_{n>0} ^3), 
\eea 
with the interactions $U_{\alpha_1 \alpha_2 \alpha_3 \alpha_4}$ introduced 
in Eq.~\eqref{eq:2interactions}.  
Minimizing this energy functional leads to  
\bea 
\lambda_0^2 &\approx& \frac{\mu - E_0 (0) }{2U_{0000}}, \\
\lambda_{n>0} &\approx& -\frac{2 U_{000n} \lambda_0^3}{E_n(0) - \mu}  . 
\eea
The ratio $\frac{\lambda_{n>0}}{\lambda_0}$ is readily given as 
\bea 
\frac{\lambda_{n}}{\lambda_0} \approx - \frac{ U_{000n}}{U_{0000}} 
	\left[ \frac{\mu - E_0(0) }{E_n(0) - \mu}\right]. 
\eea 
Physically, the  high band condensate is due to competition of 
interaction energy and lattice potential energy---the interaction favors 
an extended condensate; while the potential energy favors a condensate with 
high density at the lattice minima. 
The high band population due to interaction effects is found in various 
settings~\cite{2003_Oosten_Straten_PRA,2007_Guo_Zhang_PRA,2007_Kantian_Daley_NJPHYS,2009_Larson_Collin_PRA,2011_Dutta_Eckardt_NJP,2011_Zhou_Porto_PRAR,2011_Zhou_Porto_PRB,2011_Kaspar_Alexej_NJP,2012_Luhmann_Jurgensen_NJP,2012_Hofer_Bruder_PRA,2005_Alon_Streltsov_PRL,2011_Mering_Fleischhauer_PRA,2013_Lacki_Delande_NJP}. 

In experiments, 
one can make a large fraction of high band condensate with double-well lattices, where 
the gap between lowest two bands is typically small 
and the fraction can be measured with band mapping techniques~\cite{2001_Greiner_Bloch_PRL}. 
Experimental evidence of this phenomenon has recently been achieved~\cite{2012_Soltan_Luhmann_NatPhys}. 
Furthermore, on general argument,  the double-well lattices can have the energy gap between 
the ground $s$-band and the first excited $p$-bands significantly smaller than that between the first excited bands and higher bands (e.g., between $p$ and $d$). This mechanism suppresses the decay by 
energy conservation law, making the first excited bands effectively metastable~\cite{2008_Stojanovic_Wu_PRL}. This will be further discussed in Section~\ref{sec:experiment}.  

\medskip 

For parity-symmetric lattices, the condensate wavefunction is parity 
even---$\phi(\tbf{x}) = \phi(-\tbf{x})$, which implies $\lambda_{n_{\rm odd}} = 0$, $n_{\rm odd}$ referring 
to the parity odd bands with $w_{n_{\rm odd}} (\tbf{x}) = -w_{n_{\rm odd}} (-\tbf{x})$. 
At mean field level, parity odd bands do not contribute and $\langle b_{n_{\rm odd}} \rangle = 0$. 
However they can form pair condensate orderings---$\langle b_{n} (\tbf{r}) b_{n'} (\tbf{r})\rangle$ 
due to Gaussian fluctuations~\cite{2011_Zhou_Porto_PRAR,2011_Zhou_Porto_PRB}. 
The mean field state is given by 
$|{\rm M}\rangle = \exp\left(\int d^d \tbf{x} \phi(x) \psi ^\dag (\tbf{x}) \right) |{\rm vac}\rangle.$
The effective Hamiltonian of high band modes at Gaussian level reads 
\bea 
H_{\rm eff} &\approx& \sum_{n\neq 0 ,\tbf{k}} \left(E_n (\tbf{k}) - \mu \right) 
	{b}_n  ^\dag (\tbf{k}) {b}_n (\tbf{k}) \nn \\
&+& \sum_{nm\neq 0 ,\tbf{k}\neq 0 } [ U_{00nm} \lambda_0 ^2 {b}_n^\dag (\tbf{k}) {b}_m ^\dag (-\tbf{k}) + h.c.]. 
\eea  
From standard perturbation theory, the correction on the mean field state from high band fluctuations is 
\bea 
- \sum_{nm \neq 0, \tbf{k} } 
    \frac{U_{00nm} \lambda_0^2 {b}_n ^\dag (\tbf{k}) {b}_m^\dag (-\tbf{k}) |{\rm M}\rangle}
      {E_n(\tbf{k}) + E_m(-\tbf{k}) -2 \mu }, 
\eea 
which mediates pairings 
\bea
&& \langle b_{n , \tbf{r}} b_{m, \tbf{r}} \rangle = \frac{1}{N_s} \sum_\tbf{k} \langle {b}_n (\tbf{k}) {b}_m (-\tbf{k}) \rangle \nn \\
&& = - \int \frac{d^d \tbf{k}}{(2\pi)^d} \frac{U_{00nm} \lambda_0^2 }{E_n(\tbf{k}) + E_m(-\tbf{k}) -2 \mu  }. 
\eea 
In parity odd bands, bosons form pair condensate with 
$\langle b_{n_{\rm odd},   \tbf{r}} \rangle =0$ and  $\langle b_{n_{\rm odd}, \tbf{r}} b_{n_{\rm odd}, \tbf{r}}  \rangle \neq 0$.

\subsubsection{Three dimensional $p$-orbital BEC and frustrated orbital ordering}

For bosons loaded on $p$-bands of a three dimensional cubic lattice, the tight binding Hamiltonian is~\cite{2006_Liu_Wu_PRA} 
\bea 
H &=& \sum_{\tbf{r} \alpha \beta} 
      [t_\parallel \delta_{\alpha \beta} - t_\perp ( 1 - \delta _{\alpha \beta})  ] 
      \left( b_{\alpha, \tbf{r} + e_\nu} ^\dag b_{\beta \tbf{r}} + h.c.\right) \nn \\
&&+ \frac{U}{2} \sum_\tbf{r} \left[ n_{\tbf{r}} ^2 - \frac{1}{3} \vec{L}_\tbf{r} ^2  \right], 
\label{eq:3Hpband3d} 
\eea 
where $n$ and $\vec{L}$ are boson density and angular momentum operators 
$n_\tbf{r} = \sum_\alpha b_{\alpha \tbf{r}} ^\dag b_{\alpha \tbf{r}} $ and 
$L_{\alpha \tbf{r}} = -i \epsilon_{\alpha \beta \gamma} b_{\beta \tbf{r} } ^\dag b_{\gamma \tbf{r}}$. 
Without interaction, there are three degenerate $p$-bands and the energy minima are at 
$\tbf{Q}_x = (\pi, 0,0)$, $\tbf{Q}_y = (0, \pi, 0)$ and $\tbf{Q}_z = (0, 0, \pi)$. The degenerate 
single-particle states are $|\tbf{Q}_\alpha \rangle = {b}_\alpha ^\dag (\tbf{Q}_\alpha ) | {\rm vac}\rangle$.  
Thus any condensate wavefunction of a linear superposition of $|\tbf{Q}_\alpha \rangle$~\cite{2012_Cai_Wang_PRBR}, 
$$
|\vec{c} \rangle  = \sum_\alpha c_\alpha |\tbf{Q} _\alpha \rangle 
$$
has the same single-particle energy. Here $\vec{c} = (c_x, c_y, c_z)$ is a complex vector 
normalized to $1$, i.e., $|\vec{c}|=1$. This complex vector could be parametrized as~\cite{2006_Liu_Wu_PRA}
\bea 
\left[ 
    \begin{array}{c}
     c_x \\
     c_y \\
     c_z 
    \end{array}
\right] 
 = e^{i \varphi - i {\rm T}_\alpha \theta_\alpha } 
\left[ 
  \begin{array}{c}
   \cos (\chi) \\
   i \sin (\chi) \\
   0
  \end{array}
\right], 
\eea 
with $T_{\alpha = x,y,z}$ the generators of SO($3$) orbital rotation in the following 
matrix representation: $[T_{\alpha}]_{\beta \gamma} = -i \epsilon_{\alpha \beta \gamma}$. 

Although the SO($3$) orbital rotation is not a symmetry of the total Hamiltonian, it keeps 
the interaction term invariant because $n_{\bf r} $ and $\vec{L}_{\bf r} ^2$ are both SO($3$) scalars. 
With a condensate at the single-particle state $|\vec{c}\rangle$, 
the mean field interaction energy is readily given as~\cite{2006_Liu_Wu_PRA}
\bea 
E_{\rm int} = \frac{1}{2} U N_s n_0 ^2 \left[1 - \frac{1}{3} \sin^2 (2 \chi)\right],  
\eea 
with $n_0$ the boson occupation number per site. 
The interaction energy is minimized at $\chi = \pm \frac{\pi}{4}$. Similar to the two dimensional case, 
the time-reversal symmetry is spontaneously broken in the $p$-band condensate. 
The ground state manifold is $U(1) \times Z_2 \times SO(3)$ at mean field level. The $Z_2 \times U(1)$ degeneracy 
remains due to the symmetries of the Hamiltonian, whereas the $SO(3)$ degeneracy is an artifact of the mean field 
theory and such a degeneracy is lifted by fluctuations through an ``order by disorder'' mechanism. 
Ref.~\cite{2012_Cai_Wang_PRBR} carried out a variational comparison between the two superposition states 
$$ 
|{\rm planar} \rangle  = \frac{1}{\sqrt{2}} \left( |\tbf{Q}_x \rangle + i |\tbf{Q}_y\rangle \right),  
$$ 
and 
$$
|{\rm diag}\rangle = \frac{1}{\sqrt{3}} \left( |\tbf{Q}_x \rangle + e^{ i 2\pi/3} |\tbf{Q}_y\rangle + e^{-i 2\pi/3} |\tbf{Q}_z\rangle  \right). 
$$ 
It is found that the latter has lower energy under Bogoliubov approximation.

\begin{figure}[htp]
\includegraphics[angle=0,width=.5\linewidth]{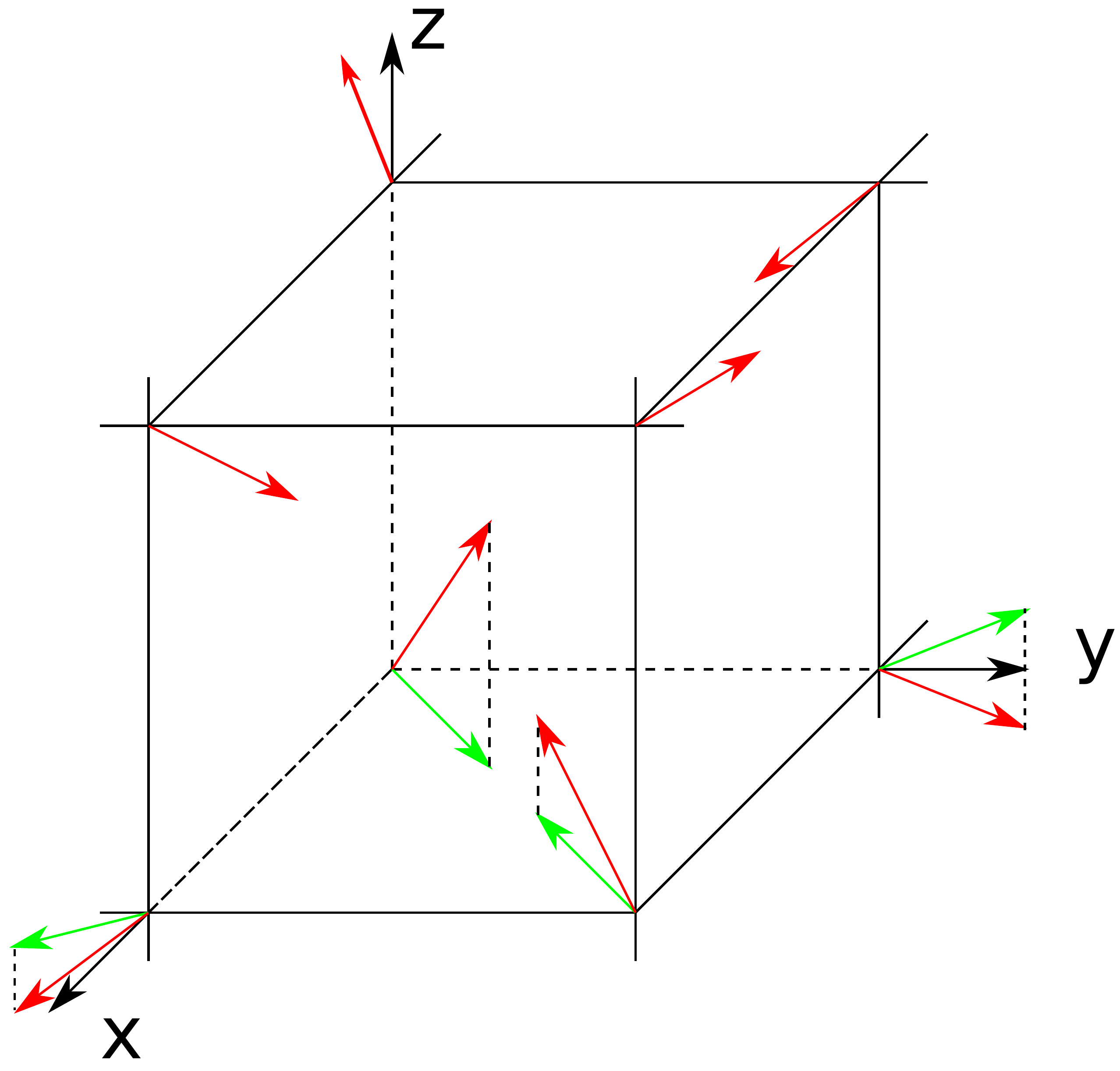}
\caption{Noncollinear angular momentum order in a 3D $p$-orbital Bose-Einstein condensate in a cubic lattice~\cite{2012_Cai_Wang_PRBR}. Red arrows indicate the polarization direction of the angular momentum $\langle \vec{L}\rangle$.  }
\label{fig:3noncollinear}
\end{figure}

One interesting consequence is that the angular momentum ($\vec{L}$) order in the 3D $p$-orbital condensate state is noncollinear, which is different from 
the 2D case. The polarization configuration of $\langle \vec{L} \rangle$ is shown in Fig.~\ref{fig:3noncollinear}. 
Such configuration exhibits non-zero chirality defined to be 
\bea 
\chi_{ijk} = \vec{L}_i \cdot (\vec{L}_j \times \vec{L}_k ), 
\eea 
where $ijk$ denote nearby three sites of the four corners of a square plaquette in a clockwise direction. 
With thermal/quantum fluctuations, the presence of such chiral order may lead to unconventional phase transitions~\cite{2013_Li_Paramekanti_NatComm}.

In a relative shallow lattice, Eq.~\eqref{eq:3Hpband3d} derived under the
harmonic approximation  would receive  significant unharmonic corrections. There Gutzwiller calculations suggest a  more exotic condensate with nematic order~\cite{2010_Collin_Larson_PRA}, which spontaneously breaks the cubic lattice symmetry.

\subsubsection{Renormalization group analysis} 
Fluctuation effects on $p$-band condensates beyond mean field theories  are studied with 
perturbative one-loop analysis~\cite{2013_Liu_Yu_PRA}, where the partition function takes 
the form 
\bea 
Z = \int D[\phi_x ^*, \phi_x, \phi_y ^*, \phi_y] e^{-S[\phi_x ^*, \phi_x , \phi_y ^*, \phi_y] } , 
\eea 
with 
\bea 
S &=& \int \frac{d\omega}{2\pi} \frac{d^2 \tbf{k}}{(2\pi)^2} 
      \sum_{\alpha} \phi_\alpha^* (\omega, \tbf{k}) 
    (-i\omega + \epsilon_\tbf{k} -r_\alpha) \phi_\alpha (\omega, \tbf{k}) \nn \\ 
&+& \int_{\omega, \tbf{k}} \left\{ 
	\sum_{\alpha} g_1 ^\alpha \phi_\alpha ^* (4) \phi_\alpha ^* (3) \phi_\alpha(2) \phi_\alpha(1) 
      \right. \nn \\
&+& g_2 \phi_y ^* (4) \phi_x ^* (3) \phi_x (2) \phi_y (1) \nn \\
&+& \left. g_3 [ \phi_x ^* (4) \phi_x ^* (3) \phi_y (2) \phi_y (1) + h.c. ]\right\} . 
\eea 
Here $\int_{\omega, \tbf{k}} = \prod_{j=1} ^4 \int \frac{d\omega}{2\pi} 
      \int \frac{d^2 \tbf{k}_j}{(2\pi)^2} (2\pi)^3 
    \delta (\tbf{k}_4 + \tbf{k}_3 - \tbf{k}_2 - \tbf{k}_1) 
    \delta (\omega_4 + \omega_3 - \omega_2 - \omega_1)$ and
$\phi_\alpha(j)$ denotes $\phi_\alpha(\omega_j, \tbf{k}_j)$. For
general lattices lacking of $C_4$ rotational symmetry, the energy potential
parameters are not equal, $r_x\neq r_y$. 
Performing a momentum-shell renormalization group (RG) analysis, the fields are split into 
fast and slow parts,  $\phi_\alpha ^> (\omega, \tbf{k})|_{\Lambda/s<|\tbf{k}| <\Lambda} $ 
and $\phi_\alpha ^< (\omega, \tbf{k}) | _{|\tbf{k}| < \Lambda/s}$. 
Following the standard Wilsonian RG procedure (integrating out 
fast modes and rescaling the effective action for the slow modes), the RG flow equations 
(or $\beta$-functions) for the potential parameters $r_\alpha$ to one-loop order are obtained to be 
\bea 
&& \frac{d\tilde{r}_x}{dl} = 2 \tilde{r}_x + 4 \tilde{g}_1 ^x \Theta (\tilde{r}_x - 1/2) + 
\tilde{g}_2 \Theta(\tilde{r}_y - 1/2), \nn \\
&& \frac{d \tilde{r}_y}{dl} = 2 \tilde{r}_y + 4 \tilde{g}_1 ^y  \Theta(\tilde{r}_y - 1/2) + 
\tilde{g}_2 \Theta(\tilde{r}_x -1/2) . 
\eea 
Here the dimensionless parameters are defined as $\tilde{r} = r m/\Lambda^2$ 
and 
$\tilde{g} = g m/(2\pi)$ and $\Theta(x)$ is the Heavyside step function. 
The RG flow equations for the quartic couplings are 
\bea 
&& \frac{d \tilde{g}_1 ^x} {dl} = -2 \tilde{g}_1 ^{x2} -2 \tilde{g}_3 ^2, \nn \\
&& \frac{d \tilde{g}_1 ^y} {dl} = -2 \tilde{g}_1 ^{y2} - 2 \tilde{g}_3^2, \nn \\
&& \frac{d \tilde{g}_2 } {dl} = - \tilde{g}_2 ^2, \nn \\ 
&& \frac{d\tilde{g}_3} {dl} = -2 \tilde{g}_3 (\tilde{g}_1 ^ x + \tilde{g}_2 ^y).  
\eea 
With bare repulsive interaction, these quartic couplings are all marginally irrelevant. 
However they could strongly modify the RG flow of $r_\alpha$ before they renormalize to zero.

In the region with $\tilde{r}_x (0) \ge \frac{1}{2}$ and $\tilde{r}_y (0) \ge \frac{1}{2}$ 
the solutions are 
\bea 
&& \tilde{r}_x (l) = e^{2l} \left[ \tilde{r}_x (0) 
  + \int _0 ^l dl' e^{-2l'} (4 \tilde{g}_1 ^x (l') + \tilde{g}_2 (l') ) \right], \nn \\
&& \tilde{r}_y (l) = e^{2l} \left[ \tilde{r}_y (0) 
  + \int _0 ^l dl' e^{-2l'} (4 \tilde{g}_1 ^y (l') + \tilde{g}_2 (l') ) \right] . \nn \\
\eea 
In this region, $\tilde{r}_x$ and $\tilde{r}_y$ quickly run to positive infinity. 
In the region with $\tilde{r}_x <\frac{1}{2}$ and $\tilde{r}_y < \frac{1}{2}$, the 
solutions are 
\bea 
&& \tilde{r}_x (l) = \tilde{r}_x (0) e^ {2l}, \nn \\ 
&& \tilde{r}_y (l) = \tilde{r}_y (0) e^{2l}, 
\eea 
from which the behaviors of RG flow are also fully determined by initial values of 
$\tilde{r}_{x,y}$. 
In other regions, one-loop corrections play more important roles in making the eventual 
values of $\tilde{r}_{x,y}$ positive or negative. Numerical studies have found interesting 
regions 
in the phase digram where $\tilde{r}_x (0)  <0$ and  $\tilde{r}_y (0) >0$ (or vice versa) 
flow to $\tilde{r}_x \to + \infty $ and $\tilde{r}_y \to + \infty$. 
Depending on the flow directions of $\tilde{r}_{x,y}$, four states can be 
identified: (1) Complex BEC ($\tilde{r}_x \to + \infty$, $\tilde{r}_y \to+ \infty$) ; 
(2) $p_x$ BEC ($\tilde{r}_x \to +\infty$, $\tilde{r}_y \to -\infty$); 
(3) $p_y$ BEC ( $\tilde{r}_x \to -\infty$, $\tilde{r}_y \to + \infty$); 
and (4) vacuum ($\tilde{r}_x \to -\infty $, $\tilde{r}_y \to -\infty$). 

The RG study sketched above does not really capture the TSOC state because 
$g_3$ flows to $0$, making an illusion that quantum fluctuations wash away the phase locking between 
$p_x$ and $p_y$ components. 
However this is not physically correct. A more careful RG 
study requires introducing $U(1)$ and $Z_2$ order parameters to characterize the fluctuation 
effects in the TSOC state.

\subsection{Mott states, orbital exchange and frustration of bosons} 
In the strongly interacting regime, bosons localize and form Mott insulator phases. 
Unlike the ``featureless'' $s$-band Mott insulators, the $p$-band Mott insulators 
have orbital degrees of freedom. Details of preparation of $p$-band Mott states including 
relaxation dynamics are studied in~\cite{2009_Challis_Girvin_PRA}. The orbital ordering is governed by the orbital exchange interactions 
which result from virtual boson tunnelings. Here we will derive the orbital super-exchange 
interactions and discuss the orbital frustrations on certain lattice geometries. 

\subsubsection{Mott states with filling factor larger than $1$.} 

The procedure to derive super-exchange interactions 
is to take the local terms as the leading part  and 
the hopping terms as perturbation. 
Consider the two dimensional $p$-band Bose gas for example. 
The local interaction is given by 
\bea 
H_U = \frac{U}{2} \left( n^2 -\frac{1}{3}L_z ^2 \right).  
\eea 
It can be verified that the angular momentum operator $L_z$ commutes with the 
local interaction, i.e., 
$$
[L_z, H_u] = 0.
$$
Thus the eigenstates of the local interaction can be chosen as states 
with definite angular momentum. 
For filling factor $\nu>1$, the degenerate eigenstates with lowest energy $=\frac{\nu^2}{3}U$ are 
\begin{align*} 
|+\rangle &= \frac{\left( b_\uparrow ^\dag \right) ^\nu }{\sqrt{\nu!}} |{\rm vac} \rangle, \\
|-\rangle &= \frac{\left( b_\downarrow ^\dag \right) ^\nu }{\sqrt{\nu!}} |{\rm vac} \rangle , 
\end{align*} 
where $b_{\uparrow/\downarrow}  = \frac{b_x  \pm i b_y  }{\sqrt{2}}$. 
The states $|+\rangle$ and $|-\rangle$ have angular momentum $+\nu$ and $-\nu$, 
respectively.  On a square lattice, the tunneling Hamiltonian in the transformed basis reads 
\bea 
H_{\rm t} = \sum_{s,s',\tbf{r}} \left[ T_{ss'} (\hat{x}) 
					  b_{s,\tbf{r}+\hat{x}} ^\dag b_{s',\tbf{r} }  + h.c. \right]  + x\leftrightarrow y, 
\label{eq:3Htun} 
\eea 
with the matrices 
\bea
T(\hat{x}) &=& \left[ 
	      \begin{array}{cc}
              \frac{t_\parallel - t_\perp}{2} & \frac{t_\parallel + t_\perp}{2} \\
	      \frac{t_\parallel + t_\perp}{2} & \frac{t_\parallel - t_\perp}{2} 
             \end{array}
	     \right],  \\
T(\hat{y}) &=& \left[ 
	      \begin{array}{cc}
              \frac{t_\parallel - t_\perp}{2} & -\frac{t_\parallel + t_\perp}{2} \\
	      -\frac{t_\parallel + t_\perp}{2} & \frac{t_\parallel - t_\perp}{2} 
             \end{array}
	     \right]. 
\eea

The low energy sub-space is spanned by the product 
states 
$$
|\{ {s} (\tbf{r}) \} \rangle 
\equiv \otimes_\tbf{r} |s(\tbf{r}) \rangle, 
$$
where $s(\tbf{r}) = \pm$ and $\tbf{r}$ runs over all lattice sites. 
All the states in this subspace have the same energy to leading order 
in $U$ and there is thus a  macroscopically huge degeneracy. 
The corrections due to the hopping term $H_t$ lift the degeneracy. 
The first order corrections vanish because $H_t$ does not connect 
any states in the low energy sub-space. The second order correction is calculated 
by the standard perturbation theory, 
\bea 
\Delta E(|\{s (\tbf{r}) \} \rangle ) 
= \sum_m \frac{|\langle m | H_t | \{s(\tbf{r}) \}\rangle|^2}
	      {E^{(0)} ( | \{ s (\tbf{r})\}\rangle ) - E^{(0)} (|m\rangle) },  
\eea 
where $| m \rangle$ is a higher energy state orthogonal to the 
product states $| \{ s (\tbf{r}\}  \rangle$, and $E^{(0)}$ is the leading order 
energy.  

Keeping only tunneling between nearest neighbors as in Eq.~\eqref{eq:3Htun}, 
$\Delta E(|\{s (\tbf{r}) \} \rangle )$ simplifies to 
\bea 
\Delta E (|\{s (\tbf{r}) \} \rangle ) 
= \sum_{\langle \tbf{r}, \tbf{r}' \rangle} 
      \Delta E(|s(\tbf{r}) s(\tbf{r}') \rangle ), 
\eea 
where $\tbf{r}$ and $\tbf{r}'$ are adjacent sites. 
Calculating the energy correction on a two-site state 
$\Delta E(|s(\tbf{r}) s(\tbf{r}') \rangle )$ is straightforward. The energy 
corrections are 
\bea 
\Delta E (|++\rangle ) &=& \Delta E(|--\rangle  )\nn \\
			&=&   
			 \frac{3}{4}\left\{ \frac{\nu (\nu+1) |t_\parallel - t_\perp|^2  }{- U } 
			  + \frac{\nu |t_\parallel + t_\perp|^2}
			      {- U (\nu+1)} \right\}, \nn \\
\Delta E (|+- \rangle ) &=& \Delta E (|-+ \rangle )\nn \\
	    &=& \frac{3}{4}\left\{  \frac{ \nu(\nu+1) |t_\parallel + t_\perp|^2 }{ - U} 
			  + \frac{\nu |t_\parallel - t_\perp|^2}{- U (\nu+1) } \right\}. \nn \\
\eea 
Then the correction $\Delta E (|\{s (\tbf{r}) \} \rangle )$ 
is given as 
\bea 
\Delta E (|\{s (\tbf{r}) \} \rangle )  
=\sum_{\langle \tbf{r}, \tbf{r}'\rangle} 
  J_\nu   s(\tbf{r}) s(\tbf{r}'), 
\label{eq:3Ecorrect}
\eea 
with 
$$J_\nu = \frac{3 \nu^2 (\nu+2)}{2(\nu+1) }\frac{t_\parallel t_\perp}{U}>0 .$$ 
Including this correction into the Hamiltonian, we get 
\bea 
\Delta \hat{H} = \sum_{\langle \tbf{r}, \tbf{r}' \rangle} J_\nu 
      {\sigma}_y (\tbf{r}) {\sigma}_y (\tbf{r}'),  
\label{eq:3orbexchange}
\eea 
where $\sigma_y$ is defined to be $\sigma_y =  \nu ^{-1} P {L}_zP $, with $P$ a projection operator 
$P = |+\rangle \langle +| + |-\rangle \langle - |$. 
The orbital super-exchange makes the staggered angular momentum 
ordering energetically favorable. 

It should be emphasized here that the energy corrections in Eq.~\eqref{eq:3Ecorrect} 
actually do not depend on the orientation of the link $\tbf{r}-\tbf{r}'$ and that 
the effective Hamiltonian in Eq.~\eqref{eq:3orbexchange} is independent of lattice 
geometries. Considering $p$-band Mott insulators on a triangle lattice, 
the effective orbital model is geometrically frustrated making both of 
ferromagnetic and antiferromagnetic correlations suppressed.

The above analysis holds in the deep lattice regime. For a
  relatively shallow lattice, the degeneracy in local Hilbert space could be
  lifted up~\cite{2010_Collin_Larson_PRA}. Treating such effects as
  perturbations, based on well established results in transverse field Ising
  models~\cite{2011_Sachdev_Book} the staggered angular momentum order in the
  Mott state is expected to be stable when the perturbations are reasonably weak
  as compared to the super-exchange. But we would like to emphasize that the
  competition of charge (atom number for neutral atoms) and spin orders in the
  shallow lattice regime may alter the above speculation and lead to potentially rich
  physics.   

\subsubsection{Mott state with filling factor $1$} 

\begin{figure}[htp]
\includegraphics[angle=0,width=.5\linewidth]{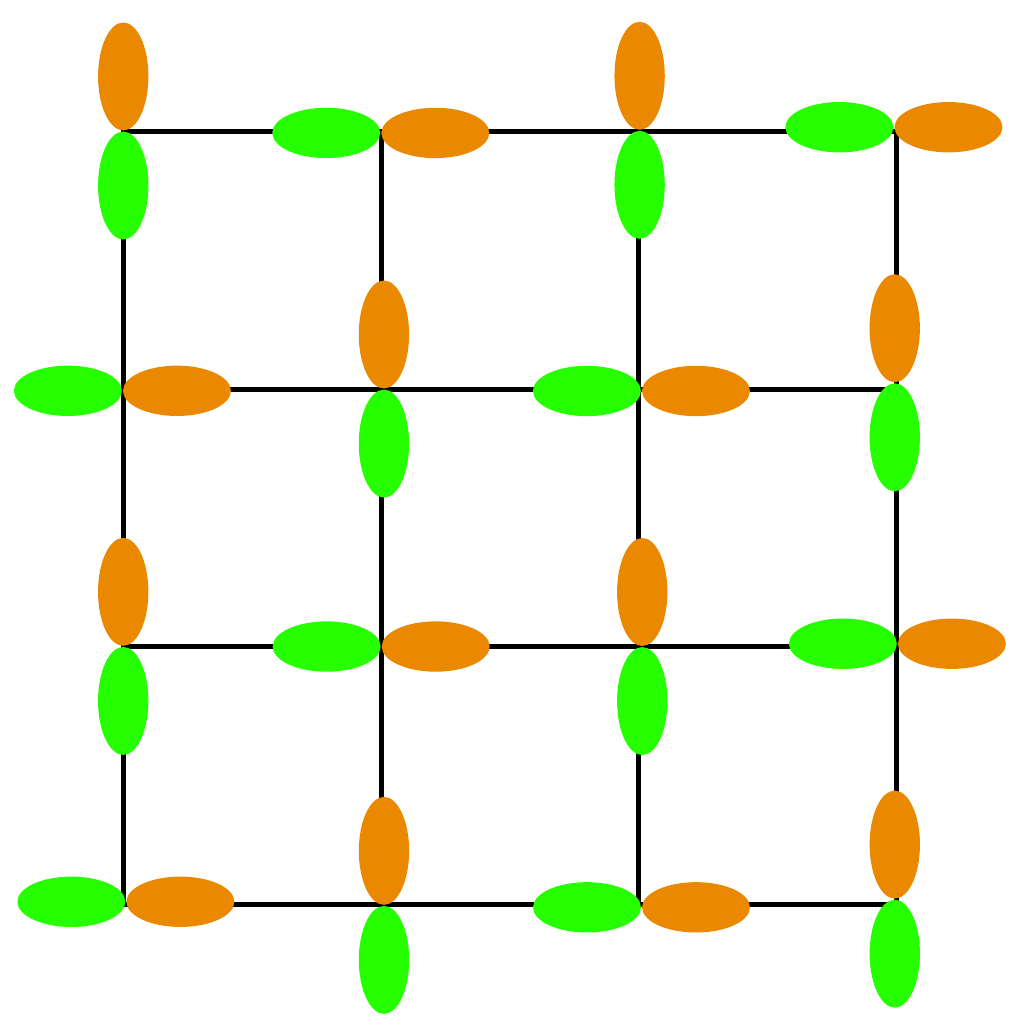}
\caption{Illustration of the alternating px/py orbital order~\cite{2011_Li_Zhao_PRA} for a $p$-band Mott insulator at unit filling. }
\label{fig:3alternatepxpy}
\end{figure}

For Mott states with filling factor $\nu = 1$, the convenient basis to calculate the super-exchange interaction 
is the $p_x$, $p_y$ basis, rather than the $p_x \pm ip_y$ basis. 
The generic form of interaction in Eq.~\eqref{eq:3genericHint} is used here.
Like deriving super-exchange for filling $\nu>1$, we  need 
to calculate the second order corrections of nearest neighbor product states---
 $|1, 0; 1, 0\rangle$, 
  $|0, 1; 0, 1\rangle$, 
$|0, 1; 1, 0 \rangle$, 
and 
   $|1, 0; 0, 1 \rangle$,   
where a notation 
\bea  
\label{eq:3orbitalfockstate} 
&& |m_x, m_y; m_x', m_y'\rangle \\
      &=& \frac{\left[b_x ^\dag (\tbf{r})\right]^{m_x} \left[b_y ^\dag (\tbf{r})\right]^{m_y} 
	      \left[b_x ^\dag (\tbf{r}')\right]^{m_x'} \left[b_y ^\dag (\tbf{r}')\right]^{m_y'} }
	     {\sqrt{m_x! m_y!m_x'! m_y' !}} |{\rm vac} \nn \rangle 
\eea  
 is adopted to save writing. 
The zeroth order energy of these four states is $U_1$. 
The higher energy virtual states that $H_t$ will couple to are 
$\frac{1}{\sqrt{2}} \left( 
 |2, 0; 0, 0\rangle 
  + 
   |0, 2; 0, 0\rangle 
\right) $, 
$\frac{1}{\sqrt{2}} \left( 
 |2, 0; 0, 0 \rangle  
  - 
  |0, 2; 0, 0\rangle 
\right) $,  
$
|1, 1; 0, 0\rangle $, 
$\frac{1}{\sqrt{2}} \left( 
 |0, 0; 2, 0\rangle 
\right.  $ + $ \left.
  |0, 0; 0, 2\rangle 
\right) $, 
$\frac{1}{\sqrt{2}} \left( 
 |0, 0; 2, 0\rangle  
  \,-\, 
   |0, 0; 0, 2\rangle 
\right) $,  and 
$
|0, 0; 1, 1\rangle $, 
with corresponding energies $2 U_1 + U_3$, $2U_1 - U_3 $, $U_1 + 2 U_2$, 
$2 U_1 + U_3$, $2U_1 - U_3 $ and $U_1 + 2 U_2$.  
For the link with $\tbf{r}' = \tbf{r} + \hat{x}$, 
the second order energy corrections are given by 
\bea 
&& \Delta E \left(
   |1, 0; 1, 0\rangle 
   \right) = -2 t_\parallel ^2 
	\left\{  \frac{1}{U_1 + U_3} + \frac{1}{U_1 - U_3} \right\} ,\nn \\
&& \Delta E \left(
    |0, 1; 0, 1\rangle \right) =0,  \nn\\
&& \Delta E \left(
    |0, 1; 1, 0\rangle \right) = 
\Delta E\left(
 |1,0; 0, 1\rangle \right) = - \frac{t_\parallel ^2 }{2 U_2} . 
\eea  
(Note that the transverse tunneling is neglected here, for the reason that 
 the longitudinal tunneling is enough to lift the degeneracy  
and is significantly stronger than the transverse one.) 
Mapping $p_x$ and $p_y$ orbitals to the pseudo-spin $1/2$ states, 
$\sigma = \uparrow$ and $\downarrow$, respectively, 
the effective  Hamiltonian on this link reads 
\bea 
H_{\rm x} =  J_1 \sigma_z (\tbf{r}) \sigma_z (\tbf{r}+\hat{x}) 
       +  M_z \left[ \sigma_z (\tbf{r}) + \sigma_z (\tbf{r}+\hat{x}) \right], 
\label{eq:3Hx}
\eea 
with $J_1 = - \frac{t_\parallel^2}{2} 
	  \left[ (U_1 + U_3)^{-1} + (U_1 - U_3)^{-1} - (2U_2)^{-1}\right] $, 
$M_z = -\frac{t_\parallel^2}{2}
	   \left[ (U_1 + U_3)^{-1} + (U_1 - U_3)^{-1} \right]. 
$
Similarly, the effective Hamiltonian for the  link $\tbf{r}' - \tbf{r} =  \hat{y}$ is 
obtained as 
\bea 
H_{\rm y} =  J_1 \sigma_z (\tbf{r}) \sigma_z (\tbf{r}+\hat{y}) 
       -  M_z \left[ \sigma_z (\tbf{r}) + \sigma_z (\tbf{r}+\hat{y}) \right].  
\eea 
Then the total effective Hamiltonian for filling $\nu =1 $ on a square lattice is  
\bea 
\Delta H = \sum_{\tbf{r}} 
	 J_1 \left[ {\sigma}_z (\tbf{r}) {\sigma}_z  (\tbf{r} + \hat{x}) 
		+  {\sigma}_z (\tbf{r}) {\sigma}_z  (\tbf{r} + \hat{y})
	    \right].  
\eea 
With $U_1 = 3U_2 = 3 U_3$, the coupling $J_1$ is positive and the ground state has  
an antiferromagnetic ordering  with alternating $p$-orbitals (see Fig.~\ref{fig:3alternatepxpy}). 
For a one-dimensional lattice, $H_{\rm x}$ makes $p_x$ orbitals favorable due to the
effective Zeeman term $M_z$ (Eq.~\eqref{eq:3Hx}). 
We mention here that including the transverse tunneling would give rise to 
even richer physics, e.g., an XYZ quantum Heisenberg model can emerge~\cite{2013_Pinheiro_Bruun_PRL}.

One key difference between filling $\nu=1$ and higher fillings is that the super-exchange interaction depends
on the orientation of the link $\tbf{r}'-\tbf{r}$, which makes the orbital frustration on 
triangle/Kagome lattices even more interesting.

\subsubsection{Phase diagram of $p$-band Bose-Hubbard model}
\begin{figure}[htp]
\includegraphics[angle=0,width=.5\linewidth]{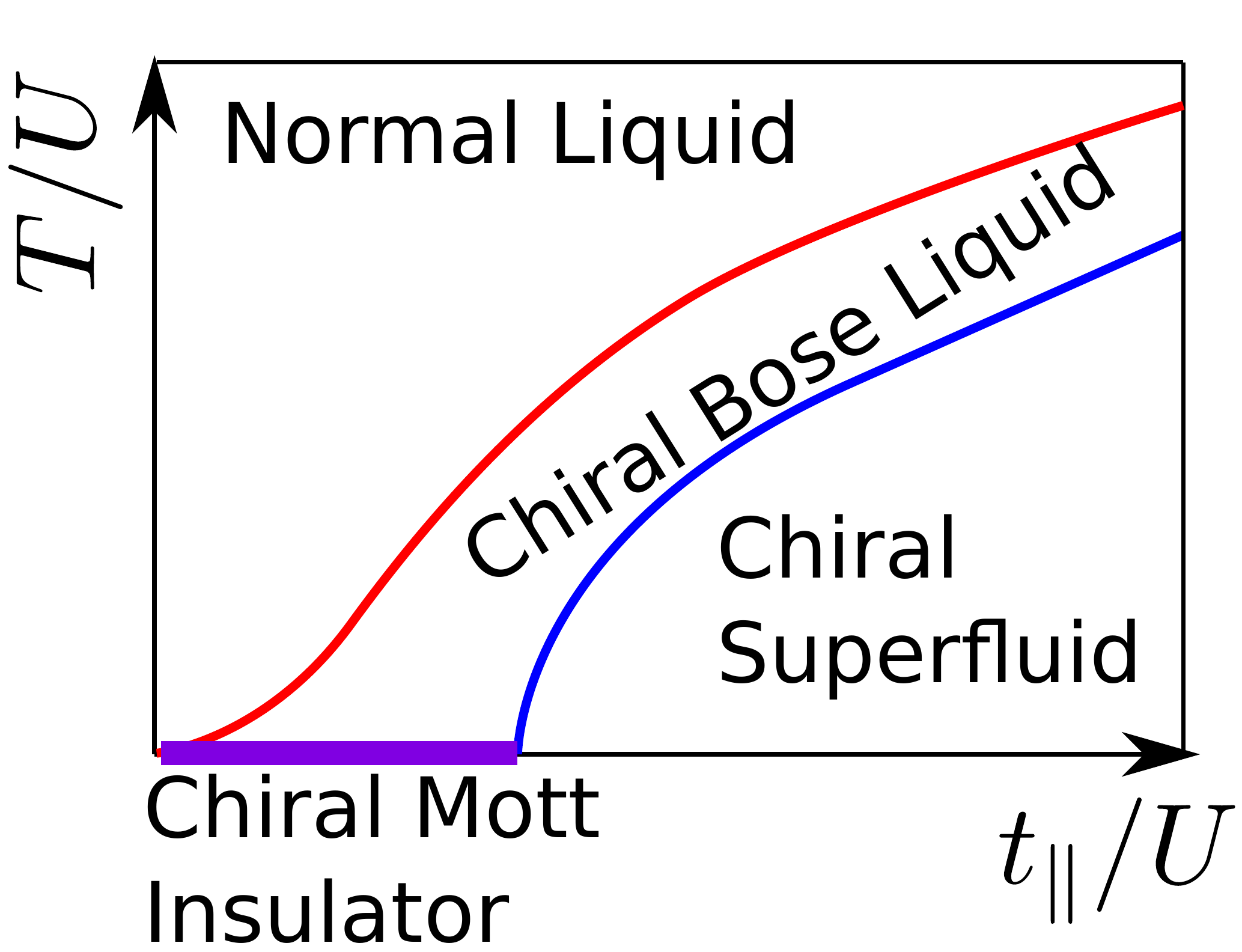}
\caption{ The schematic phase diagram of the two dimensional $p$-band Bose-Hubbard model with filling factor $\le 2$. 
The Chiral Mott and superfluid states have staggered angular momentum ordering. At zero temperature there is a 
quantum phase transition between the chiral Mott and superfluid states. At finite temperature, there is a chiral 
Bose liquid state which has angular momentum order but no superfluidity. Upon heating, the chiral superfluid undergoes 
a Kosterlitz-Thouless transition into the chiral Bose liquid, which subsequently undergoes an Ising transition at a 
higher temperature into a normal Bose liquid~\cite{2013_Li_Paramekanti_NatComm}.}
\label{fig:3pbandphasediag}
\end{figure}

The phase diagram of $p$-band Bose-Hubbard model in a two dimensional square lattice is studied by quantum~\cite{2013_Hebert_Cai_PRB} and 
classical~\cite{2013_Li_Paramekanti_NatComm} Monte-Carlo simulations. For filling of two particles per site or higher, a second order 
quantum phase transition from antiferromagnetic Mott state to the TSOC state is found at zero temperature in the quantum Monte-Carlo study as well as 
in Gutzwiller approach~\cite{2009_Larson_Collin_PRA,2010_Collin_Larson_PRA,2012_Martikainen_Larson_PRA}. 
In the weakly interacting regime at finite temperature, the fluctuations are modeled by a phase-only model studied by 
the classical Monte-Carlo. It is found that the TSOC state develops a two-step phase transition to the normal state, a Kosterlitz-Thouless 
transition followed by a higher temperature Ising transition. Sandwiched between the two transitions 
is a time-reversal symmetry breaking non-superfluid intermediate state. 
By combining the numerical results from Monte Carlo in the weak coupling regime and the analytical exact result from mapping the Mott limit of the $p$-band model to the orbital equivalent of  the Onsager Ising model~\cite{1944_Onsager_PR}, the phase diagram for $p$-band Bose-Hubbard model 
in two dimensions is proposed (Fig.~\ref{fig:3pbandphasediag}).  
We would like to mention here that strong correlation effects may give rise to exotic intermediate phases between p-band Mott insulator and superfluid states~\cite{2007_Xu_Fisher_PRB}. 

\subsection{Interacting $p$-orbital fermions}

\subsubsection{Nested Fermi surface---FFLO state} 
Searches for superconducting Fulde-Ferrell-Larkin-Ovchinnikov (FFLO) phases with 
spatially varying order parameters have been attracting 
tremendous interest 
in both atomic gases and electronic materials. 
It appears that 
the parameter window for this novel state to occur in two or three dimensions is quite 
narrow for conventional settings~\cite{2010_Radzihovsky_Sheehy_RPP}. 
(In an optical lattice, the spin imbalance window to reach the FFLO state is larger but only near half filling~\cite{2010_Loh_Trivedi_PRL}.)  
In one dimension the parameter regime is considerably larger and 
much progress has been made to find FFLO phases~\cite{2001_Yang_PRB,2007_Orso_PRL, 2007_Hu_Liu_PRL,2007_Guan_Batchelor_PRB,
2007_Parish_Baur_PRL,2007_Feiguin_Heidrich_PRB,2008_Casula_Ceperley_PRA,2009_Kakashvili_Bolech_PRA,2010_Liao_Rittner_Nature}. Nonetheless, long range order is prohibited due to 
fluctuations in one dimension, which makes it challenging to observe FFLO states even in one dimension.

\begin{figure}[htp]
\includegraphics[angle=0,width=.5\linewidth]{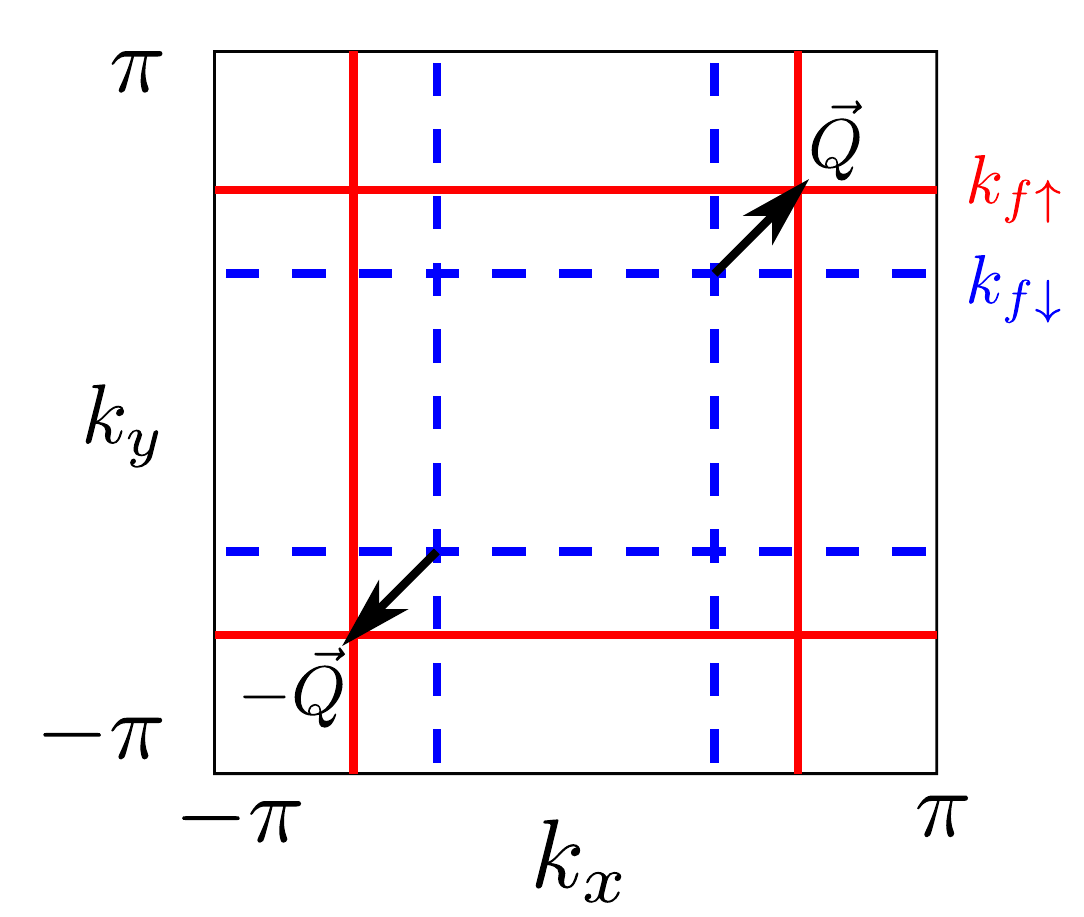}
\caption{Nesting of $p$-orbital Fermi surfaces on a square lattice~\cite{2011_Cai_Wang_PRA}.
The Fermi surfaces of $p_x$ ($p_y$) orbitals are vertical (horizontal) lines. Solid and dashed lines are for the minority and 
majority hyperfine states, respectively. Fermi surfaces are perfectly matched as $t_\perp/t_\parallel \to 0$, 
with pairing momenta $\pm \tbf{Q} =\pm (\delta k_f, \delta k_f)$, 
with $\delta k_f = k_{f\uparrow} - k_{f\downarrow}$. By lattice rotation symmetry, $\tbf{Q}' = (\delta k_f, -\delta k_f)$ is the other choice 
of pairing momentum.  }
\label{fig:3porbfermion}
\end{figure}

Due to intrinsic anisotropy of $p$-orbital wavefunctions, FFLO states in $p$-orbital fermion systems 
are found to occur in a wide window in two dimensions or even in three dimensions~\cite{2011_Cai_Wang_PRA}, 
not restricted to half filling. 
Here, we shall focus on a two dimensional square lattice, where the tunneling Hamiltonian is~\cite{2011_Cai_Wang_PRA}
\bea 
H_{0}  &=& t_\parallel \sum_{\tbf{r}, \sigma} 
    \left[  c_{x, \sigma,\tbf{r} } ^\dag c_{x, \sigma, \tbf{r}+\hat{a}_x} + 
	     c_{y, \sigma,\tbf{r} } ^\dag c_{y, \sigma, \tbf{r}+\hat{a}_y}
    + h.c.  \right ] \nn \\
    && - \mu \sum_{\tbf{r}, \sigma} n_\sigma(\tbf{r}) 
      - \frac{h}{2} \left( n_\uparrow (\tbf{r}) - n_\downarrow (\tbf{r})  \right), 
\eea 
with $c_{x, \sigma}$ the fermionic annihilation operators for $p_x$ ($p_y$) orbital with pseudo-spin 
$\sigma = \uparrow/\downarrow$ and  $n_\sigma = \sum_\alpha c_{\alpha, \sigma} ^\dag c_{\alpha, \sigma}$ 
the density operators for spin $\sigma$. The transverse tunneling $t_\perp$ ($\ll t_\parallel$) is neglected for simplicity and 
in presence of spin-imbalance, this leads to perfect nesting of $p$-orbital Fermi surfaces (see Fig.~\ref{fig:3porbfermion}). 

In atomic gases the pseudo-spin components are hyperfine states. The interactions between them maybe 
engineered by $s$-wave Feshbach Resonance. With the Feshbach Resonance, a matured technique in experiments, 
the induced interactions are given as~\cite{2010_Zhang_Hung_PRA2}
\bea 
H_{\rm int} &=& U\sum_\tbf{r} \left[ n_{x\uparrow,\tbf{r}} n_{x \downarrow, \tbf{r}} +  n_{y\uparrow,\tbf{r}} n_{y \downarrow, \tbf{r}} \right] \nn \\
      && - \sum_\tbf{r} J \left[ \vec{S}_{x,\tbf{r}}  \cdot \vec{S}_{y, \tbf{r}}  - \frac{1}{4} n_{x,\tbf{r}} n_{y, \tbf{r}}  \right] \nn \\ 
      && + \sum_\tbf{r} V_{xy} \left[ c_{x\uparrow, \tbf{r}} ^\dag c_{x\downarrow, \tbf{r}} ^\dag c_{y \downarrow, \tbf{r}} c_{y \uparrow, \tbf{r}} + h.c.  \right]. 
\label{eq:pbandint} 
\eea 
At tree level, $J$, $V_{xy} $ and $U$ are related: $J = \frac{2U}{3}$, $V_{xy} = \frac{U}{3}$. 
With attractive interaction, $U<0$, the induced Cooper parings are 
$\Delta_{x, \tbf{r}} = \langle c_{x\uparrow, \tbf{r}} c_{x \downarrow, \tbf{r}}  \rangle$,
$\Delta _{y, \tbf{r}} = \langle c_{y\uparrow, \tbf{r}} c_{y \downarrow, \tbf{r}}  \rangle$, 
$\Delta _{xy, \tbf{r}} = \langle c_{x\uparrow, \tbf{r}} c_{y\downarrow, \tbf{r}} \rangle$, 
and 
$\Delta _{yx, \tbf{r}} = \langle c_{y\uparrow, \tbf{r}} c_{x\downarrow, \tbf{r}} \rangle$. 
The BCS mean field Hamiltonian is 
\bea 
H_{\rm M} &=& H_0 \nn \\
      &-& U \sum_\tbf{r} \left[ 
		\Delta_{x,\tbf{r}} c_{x\uparrow, \tbf{r}} ^\dag c_{x\downarrow, \tbf{r}} ^\dag 
		+\Delta_{y, \tbf{r}} c_{y \uparrow, \tbf{r}} ^\dag c_{y \downarrow, \tbf{r}}^\dag  + h.c.\right] \nn \\
      &+& \frac{1}{2}J \sum_\tbf{r} 
		\left\{ \Delta_{xy,\tbf{r}} c_{x\downarrow, \tbf{r} } ^\dag c_{y\uparrow, \tbf{r} }^\dag 
		      + \Delta_{yx,\tbf{r}} c_{x\uparrow, \tbf{r}} ^\dag c_{y \downarrow, \tbf{r}} ^\dag \right. \\ 
      && \,\,\,\,\,\,\,\,\,\left. 
		      + \Delta_{yx,\tbf{r}} c_{y \uparrow, \tbf{r}} ^\dag c_{x\downarrow,\tbf{r}} ^\dag 
		      + \Delta_{xy,\tbf{r}} c_{x\uparrow, \tbf{r}} ^\dag c_{y \downarrow,\tbf{r}} ^\dag +h.c. \right\}\nn \\
      & -&  V_{xy} \sum_\tbf{r} 
		\left[  \Delta_{y,\tbf{r}} c_{x\uparrow,\tbf{r}} ^\dag c_{x\downarrow, \tbf{r}} ^\dag 
			+ \Delta_{x, \tbf{r}} c_{y\uparrow, \tbf{r}}^\dag c_{y\downarrow, \tbf{r}} ^\dag  + h.c. \right]\nn . 
\eea 
The last term $V_{xy}<0$ locks the phase difference between $\Delta_x$ and $\Delta_y$ at $0$, which plays a central role 
in making the superconducting coherence two-dimensional. 
Solving the gap equation numerically yields 
\bea 
&& \Delta_{xy} , \Delta_{yx} \approx 0, \\
&& \Delta_{x,\tbf{r}} = \Delta_{y, \tbf{r}} = |\Delta | \cos \left(\tbf{Q} \cdot \tbf{r}\right) .
\eea
The mean field phase diagram has been mapped out in Ref.~\cite{2011_Cai_Wang_PRA}. The parameter regime supporting FFLO states is considerably large 
in spin imbalanced $p$-band fermions.

\subsubsection{Nested Fermi surface---density stripes}

\begin{figure}[htp]
\includegraphics[angle=0,width=\linewidth]{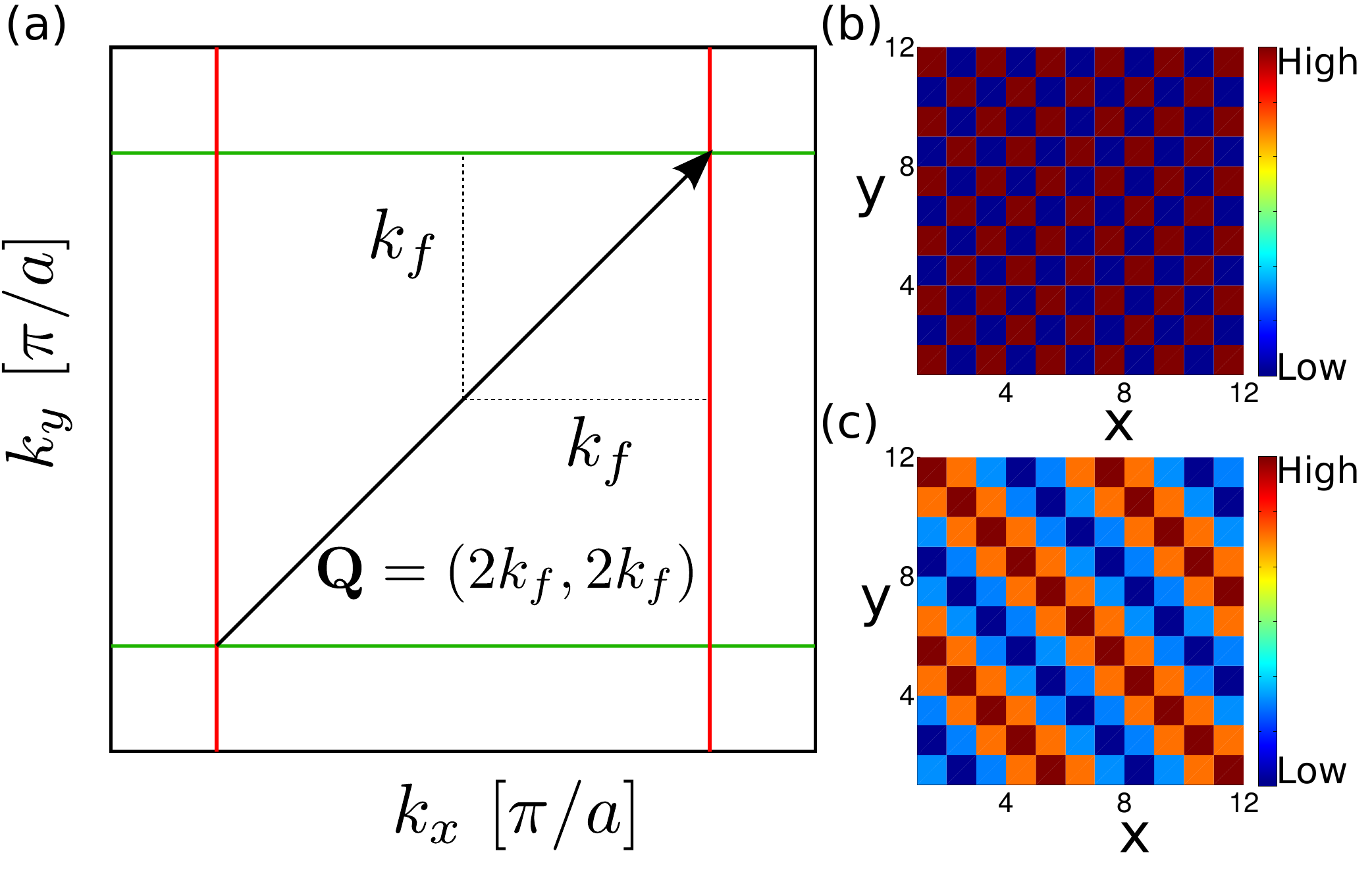}
\caption{Fermi surface nesting and density waves of spinless fermions on $p$-orbital bands~\cite{2012_Zhang_Li_PRA}. 
(a) shows the Fermi surface nesting. Red (dark gray) and green (light gray) solid curves indicate Fermi surfaces of $p_x$ and $p_y$ orbital bands, 
respectively. The solid arrow shows the ($2k_{\rm F}, 2k_{\rm F}$) momentum of particle-hole pairing simultaneously satisfying the nesting condition for both $p_x$ and $p_y$ bands. (b) shows the checkerboard density pattern at half filling. (c) shows  the density pattern of the striped CDW/ODW phase lower than half-filling. 
}
\label{fig:3porbspinlessfermi}
\end{figure}

Besides the superconducting stripes in FFLO states, $p$-orbital fermions also naturally support 
density stripe orders, again from Fermi surface nesting (Fig.~\ref{fig:3porbspinlessfermi}). 
Stripe and checkerboard orders are found in a system of {\it spinless} fermions loaded up to $p$-orbital bands~\cite{2012_Zhang_Li_PRA}, 
where the Hamiltonian takes a simple form 
\bea 
H &=& \sum_{\tbf{r} \alpha \beta} 
      \left(t_{\alpha \beta} c_{\alpha, \tbf{r} + \hat{a}_\beta} ^\dag c_{\alpha, \tbf{r} } + h.c. \right) \nn \\
   && -\mu \sum_{\tbf{r} \alpha} n_{\alpha \tbf{r}} 
 + g\sum_\tbf{r} n_{x, \tbf{r}} n_{y, \tbf{r}} , 
\label{eq:3Ham2dspinless}
\eea 
with $t_{\alpha \beta} = [t_\parallel \delta_{\alpha \beta} - t_\perp (1-\delta_{\alpha \beta}) ] $. 
Fermi surface nesting of this system is pictorially illustrated in Fig.~\ref{fig:3porbspinlessfermi}. Fermi surfaces of $p_x$ and $p_y$ bands are approximately perpendicular to each other, which greatly suppresses the Cooper instability. The reason is 
that for the spinless case, the onsite interaction can lead to only cross-orbital Cooper pairing that is antisymmetric in px and py orbitals. The nearly orthogonal geometry of the two Fermi surfaces now makes it impossible to condense such Cooper pairs at a single center of mass momentum.
In contrast, each $p_x$ ($p_y$) particle-hole pair in the density channel composed of one particle and one hole within the $p_x$ ($p_y$) band benefits from the fermi surface nesting. To simultaneously condense particle-hole pairs in each orbital band, the wave-vector  
$$
\tbf{Q}_ {1,2} = (2 k_{\rm F} , \pm 2 k_{\rm F}) 
$$
is most favorable (see Fig.~\ref{fig:3porbspinlessfermi}).

To characterize the Fermi surface nesting effect observed in Fig.~\ref{fig:3porbspinlessfermi}, one can look at the 
density-density correlations, which can be calculated by the field theory with partition function  
$Z = {\rm Tr} e^{-\beta H} = \int D \left(\psi^*_\alpha(\tbf{r}, \tau) \psi_\alpha(\tbf{r}, \tau) \right)
      \exp(-S_F)$, and the action  
\bea 
S_F&=& \int d\tau \sum_{\mathbf{r},\alpha} \psi^\ast_\alpha(\mathbf{r},\tau) (\partial_\tau -\mu)\psi_\alpha(\mathbf{r},\tau)\nonumber\\
&& +\sum_{\mathbf{r} \alpha \beta}t_{\alpha \beta}
(\psi^{\ast}_{\alpha}(\mathbf{r}+{\hat{a}_{\beta}},\tau)
\psi_{\alpha}(\mathbf{r},\tau)+h.c.)\nonumber\\
&&+g\sum_{\mathbf{r}} \psi^\ast_x(\mathbf{r},\tau) \psi^\ast_y(\mathbf{r},\tau) \psi_y(\mathbf{r},\tau)\psi_x(\mathbf{r},\tau).
\eea 
The density fields are defined as 
$\rho_\alpha (\tbf{r}, \tau) = \psi^* (\tbf{r}, \tau) \psi (\tbf{r}, \tau)$, and 
density-density correlations are given by 
$$\Pi _{\alpha \beta} (q) = \frac{T}{N_s} \langle \rho_{\alpha, q} \rho_{\beta, q} \rangle,$$
with 
$q = (\tbf{q}, i\omega)$, 
$$\rho_{\alpha q} = \sum_\tbf{r} \int d\tau \rho (\tbf{r}, \tau) e^{i\tbf{q}\cdot \tbf{r} - i\omega \tau} 
= \sum_{ k}\tilde{ \psi }_\alpha ^\dag (k+q) \tilde{\psi}_\alpha (k), $$ 
and $\tilde{\psi}_\alpha (k) = \frac{1}{\sqrt{\beta N_s} }\int d\tau \sum_\tbf{r} \psi_\alpha (\tbf{r}, \tau) e^{- i\vec{k} \cdot\tbf{r} + i\omega \tau}$. 
It is useful to split  into two channels---number density ($\rho_+$) and orbital density ($\rho_-$): $\rho_\pm (q) = \rho_{x, q} \pm \rho_{y, q} $. 
The correlations in these two separate channels are defined as
$$
\Pi_\pm = \frac{T}{N_s} \langle \rho_\pm (q) \rho_\pm (-q) \rangle. 
$$
Summing up ring diagrams under random phase approximation (RPA), 
the correlations are obtained as~\cite{2012_Zhang_Li_PRA}
\bea 
\Pi_\pm (\tbf{Q}, 0) = \frac{2 \chi^0} {1 \pm g \chi^0}, 
\eea 
with 
$\chi^0$ given by 
$$
\chi^0 = D (E_F) \ln\left( \frac{\omega_D}{T} \right), 
$$
in the limit of $t_\perp \to 0$. Here 
$D(E_F)$ is the density of states near the Fermi surface and $\omega_D$ is some energy cutoff in the field theory. 
Due to the logarithmic divergence in $\chi^0$, any arbitrarily weak attractive (repulsive) interaction $g<0$ ($g>0$) induce 
divergence of $\Pi_+$ ($\Pi_-$) at sufficiently low temperature. The divergence of $\Pi_+$ and $\Pi_-$ indicates long range ordering 
of charge density wave (CDW) and orbital density wave (ODW), respectively. Transitions to these density waves are studied with mean field theory, where 
the Hamiltonian is approximated by 
\bea 
H_{\rm MF} =  \sum_{\tbf{r} \alpha \beta} t_{\alpha \beta}
      \left( c_{\alpha, \tbf{r} + \hat{a}_\beta} ^\dag + h.c. \right) 
   -\mu \sum_{\tbf{r} \alpha} n_{\alpha \tbf{r}} \nn \\ 
+ g \sum_\tbf{r} \left( n_{x,\tbf{r}} M_{y, \tbf{r}} + n_{y, \tbf{r}} M_{x,\tbf{r}} - M_{x,\tbf{r}} M_{y, \tbf{r}}  \right),  
\eea 
with $M_{\alpha, \tbf{r}} = \langle n_{\alpha, \tbf{r}} \rangle$. 
Self-consistent mean field calculations confirm that 
repulsive and attractive interactions favor CDW and ODW, respectively.  
The density patterns of these density waves are shown in Fig.~\ref{fig:3porbspinlessfermi}. 

The order parameter for the charge density wave phase is introduced by 
$$
\rho (\tbf{r}) = [\phi_1 e^{i\tbf{Q}_1 \cdot \tbf{r}} + \phi_2 e^{i\tbf{Q}_2\cdot \tbf{r}} + c.c.] + {\rm const},  
$$
where $\phi_1$ and $\phi_2$ are complex valued fields slowly varying in space. 
The phenomenological free energy reads
\bea 
\label{eq:3Freeenergypdensitywave} 
&& F = \\ 
&& \int d^2 \tbf{r} \sum_{j = 1,2} \left( K|\vec{\nabla} \phi_j|^2 + r |\phi_j|^2 + u|\phi_j|^4 + v|\phi_1|^2 |\phi_2|^2  \right). \nn 
\eea 
Here,  incommensurate filling is assumed and the theory has an emergent $U(1) \times U(1)$ symmetry; otherwise 
the terms such as $\left(\phi_j ^p + c.c.\right)$ are allowed and the theory has lower symmetry.

The effective couplings, $K$, $r$, $u$, and $v$ in Eq.~\eqref{eq:3Freeenergypdensitywave} have been 
connected to microscopic parameters by field theory calculations~\cite{2012_Zhang_Li_PRA}.  
At low temperature, we have ($r<0$, $u>2v$), and the system is in a striped CDW phase with
the wavevector $\tbf{Q}_1$ or $\tbf{Q}_2$ spontaneously chosen. 
Assuming $\tbf{Q}_1$ is spontaneously chosen 
there is an algebraic long range order in $\phi_1$, i.e., $
\langle \phi_1^* (\tbf{r}) \phi_1 (\tbf{r}')\rangle \propto \frac{1}{|\tbf{r}-\tbf{r}'|^\gamma}. 
$
At higher temperature it is found that the striped CDW phase first melts to a nematic phase through an Ising transition and 
then to normal through a Kosterlitz-Thouless transition.

\subsubsection{Strongly correlated orbital models}
At half filling $p$-orbital fermions described by the Hamiltonian in Eq.~\eqref{eq:3Ham2dspinless} exhibits a Mott transition with strong repulsive interaction, which  is studied in Ref.~\cite{2008_Zhao_Liu_PRL,2008_Wu_PRL2}. In the fermionic Mott state, like in the bosonic case, fermions are localized on each lattice site. As a result, the low energy physics is described by an effective model of super-exchange interactions. 

Considering a link $<${\bf r}, {\bf r}'$>$ the super-exchange interactions are determined by energy corrections on the states 
$
   |1,0; 1, 0\rangle 
   $, 
$  
 |0, 1; 0, 1\rangle 
 $, 
$
 |0, 1; 1, 0\rangle 
 $ 
and 
$
  |1,0; 0, 1\rangle
  $,  
where a notation is taken from Eq.~\eqref{eq:3orbitalfockstate} with the bosonic operators 
$b_{\alpha}$ replaced  by fermionic ones $c_{\alpha}$.  
Suppose this link is in the $x$ direction, the tunneling is then 
$H^x_{\rm t} = -t_\parallel c_{x, \tbf{r}} ^\dag  c_{x, \tbf{r}'} + h.c. $,
with the transverse tunneling neglected. 
From standard perturbation theory, the energy corrections due to virtual fermion fluctuations are 
\bea 
&& \Delta E \left(
 |0, 1; 0, 1\rangle \right)  = 
  \Delta E \left(
    |1, 0; 1, 0\rangle \right)  = 0, \\
&& \Delta E \left(
   |1, 0; 0, 1\rangle \right)  = 
  \Delta E \left(
   |0, 1; 1, 0 \rangle \right)  =  -\frac{t_\parallel ^2}{U}. 
\eea 
Mapping $p_x$ ($p_y$) to pseudo-spin $\uparrow$ ($\downarrow$) states, the super-exchange interactions are given in a compact form as 
\bea 
h_{\rm eff} ^x = J_z \sigma_z (\tbf{r}) \sigma_z (\tbf{r}')  + const,  
\eea 
with $J_z = \frac{t_\parallel ^2}{2 U}$. 
Rotating $p$-orbitals by an angle $\theta$,  we have the following transformation
\bea 
\left( \begin{array}{c}
        c_x \\ 
        c_y 
       \end{array}
\right) \to 
{\cal U} (\theta) 
\left( \begin{array}{c}
        c_x \\ 
        c_y 
       \end{array}
\right), 
\eea 
with 
$$
{\cal U} (\theta )  = 
\left( 
\begin{array}{cc}
 \cos \theta & \sin\theta \\
 -\sin \theta & \cos\theta
\end{array}
\right).
$$ 
For a link oriented at an angle $\theta$ with respect the $x$ axis, the super-exchange interaction reads as 
\bea 
h_{\rm eff} ^ {\theta} = J_z   \tilde{\sigma}_z (\tbf{r}) \tilde{\sigma}_z (\tbf{r}'), 
\eea 
with 
$\tilde{\sigma}_z = {\cal U} ^\dag (\theta) \sigma_z {\cal U} (\theta) 
= \sin (2\theta) \sigma_x + \cos(2\theta) \sigma_z$. 

\begin{figure}[htp]
\includegraphics[angle=0,width=0.6\linewidth]{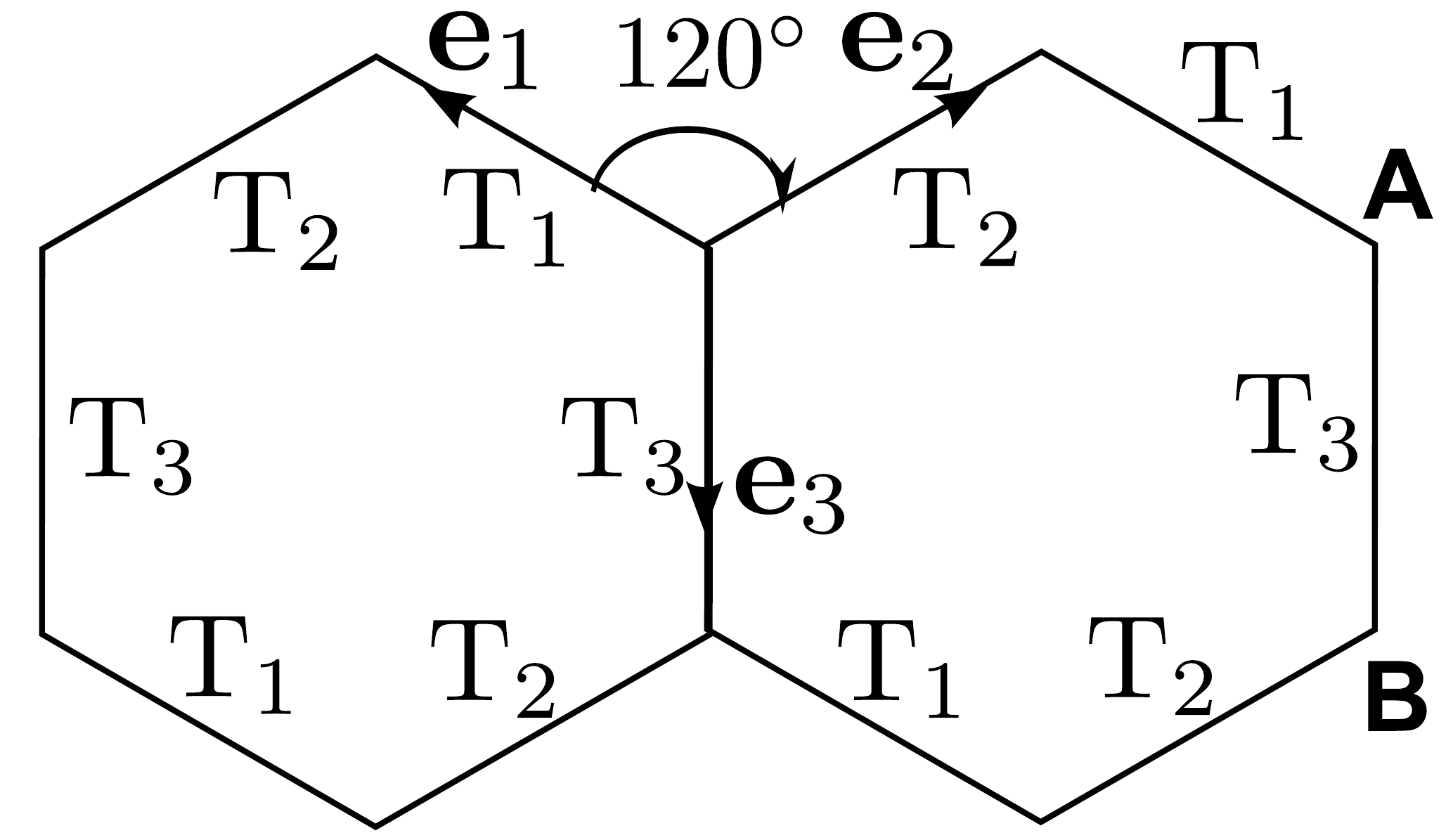}
\caption{Illustration of quantum $120^\circ$ model on a honeycomb lattice. $T_1$ ,$T_2$ and $T_3$ denote three different super-exchange interactions. }
\label{fig:3porb120}
\end{figure}

On a square lattice, the Hamiltonian describing the orbital order is 
\bea 
H_{\rm eff}^{\rm sq} = J_z \sum_{\tbf{r}} 
      \sigma_z (\tbf{r}) \sigma_z (\tbf{r}+ \tbf{a}_x) 
      + \sigma_z (\tbf{r}) \sigma_z (\tbf{r}+ \tbf{a}_y), 
\eea 
which has the same form as the $p$-band Mott insulator of bosons. This Hamiltonian supports an alternating $p_x/p_y$ order as shown in Fig.~\ref{fig:3alternatepxpy}. 
On a honeycomb lattice, the super-exchange Hamiltonian is
\bea 
H_{\rm eff} ^{\rm hc} = J_z \sum_{\tbf{r}, j} 
      T_j (\tbf{r}) T_j (\tbf{r} + \tbf{e}_j), 
\label{eq:3Ham120}
\eea 
with 
\bea 
T_1 &=& -\frac{\sqrt{3}}{2} \sigma_x + \frac{1}{2} \sigma_z, \nn \\
T_2 &=& \frac{\sqrt{3}}{2} \sigma_x + \frac{1}{2} \sigma_z, \nn \\ 
T_3 &=& -\sigma_z. 
\eea 
Here the summation $\sum_\tbf{r}$ includes one set of the `A' sublattices~(Fig.~\ref{fig:3porb120}). This model, dubbed quantum $120^\circ$ 
model~(\onlinecite{2008_Zhao_Liu_PRL}; \onlinecite{2007_Wu_Bergman_PRL,2008_Wu_PRL2}), is geometrically frustrated. The complication of this model originates precisely 
from the spatial nature of orbitals, which makes orbital degrees of freedom drastically different from real spins. 

In three dimensions on a diamond lattice, the $p$-orbital exchange interaction leads 
to an exact orbital Coulomb phase characterized by ice rules and emergent gauge structures~\cite{2011_Chern_Wu_PRE}. 

\subsubsection{Anti-Ferromagnetic phases of spinor $p$-orbital fermions }  
As motivated by understanding the role of magnetism in high temperature superconductors, studies of antiferromagnetic transitions in $s$-band fermions attracted tremendous interest, but the transition temperature is still out of reach for current cooling techniques. One way to improve the transition temperature could be provided by considering $p$-band fermions. The antiferromagnetic transition of spin-$1/2$ fermions loaded in $p$-bands of a 3D  cubic lattice are studied in~\cite{2008_Wu_Zhai_PRB}, where half filling 
(three fermions per lattice site) is assumed. The Hamiltonian describing such a system is $H = H_0 + H_{\rm int}$, with 
$$
H_0 = \sum_{\tbf{r} \alpha \beta} 
      \left(t_{\alpha \beta}  c_{\alpha, \tbf{r} + \hat{a}_\beta} ^\dag  c_{\alpha, \tbf{r}}  + h.c. \right), 
$$ 
and 
\begin{widetext} 
$$
H_{\rm int} = \sum_{\tbf{r}} 
      \left\{ U \sum_\alpha n_{\tbf{r},  \alpha, \uparrow} n_{\tbf{r},  \alpha, \downarrow} 
	+ W \sum_{\alpha \neq \beta} \left( n_{\tbf{r},  \alpha, \uparrow} n_{\tbf{r},  \beta, \downarrow} 
 + c_{\tbf{r},  \alpha, \uparrow} ^\dag c_{\tbf{r}, \beta, \downarrow} ^\dag c_{\tbf{r}, \alpha, \downarrow} c_{\tbf{r},  \beta, \uparrow} 
		    + c_{\tbf{r},  \alpha, \uparrow} ^\dag c_{\tbf{r},  \alpha, \downarrow} ^\dag c_{\tbf{r},  \beta, \downarrow} c_{\tbf{r},  \beta, \uparrow}\right) \right\}. 
$$
\end{widetext} 

With strong repulsion, we can project to the low-energy subspace determined by $H_{\rm int}$; projecting out high energy subspace will contribute to super-exchange interactions. The low-energy states are 
the four degenerate components of total spin-$3/2$: $|\uparrow \uparrow\uparrow\rangle$,  
$\frac{1}{\sqrt{3}} \left( 
    |\downarrow \uparrow \uparrow \rangle 
  + |\uparrow \downarrow \uparrow \rangle 
  + |\uparrow \uparrow \downarrow \rangle \right), $ 
$\frac{1}{\sqrt{3}} \left( 
    |\downarrow \downarrow \uparrow \rangle 
  + |\uparrow \downarrow \downarrow \rangle 
  + |\downarrow \uparrow \downarrow \rangle \right), $ 
and $|\downarrow \downarrow \downarrow \rangle, $  
where a notation $|s_1 s_2 s_3 \rangle = c_{x, s_1} ^\dag c_{y, s_2} ^\dag c_{z, s_3} ^\dag |{\rm vac}  \rangle $ is used. 
This is  manifestation of the Hund's rule. 
It is quite involved to perform the quantum mechanical second order perturbation theory here.
A more elegant way is to take the  Brillouin-Wigner approximation where the super-exchange interaction is given by 
\bea 
H_J = -P_G H_0 P_e H_{\rm int} ^{-1} P_e H_0 P_G. 
\eea 
Here $P_e$ and $P_G$ mean projections onto excited and low-energy subspaces, respectively. 

The calculations are greatly simplified by two observations~\cite{2008_Wu_Zhai_PRB}. Firstly, the hopping processes only take place within the same orbital. Secondly, all terms in  $H_0$ acting on the low-energy subspace create eigenstates of $H_{\rm int}$ with the same excitation energy $U+ 2 W$. Then the super-exchange Hamiltonian $H_J$ on a link 
$<\tbf{r}, \tbf{r}'= \tbf{r} + \hat{x} >$ is given by 
$$ 
 \sum_\alpha \frac{2 |t_{\alpha, x}|^2 }{U+ 2W} 
P_G \left( \sum_{s s'} c_{\tbf{r}, \alpha, s} ^\dag c_{\tbf{r}, \alpha, s'} 
c_{\tbf{r}', \alpha, s'} ^\dag c_{\tbf{r}', \alpha, s}  \right) P_G. 
$$ 
By symmetrizing $c_{\tbf{r}, \alpha, s} ^\dag c_{\tbf{r}, \alpha, s'}$,  one can show 
$$
P_G c_{\tbf{r}, \alpha, s} ^\dag c_{\tbf{r}, \alpha, s'} P_G 
= \frac{1}{3} P_G \sum_\beta 
  \left( c_{\tbf{r}, \beta, s} ^\dag c_{\tbf{r}, \beta, s'}  \right) P_G. 
$$ 
Then the super-exchange Hamiltonian is obtained to be 
\bea 
H_J = J \sum_{<\tbf{r}, \tbf{r}'>} \vec{S}_\tbf{r}\cdot \vec{S}_{\tbf{r}'}, 
\eea 
with $\vec{S}_\tbf{r} = \frac{1}{2} P_G 
    \sum_{\alpha s s'} c_{\tbf{r}, \alpha, s} ^\dag \vec{\sigma}_{s s'} c_{\tbf{r}, \alpha, s'} P_G$, 
and the effective coupling 
\be 
J = 4 (t_\parallel ^2 + 2 t_\perp ^2 ) /(9U + 18 W) >0. 
\ee 
The effective description is an isotropic spin-$3/2$ Heisenberg model.  
%
We remind the reader that this is the model for half filling, with the full Hilbert space of each site being spanned by three $p$-orbitals and two spins. Hund's rule reduces the low energy subspace to the total spin-$3/2$ space. 
The ground state of the system  thus has an antiferromagnetic long range order. 
This antiferromagnetic order is destroyed by thermal fluctuations when the temperature is above N$\acute{\text {e}}$el temperature $\sim J$.

\subsection{Topological bands and nontrivial orbital states}
\label{sec:TopologicalPhase}
In optical lattice experiments considerable efforts have been made to create topological bands. 
Neutral atoms loaded in such bands would experience effective magnetic fields due to non-trivial 
Berry curvatures. 
These experimental developments are motivated by consideration of novel 
quantum many-body states such as quantum Hall states and topological insulators/superconductors. 
While previous experiments largely focused on manipulating different  hyperfine states of atoms 
with synthetic gauge fields, 
recent theoretical studies~\cite{2010_Liu_Liu_PRA,2011_Sun_Liu_NatPhys,2011_Sun_Gu_PRL,2013_Li_Zhao_NatComm,2014_Liu_Li_NatComm,2014_Dutta_Lewenstein_NJP,2014_Dutta_Przysiezna_PRA,2015_Liu_Li_arXiv2,2015_Yin_Baarsma_arXiv} point 
to alternate ways to achieve topological bands by considering high orbital states 
in the optical lattices of non-standard geometry.

\subsubsection{Topological $sp$-orbital ladder} 
\begin{figure}[htp]
\includegraphics[angle=0,width=.6\linewidth]{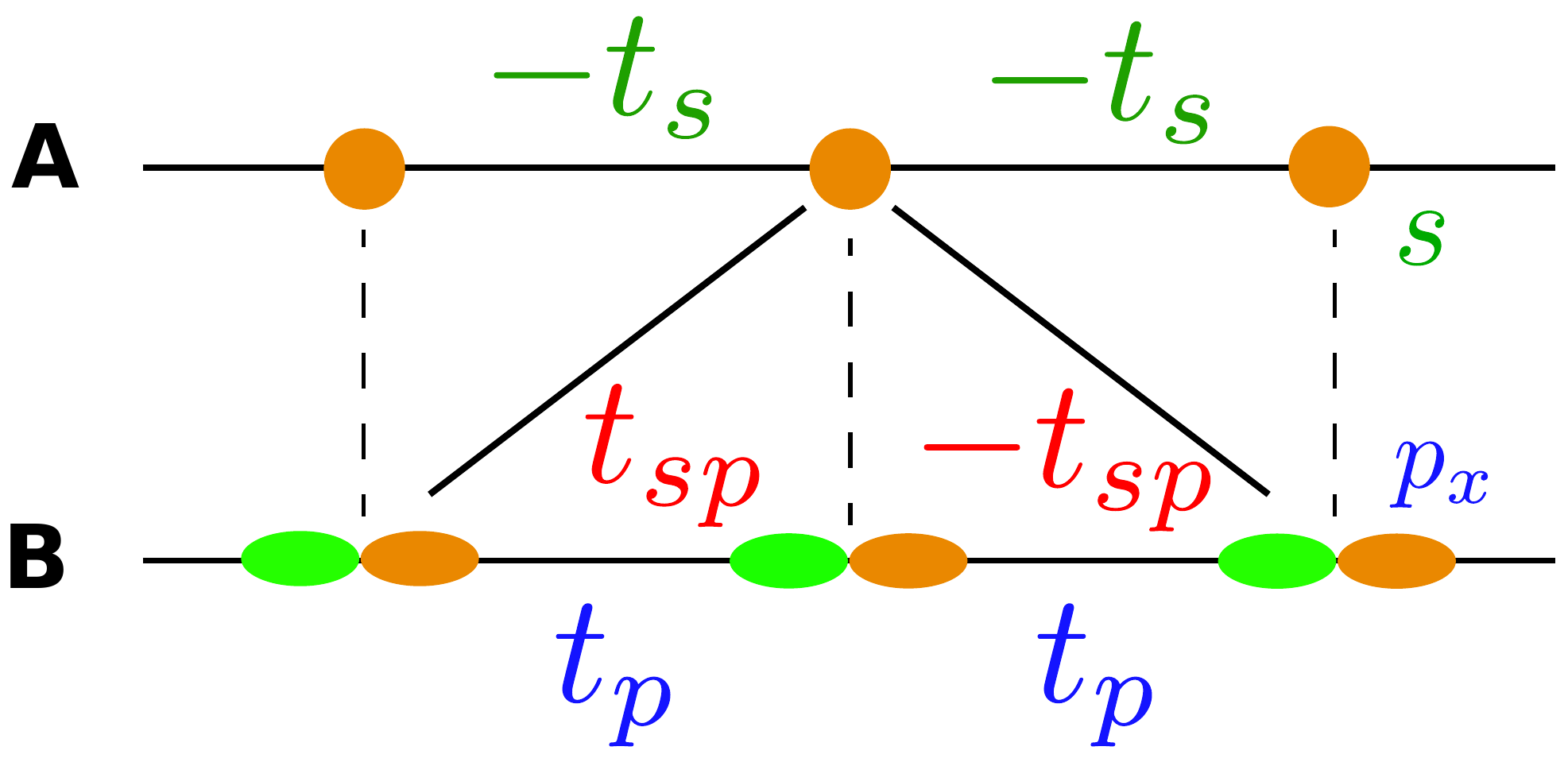}
\caption{A one dimensional $sp$-orbital ladder~\cite{2013_Li_Zhao_NatComm}. This ladder consists of two chains, A and B. The $s$ orbitals of A chain 
are level with the $p_x$ orbitals of B chain. The inter-orbital tunneling has a `$\pm$' staggering sign as shown. 
}
\label{fig:3spladder}
\end{figure}

A one dimensional ladder composed of two chains of $s$ and $p$-orbitals is shown in Fig.~\ref{fig:3spladder}. The two orbitals are level in energy, and other lower orbitals are energetically separated with a large gap, and thus can be neglected when considering the $sp$ ladder. 
The Hamiltonian describing this orbital ladder system is given by 
\bea
H_0= \textstyle \sum_j
                C_j ^\dag
                \left[
                \begin{array}{cc}
                -t_s & -t_{sp} \\
                t_{sp} & t_p
                \end{array}
                \right]
                C_{j+1}
                +h.c. 
       - \sum_j \mu    C_j ^\dag C_j,
\label{eq:ham}
\eea
where $C^\dagger_j = \left[ a_s^\dagger (j), a^\dagger_{p_x} (j) \right]$, 
{with $a_s^\dagger (j) $ and $a^\dagger_{p_x} (j) $ being fermion creation operators 
for the $s$- and $p_x$-orbitals on the A and B chain respectively.} The
relative sign of the hopping amplitudes is fixed by parity symmetry of the
$s$ and $p_x$ orbital wave functions. As depicted in Fig.~\ref{fig:3spladder},
the hopping pattern plays a central role in producing a topological phase. 
With a proper global gauge choice, $t_s$, $t_p$ and $t_{sp}$ are all positive. 
Focusing on half filling with chemical potential $\mu = 0$, the Hamiltonian is particle-hole symmetric under transformation $C_j \to (-1)^j C_j ^\dag$. Heuristically, topologically non-trivial band structure of the $sp$-orbital ladder may be speculated by rewriting the staggered quantum tunneling as 
$$
t_{sp} \sum_j \left[ C_j ^\dag \left( -i \sigma_y \right) C_{j+1}  + h.c. \right] 
$$
resembling the spin-orbit interactions when the $s$ and $p$ orbitals are mapped to pseudo-spin ($1/2$) states. The physics of the $sp$ orbital ladder is also connected to the more familiar frustrated ladder with magnetic $\pi$-flux, but the $sp$-ladder appears much easier to realize in optical lattice experiments. 

In the momentum space, the Hamiltonian takes a suggestive form 
\be
\mathcal{H} (k ) = h_0 (k) \mathbb{1} + \vec{h}(k) \cdot \vec{\sigma},
\label{eq:bulkhk}
\ee
where $h_0 (k) =(t_p -t_s) \cos(k)$,
$h_x =0$,
$h_y (k) =  2t_{sp} \sin(k)$ and
$h_z (k)  = -(t_p + t_s) \cos(k)$. Here, $\mathbb{1}$ is the unit matrix,  
and $\sigma_x$, $\sigma_y$ and $\sigma_z$ are Pauli matrices in the two-dimensional
orbital space. 
The energy spectrum consists of two branches, 
$$
E_\pm (k) = h_0 (k) \pm \sqrt{h_y ^2 (k) + h_z^2 (k) }.  
$$ 
An interesting limit is that when $t_s =t_p = t_{sp}$, the two bands are both completely flat. 
As the momentum $k$ is varied from $-\pi$ through $0$ to $+\pi$, crossing the entire Brillouin zone, the direction of the vector $\vec{h} (k)$ winds an angle of $2\pi$. In the notation of Ref.~\cite{2012_Wen_PRB}, the $sp$-orbital ladder belongs to the symmetry group $G_{++} ^{-+} (U, T, C)$, as it has both particle-hole and time-reversal symmetries in addition to the usual charge $U(1)$ symmetry. At half filling, it is characterized by an integer topological invariant, in this case the winding number $1$. 

A manifestation of the nontrivial band topology is existence of edge states. It is easiest to show the edge states in the flat band limit, $t_s = t_p = t_{sp} \equiv t $, by introducing auxiliary operators, 
$\phi_\pm (j) = [a_p (j) \pm a_s (j)] /\sqrt{2}$. Then the Hamiltonian only contains coupling between $\phi_+$ and $\phi_-$ of nearest neighbors, 
$$
H_0  \to 2 t \sum_j \phi_-^\dag (j) \phi_+ (j+1) + h.c. 
$$ 
One sees immediately that the edge operators $\phi_+ (1)$ and $\phi_- (N_s) $  are isolated from the bulk, i.e., decoupled from the rest of the system. These modes describe two edge states at zero energy.  
Away from the flat band limit, the wavefunctions of the edge states analytically constructed in Ref.~\cite{2013_Li_Zhao_NatComm} are found not to confine strictly at the ends, but instead decay exponentially with a characteristic length scale 
$$\xi  = 2/\log \left( | ( \sqrt{t_s t_p} + t_{sp} ) /( \sqrt{t_s t_p } -t_{sp} ) | \right). $$
Here, recall that the implicit length unit is the lattice constant along the ladder leg direction. 
For $t_{sp} = \sqrt{t_s t_p}$, which includes the flat band limit, the decay length $\xi$ vanishes and we have sharply confined edge states. 
 
A topological phase transition to a trivial insulator state can be driven by inducing a coupling between 
$s$ and $p$ orbitals, $\Delta H = \Delta_y \sum_j C_j^\dag \sigma_y C_j$, which can be engineered by rotating the atoms locally on each site~\cite{2010_Gemelke_Sarajlic_arXiv}. 
For the coupling strength $\Delta_y$ greater than some critical value $\Delta_y ^c$, Berry phase vanishes 
and the system becomes a trivial band insulator, and the zero energy edge states disappear. Such a phase transition can be detected by measuring the density correlation between two ends in experiments.

Regarding practical experimental realizations, careful treatments of band structures and Wannier functions are required as the details of tight binding models could receive significant corrections beyond harmonic approximations (Eq.~\eqref{eq:HarmonicTunneling})~\cite{2014_Ganczarek_Modugno_PRA}.  
One controllable way to couple $s$ and $p$ orbitals is to use a one dimensional shaking 
lattice~\cite{2012_Sowinski_PRL,2013_Lachi_Zakrzewski_PRL,2014_Zhang_Zhou_PRA,2015_Zhang_Lang_arXiv,2015_Strater_Eckardt_PRA,2015_Przysiezna_Dutta_NJP,2015_Dutta_Przysiezna_SR}, 
which  has recently been realized in experiments~\cite{2011_Fort_Fabbri_JPCS,2013_Parker_Ha_NatPhys,2015_Khamehchi_Qu_arXiv,2015_Weinberg_Olschlager_PRA,2015_Niu_Hu_OptExp}. 
The other way to systematically control the $sp$-orbital coupling is to consider a noncentrosymmetric lattice where the coupling can be turned on and off by manipulating inversion symmetries~\cite{2015_Liu_Li_arXiv2}.

\subsubsection{Topological semimetal from mixing $p$ and $d$ orbitals}
\begin{figure}[htp]
\includegraphics[angle=0,width=\linewidth]{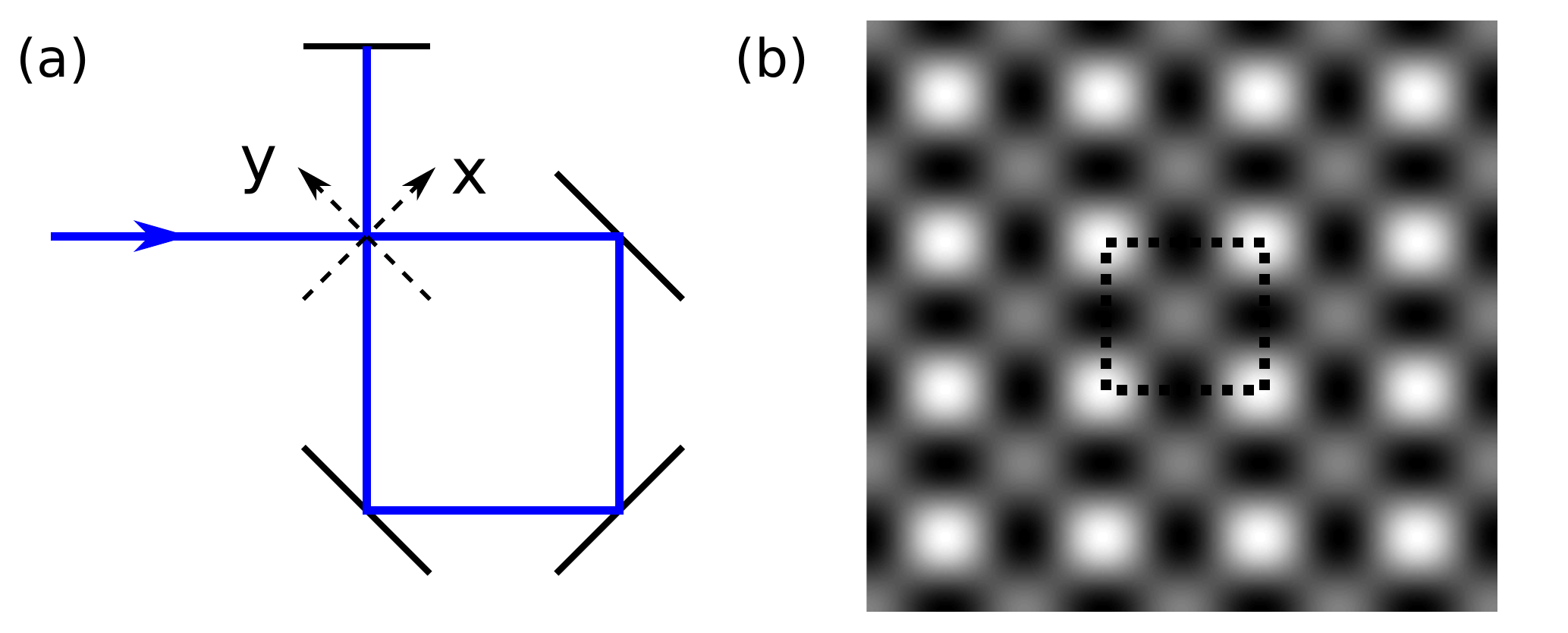}
\caption{Optical lattice realization of the topological semimetal~\cite{2011_Sun_Liu_NatPhys}. (a), the experimental setup to realize the lattice potential in Eq.~\eqref{eq:3TopPotential} 
for $V_2/V_1 \ge 1/2$. The linear polarization of the incident monochromatic light beam (solid blue line) encloses an angle $\alpha$ with respect to the direction normal to the drawing 
plane. (b), the optical potential for $V_1 = 2.2 E_R$, and $V_2 = 3.4 E_R$. The darker (lighter) regions represent areas where the potential is low (high). The dashed line marks one unit cell of the lattice. 
}
\label{fig:3TopSemimetal}
\end{figure}

We now turn to two dimensions and study how degeneracy of higher orbital bands may give rise to topological phases~\cite{2010_Liu_Liu_PRA,2011_Sun_Liu_NatPhys}. Consider a double-well optical lattice of the configuration shown in Fig.~\ref{fig:3TopSemimetal}~\cite{2011_Sun_Liu_NatPhys}.  By the space group symmetry ($D_4$) of the lattice, the two $p$-orbital states ($p_x$ and $p_y$) are degenerate with the lowest $d$-orbital (i.e., $d_{x^2-y^2}$) at high symmetry points in the momentum space.  The lattice configuration is found to exhibit degenerate $p$ and $d$ orbitals. 
Considering a square lattice with three orbitals on each site ($p_x$, $p_y$ and $d_{x^2 -y^2}$), the Hamiltonian of the tight binding model takes the following form~\cite{2011_Sun_Liu_NatPhys}
\bea 
 H_0 &=& \delta \sum_\tbf{r} d_\tbf{r} ^\dag d_\tbf{r} -t _{dd} \sum_\tbf{r} 
    \left( 	d_\tbf{r} ^\dag d_{\tbf{r} + \tbf{a}_x } + d_\tbf{r} ^\dag d_{\tbf{r} + \tbf{a}_y} + h.c. \right) \nn \\ 
    && + t_{\parallel} \sum_\tbf{r} 
      \left( 	p_{x, \tbf{r}} ^\dag p_{x, \tbf{r} + \tbf{a}_x} 
	+ p_{y, \tbf{r}} ^\dag p_{y, \tbf{r} + \tbf{a}_y} + h.c.\right)  \nn \\
    && - t_\perp \sum_\tbf{r} 
	\left( p_{x, \tbf{r}} ^\dag p_{x, \tbf{r} + \tbf{a}_y }  
	  + p_{y, \tbf{r}} ^\dag p_{y, \tbf{r} + \tbf{a}_x} + h.c. \right) \nn \\ 
    && + t_{pd} \sum_\tbf{r} 
	\left( d_\tbf{r} ^\dag p_{x, \tbf{r}+ \tbf{a}_x} - p_{x, \tbf{r}} ^\dag d_{\tbf{r}+ \tbf{a}_x} \right. \nn \\ 
    &&\,\,\,\,\,\,\, \left. + d_\tbf{r} ^\dag p_{y, \tbf{r} + \tbf{a}_y }    - p_{y, \tbf{r}} ^\dag d_{\tbf{r} + \tbf{a}_y} 
	  + h.c. \right), 
\label{eq:3TopSemimetal} 
\eea 
where $\tbf{a}_x$ ($\tbf{a}_y$) is the lattice vector in $x$ ($y$) direction, and $p_{x,\tbf{r}} $, $p_{y,\tbf{r}}$ and $d_\tbf{r} $ are fermionic annihilation operators for $p_x$, $p_y$  and $d_{x^2 -y^2}$ orbitals at site $\tbf{r}$. The amplitudes of tunneling between these orbitals at nearby sites are $t_{dd}$, $t_\parallel$, $t_\perp$, and $t_{pd}$. With a proper gauge choice, these tunneling amplitudes are all positive. Here a point group $D_4$ and time-reversal symmetries have  been assumed. This tight binding Hamiltonian can  be realized by a double-well 
optical lattice potential 
\bea 
V(x,y) &=& -V_1 [\cos (k x) + \cos (k y) ] \nn \\ 
&& + V_2 [\cos (k x + k y) + \cos (kx - ky) ] . 
\label{eq:3TopPotential} 
\eea 
A typical configuration and the experimental protocol to realize it are shown in Fig.~\ref{fig:3TopSemimetal}~\cite{2011_Sun_Liu_NatPhys}.  By the point group symmetry ($D_4$) of the lattice, the two $p$-orbital states ($p_x$ and $p_y$) are degenerate at high symmetry points in the momentum space. By dialing the relative strength of $V_1$ and $V_2$, the two $p$-orbitals may be tuned to degeneracy with the lowest $d$-orbital (i.e., $d_{x^2-y^2}$). That corresponds to the control of the value of the band gap $\delta$ in Eq.~\eqref{eq:3TopSemimetal}.
Band structure calculation has confirmed that the relevant physics is captured by the tight binding model~\cite{2011_Sun_Liu_NatPhys}. 

In the momentum space, the tight binding Hamiltonian becomes, 
\be
H_0 = \sum_\tbf{k} \left( d_\tbf{k} ^\dag, p_{x, \tbf{k}} ^\dag, p_{y, \tbf{k}} ^\dag \right) {\cal H}(\tbf{k}) 
\left( \begin{array}{c}
	d_\tbf{k} \\ 
	p_{x, \tbf{k} } \\
	p_{y, \tbf{k}}  
	\end{array} 
\right) ,
\label{eq:3pdham}  
\ee   
with ${\cal H} ({\bf k}) $ given by 
\begin{widetext} 
$$
	\left( 
	      \begin{array}{ccc} 
	       -2t_{dd} ( \cos k_x + \cos k_y ) + \delta	& 2 it_{pd} \sin k_x 	& 2i t_{pd} \sin k_y  \\ 
	      -2 it_{pd} \sin k_x	& 2 t_\parallel \cos k_x -2 t_\perp \cos k_y & 0 	\\ 
	      -2i t_{pd} \sin k_x 	&0	& 2 t_\parallel \cos k_y -2 t_\perp \cos k_x
	      \end{array}
	    \right).  
$$
\end{widetext} 
Depending on the value of the energy difference $\delta$, there are two types of band structures for this model. For $\delta > 4 t_{dd} + 2 t_\parallel -2 t_\perp$, $d_{x^2 -y^2}$ orbitals are weakly hybridized with $p$-orbitals; for $0<\delta < 4 t_{dd} + 2 t_\parallel -2 t_\perp$, the orbitals are strongly hybridized. For the latter case, a band touching point between the top and middle bands shows up at $(k_x = 0, k_y =0)$ ($\Gamma$ point). This band touching point
has non-trivial topological property, which is characterized by the Berry flux defined as the contour integral of the Berry connection in the momentum space, 
$$ \gamma_n  = \oint _{\cal C} d \tbf{k} \cdot \tbf{A}_n (\tbf{k}), $$  
with $n$ the band index, ${\cal C}$ a close contour enclosing the band-touching point, and the Berry connection $\tbf{A}_n (\tbf{k}) = i \langle u_\tbf{k} | \partial_\tbf{k} |u_\tbf{k} \rangle$, where $|u_\tbf{k} \rangle$ is the eigenstate of the Hamiltonian ${\cal H}(\tbf{k})$. 
The Berry flux $\gamma_n$ is quantized to an integer multiplied by $2\pi$, and only two cases $\gamma_n = 0$ or $\pi$ are distinguishable without any symmetry requirement due to the gauge choice in $|u_n (\tbf{k}) \rangle $. However, with space-inversion symmetry, we can restrict $I |u_n (\tbf{k}) \rangle = |u_n (-\tbf{k} ) \rangle$, with $I$ the space-inversion operator.  The Berry flux then becomes well defined up to $\mod 4 \pi$~\cite{2011_Sun_Liu_NatPhys}. For the band touching point considered here, $\gamma_n$ is $2\pi$, and this band touching is topologically protected (in presence of symmetry). Filling fermions up to such a touching point  gives rise to a topological semimetal. 

A more illuminating way to show the topological protection  is to construct an effective two band Hamiltonian in the vicinity of $\Gamma$ point. Near this point, the $d_{x^2 -y^2}$ orbital band is far below in energy and can thus be eliminated. With standard perturbation theory, the effective Hamiltonian is given to second order as~\cite{2011_Sun_Liu_NatPhys}
$$
{\cal H} _{\rm eff} = 
	\left( \begin{array}{cc} 
		{\cal H} _{22} & {\cal H}_{23} \\ 
		{\cal H}_{32} & {\cal H}_{33} 
		\end{array} 
	\right) 
	- \frac{1}{H_{11} -\mu} 
	\left( \begin{array}{cc}
		{\cal H}_{21} {\cal H}_{12}	& {\cal H}_{21} {\cal H}_{13} \\ 
		{\cal H}_{31}{\cal H}_{12} & {\cal H}_{31} {\cal H}_{13}  
		\end{array} 
	\right) , 
$$
with $\mu$ the chemical potential of the topological semimetal. 
Further expanding momentum around $0$, the effective Hamiltonian takes the following form 
\bea
{\cal H}_{\rm eff} &=& \frac{t_1 + t_2}{2} (k_x ^2 + k_y^2) \mathbb{1} \nn \\
	&+&  2 t_3 k_x k_y \sigma_x 
	+\frac{t_1 - t_2} {2} (k_x^2 -k_y^2) \sigma_z, 
\eea 
where $t_1 = t_\parallel + \frac{4 t_{pd} ^2 }{ 2t_\parallel - 2t_\perp + 4 t_{dd} - \delta }$, 
$t_2 = -t_\perp$, and $t_3 = \frac{2 t_{pd} ^2 }{ 2t_\parallel - 2t_\perp + 4 t_{dd} - \delta }$.  
The absence of $\sigma_y $ component is protected by time-reversal and space-inversion symmetries. 
The energy gap near $\Gamma$ point is $2 |\vec{h}|$, with $\vec{h}$ a planar vector 
$\vec{h} ({\bf k}) = (2t_3 k_x k_y, \frac{t_1 - t_2}{2} (k_x ^2 -k_y^2) )$. The vector $\vec{h}$ forms a vortex configuration with winding number $2$ in the momentum space. At the vortex core 
(the $\Gamma$ point ${\bf k} = (0,0)$), 
it is guaranteed that $\vec{h} =0$, which means the degeneracy (or band touching) point is topologically protected. 

A question naturally arising is whether the required time-reversal and space inversion symmetries can spontaneously break at low temperature. Renormalization group analysis~\cite{2009_Sun_Yao_PRL,2011_Sun_Liu_NatPhys} points to the spontaneous symmetry breaking of time-reversal, and a state with angular momentum order $\langle i p_x ^\dag p_y + h.c. \rangle$ is stabilized at low temperature, 
if the interaction is repulsive. 
Taking this into the effective Hamiltonian, a gap opens at $\Gamma$ point. As a result, the topological semimetal gives way to an insulator state at low temperature. This insulator is topologically non-trivial with finite Chern number. 
If the bare interaction is attractive, the renormalization equation shows that it flows to the fixed point of zero (usually called a marginally irrelevant term). In other words, the topological semimetal phase is stable against any  attractive interaction in the perturbative renormalization group sense.

\subsubsection{Nearly flatbands with nontrivial topology} 
In the system described by Eq.~\eqref{eq:3pdham} at low temperature, the developed angular momentum order generates an additional coupling between two $p$ orbitals, 
$$
\Delta H =  \sum_\tbf{k}  i \Delta p_{x,\tbf{k}} ^\dag p_{y, \tbf{k}} + h.c. , 
$$
which breaks time-reversal symmetry and thus allows the Chern number to be non-trivial. With the parameter choice $\delta = -4 t_{dd} + 2 t_\parallel + \Delta - 2 t_\parallel \Delta /(4 t_\parallel + \Delta ) $ and $t_\perp = t_\parallel \Delta /(4 t_\parallel + \Delta) $, the energies of the top band at $\Gamma $ and $M$ points are equal~\cite{2011_Sun_Gu_PRL}.  Varying $\Delta $ with $t_{dd} = t_{pd} = t_\parallel =t $ fixed, they found that the ratio of the bandwidth/band gap is minimized ($\approx 1/20$) at $\Delta/t = 2.8$ for the top band. The top and bottom bands carry opposite Chern numbers $\pm 1$, and are thus topologically non-trivial while the middle band is topologically trivial. Such nearly flatbands with nontrivial topology mimic the Landau levels of 2D electron gas in strong magnetic fields. The flatness is crucial to reach fractional topological states in lattice models. Further numerical investigations have shown that fractional quantum Hall states are supported when the flatbands are filled at certain fractional filings~\cite{2011_Wang_Gu_PRL,2011_Sheng_Gu_NatComm}.

\subsection{Numerical calculations on lifetime and stability} 
Earlier discussion of the lifetime of $p$-orbital BEC based on Fermi's golden rule calculation largely relies on single-particle picture, and may underestimate many-body effects. Numerical studies based upon Gross-Pitaevskii approach (Eq.~\eqref{eq:3EGP})~\cite{2011_Martikainen_PRA,2013_Xu_Chen_PRA} indeed find interesting phenomena beyond the scope of Fermi's golden rule treatment. To study the TSOC superfluid in continuous space, a variational condensate wavefunction is taken, 
\bea 
\psi_0(\tbf{x}) = \sum_{\tbf{K}} \left[  X_\tbf{K} e^{i  ( \tbf{Q}_x + \tbf{K}) \cdot \tbf{x} } + i  Y _{\tbf{K}}  e^{i  ( \tbf{Q}_y + \tbf{K}) \cdot \tbf{x} } \right], 
\label{eq:3variationansatz} 
\eea 
with $\tbf{K}$  the reciprocal lattice vectors, $X_\tbf{K}$ and $Y_\tbf{K}$ the variational parameters, and $\tbf{Q}_x$ ($\tbf{Q}_y$) the minima for the $p_x$ ($p_y$) band. This wavefunction is superposed of two Bloch functions, and it breaks lattice translation symmetry, as required to describe the TSOC superfluid. 
The key features of TSOC superfluid
which are time-reversal symmetry breaking and staggered orbital current, 
as predicted based on tight binding models, 
are confirmed in the numerical calculations for continuous space~\cite{2013_Xu_Chen_PRA}. 

The stabilities of the TSOC superfluid state are investigated within the time-dependent GP equation, 
\bea 
i \partial_{\tau} \psi (\tbf{x}, {\tau} ) = \left[ -\frac{\vec{\nabla} ^2 }{2M} + V (\tbf{x} ) + g|\psi|^2 \right] \psi( \tbf{x}, {\tau}). 
\eea  
Rewriting $\psi(\tbf{r}, {\tau} )$ into condensate and fluctuation parts, 
$$
\psi(\tbf{x}, {\tau} ) = \psi_0 (\tbf{x} , {\tau}) + u_\tbf{q} (\tbf{x}, {\tau}) e^{i \tbf{q} \cdot \tbf{x}} + v_\tbf{q} ^* (\tbf{x}, {\tau} ) e^{-i \tbf{q} \cdot \tbf{x}} 
$$
the time-dependent GP equation determines the dynamics of fluctuations~\cite{2013_Xu_Chen_PRA}
\bea 
i\partial_{\tau} \left ( 
			\begin{array}{c}
			u_\tbf{q} (\tbf{x}, {\tau} )  \\ 
			v_\tbf{q} (\tbf{x}, {\tau} ) 
			\end{array} 
			\right) 
		= \sigma_z {\cal K} _\tbf{q} 
		\left( 
			\begin{array}{c} 
			u_\tbf{q} (\tbf{x}, {\tau} ) \\ 
			v_\tbf{q} (\tbf{x}, {\tau} ) 
			\end{array} 
			\right), 
\label{eq:3dynamicsuv}
\eea 
with 
\bea 
 {\cal K} _\tbf{q}  &=& \left( 
	\begin{array}{cc} 
	{\cal L} (\tbf{q})   & g \psi_0 ^2 \\ 
	g\psi_0 ^{*2} &	{\cal L}(\tbf{q}) 
	\end{array} 
	\right), \\ 
{\cal L} (\tbf{q})  &=&  -\frac{ \left( \vec{\nabla} + i\tbf{q} \right) ^2}{2M}  + V(\tbf{x} ) + 2 g |\psi_0|^2. \nn 
\eea 
Note that the vector $\tbf{q}$ is the lattice momentum after doubling periods to make $\tbf{q}$ a good quantum number, and that 
$u_\tbf{q}$ and $v_\tbf{q}$ are periodic---$u_\tbf{q} ( \tbf{x} + 2 \tbf{a}_x ) = u_\tbf{q} ( \tbf{x} + 2 \tbf{a}_y ) = u_\tbf{q} (\tbf{x}) $,  
$v_\tbf{q} ( \tbf{x} + 2 \tbf{a}_x) =v_\tbf{q} ( \tbf{x} + 2 \tbf{a}_y) = v_\tbf{q} ( \tbf{x} ) $. 
The eigenvalues of $\sigma_z {\cal K}_\tbf{q}$ determine the Bogoliubov spectra, which are studied for square 
and checkerboard lattices. The fluctuations would grow in time if the eigenvalues are imaginary, leading 
to dynamical instability. 
This instability is cross checked by simulating real time dynamics in the continuous space where the optical lattice is treated exactly by a periodic potential~\cite{2013_Xu_Chen_PRA}, beyond the standard tight-binding model approximation.

For a square lattice, the TSOC superfluid state is found to be dynamically unstable unless the interaction 
strength is extremely weak. In presence of dynamical instability, the lifetime of the TSOC superfluid state 
in a simple square lattice could be tens of milliseconds, rendering that such a state is experimentally 
unreachable for the simple square lattice. 
This conclusion is fully consistent with the early experimental finding of a relatively fast decay of the 
$p$-orbital atoms in a quasi-1D lattice system~\cite{2007_Muller_Folling_PRL}.  
In contrast,  for the checkerboard lattice as used in 
experiments~\cite{2011_Wirth_Olschlager_NatPhys,2011_Olschlager_Wirth_PRL}, when the lattice is not 
too shallow and the interaction is not too strong, the TSOC superfluid state is shown numerically to be dynamically stable. 
This is consistent with the long lifetime as observed in experiments. Similar improvement with superlattices 
is also found in one dimension~\cite{2011_Martikainen_PRA}.  When the interaction is stronger 
than some critical value, the TSOC superfluid is no longer dynamically stable even for the checkerboard 
lattice. 
Based on the dynamical stability, a phase diagram is predicted in Ref.~\cite{2013_Xu_Chen_PRA}, which is consistent with experimental observations.


Another way to understand the dynamical instability is to look at the energy cost for fluctuations 
$u_\tbf{q}$, $v_\tbf{q}$, which takes the following form~\cite{2001_Wu_Niu_PRA}, 
\bea 
\delta E_\tbf{q} = \int d^2 \tbf{x} 
      \left(u_\tbf{q} ^* ({\bf x} ),  v_\tbf{q} ^* ({\bf x}) \right) {\cal K}_\tbf{q} 
    \left( 
      \begin{array}{c} 
       u_\tbf{q} ({\bf x}) \\ 
       v_\tbf{q} ({\bf x} ) 
      \end{array}
   \right) 
\eea 
The fact that the eigenvalues of $\sigma_z {\cal K}_\tbf{q}$ are imaginary implies the matrix  ${\cal K}_\tbf{q}$ 
is not positive definite (although the reverse may not be true), 
which means that the variational ansatz in Eq.~\eqref{eq:3variationansatz}  is 
not a stable saddle point of the GP energy functional.  
This in principle indicates tendency of forming 
some crystalline ordering~\cite{2011_Li_Liu_PRAR}.

The other type of instability is Landau instability for the reason that there are always
Bogoliubov modes causing the free energy to be negative for p-orbital BEC, which means the state is a local
saddle point that can decay into the lowest s-band.   
However this instability is less important than 
the dynamical instability within the lifetime of experiments. The time scale for Landau instability to destroy 
the $p$-orbital BEC is estimated to be $500$ms while it is found to be around $10$ms in numerical simulations 
for dynamical instability. Although the $p$-orbital BEC is not strictly a metastable state due to Landau instability, 
it is fairly stable within the experimentally relevant time-scale. 
In the checkerboard lattice experiment~\cite{2011_Wirth_Olschlager_NatPhys} where each lattice site actually represents an elongated tube in the third direction, the dynamical phenomena are even richer. For example, a collision process with two atoms decaying into the lowest band is allowed as the energy could be released to the kinetic motion in the third direction~\cite{2013_Paul_Tiesinga_PRA}.

The dynamical instability of excited band condensate in a double-well lattice has also been studied in detail, and 
the loop structure in Bogoliubov spectra is found to be correlated with the dynamical instability~\cite{2012_Hui_Barnett_PRA}.

%% file: 4ExpProbes.tex
\section{Experimental probes and novel lattices} 

The theoretical discovery of richness of many-body physics with $p$-orbital
atoms has motivated considerable experimental efforts  in recent years.  So
  far the experiments have been done only for bosonic atoms.  It
  has been demonstrated  in a checkerboard optical lattice that the chiral
$p+ip$ Bose-Einstein condensate  gives rise to nontrivial quantum
interference. In this section, we will review the experimental challenges to
detect the chiral order, the recent proposals in theory and attempts in
experiment, and the current status.

\label{sec:experiment} 

\subsection{Early experimental observations of higher bands in a cubic lattice}
 
Coherent bosonic cold atoms were observed in the higher bands of an optical lattice in the pioneering experiments of accelerating lattices~\cite{2005_Browaeys_Haffner_PRA} and  of cross-band Raman transitions~\cite{2007_Muller_Folling_PRL}.

\begin{figure}[htp]
\includegraphics[angle=0,width=\linewidth]{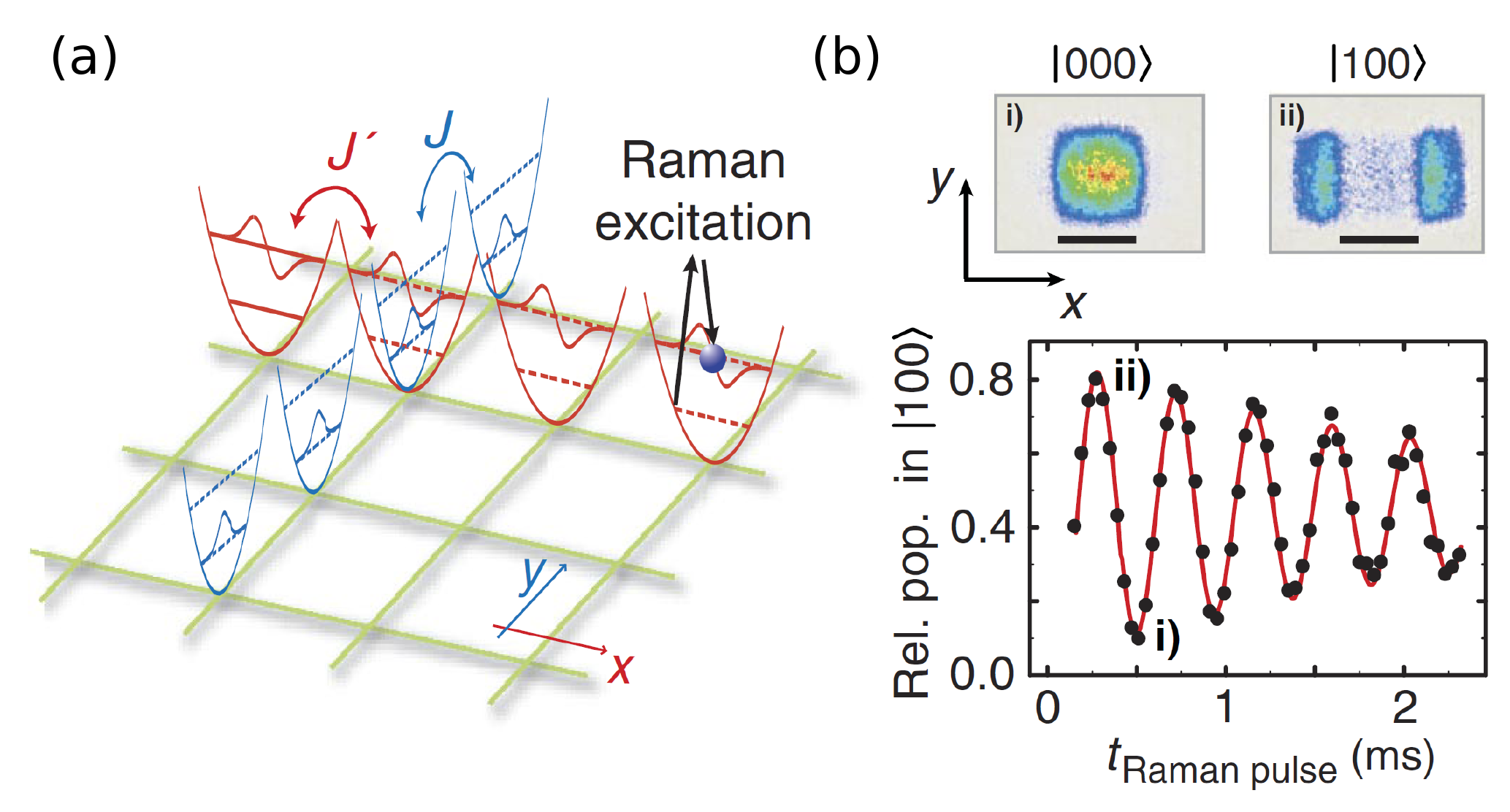}
\caption{Population of higher orbitals with Raman transition~\cite{2007_Muller_Folling_PRL}. 
(a), schematic  of stimulated Raman transitions from $s$- to $p$-wave orbital. (b), the population of the 
lowest (i) and first excited band (ii) measured by time-of-flight techniques. Rabi 
oscillations between 
the $s$- and $p$-wave orbital demonstrate the coherent coupling. }
\label{fig:4BlochExp}
\end{figure}

In the experiment of~\onlinecite{2007_Muller_Folling_PRL}, 
the sample is prepared by first loading a Bose-Einstein condensate of $^{87}$Rb atoms into a deep symmetrically simple cubic 3D optical lattice formed by three 
far detuned laser standing waves. For this deep lattice, it can be treated as an array of 3D harmonic oscillators with discrete vibrational levels, which can be labeled as $|m_x m_y m_z\rangle$ with $m_j$ the vibrational quantum number along the $j$ axis. Population transfer in these orbital levels can be controlled using a stimulated two-photon Raman process with propagating laser beams along the $x$ axis (see Fig.~\ref{fig:4BlochExp}), which provides an inter-orbital coupling 
$$ 
\Omega_{\rm eff} |m_x ' m_y m_z \rangle \langle m_x , m_y m_z | , 
$$ 
with $\Omega_{\rm eff}$ the effective Rabi frequency. 
The experiment restricts the Raman coupling to the lowest Bloch bands and demonstrates orbital transition from the $|000\rangle$ state ($s$-orbital) to $|100\rangle$ ($p_x$-orbital). 
Rabi oscillations between the two orbitals have been observed.
A maximal transfer efficiency of nearly $80\%$ is achieved. 

The decay of atoms into the lowest orbital due to collisional events has also been measured. 
The lifetime was found to be $10-100$ times longer than the tunneling scale. Emergence of coherence 
compatible with a Bose-Einstein condensation to a nonzero momentum state has been 
seen; yet the  experimental system was anisotropic and the predicted $p_x+ip_y$-wave condensate was not studied for  the absence of $p_x$ and $p_y$ orbital symmetry.

\subsection{Observation of high-band condensation in a checkerboard lattice}
\begin{figure}[htp]
\includegraphics[angle=0,width=.8\linewidth]{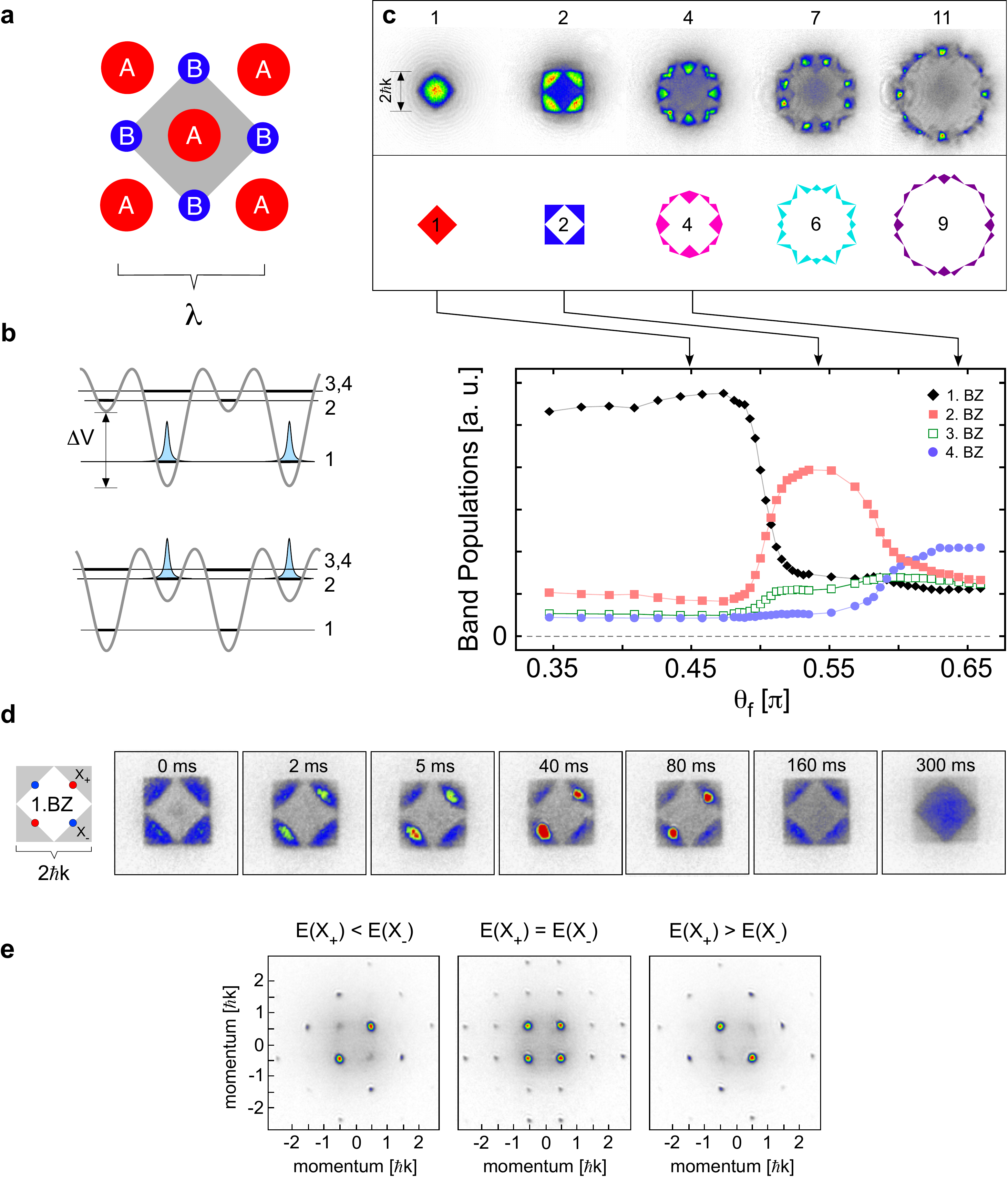}
\caption{
Population of excited bands (figure provided  by A. Hemmerich as courtesy). {\bf a}, the checkerboard lattice with 
two sublattices $A$ and $B$. {\bf b}, the experimental sequence to populate excited bands versus the final value of 
$\theta$ (see Eq.~\eqref{eq:4HemmerichExp}) in step $2$ in {\bf b}. {\bf c}, 
the populations of higher bands with varying $\theta_f$. The upper panel illustrates momentum distributions in different Brillouin zones (top row), and their dependence on $\theta_f$. {\bf d} shows the condensation in the $X_{+}$  point and the band relaxation to the first BZ after long times. In (e) three momentum spectra are shown, with the middle one corresponding to the interesting case of equal populations in $X_{+}$ and $X_{-}$. 
Original results in a different form were published in~\cite{2011_Wirth_Olschlager_NatPhys}. 
}
\label{fig:4HemmerichExp}
\end{figure}

After the early observation of higher band population~\cite{2007_Muller_Folling_PRL,2009_Johnson_Tiesinga_NJP},   
long-lived Bose-Einstein condensate in the high-bands was not achieved 
until the groundbreaking experiment~\cite{2011_Wirth_Olschlager_NatPhys}. In this experiment, a square optical lattice, 
composed of two classes ($A$ and $B$) of (tube-shaped) lattice sites is used (see Fig.~\ref{fig:4HemmerichExp}). 
Formed by two standing waves oriented along the $x$ and $y$ axes with polarization along the $z$ 
axis, the lattice potential is 
\bea 
V(x,y) &=& -\frac{V_0}{4} \left| \eta [ (\hat{z} \cos(\alpha) + \hat{y} \sin (\alpha))e^{ikx} + \epsilon \hat{z} e^{-ikx}    ] \right. \nn \\ 
 && \left. + e^{i\theta} \hat{z} (e^{iky} + \epsilon e^{-iky} )\right| ^2, 
\label{eq:4HemmerichExp} 
\eea  
where $\eta \approx 0.95$  accounts for a small difference in the powers directed 
to interferometer branches, $\epsilon \approx 0.81$ accounts for the imperfect retro-reflections, 
and the angle $\alpha$ permits tunability of anisotropy in the $x$-$y$ plane. 
An isotropic $p$-band with degenerate band minima arises when $\cos(\alpha ) \approx \epsilon$ (or 
$\alpha = \alpha_{\rm iso} \approx \pi/5$). The controllability of the phase difference $\theta$ allows to adjust the 
relative depth of potentials at $A$ and $B$ sites, which is crucial in this experiment to populate higher bands. 
For $\theta < \pi/2$ the $A$ sites are shallower than the $B$ sites 
and vice versa. 

Initially a Bose-Einstein condensate of rubidium ($^{87}$Rb) atoms is prepared and the lattice potential is 
adiabatically turned on with $\theta = 0.38\pi$ such that $B$-sites are much lower than $A$. A lowest band lattice 
Bose-Einstein condensate is thus created with most of atoms confined in $B$ sites. Then $\theta$ is rapidly increased 
to a final value $\theta_f >\pi/2$ such that the $s$-orbitals in the $B$ sites are level with the $p$-band of the lattice 
in energy. In doing so, atoms are efficiently transfered to the $p$-band. Since this preparation procedure is abrupt, 
the prepared state is not immediately a condensate state but rather an incoherent state, in which the atomic distribution 
in the Brillouin zone is fairly uniform. Surprisingly, 
after some holding time around $10$ms, sharp peaks arise at $p$-band minima and the $p$-band Bose-Einstein condensate 
spontaneously emerges. 
In theory the emergence of phase coherence is beyond the scope of Gross-Pitaevskii approach, 
and can be studied by constructing a quantum rotor model~\cite{2012_Sau_Wang_PRA}, 
where the dynamics is well captured by the truncated Wigner approximation~\cite{2002_Polkovnikov_Sachdev_PRA}.  

The $p$-band condensate is not a ground state of the system 
but a metastable state;
 decaying  into the lowest band is unavoidable. 
In this checkerboard lattice, the band gap between $p$-band and the lowest band is 
largely mismatched with the gap between $p$-band and the higher band, and Fermi's golden rule calculation 
(see Sec.~\ref{sec:fermigoldenlifetime}) predicts a significant improvement of stability. In the experiment, the lifetime 
of the $p$-band condensate could reach $100$ms or longer. 

From the measurements of momentum distribution, the experimental evidence of $p$-band condensate is conclusive. However 
there is no direct evidence for the orbital ordering in the TSOC state as predicted in theory. As a step further, a 
phase diagram is mapped out with varying $\alpha$ (controlling the anisotropy) and the phase diagram is quantitatively 
consistent with theoretical predictions~\cite{2013_Olschlager_Kock_NJPHYS}. 
The remarkable consistency of experimental 
observations with theories strongly suggests the $p$-band condensate be a TSOC state. 
Yet, direct evidence of the orbital order 
requires further experimental investigation.

Population of even higher bands, say $f$-bands, is also achieved in this checkerboard 
lattice~\cite{2011_Olschlager_Wirth_PRL} thanks to the tunability of relative depth 
between two sublattices. Similar procedure was implemented as in preparing the $p$-band condensate. 
The resulting $f$-band condensate also has a complex nature. The condensate wavefunction locally 
resembles the superposition $\psi_{[3,0]} \pm i \psi_{[0,3]}$ of eigenfunctions $\psi_{[n,m]}$ of a 2D 
harmonic oscillator with $n$ and $m$ oscillator quanta in $x$ and $y$ directions, which has a spatial 
$(2 x^3 -3x) \pm i (2y^3 -3y)$ dependence locally. The complex $f$-band condensate emerges from the same 
mechanism as the TSOC state of the $p$-band, 
namely maximizing the local angular momentum. 

Besides the way of loading atoms into the excited bands demonstrated in the checkerboard lattice, there are other 
possibilities, for example by Bloch oscillation techniques~\cite{2011_Larson_Martikainen_PRA,2012_Tarruell_Greif_Nature} 
or by vibrating lattices~\cite{2012_Sowinski_PRL,2013_Lachi_Zakrzewski_PRL}.

\subsection{Early experimental realization of double-well lattices} 
Observations of higher bands in optical lattices are achieved in the early experiments 
manipulating double-well lattices, which were largely motivated by implementing 
coherent control of quantum degrees of freedom~\cite{2006_Sebby-Strabley_Anderlini_PRA,2007_Anderlini_Lee_Nature,2008_Trotzky_Cheinet_Science,2008_Lundblad_Lee_PRL,2008_Cheinet_Trotzky_PRL}.

\begin{figure}[htp]
\includegraphics[angle=0,width=\linewidth]{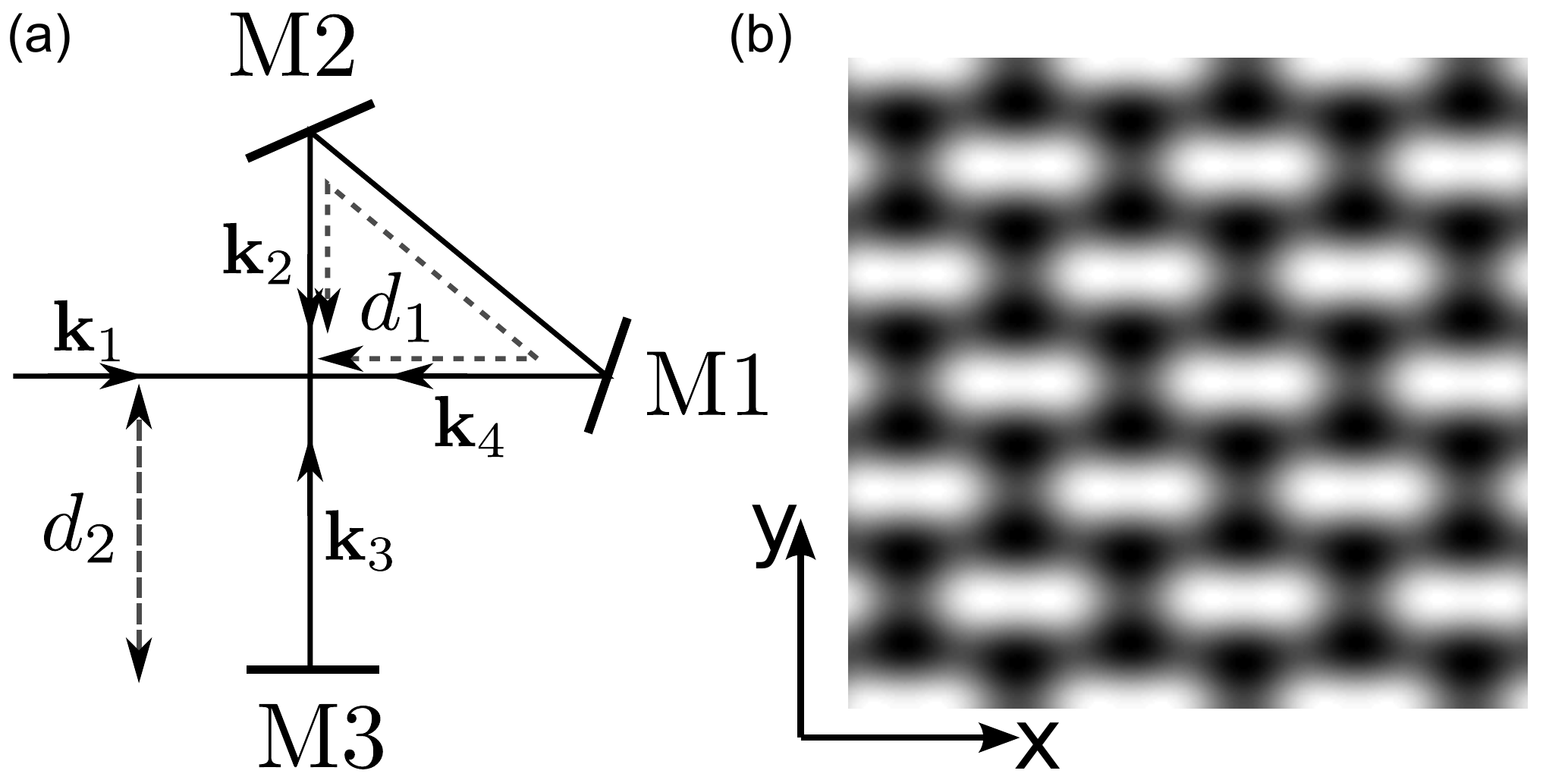}
\caption{ Laser beams to generate a double-well lattice~\cite{2006_Sebby-Strabley_Anderlini_PRA}. (a) shows the laser setup. The 
incoming beam with wave vector ${\bf k}_1$ is reflected by mirrors M1 and M2 and after traveling distance $d_1$ returns to the cloud 
with a wave vector ${\bf k}_2$. The beam is then retro-reflected by M3 and returns with a wave vector ${\bf k}_3$, having traveled with 
an additional distance $2d_2$. (b) shows the generated double-well lattice with $I_{z,0}/I_{xy,0} = 0.4$, $\phi_{xy} - \phi_z = \pi/2$ and 
$\theta_{xy} - \theta_z = -\pi/2$ (see text). The darker (lighter) regions represent areas where the potential is low (high).  }
\label{fig:4Portolasersetup}
\end{figure}

Here we use the experiment~\cite{2006_Sebby-Strabley_Anderlini_PRA} to demonstrate how 
the higher bands are populated in double-well lattices and what consequent observables are achieved. 
This double-well lattice is a two 
dimensional lattice formed by 
superimposing two lattices with orthogonal polarizations. Having a laser setup as shown in 
Fig.~\ref{fig:4Portolasersetup}(a), the electric field generated by the four laser beams is 
${\rm Re} [\vec{E}(x,y)] e^{i\omega t} $, with 
\bea 
\vec{E} (x,y) = E  \left( e^{i k x} + e^{i (2\theta + 2\phi -kx)} \right) \hat{e}_1 \nn \\ 
		+E \left( e^{i(\theta -k y)} + e^{i (\theta + 2\phi + k y)} \right) \hat{e}_2, 
\eea 
where $k = 2\pi/\lambda$ ($\lambda$ is the wavelength of the laser light), 
$\theta = k d_1 + \delta \theta$, and $\phi = k d_2 + \delta \phi$ (the extra phase shifts 
$\delta \theta$ and $\delta \phi$ are polarization dependent and can be controlled in experiments).
We have neglected several imperfections such as imperfect alignment and reflections for simplicity here. 
In experiments these imperfections could cause technical challenges. 
For light polarizations being all in plane such that $\hat{e}_1 = \hat{y}$, $\hat{e}_2 = \hat{x}$, we have 
a laser intensity field 
\bea 
&& I_{xy} (x,y)/I_{xy,0}  \\ 
&&= 2 \cos(2kx - 2\theta_{xy} - 2\phi_{xy} ) + 2 \cos(2ky  + 2\phi_{xy}) + 4 \nn, 
\eea 
with subscripts in $\theta $ and $\phi$ specifying the polarization dependence. 
For the out-of-plane case, $\hat{e}_1 = \hat{e}_2 = \hat{z}$, the laser intensity field is 
\bea 
&& I_z(x,y) /I_{z,0} \\ 
&& = 16 \left[ \cos(\frac{k}{2} (x+y) - \frac{\theta_z}{2} \right] ^2 
		   \left[ \cos(\frac{k}{2} (x-y) - \frac{\theta_z}{2} - \phi_z \right]^2. \nn 
\eea 
The laser field creates an optical potential $V(x,y) \propto (I_{xy} (x,y) + I_z (x,y))$. 
With in-plane and out-plane polarized laser beams combined, 
a double-well lattice can be created (Fig.~\ref{fig:4Portolasersetup}(b)). 

Ground state can be achieved by adiabatically loading atoms into the lattice. 
For the double-well lattice, different from simple Bravais lattices,  the band gap could be 
very small compared with the energy scale $\hbar T_{\rm load}^{-1} $, with $T_{\rm load}$ the 
loading time. 
Then the Landau-Zenner transitions across the lowest and first excited bands can be significant. 
The population of the first excited band causes the oscillations in the  momentum distribution 
measured in time-of-flight, which are observed in experiments. 

The relation between the observed oscillations in the momentum distribution and the population of the excited 
band can be quantified by constructing a two-band model, 
\bea 
H = \sum_{\tbf{r}, \tbf{r}'} \phi_\tbf{r} ^\dag T_{\tbf{r} \tbf{r}'} \phi_{\tbf{r}'}, 
\eea 
with $\phi_\tbf{r} = [\phi_{A, \tbf{r}} , \phi_{B, \tbf{r}} ]^T$ where 
$\phi_A$ and $\phi_B$ are annihilation operators for the localized orbitals,
 $w_A (\tbf{x} - \tbf{r})$ and $w_B (\tbf{x}-\tbf{r})$, 
in the two sub-wells at site ${\bf r}$ in the double-well lattice. 
In momentum space, the Hamiltonian then reads 
$
H = \sum _\tbf{k} \phi^\dag (\tbf{k}) {\cal H }(\tbf{k}) \phi(\tbf{k}), 
$
with $\phi(\tbf{k})$ Fourier transform of $\phi_\tbf{r}$. 
After loading bosonic atoms into the lattice, the condensate is a superposition of the 
ground state and excited state at lattice momentum $\tbf{k}=0$, 
$$
|\psi \rangle = \psi_g |g \rangle + \psi_e |e\rangle.  
$$ 
Writing ${\cal H} (\tbf{0})$ as 
$$
{\cal H} (\tbf{0}) = h_0 \mathbb{1} + h_x \sigma_x + h_y \sigma_y, 
$$ 
the dynamics of the state $|\psi\rangle$ is given as 
$|\psi (t) \rangle  = \psi_g e^{i \Delta t/2} |g\rangle + \psi_e e^{-i\Delta t/2} |e\rangle$, 
with $\Delta = 2 \sqrt{h_x ^2 + h_z ^2}$. In terms of $\phi_{A,B} (\tbf{k})$ basis, we have 
\begin{align*}
|\psi(t) \rangle &= \left\{ 
\left[ \psi_e e^ {-i\Delta t/2} \cos (\gamma /2) - \psi_g e^{i\Delta t/2} \sin(\gamma/2) \right] 
\phi_A ^\dag (0)\right.  \\
&\left. + \left[ \psi_e e ^{-i\Delta t/2} \sin(\gamma/2) + \psi_g e^{i\Delta t/2} \cos(\gamma/2) \right] 
\phi_B ^\dag (0) \right\} |0\rangle,   
\end{align*}
with $\gamma$ the polar angle of the vector $(h_z, h_x)$. The momentum distribution is then given 
as 
$$
 n(\tbf{k}) = const   
 + 2  {\rm Re} \left[ \psi_e ^* \psi_g e^{i\Delta t} 
	  (\tilde{w}_A ^* (\tbf{k}) \cos(\theta/2) + \tilde{w}_B ^* (\tbf{k})  \sin(\theta/2) ) 
	\nn \right. 
	  \left. (\tilde{w}_B (\tbf{k}) \cos(\theta/2) + \tilde{w}_A (\tbf{k})  \sin(\theta/2) ) \right] 
$$
where $\tilde{w}_{A,B} (\tbf{k}) $ is the Fourier transform of 
$\sum_\tbf{r} w_{A,B} (\tbf{x} - \tbf{r})$.  The population 
fraction of the excited band could thus be extracted from  the dynamical evolution 
of momentum distribution.

Although the above discussions were restricted to the setup in the 
experiment~\cite{2006_Sebby-Strabley_Anderlini_PRA}, the coherent oscillation in time-of-flight is a generic phenomenon 
when a superposed state of ground and excited bands is prepared. And indeed similar oscillations are observed in other double-well 
lattices as well~\cite{2007_Anderlini_Lee_Nature,2007_Muller_Folling_PRL,2008_Trotzky_Cheinet_Science}.

\subsection{Theoretical understanding of experiments}
Early theoretical studies of $p$-band condensates focus on the case with the point group $D_4$ symmetry. 
For the lattice potential realized in the experiment of Hamburg (Eq.~\eqref{eq:4HemmerichExp}),
the point group symmetry is 
maintained only for the ideal case $\epsilon = 1$ and $\alpha = 0$, where the potential reduces to 
$V = -V_0 \left(\eta^2 \cos^2 k x + \cos^2 ky +2 \eta \cos \theta \cos k x \cos ky  \right) $. 
For the realistic situation with $\epsilon <1$, the $D_4$ 
symmetry is thus broken and only reflection symmetry with respect 
to the $x$-axis is preserved. The asymmetry could be partially compensated by setting $\alpha = \alpha_{\rm iso}$, 
for which the potential reads $V = -V_0 \epsilon \left[ \eta ^2 \epsilon \cos^2 kx + \cos^2 ky \right] 
- V_0 \epsilon \eta \cos kx \left[\cos(ky + \theta) + \epsilon^2 \cos(ky-\theta) \right] $. 
The consequences of asymmetry are studied in detail in~\cite{2011_Cai_Wu_PRA,2012_Shchesnovich_PRA}. 

The band structure is calculated by plane-wave expansion~\cite{2011_Cai_Wu_PRA}. 
The reciprocal lattice lattice vectors are defined as 
$\tbf{G}_{m,n} = m \tbf{b}_1 + n \tbf{b}_2$, with $\tbf{b}_{1,2} = (\pm \pi/a, \pi/a)$ ($a$ the lattice constant). 
Taking the single-particle Hamiltonian $H_0 = -\hbar^2 \vec{\nabla}^2/(2M) + V(\tbf{x})$, the diagonal matrix elements 
are 
$\langle \tbf{k} + \tbf{G}_{m,n} | H_0 | \tbf{k} + \tbf{G}_{m,n} \rangle 
= E_r \left\{ [ak_x/\pi + (m-n) ]^2 + [ak_y/\pi + (m+n) ]^2 \right\}, 
$ 
with $E_r$ the single-photon recoil energy, and the off-diagonal matrix elements are 
\bea 
&& \langle \tbf{k} | H_0 | \tbf{k} + \tbf{G}_{\pm 1, 0} \rangle =
    -\frac{V_0}{4} \eta \epsilon ( \cos \alpha e^{\mp i \theta} + e^{\pm i\theta} ), \nn \\ 
&& \langle \tbf{k} |H_0 | \tbf{k} + \tbf{G}_{0, \pm 1} \rangle = -\frac{V_0}{4} \eta 
      (\cos(\alpha) e^{\pm i\theta} + \epsilon ^2 e^{\mp i\theta} ), \nn \\ 
&& \langle \tbf{k} | H_0 | \tbf{k} + \tbf{G}_{\pm 1, \mp 1}  \rangle = -\frac{V_0}{4} \epsilon \eta ^2 \cos \alpha, \\
&& \langle \tbf{k} | H_0 | \tbf{k} + \tbf{G}_{\pm 1, \pm 1} \rangle = -\frac{V_0}{4} \epsilon \cos \alpha. \nn 
\eea 
There are four time-reversal invariant points in the Brillouin zone, 
$O = (0,0)$, $X_{\pm} = (\pm \frac{\pi}{2 a} , \frac{\pi}{2a} )$, and $M  = (\frac{\pi}{a}, \frac{\pi}{a})$, at which 
the Bloch functions are real valued. The band spectra are symmetric at these points, and consequently 
$\partial_\tbf{k} \varepsilon (\tbf{k}) =0$, which means that they are saddle points in the band structure. For the choice 
$\alpha = \alpha_{\rm iso}$, the second band has double degenerate minima at $X_+$ and $X_-$. 
 For $\alpha<\alpha_{\rm iso}$ ($\alpha > \alpha_{\rm iso}$) , $X_+$ ($X_-$) becomes the unique band minimum. 

To investigate the interaction effects, the Gross-Pitaevskii equation 
\bea 
\left\{ -\frac{\hbar^2 \vec{\nabla}^2}{2M} + V_{\rm eff} (\tbf{x}) \right\} \Psi (\tbf{x}) = E \Psi(\tbf{x}),
\eea 
with $V_{\rm eff} (\tbf{x}) = V(\tbf{x}) + g\rho |\Psi(\tbf{x})|^2 $, is solved self-consistently by assuming the 
condensate wavefunction is a superposition of Bloch functions at $X_{\pm}$, 
\bea 
\Psi(\tbf{x}) = \cos (\delta) \psi_{X_+} (\tbf{x}) + \sin(\delta) e^{i\phi} \psi_{X_-} (\tbf{x}) . 
\eea 
The Bloch functions $\psi_{X_{\pm}}$ have nodal lines in space, while the variational condensate wavefunction 
could avoid nodal lines by having complex values (with $\delta \neq 0$ or $\pi/2$, and $\phi \neq 0$). 
The complex solution is spatially more uniform and thus more favorable by 
interactions, but at the same time costs more kinetic energy when $\alpha \neq \alpha_{\rm iso}$. 

The competition between interactions and anisotropy leads to an interesting phase diagram containing 
two real and one complex states of Bose-Einstein condensation. 
The Gross-Pitaevskii approach finds second order transitions at zero temperature~\cite{2011_Cai_Wu_PRA}. 
The phase transitions can be understood within a Ginzburg-Landau theory, 
\bea 
&& F = -r_1 |\psi_+ |^2  -r_2 |\psi_-|^2 + g_1 |\psi_+|^4 + g_2 |\psi_-|^4 \nn \\
&&  + g_3 |\psi_+ | ^2 |\psi_-|^2 
  + g_4 ( \psi_+ ^{*2} \psi_- ^2 + c.c.), 
\eea 
with $\psi_{\pm}$ describes the condensate component at $X_{\pm} $. 
The Umklapp term $g_4>0$ favors the complex state. 
Assuming $r_1, r_2$, and $g_3 -2 g_4$ $>0$, the complex state occurs in the regime 
\be
\frac{g_3 -2 g_4}{2 g_2} < \frac{r_1}{r_2} < \frac{2 g_1 }{g_3 - 2 g_4}. 
\ee
The predicted phase diagram is confirmed in the experiment~\cite{2013_Olschlager_Kock_NJPHYS}. 


\subsection{Measurement of orbital orders by quench dynamics}
\label{sec:measureorbital}

Direct measurement of orbital ordering, namely the staggered angular momentum, was thought to be 
an experimental challenge, which motivates a theoretical proposal of using quench dynamics~\cite{2013_Li_Paramekanti_NatComm}. The key idea could be understood by drawing an analogy 
between the two orbital states at each site $(p_x, p_y)$, and a pseudospin-$1/2$ degrees of freedom 
($\uparrow$, $\downarrow$). In this analogy, the $p_x\pm i p_y$ state corresponds to a pseudospin pointing 
along the $y$ direction in spin space. Applying a `magnetic field' along the $x$ direction to this pseudospin 
should then induce Larmor precession, leading to periodic oscillations of the $z$-magnetization, corresponding 
to the population imbalance between two $p$-orbitals, $\Delta N = N(p_x) - N (p_y)$. Here we consider a square lattice. We can take a certain initial state 
and then quickly turn  on a strong `magnetic field'
\be
H_{\rm mag} = \sum_\tbf{r} (-1)^{r_x + r_y} \lambda (\tbf{r}) 
    \left[ b_x ^\dag (\tbf{r}) b_y (\tbf{r}) + h.c. \right]
\ee 
at time ${\tau}  =0$. For simplicity, the `magnetic field' is assumed to be strong enough to 
completely dominate  the short-time dynamics. If initially a staggered superposition 
$p_x \pm e^{i\theta} p_y$ is prepared, all local Larmor precessions add up to produce a 
macroscopic oscillation in the orbital imbalance $\Delta N$. This imbalance 
evolves within a Heisenberg picture as 
\bea 
\frac{d \Delta N (\tbf{r}, {\tau} ) } {d{\tau} } 
&=& -i [\Delta N(\tbf{r}, {\tau} ), H_{\rm mag} ] \nn \\ 
&=& -2 \lambda (\tbf{r}) { L}_z ^{\rm stag} (\tbf{r}, {\tau} ), 
\eea 
with $L_z ^{\rm stag}$ the staggered angular momentum operator, whose time evolution is described by 
\bea 
\frac{d L_z ^{\rm stag} } {d{\tau} }  = 2 \lambda (\tbf{r}) \Delta N (\tbf{r}, {\tau} ) . 
\eea
This leads to oscillations in $\langle \Delta N (\tbf{r}, {\tau} ) \rangle$,  
\bea 
&& \langle \Delta N (\tbf{r}, {\tau} ) \rangle \nn \\
&=& \langle \Delta N (\tbf{r}, 0) \rangle  \cos (2 \lambda(\tbf{r}) {\tau} ) 
- \langle L_z ^{\rm stag} (\tbf{r}, 0) \rangle \sin ( 2 \lambda (\tbf{r}) {\tau} ) \nn \\
&\equiv& A (\tbf{r}) \cos (2 \lambda(\tbf{r}) {\tau}  + \phi(\tbf{r}) ), 
\eea 
where $\langle \Delta N (\tbf{r}, 0) \rangle$ and $\langle L_z ^{\rm stag} (\tbf{r}, 0) \rangle$ 
denote the  orbital imbalance and staggered angular momentum for the initial state. 
The trigonometric form of this time-dependent equation thus defines the quantities 
$A({\bf r} )$ and $\phi({\bf r})$, ready to compare with the experimental measurement of $\Delta N$.

Neglecting spatial inhomogeneity in $\lambda (\tbf{r})$ and $\phi (\tbf{r})$, we can set 
$\lambda (\tbf{r}) = \lambda$ and $\phi(\tbf{r}) = \phi$, and extract 
the initial angular momentum order from the amplitude $A$ and the phase shift $\phi$ in the dynamics 
of the spatially averaged orbital imbalance 
$\overline{ \langle \Delta N ({\tau} ) \rangle } = 1/N_s \sum_\tbf{r} \Delta N (\tbf{r}, {\tau} )$. The 
coefficient $\lambda$ can be read off from the oscillation period ${\tau} _{Q} \equiv \pi / \lambda$. 
The orbital population imbalance can be measured directly in time-of-flight experiments. 

For a $C_4$ symmetric initial state with non-zero staggered angular momentum, 
but no orbital imbalance, 
$\overline{ \langle \Delta N ({\tau} ) \rangle} $ is expected to oscillate with a non-zero amplitude and phase 
shift $\phi = \pm \pi/2$ whose sign will fluctuate from realization to realization. By contrast 
$\overline{ \langle \Delta N ({\tau} ) \rangle} = 0$ should be observed for a completely disordered state. The amplitude of the signal 
is thus a direct measure of the staggered angular momentum order parameter. 

In the case that $C_4$ symmetry is explicitly broken as achieved in the recent experiment, a state with an initial 
orbital imbalance but no angular momentum order would exhibit oscillations with a finite amplitude but no phase shift, 
i.e., $\phi =0$. In contrast, for a state with angular momentum order, the spontaneous time-reversal symmetry breaking yields 
a finite phase shift $\phi \neq 0$, which would vanish in a singular fashion as we tune from the angular momentum ordered 
to disordered regime through a second order phase transition. 

The required coupling $H_{\rm mag}$ can be engineered by adding a quench potential $V_{\rm mag}(\tbf{x})$ 
modulated in the $(1,1)$ direction with respect to the original lattice potential. The add-on potential generates a coupling 
between $p_x$ and $p_y$ orbitals
\bea 
\epsilon (\tbf{r})  \approx \frac{\hbar}{4 m \omega_0} 
    \frac{\partial^2 V_{\rm mag} 
      \left(\tbf{r} + l \left[ \frac{\tbf{a}_x + \tbf{a}_y}{\sqrt{2}} \right] \right) }{a^2 \partial  l^2 }|_{l\to 0} , 
\eea 
where $\omega_0$ is the harmonic oscillator frequency of the lattice wells hosting the $p$-orbitals and 
$a = |\tbf{a}_x | = |\tbf{a}_y| $ is the lattice constant.
The above estimate for the coupling strength is valid in the tight binding regime when the quench potential 
is weak as compared with the original optical lattice. 
Without loss of generality, one may consider an 
add-on optical potential of the form 
\bea 
V_{\rm mag} (\tbf{x}) = -\Gamma \cos ^2 \left( \frac{2 \nu+1}{4} (\tbf{K}_x + \tbf{K}_y) \cdot \tbf{x} \right), 
\eea 
with some integer $\nu \ge 0$, a positive amplitude $\Gamma$, and $({\bf K}_x$, ${\bf K}_y)$ denoting the primitive vectors 
of the reciprocal lattice. 
This potential leads to a $p_x$/$p_y$ coupling 
\bea 
\epsilon (\tbf{r}) = \frac{E_{\rm r}}{\hbar \omega_0} \frac{\Gamma}{4} 
    (2 \nu + 1)^2 (-1)^{r_x + r_y}, 
\eea 
with $E_r$ the photon recoil energy with wave number $1/2|{\bf K}_x + {\bf K}_y |$. 
The staggering factor in the engineered coupling is crucial to probe the staggered angular momentum order. 

This quench proposal brings other interesting possibilities in addition to providing a method to probe orbital order. For instance, one can simulate spin dynamics in solid state materials by studying 
orbital dynamics of $p$-band bosons. 
One advantage about orbital dynamics is that engineering artificial 
effective magnetic fields is intrinsically easier due to the the spatial nature of orbital degrees of freedom than engineering real staggered magnetic fields.

\subsection{Measurement of the complex phase by Raman transitions}

There is another proposed scheme to measure the inter-orbital phase coherence in $p_x \pm ip_y$ superfluid by 
Raman transition~\cite{2012_Cai_Duan_PRA}. In the $p_x \pm ip_y$ superfluid, condensation takes place at 
$X_+$ and $X_-$ and the condensate state is 
$ |\Psi \rangle \propto \left( b_{X_+} ^\dag + e^{i\theta} b_{X_-} ^\dag \right) ^ {N} |0\rangle$ in general. 
The idea is to transform the phase coherence to number difference in momentum space. With a Raman operation, bosons 
in the original condensate can be transfered to a state with 
\bea 
&& b_{X_+} ' = \frac{1}{\sqrt{2}} ( b_{X_+} - i e^{i\phi}  b_{X_-}  ), \nn \\
&& b_{X_-} ' = \frac{1}{\sqrt{2}} ( b_{X_-} - i e^{-i\phi} b_{X_+} ). 
\label{eq:4Ramantransfer} 
\eea 
With $\phi = 0$, 
the phase coherence in the $p_x \pm ip_y$ state is then transformed as  
\bea 
\langle i b_{X_+} ^\dag b_{X_-} + h.c. \rangle 
 = \langle b_{X_-} ^{'\dag} b_{X_-}' - b_{X_+} ^{'\dag} b_{X_+} '  \rangle 
 \equiv \delta n', 
\eea 
which can be extracted in time-of-flight experiment. 

\begin{figure}[htp]
\includegraphics[angle=0,width=.5\linewidth]{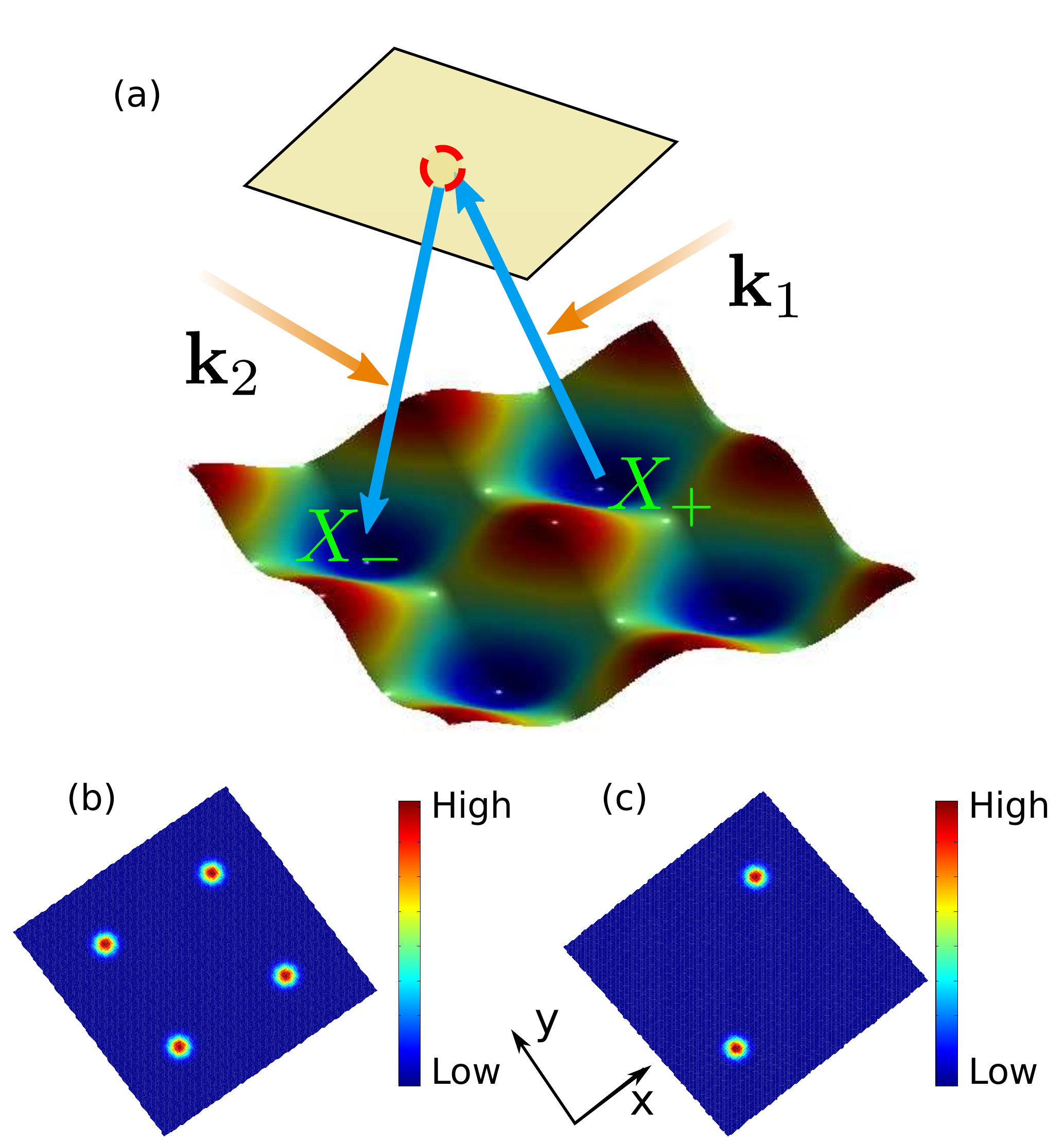}
\caption{ Illustration of proposed Raman scheme to detect the complex orbital order in $p_x \pm ip_y$ superfluid~\cite{2012_Cai_Duan_PRA}.
(a) shows the Raman pulses with different propagating directions to build up  momentum transfer between bosons at $X_+$ and $X_-$. 
(b) and (c) show the time-of-flight imaging after Raman transition for the complex coherent $p_x \pm ip_y$ state and incoherent 
mixing of $p_x$ and $p_y$ condensates, respectively. } 
\label{fig:4RamanProbe}
\end{figure}

The required Raman transition can be implemented by two traveling-wave laser beams along different directions with 
corresponding wave vector $\tbf{k}_{1,2}$ and frequency $\omega_{1,2}$~\cite{2006_Duan_PRL}. 
These laser beams induce an effective Raman Rabi frequency with 
a spatially varying phase $\Omega (\tbf{x}, t) = \Omega_0 e^{ i ( \delta \tbf{k} \cdot \tbf{x} -\delta \omega t + \phi) }$, where 
$\delta \tbf{k} = \tbf{k}_1 - \tbf{k}_2$, $\delta \omega = \omega_1 - \omega_2$, and $\phi$ is the relative phase between the two laser 
beams (see Fig.~\ref{fig:4RamanProbe}(a)).  The effective Hamiltonian for the Raman process is described by 
\bea 
H_R = \int d \tbf{x} \Omega (\tbf{x}, t) \phi ^\dag (\tbf{x})  \phi (\tbf{x}) + h.c.,  
\eea 
where $\phi(\tbf{x})$ is the boson annihilation operator in continuous space. 
The generated spatially dependent potential couples the two condensate components at the two momentum points 
[in the Hamburg experiment~\cite{2011_Wirth_Olschlager_NatPhys} 
${X}_{\pm} = (\pm \pi/2, \pi/2) $, requiring $\delta \tbf{k} = X_+ -X_- = (\pi, 0)$]. 

To avoid complications of  interband transitions (with band gap $\Delta$) 
and dynamics caused by tunnelings ($t$), an optimal choice for the Raman coupling 
strength is 
$ t \ll  \hbar \Omega_0 \ll \Delta$. For the experimental situation, 
the Raman coupling strength should be chosen to be $\Omega_0 \approx 2 \pi \times 0.5$kHz.  Thus 
the required duration of the Raman pulse is around $1$ms. 
To get efficient Raman operation, the frequency $\delta \omega$ should match the energy difference between the initial and final states which 
is around a few Hz. Therefore the phase accumulation $\delta \omega t$ within the duration of Raman pulse is negligible. With this approximation 
the Raman coupling is simplified to be 
\bea 
H_R \approx \sum_{\tbf{k}} e^{i\phi} \lambda (\tbf{k}) b_{\tbf{k} + \delta \tbf{k}}^\dag b_{\tbf{k}} + h.c. 
\eea 
Here  $\lambda (\tbf{k})$ is the $\tbf{k}$ dependent effective coupling, which can be calculated from the Bloch functions. 
For the Hamburgh experiment, 
it is estimated that $\lambda(X_\pm ) \approx 0.98 \Omega_0 \equiv \lambda$. 
Choosing the duration of the Raman pulse to be $\lambda \delta t = \pi/4$, 
the required state transfer in Eq.~\eqref{eq:4Ramantransfer} is achieved. 
The resultant density difference is 
\bea 
\delta n' = \langle i e^{i\phi} b_{X_+ } ^\dag b_{X_-} + h.c. \rangle.
\eea 
For the $p_x \pm ip_y$ superfluid, the density difference would be $\delta n' \propto \cos (\phi)$. 
With $\phi = 0$, $\delta n' = \langle i  b_{X_+ } ^\dag b_{X_-} + h.c. \rangle$ represents the 
order parameter of the complex orbital ordering (Fig.~\ref{fig:4RamanProbe}).

\subsection{Interference measurement of the complex phase} 

\begin{figure}[htp]
\includegraphics[angle=0,width=.8\linewidth]{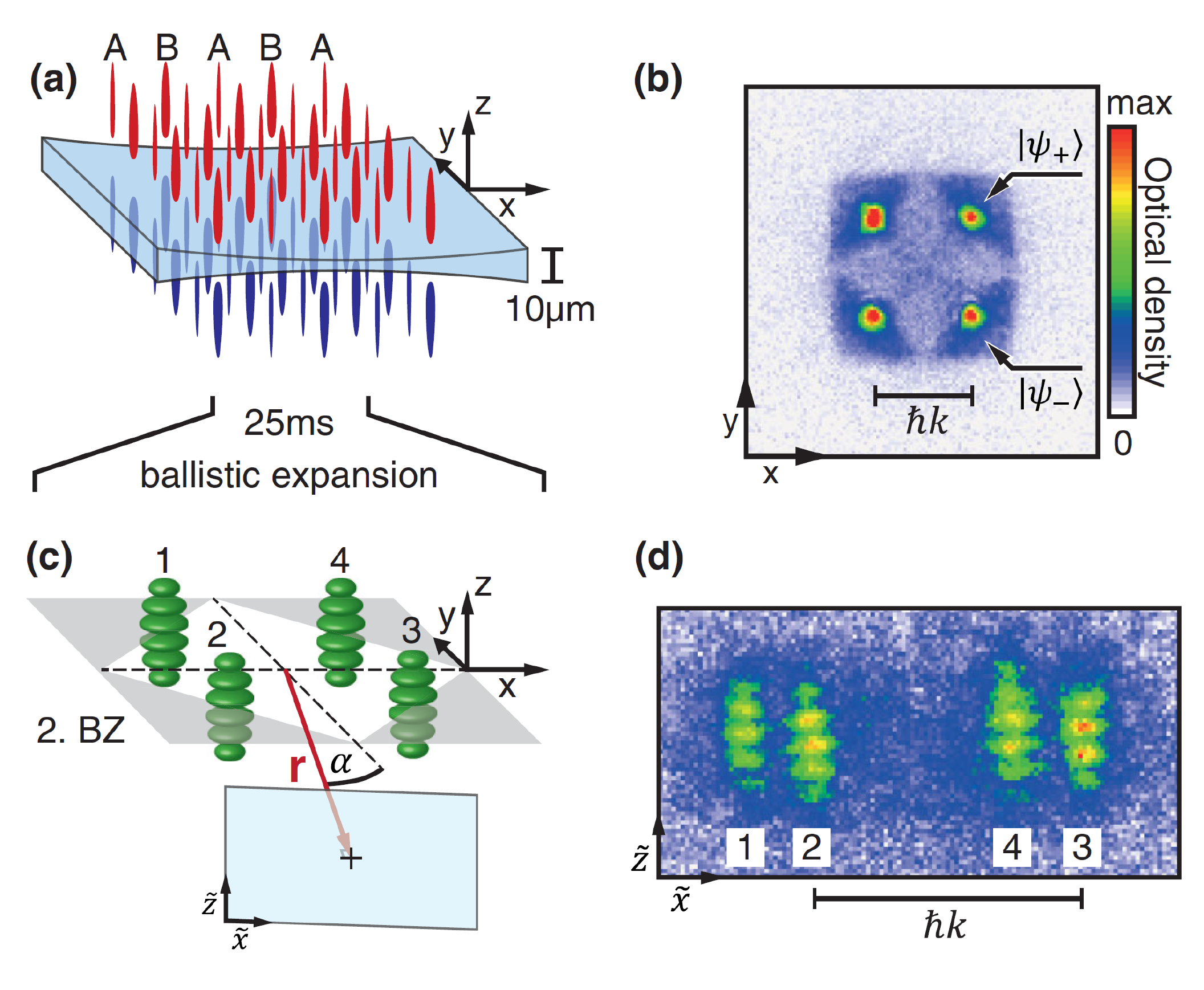}
\caption{
Interference measurement of inter-orbital coherence in the $p_x + i p_y$ superfluid~\cite{2015_Kock_Olschlager_PRL}. (a) shows the experimental protocol to prepare two copies of lattice condensates (red and blue). (b) shows the momentum distribution for the $p_x + ip_y$ superfluid. (c) shows the atomic spatial distribution after ballistic expansion of the two condensates. The four Bragg peaks are labeled by $1$-$4$. (d)  shows the experimental observation of the interference pattern of the four Bragg peaks. The interference structure is along the $z$ direction.
} 
\label{fig:4HemmerichInterference}
\end{figure}

In a recent experiment~\cite{2015_Kock_Olschlager_PRL}, that generalizes the idea of Young's double slits, an interference measurement has been implemented to detect the inter-orbital phase coherence in the $p_x +ip_y$ superfluid. In this experiment, two independent copies of the lattice condensates 
are prepared with the experimental setup as illustrated in 
Fig.~\ref{fig:4HemmerichInterference}. The condensates are simultaneously prepared in the second band in two spatially 
separated regions of the lattice. After the state preparation, all potentials are switched off. The zeroth-order Bragg 
peaks observed in the $xy$-plane carry interference patterns in the $z$ direction due to overlapping 
contributions from 
the condensates originally separate in space. 
In the simplified picture approximating the two condensates by two  point sources, the wave length of the density grating in the interference is $\lambda_z = \frac{2\pi \hbar t_{\rm TOF}}{m d_z}$, with $t_{\rm TOF}$ the time of ballistic expansion, $d_z$ the spatial separation of the two condensates. This estimate is quantitatively consistent with experimental results. 

In the ballistic expansion, the Bragg peaks (labeled by $1$, $2$, $3$ and $4$ in Fig.~\ref{fig:4HemmerichInterference}) yield the Fourier components of the condensate wavefunction, and we can associate a phase for each component, $\theta_{j=1,2,3,4}$. Since the spatially separate condensates are decoupled, they carry different phases, $\theta_j$ and $\theta_j'$. From the relative phase $\Delta \theta_j = \theta_j - \theta_j'$, we can introduce $\Delta \theta_{i,j} = \Delta \theta_i - \Delta \theta_j$, which directly determines the correlation among the interference patterns in the Bragg peaks. If $\Delta \theta_{i,j} = 0$ ($\pi$), the density patterns of the $i$th and $j$th peak are positively (negatively) correlated. The interference patterns obtained in experiments yield that $\Delta \theta_{1,3} = \Delta \theta_{2,4} = 0$, 
over $420$ independent realizations, 
and that $\Delta \theta_{1,2} = \Delta \theta_{1,4} = \Delta \theta_{2,3} = \Delta \theta_{3,4}$, and their value spontaneously chooses $0$ or $\pi$. The interference measurement unambiguously tell that the phase of different momentum components is indeed correlated. 
To the best of our knowledge, the experiment~\cite{2015_Kock_Olschlager_PRL} appears to be the first phase sensitive measurement which poses an important constraint on the nature of $p$-orbital Bose Einstein condensates. It is desirable that future experiments can directly probe the phase lock between the condensate components at two band minima, corresponding to the $X_+$ and $X_-$ points in the paper~\cite{2015_Kock_Olschlager_PRL}.

%% file: 5Discussion.tex
\section{Discussion and outlook}

\subsection{Orbital physics in electronic materials} 
The crystal structure of the atomic ions in solids provide confining potential
for electrons due to strong Coulomb force.  Electrons in solids are usually
nearly localized on atomic ions and the resulting orbital wavefunctions (or the
shape of the electron cloud) are determined by the strong confining
potential. This orbital degree of freedom is of great importance in correlated
materials such as transition metal
oxides~\cite{2000_Tokura_Nagaosa_Science}. Many intriguing phenomena such as
metal-insulator transitions and colossal magnetoresistance can be attributed (or
partially attributed) to the interplay of $d$-orbitals with charge and spin
degrees of freedom.

\begin{figure}[htp]
\includegraphics[angle=0,width=.6\linewidth]{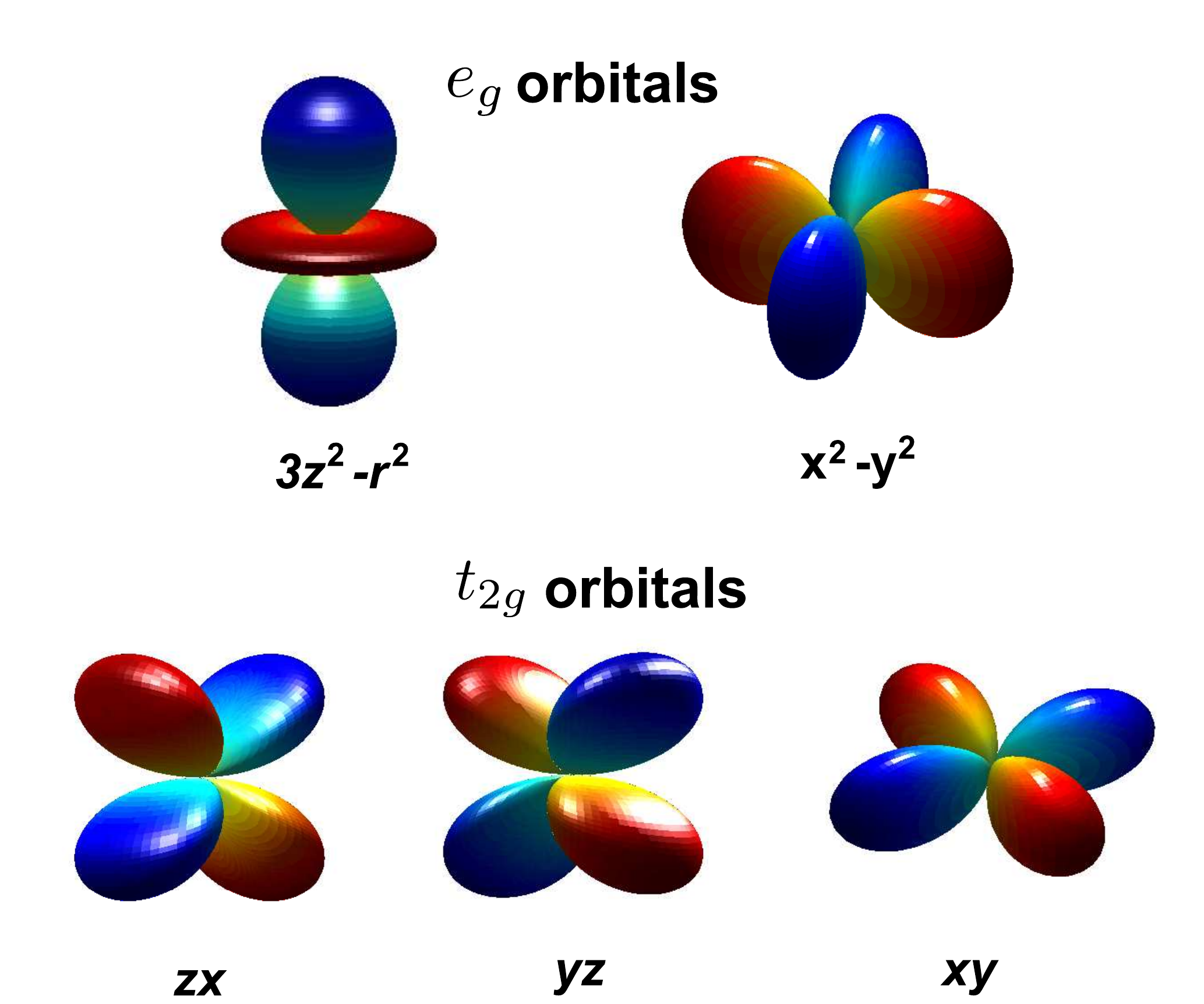}
\caption{ Five $d$-orbitals. In the presence of crystal field, the orbital degeneracy splits into  two groups, $e_g$ and $t_{2g}$.}
\label{fig:5dorbitals}
\end{figure}

Considering a transition-metal oxide material with perovskite crystal structure, $d$-orbital electrons localized on 
the transition-metal atom are surrounded by six oxygen ions $O^{2-}$, which give rise to crystal field and 
consequent energy splitting of the $d$-orbitals. Orbital wavefunctions pointing towards the negative-charged oxygen ions 
(the $e_g$ orbitals, $d_{x^2-y^2}$ and $d_{3z^2 -r^2}$) have higher energy compared with those pointing in other orientations 
(the $t_{2d}$ orbitals, $d_{xy}$, $d_{yz}$ and $d_{xz}$) due to Coulomb repulsion (see Fig.~\ref{fig:5dorbitals}). 
The spatial nature of orbital makes it intrinsically attached to the crystal fields, even in the absence of the relativistic 
spin-orbital interaction, and this intrinsic coupling of orbital degree of freedom to crystal fields and the resultant crystal 
symmetry make it distinct from real spins.
When orbitals are modeled as pseudo-spins,  the model Hamiltonian is in general lack of $SU(2)$ symmetry. Consider a 
typical Mott insulator LaMnO$_3$ as an example. A neutral Mn atom has an electron configuration $3d^54s^2$. Losing 
three electrons, Mn$^{3+}$ in this material has four electrons in those five $d$-orbitals. From Hund's rule, the spins are 
aligned ferromagnetically, and there are thus two possibilities for $e_g$ orbitals with either $d_{x^2-y^2}$ or $d_{3z^2 - r^2}$ being 
occupied. This represents the orbital degree of freedom in this Mott insulator, which can be modeled as pseudo-spins $T_{x,y,z}$. 
The model Hamiltonian is 
$$
H = \sum_{\tbf{r} \tbf{r}'} J_{\tbf{r}  \tbf{r}'} ^{\alpha \beta} T_\alpha (\tbf{r}) T_\beta (\tbf{r}'), 
$$
which is typically not $SU(2)$ symmetric. 
With a long range orbital order, spin magnetism would be strongly affected by so called Jahn-Teller effect~\cite{1937_Jahn_Teller_RSLPSA}. 

Most $p$-orbital solid state materials, for example the semiconducting silicon and graphene, are actually weakly correlated. 
However, recent studies in one oxide heterostructure LAO/STO have found that correlated physics such as ferromagnetism emerges from 
the effective 
$p$-orbitals, where $p_x$ and $p_y$ are mimicked by $d_{xz}$ and $d_{yz}$ orbitals 
(the degeneracy with $d_{xy}$ orbital is broken due to lack of out-of-plane inversion symmetry at the interface).  
In $d$-orbital systems, correlated physics usually emerges due to large Hubbard $U$ interaction because of the tight 
confinement of these orbitals.  The emergence of correlated physics in $p$-orbital systems on the other hand 
could be attributed to a different origin, which is the quasi-one 
dimensionality~\cite{2013_Chen_Balents_PRL,2013_Li_Lieb_PRL}. In one dimension at low filling, 
the magnetic susceptibility diverges as $\chi_{1d} \sim 1/\rho^2$, where $\rho$ is the occupation number per site. Even 
for infinitesimal interaction $U$, there is a strong interaction effect: the ratio of the interacting to free fermion 
susceptibility diverges, $\chi_{1d}/\chi_{ff} \to \infty$ for $\rho \to 0$. 
A general result for the free energy (per site) versus magnetization at low density is obtained to be 
\bea 
F = 2 \rho J_{\rm eff} F_1 \left[ \frac{M}{\rho} , \frac{k_B T}{J_{\rm eff}} \right] 
  - J_H M^2, 
\eea 
where $M$ is the magnetization (per site), $J_H$ is the Hund's rule coupling, $J_{\rm eff}$ is the effective antiferromagnetic coupling, 
and $F_1 [m,t]$ is the free energy per site of the one-dimensional antiferromagnetic chain, 
with reduced magnetization $m$ and temperature $t$ 
(this is known from thermodynamic Bethe ansatz).  
The effective coupling $J_{\rm eff}$ is reasonably conjectured to 
scale as $J_{\rm eff} \propto \rho^3$~\cite{2013_Chen_Balents_PRL}. 
From the free energy, the Hund's energy is dominant and favors a ferromagnetic state with sufficiently low density for arbitrarily 
weak Hund's coupling $J_H$. A rigorous work~\cite{2013_Li_Lieb_PRL}  studies the higher filling regime (but assumes
no double occupancy), where a ferromagnetic ground state for $p$-orbital fermions is proved based on transitivity 
and non-positivity of the many-body Hamiltonian.  
Further studies are required to find out the boundary of 
the ferromagnetism in $p$-orbital fermions. 

\subsection{Synthetic orbital matter and material design}
In material science, design of materials for applications is an important
subject. Recent developments involve engineering heterostructures with hybrid
materials. For example oxide heterostructures such as LaAlO$_3$/SrTiO$_3$ and
GdTiO$_3$/SrTiO$_3$ have been created and extensively studied. While the
properties of many materials can be calculated within the density functional
theory (DFT), this approach fails for ones of strong correlation for which
$d$-orbital electrons typically play an important role.  At the same time, these
strongly correlated materials could have fascinating properties including
important applications. High $T_c$ superconductivity belongs to this
class. Lots of efforts have been made in searching for materials with higher
$T_c$, but there is no real improvement in the last two decades.  Lack of
reliable tools in predicting $T_c$ leaves the design of high $T_c$
superconductivity essentially to empirical trials, which are costly in 
both time and materials. Developing new tools to simulate
strongly correlated materials by incorporating correlation effects in DFT has
triggered tremendous interest but appears to be very challenging.

To address the challenge of simulating correlated $d$-orbital electrons 
in classical computers, one alternative way is to create synthetic orbital matter with optical lattices and take it as a quantum orbital simulator. 
With this optical-lattice-based quantum orbital simulator, the ultimate procedure for material design would 
be---(1) conceive a particular design of materials; (2) determine the orbital configuration of the imagined material by 
quantum chemistry; and (3) apply cold atoms in optical lattices to simulate the properties. In such a way, we could explore the imagined quantum materials for desired properties,  bypassing the often tedious chemical process of really fabricating them from electronic compounds. This would significantly speed up the material design 
and should help improve key quantities of great interest, for instance, the value of critical temperature $T_c$ of superconductivity in future.   
Although the optical lattice experiment is still 
at a very early stage, with future developments, synthetic orbital matter in optical lattices could be extremely 
helpful to the design of real materials.

Finally, we  would like to  point out that orbital degrees of freedom are found to play an important  role for    a vast majority of intriguing electronic quantum materials that condensed matter physicists have found since 1970s.   Magnetic materials of spin only are an important class of systems that have been studied with great progress and remain to pose new challenges, such as frustrated magnets possibly showing spin liquid phases.  In fact, the spin-only systems represent a small fraction of the world of real materials.  Furthermore, past theoretical studies predicted exotic phenomena for model systems that have no spin but only orbital degrees of freedom. Such hypothetical models, which previously might have seemed too special and excessive, now become readily realizable with optical lattices.   On this regard, using higher orbital bands of the optical lattice appears to open up a new front to explore orbital physics, both for understanding the electronic systems and for exploring artificial quantum orbital-only models that have no prior analogue in solids.

\subsection{Many-body dynamics of high orbital atoms}
Coherent dynamics across different bands has been observed in many 
experiments~\cite{2003_Lasinio_Morsch_PRL,2006_Sebby-Strabley_Anderlini_PRA,2007_Anderlini_Lee_Nature,
2008_Trotzky_Cheinet_Science,2008_Cheinet_Trotzky_PRL,2013_Zhai_Yue_PRA,2015_Hu_Niu_arXiv}. In particular 
the recent experiments~\cite{2013_Zhai_Yue_PRA,2015_Hu_Niu_arXiv} have demonstrated fast coherent controllability of orbital degrees of freedom. 
These experimental developments open up possibilities of studying many-body dynamics of high orbital, where 
the observed Rabi-like oscillations between different bands can be affected by interaction. One particular 
example would be orbital Josephson effect, 
which has been studied for double-well potentials~\cite{2011_Garcia-March_Dounas-Frazer_PRA,2012_Garcia-March_Dounas-Frazer_FrontPhys,2014_Gillet_Garcia-March_PRA,2015_Garcia-March_Carr_PRA}. This effect has also been seen in numerical simulations of a dynamical procedure, proposed to detect the $p+ip$ BEC~\cite{2012_Cai_Duan_PRA,2013_Li_Paramekanti_NatComm}.

The orbital Josephson effect is expected to be generic for various experimental setups for high orbital atoms. Here we consider  
the specific setup proposed to probe the complex order (see Sec.~\ref{sec:measureorbital}). 
Assuming all atoms condense, the dynamics 
is then approximately captured by a two-mode Hamiltonian, 
\bea 
H &=& \lambda b_{{\bf K}_1} ^\dag b_{{\bf K}_2} + h.c. \nn \\
&+& g_1 \left( b_{{\bf K}_1} ^\dag b_{{\bf K}_1} ^\dag b_{{\bf K}_1}  b_{{\bf K}_1} + {\bf K}_1 \to {\bf K}_2 \right) 
+ g_2 b_{{\bf K}_1} ^\dag b_{{\bf K}_1} b_{{\bf K}_2} ^\dag b_{{\bf K}_2} \nn \\
&+& g_3 \left( b_{{\bf K}_1} ^\dag b_{{\bf K}_1}^\dag b_{{\bf K}_2} b_{{\bf K}_2} + h.c. \right), 
\eea  
where $b_{{\bf K}_{1,2}}$ are the two condensed modes and the last term $g_3$ is a Umklapp process. 
Following the treatment of Josephson effect developed for double-well Bose-Einstein condensates~\cite{1997_Smerzi_Fantoni_PRL,1998_Zapata_Sols_PRA}, 
the dynamical 
state could be approximated by 
\bea 
|\Psi (t) \rangle = \frac{1}{\sqrt{N!}} \left( \psi_1(t) b_{{\bf K}_1} ^\dag + \psi_2 (t) b_{{\bf K}_2} ^\dag  \right) ^N |0\rangle.
\eea 
The corresponding time-dependent Gross-Pitaevskii equation is~\cite{2012_Cai_Duan_PRA}
\bea 
i\partial_t \psi_1 (t) &=& \lambda \psi_2 (t) + (2 g_1 |\psi_1|^2 + g_2 |\psi_2| ^2 ) \psi_1 
+ 2 g_3 \psi_1 ^* \psi_2 ^2 \,, \nn \\
i\partial_t \psi_2 (t) &=& \lambda \psi_1 (t) + (2 g_1 |\psi_2|^2 + g_2 |\psi_1| ^2 ) \psi_2
+ 2 g_3 \psi_2 ^* \psi_1 ^2\,. \nn \\
\eea 

To make the dynamics more physical, one can rewrite the wavefunctions $\psi_j (t)$ in terms 
of densities and  phases as 
\begin{align*}
 & \psi_1 \to \sqrt{\rho_1} e^{i\theta_1}, \nn \\
& \psi_2 \to \sqrt{\rho_2} e^{i\theta_2}.  \nn 
\end{align*} 
The equation of motion is most easily derived by constructing the Lagrangian, which 
takes the form, 
$$
L = -\rho_1 \partial_t \theta_1 - \rho_2 \partial_t \theta_2  
- \left\{  2\lambda\sqrt{\rho_1 \rho_2} \cos(\theta_2 -\theta_1) 
+ 2 g_3 \rho_1 \rho_2 \cos(2(\theta_2 - \theta_1)) \right.   
 +\left.  g_1 (\rho_1 ^2 + \rho_2 ^2 ) + g_2 \rho_1 \rho_2  \right\}. 
$$ 
From Euler-Lagrangian equations, 
\begin{align*} 
& \partial_t \rho_j = - \frac{\partial L} {\partial \theta_j } ,  \\
& \frac{\partial L}{\partial \rho_j} =0, 
\end{align*} 
one gets 
\begin{align*}
& \partial_t \rho_1 = - \partial_t \rho_2 = 2 \lambda \sqrt{\rho_1 \rho_2} \sin(\theta_2 -\theta_1) \\  
& \,\,\, \,\,\, \,\,\, \,\,\, \,\,\, \,\,\, \,\,\, \,\,\, 
 + 4 g_3 \rho_1 \rho_2 \sin(2 (\theta_2 - \theta_1)),  \\
& \partial_t \theta_1 = -\lambda \sqrt{\frac{\rho_2}{\rho_1}} \cos(\theta_2 - \theta_1) 
      - 2 g_3 \rho_2 \cos(2 (\theta_2 - \theta_1) )  \\
& \,\,\, \,\,\, \,\,\, \,\,\, \,\,\, \,\,\, \,\,\, \,\,\, 
      - 2 g_1 \rho_1 - g_2 \rho_2,  \\
& \partial_t \theta_2 =  -\lambda \sqrt{\frac{\rho_1}{\rho_2}} \cos(\theta_1 - \theta_2) 
      - 2 g_3 \rho_1 \cos(2 (\theta_1 - \theta_2) ),  \\
& \,\,\, \,\,\, \,\,\, \,\,\, \,\,\, \,\,\, \,\,\, \,\,\, 
      - 2 g_1 \rho_2 - g_2 \rho_1.  
\end{align*} 
To make a direct connection to Josephson effects, the number imbalance and phase difference are defined to be 
 $z = \rho_1 - \rho_2$ and $\phi = \theta_1 - \theta_2$, whose dynamical evolution is governed by  
\bea 
&& \partial_t z = 2 \left(\lambda \sqrt{1-z^2} \sin(\phi) + g_3 (1-z^2) \sin(2\phi) \right) \,, \nn \\ 
&& \partial_t \phi = (2 g_1 - g_2 ) z - \frac{2\lambda z } {1-z^2} \cos(\phi) - 2 g_3 z \cos(2 \phi).  
\label{eq:5dynamics}
\eea 
Compared with Josephson effects in double-well Bose-Einstein condensates~\cite{1997_Smerzi_Fantoni_PRL,1998_Zapata_Sols_PRA}, 
the key difference is that here we have $\sin (2 \phi)$ and $\cos (2\phi)$) terms 
which are generated by the Umklapp process $g_3$. In the ground state, these terms give rise to 
the spontaneous time-reversal symmetry breaking.

In the noninteracting limit, $g_{1,2,3} \to 0$, the Rabi-like oscillation with frequency 
$2\lambda$ is easily recovered. In the linear regime, $|z| \ll 1$, 
the dynamics in $z$ and $\phi$ is simplified to 
\begin{align*}
& \partial_t z \approx (2 \lambda + 4 g_3) \delta \phi , \\
& \partial_t \delta \phi \approx (2 g_1 - g_2 -2 \lambda - 2 g_3) z, 
\end{align*} 
assuming $\phi \ll 2 \pi$.  
This gives rise to oscillatory dynamics with a frequency 
\bea 
\omega_{\rm real} = \sqrt{(2\lambda + 4 g_3) (2 \lambda - 2g_1 + g_2 +2 g_3)}, 
\eea 
which is the Josephson frequency for a real superposition state $p_x + p_y$. 
For the complex superposition $p_x \pm ip_y$, 
expressing $\phi$ in terms of fluctuation field $\delta \phi$, 
 $\phi \to \frac{\pi}{2} + \delta \phi$ ($\delta \phi \ll 2\pi$), 
the linear dynamics is 
\begin{align*}
& \partial_t z \approx -2 g_3 (\delta \phi - \frac{\lambda}{g_3} ) ,  \\
& \partial_t \delta \phi \approx (2g_1 - g_2 + 2 g_3) z , 
\end{align*}  
which predicts a Josephson frequency 
\bea 
\omega_{\rm complex}  = \sqrt{ 2 g_3 ( 2 g_1 -g _2 + 2 g_3)}, 
\eea 
with $\lambda/g_3$ assumed to be small. 
In the Josephson effects, the frequency is different from that in non-interacting Rabi oscillations. 
This frequency difference is also seen in the numerical simulations based on Gross-Pitaevskii equations~\cite{2012_Cai_Duan_PRA} 
and Gutzwiller methods~\cite{2013_Li_Paramekanti_NatComm}.

The nonlinear effects of dynamics in Eq.~\eqref{eq:5dynamics} are expected to be more interesting, because of the $\sin(2\phi)$ term, than the usual Josephson physics of double-wells.
For example the analogy of self-trapping effect in double-wells would certainly exist in this orbital setting, and 
very likely would lead to new possibilities beyond the standard double-well Josephson effect. Details of such orbital Josephson effects call for 
further theoretical and experimental investigations.

\subsection{Relation to spin-orbit coupled quantum gases}
Orbital degree of freedom can certainly be mapped to pseudo-spins. In doing so, spin-orbit couplings 
of certain types usually arise naturally due to the spatial 
nature of orbitals~\cite{2010_Liu_Liu_PRA,2012_Sun_Jackeli_PRB,2011_Sun_Liu_NatPhys,
2013_Li_Zhao_NatComm,2014_Belemuk_Chtchelkatchev_arXiv,2014_Liu_Li_NatComm,2015_Zhou_Zhao_PRL}. 
The tunneling Hamiltonian of orbital models mixes different orbitals. In particular 
mixing of different parities could lead to non-trivial effective spin orbit couplings and 
consequent topological properties. Mixing of $s$ and $p$-orbitals in a 
ladder system~\cite{2013_Li_Zhao_NatComm} closely mimics the one dimensional 
spin orbital coupling recently engineered in cold gases by Raman 
transitions~\cite{2011_Lin_Spielman_Nature,2013_Galitski_Nat}. Such $sp$ orbital mixing 
is recently achieved in a shaken lattice experiment of $^{133}$Cs Bose-Einstein condensates~\cite{2013_Parker_Ha_NatPhys} and a similar 
band structure with double minima like the spin-orbit coupled case is indeed obtained. 
Mixing of $p$ and $d$-orbitals gives rise to the phases of topological semimetal and topological insulator~\cite{2011_Sun_Liu_NatPhys}. 
One recent work shows that mixing of $p$-orbitals in spin imbalanced fermions leads to topological superconductivity
with novel features~\cite{2014_Liu_Li_NatComm}. 
We note however that some of these novel predictions are made for fermionic species of atoms, whereas the high band experiments have been explored only for bosons so far as to this time.  Further experimental developments are expected. 

With strong repulsion, particles could form Mott states with  the charge degrees of freedom frozen. 
The orbital ordering in Mott states is then described by super-exchange interactions of orbitals, which 
typically depend on the orientation of links. 
This orientation dependent orbital super-exchange gives rise to novel pseudo-spin  models such as quantum $120^0$ model~(\onlinecite{2008_Zhao_Liu_PRL};  \onlinecite{2008_Wu_PRL2}) 
(see Eq.~\eqref{eq:3Ham120}). 

With spin-orbit couplings, many interesting quantum phases such as skyrmions and topological states have been investigated. The connection of orbital physics to spin-orbit coupling suggests possibilities of novel orbital states. One reason to study spin-orbit coupled physics in  orbital systems (with atoms loaded into higher bands) is 
that there appears no additional heating in this system, in contrast with the heating challenge faced by engineered spin-orbital couplings by the advanced Raman laser technique. 
In this regard, orbital physics provides an alternative platform to investigate
spin-orbit coupled phenomena, which is a direction worth future exploration. 

\subsection{Periodic driving induced  orbital couplings}

In recent optical lattice experiments~\cite{2013_Parker_Ha_NatPhys,2013_Struck_Weinberg_NatPhys,2013_Aidelsburger_Atala_PRL,2013_Miyake_Siviloglou_PRL,2014_Jotzu_Messer_Nature,2015_Niu_Hu_OptExp,2015_Weinberg_Olschlager_PRA}, periodically driven systems have been developed with a motivation to create exotic atomic phases. In such systems time reversal symmetry is explicitly broken. With the driving frequency matching band gaps,  energetically separated orbital bands can be efficiently coupled. 

Here we use one  example to demonstrate the key idea of using lattice shaking to induce/control orbital couplings. Consider a  one dimensional shaking lattice as implemented in experiments~\cite{2013_Parker_Ha_NatPhys}. The time-dependent optical potential of this lattice reads 
\be 
V(x,t ) = V_0 \cos \left [ k (x - x_0 (t)) \right ], 
\ee 
with $x_0(t)$ a periodic function, $x_0 (t)   = X_0 \sin (2\pi t/T)$. Taking $X_0 = 0$, we have a static lattice potential where $s$ and $p$ orbital bands are decoupled and well separated by an energy gap.  
With weak driving, we have 
$ V(x,t ) \approx V_0 \left[ \cos (k x) +  k x_0 (t) \sin (kx) \right]$. The time-dependent term introduces an effective coupling between $s$ and $p$ orbitals, approximately given by 
\be 
\lambda_{sp} = kV_0 x_0 (t) \int dx  \sin (kx) w_s ^* (x) w_p (x),  
\ee 
with $w_\nu(x)$ the orbital wavefunction. With frequency $2 \pi/T$ matching the band gap, the system is approximately described by a static two-band model with $s$ and $p$ orbitals coupled, under a rotating wave approximation.

It appears natural to engineer orbital couplings by lattice modulation/shaking techniques. But the problem is that heating effects are fundamentally unavoidable in periodically driven quantum systems. Since periodic driving breaks time translational symmetry, energy is no longer a conserved quantity. It follows that driven systems (assuming ergodicity)  at long time would necessarily be described by infinite temperature ensemble. Nonetheless, there could be long lifetime transient states that manifest interesting topological features. This requires more careful treatment of quantum dynamics than just solving for the ground states of effective static Hamiltonians.  One way out is to combine with dissipation. Driven-dissipative orbital models may exhibit steady quantum many-body states with interesting topological properties. This is worth future exploration.

\subsection{Open questions}

For bosons, firstly, it remains open how to experimentally reach the Mott
  insulator phases of the $p$-band and study the $p$-band superfluid-Mott
  insulator transition. The current experiments at Hamburg are performed with a
  two-dimensional checkerboard lattice and a relatively
  shallow harmonic trap in the third dimension. Introducing an additional
  optical lattice potential in the third dimension is required to access the
  Mott regime. Unfortunately that would also increase the on-site interaction
  between $p$-orbital bosons, which leads to faster decay \cite{2015_Hemmerich_private}.    

Secondly, it is intriguing to find out what type of new topological defects,
other than vortices, may possibly occur in the staggered $p_x\pm ip_y$-orbital Bose-Einstein
condensate.  The state breaks not only U(1) but also other interesting symmetries
that are usually not broken in other conventional Bose condensates, including for
example, time-reversal, lattice translational and rotational symmetries. On the
general ground of broken symmetries, new classification of topological defects is
expected but remains unknown. 

For  fermions, the stability of the $p$ and higher orbital bands is protected by Fermi statistics,
if the experimental system is prepared with the lowest ground band being
completely filled, as opposed to the method of band population 
inversion~\cite{2011_Wirth_Olschlager_NatPhys,2012_Olschlager_Wirth_PRL,2013_Olschlager_Kock_NJPHYS,2015_Kock_Olschlager_PRL,2007_Muller_Folling_PRL}. 
Nevertheless, this approach
would require a higher density of fermions, which in turn requires a higher
efficiency of cooling fermions down to degeneracy. The recent breakthrough in the Rice
experiment of fermions on lattice~\cite{2015_Hart_Duarte_Nature} is promising for
studying the higher orbital bands.

%% file: 6Appendix.tex
\section{Tree level estimate of couplings in effective field theory for $p$-orbital bosons}
\label{sec:EFTcouplings} 
In this appendix, the coupling constants in the effective field theory (Eq.~\eqref{eq:3HpbandEFT}) 
are related to a microscopic model. 
We start with the contact interaction for a 3D Bose gas, which reads 
\be
V_{\rm int} =  \frac{2\pi a_s \hbar^2}{m} \int d^3 \textbf{x} \psi ^\dag (\textbf{x}) \psi (\textbf{x}) \psi^\dag (\textbf{x}) \psi(\textbf{x}), 
\ee
where $\psi (\textbf{x})$ is a bosonic field operator, $m$ is the mass of atoms and $a_s$ is the 3D scattering length. 
With bosons loaded into the $p$-band of a 2D lattice that has band minima at $\tbf{Q}_x  = (\pi,0)$ and $\tbf{Q}_y = (0,\pi)$, 
the field operator is expanded by the low energy modes as~\cite{2013_Li_Paramekanti_NatComm}
\bea 
\psi(\textbf{x}) &=& \int ^\Lambda \frac{d ^2 \tbf{q}}{(2\pi)^2} e^{i \tbf{q} \cdot {\textbf{x}} } 
   \left\{ 
	  {b}   _{ \tbf{Q}_x + \tbf{q} } e^{i \tbf{Q}_x  \cdot \textbf{x} } u_{\tbf{Q}_x + \tbf{q}} (\textbf{x}) \right. \nn \\
&&\left. 	 
+ {b}  _{ \tbf{Q}_y + \tbf{q} } e^{i \tbf{Q}_y \cdot \textbf{x}} u_{\tbf{Q}_y + \tbf{q}} (\textbf{x} ) 
      \right\}, 
\eea 
where ${b}_ {\tbf{Q}_\alpha +\tbf{q}}$ is the annihilation operator for a Bloch mode near the band minimum $\tbf{Q}_\alpha$ 
and $u_{\tbf{Q}_\alpha + \tbf{q}} (\textbf{x})$ is the corresponding periodic Bloch wavefunction. 
At tree level, the high energy modes are integrated out and the resulting renormalization of the low energy theory is neglected. 
Then the interaction is written in terms of these low energy modes as 
\begin{widetext} 
\bea 
\textstyle V_{\rm int} &=& \textstyle 
	\frac{2\pi a_s \hbar^2}{m}  \int d^3 \textbf{x} \left[ \prod _{j =1}^4  \frac{d ^2 \tbf{q}_j }{(2\pi)^2}\right] 
	      e^{-i (\tbf{q}_1 + \tbf{q}_3 - \tbf{q}_2 -\tbf{q}_4) \cdot \textbf{x}} \nn \\ 
	&& \textstyle \left\{ \textstyle {b}   ^\dag _{ \tbf{Q}_x + \tbf{q}_1 } {b}  _{ \tbf{Q}_x + \tbf{q}_2 }  
		{b}  ^\dag _{ \tbf{Q}_x + \tbf{q}_3 } {b}_{    \tbf{Q}_x + \tbf{q}_4 }  
	      u_{\tbf{Q}_x + \tbf{q}_1} ^* (\textbf{x}) u_{\tbf{Q}_x + \tbf{q}_2}  (\textbf{x}) u_{\tbf{Q}_x + \tbf{q}_3} ^* (\textbf{x}) 
		u_{\tbf{Q}_x + \tbf{q}_4} (\textbf{x}) 
	       + \tbf{Q}_x \to \tbf{Q}_y 
	\right. \nn \\
       && 4\times  {b}   ^\dag _{ \tbf{Q}_x + \tbf{q}_1 } {b}  _{ \tbf{Q}_x + \tbf{q}_2 } {b}  ^\dag _{ \tbf{Q}_y + \tbf{q}_3 } {b}  _{\tbf{Q}_y + \tbf{q}_4}   
	      u_{\tbf{Q}_x + \tbf{q}_1} ^* (\textbf{x}) u_{\tbf{Q}_x + \tbf{q}_2}  (\textbf{x}) u_{\tbf{Q}_y + \tbf{q}_3} ^* (\textbf{x}) 
		u_{\tbf{Q}_y + \tbf{q}_4} (\textbf{x}) \nn \\
      && \left. 
	 {b}   ^\dag _{ \tbf{Q}_x + \tbf{q}_1 } {b}  _{\tbf{Q}_y + \tbf{q}_2} {b}  ^\dag _{\tbf{Q}_x + \tbf{q}_3} {b}  _{ \tbf{Q}_y + \tbf{q}_4}   
	      u_{\tbf{Q}_x + \tbf{q}_1} ^* (\textbf{x}) u_{\tbf{Q}_y + \tbf{q}_2}  (\textbf{x}) u_{\tbf{Q}_x + \tbf{q}_3} ^* (\textbf{x}) 
		u_{\tbf{Q}_y + \tbf{q}_4} (\textbf{x}) 
	 + \tbf{Q}_x \leftrightarrow \tbf{Q}_y  \right\}. \nn \\ 
\eea 
We can rewrite $\textbf{x} = \tbf{R} + \textbf{x}'$, where $\tbf{R}$ is the position vector of lattice sites and $\textbf{x}'$ 
centers over one unit cell. 
With $\sum_\tbf{R} e^{-i (\tbf{q}_1 + \tbf{q}_3 - \tbf{q}_2 -\tbf{q}_4) \cdot \tbf{R} } = \frac{ (2\pi)^2}{a^2}
\delta (\tbf{q}_1 + \tbf{q}_3 - \tbf{q}_2 - \tbf{q}_4)$ ($a$ is the lattice constant), we get  
\bea 
 V_{\rm int} = \int \left[ \prod_{j =1}^4  \frac{d ^2 \tbf{q}_j }{(2\pi)^2}\right] 
&&       (2\pi)^2 \delta (\tbf{q}_1 + \tbf{q}_3 - \tbf{q}_2 - \tbf{q}_4)   
 \left\{ \sum_{\alpha,\beta=x,y}g_{\alpha \beta} (\tbf{q}_1, \tbf{q}_2, \tbf{q}_3)  
	{b}   ^\dag _{ \tbf{Q}_\alpha + \tbf{q}_1 } {b}  _{\tbf{Q}_\alpha + \tbf{q}_2}  {b}  ^\dag _{\tbf{Q}_\beta + \tbf{q}_3} {b}_  {\tbf{Q}_\beta + \tbf{q}_4}
  \right.  \nn \\
&& \left. 
g_3 (\tbf{q}_1, \tbf{q}_2, \tbf{q}_3)  
	\left[ {b}   ^\dag _{\tbf{Q}_x+ \tbf{q}_1}  {b}  ^\dag _{\tbf{Q}_x + \tbf{q}_3} {b}_  {\tbf{Q}_y + \tbf{q}_2} {b}   _{\tbf{Q}_y + \tbf{q}_4}  
	+ \tbf{Q}_x \leftrightarrow \tbf{Q}_y \right] \right\}, 
\eea 
where 
\bea 
g_{xx} (\tbf{q}_1, \tbf{q}_2, \tbf{q}_3) &=& \frac{2\pi a_s \hbar^2}{ma^2} \int d^3 \textbf{x}' 
      u_{\tbf{Q}_x + \tbf{q}_1} ^* (\textbf{x}') u_{\tbf{Q}_x + \tbf{q}_2}  (\textbf{x}') u_{\tbf{Q}_x + \tbf{q}_3} ^* (\textbf{x}') 
		u_{\tbf{Q}_x + \tbf{q}_1 + \tbf{q}_3 - \tbf{q}_2} (\textbf{x}') , \nn\\
g_{yy}  (\tbf{q}_1, \tbf{q}_2, \tbf{q}_3) &=& \frac{2\pi a_s \hbar^2}{ma^2} \int d^3 \textbf{x}' 
      u_{\tbf{Q}_y + \tbf{q}_1} ^* (\textbf{x}') u_{\tbf{Q}_y + \tbf{q}_2}  (\textbf{x}') u_{\tbf{Q}_y + \tbf{q}_3} ^* (\textbf{x}') 
		u_{\tbf{Q}_y + \tbf{q}_1 + \tbf{q}_3 - \tbf{q}_2} (\textbf{x}') , \nn \\ 
g_{xy} (\tbf{q}_1, \tbf{q}_2, \tbf{q}_3) &=& g_{yx} (\tbf{q}_3, \tbf{q}_1 + \tbf{q}_3 - \tbf{q}_2, \tbf{q}_1 )
 \nn \\ 
&=& \frac{4\pi a_s \hbar^2}{ma^2} \int d^3 \textbf{x}' 
	      u_{\tbf{Q}_x + \tbf{q}_1} ^* (\textbf{x}') u_{\tbf{Q}_x + \tbf{q}_2}  (\textbf{x}') u_{\tbf{Q}_y + \tbf{q}_3} ^* (\textbf{x}') 
		u_{\tbf{Q}_y + \tbf{q}_1 + \tbf{q}_3 - \tbf{q}_2} (\textbf{x}') , \nn \\ 
g_3 (\tbf{q}_1, \tbf{q}_2, \tbf{q}_3) &=& \frac{4\pi a_s \hbar^2}{ma^2} \int d^3 \textbf{x}' 
      u_{\tbf{Q}_x + \tbf{q}_1} ^* (\textbf{x}')  u_{\tbf{Q}_x + \tbf{q}_3} ^* (\textbf{x}')  u_{\tbf{Q}_y + \tbf{q}_2}  (\textbf{x}')
		u_{\tbf{Q}_y + \tbf{q}_1 + \tbf{q}_3 - \tbf{q}_2} (\textbf{x}'). 
\eea 
\end{widetext} 
Neglecting the momentum dependence of $g_{\alpha \beta}$ and $g_3$, the derived couplings simplify to 
\bea 
&& g_{xx} = g_{yy} = \frac{2\pi a_s \hbar^2}{m a^2} \int d ^3 \tbf{x}' |u_{\tbf{Q}_x} (\tbf{x}') |^4, \nn  \\ 
&& g_{xy} = g_{yx} = \frac{4\pi a_s \hbar^2}{m a^2} \int d ^3 \tbf{x}' |u_{\tbf{Q}_x} (\tbf{x}') |^2  |u_{\tbf{Q}_y} (\bf{x}') |^2, \nn \\ 
&& g_3 =  \frac{2\pi a_s \hbar^2}{m a^2} \int d ^3 \tbf{x}' u_{\tbf{Q}_x} (\tbf{x}') ^{*2}  u_{\tbf{Q}_y} (\tbf{x}') ^{2} , \nn 
\eea

The calculation of $K_\parallel$ and $K_\perp$ is straightforward at tree level, and they are estimated to be 
$K_\parallel =- \frac{1}{2a^2} \frac{\partial^2 }{\partial k_x ^2} E_p (\tbf{k}) |_{\tbf{k}\to \tbf{Q}_x} $, and 
$K_\perp =- \frac{1}{2a^2} \frac{\partial^2 }{\partial k_y ^2} E_p (\tbf{k}) |_{\tbf{k}\to \tbf{Q}_x}$, 
with $E_p ({\bf k})$ the dispersion of the $p$-band.

%% file: orbreview-rpp.bbl
\begin{thebibliography}{169}%
\makeatletter
\providecommand \@ifxundefined [1]{%
 \@ifx{#1\undefined}
}%
\providecommand \@ifnum [1]{%
 \ifnum #1\expandafter \@firstoftwo
 \else \expandafter \@secondoftwo
 \fi
}%
\providecommand \@ifx [1]{%
 \ifx #1\expandafter \@firstoftwo
 \else \expandafter \@secondoftwo
 \fi
}%
\providecommand \natexlab [1]{#1}%
\providecommand \enquote  [1]{``#1''}%
\providecommand \bibnamefont  [1]{#1}%
\providecommand \bibfnamefont [1]{#1}%
\providecommand \citenamefont [1]{#1}%
\providecommand \href@noop [0]{\@secondoftwo}%
\providecommand \href [0]{\begingroup \@sanitize@url \@href}%
\providecommand \@href[1]{\@@startlink{#1}\@@href}%
\providecommand \@@href[1]{\endgroup#1\@@endlink}%
\providecommand \@sanitize@url [0]{\catcode `\\12\catcode `\$12\catcode
  `\&12\catcode `\#12\catcode `\^12\catcode `\_12\catcode `\%12\relax}%
\providecommand \@@startlink[1]{}%
\providecommand \@@endlink[0]{}%
\providecommand \url  [0]{\begingroup\@sanitize@url \@url }%
\providecommand \@url [1]{\endgroup\@href {#1}{\urlprefix }}%
\providecommand \urlprefix  [0]{URL }%
\providecommand \Eprint [0]{\href }%
\providecommand \doibase [0]{http://dx.doi.org/}%
\providecommand \selectlanguage [0]{\@gobble}%
\providecommand \bibinfo  [0]{\@secondoftwo}%
\providecommand \bibfield  [0]{\@secondoftwo}%
\providecommand \translation [1]{[#1]}%
\providecommand \BibitemOpen [0]{}%
\providecommand \bibitemStop [0]{}%
\providecommand \bibitemNoStop [0]{.\EOS\space}%
\providecommand \EOS [0]{\spacefactor3000\relax}%
\providecommand \BibitemShut  [1]{\csname bibitem#1\endcsname}%
\let\auto@bib@innerbib\@empty
\bibitem [{\citenamefont {Aidelsburger}\ \emph {et~al.}(2013)\citenamefont
  {Aidelsburger}, \citenamefont {Atala}, \citenamefont {Lohse}, \citenamefont
  {Barreiro}, \citenamefont {Paredes},\ and\ \citenamefont
  {Bloch}}]{2013_Aidelsburger_Atala_PRL}%
  \BibitemOpen
  \bibfield  {author} {\bibinfo {author} {\bibnamefont {Aidelsburger},
  \bibfnamefont {M.}}, \bibinfo {author} {\bibfnamefont {M.}~\bibnamefont
  {Atala}}, \bibinfo {author} {\bibfnamefont {M.}~\bibnamefont {Lohse}},
  \bibinfo {author} {\bibfnamefont {J.~T.}\ \bibnamefont {Barreiro}}, \bibinfo
  {author} {\bibfnamefont {B.}~\bibnamefont {Paredes}}, \ and\ \bibinfo
  {author} {\bibfnamefont {I.}~\bibnamefont {Bloch}}} (\bibinfo {year}
  {2013}),\ \href {\doibase 10.1103/PhysRevLett.111.185301} {\bibfield
  {journal} {\bibinfo  {journal} {Phys. Rev. Lett.}\ }\textbf {\bibinfo
  {volume} {111}},\ \bibinfo {pages} {185301}}\BibitemShut {NoStop}%
\bibitem [{\citenamefont {Alon}\ \emph {et~al.}(2005)\citenamefont {Alon},
  \citenamefont {Streltsov},\ and\ \citenamefont
  {Cederbaum}}]{2005_Alon_Streltsov_PRL}%
  \BibitemOpen
  \bibfield  {author} {\bibinfo {author} {\bibnamefont {Alon}, \bibfnamefont
  {O.~E.}}, \bibinfo {author} {\bibfnamefont {A.~I.}\ \bibnamefont
  {Streltsov}}, \ and\ \bibinfo {author} {\bibfnamefont {L.~S.}\ \bibnamefont
  {Cederbaum}}} (\bibinfo {year} {2005}),\ \href {\doibase
  10.1103/PhysRevLett.95.030405} {\bibfield  {journal} {\bibinfo  {journal}
  {Phys. Rev. Lett.}\ }\textbf {\bibinfo {volume} {95}},\ \bibinfo {pages}
  {030405}}\BibitemShut {NoStop}%
\bibitem [{\citenamefont {Anderlini}\ \emph {et~al.}(2007)\citenamefont
  {Anderlini}, \citenamefont {Lee}, \citenamefont {Brown}, \citenamefont
  {Sebby-Strabley}, \citenamefont {Phillips},\ and\ \citenamefont
  {Porto}}]{2007_Anderlini_Lee_Nature}%
  \BibitemOpen
  \bibfield  {author} {\bibinfo {author} {\bibnamefont {Anderlini},
  \bibfnamefont {M.}}, \bibinfo {author} {\bibfnamefont {P.~J.}\ \bibnamefont
  {Lee}}, \bibinfo {author} {\bibfnamefont {B.~L.}\ \bibnamefont {Brown}},
  \bibinfo {author} {\bibfnamefont {J.}~\bibnamefont {Sebby-Strabley}},
  \bibinfo {author} {\bibfnamefont {W.~D.}\ \bibnamefont {Phillips}}, \ and\
  \bibinfo {author} {\bibfnamefont {J.~V.}\ \bibnamefont {Porto}}} (\bibinfo
  {year} {2007}),\ \href {http://dx.doi.org/10.1038/nature06011} {\bibfield
  {journal} {\bibinfo  {journal} {Nature}\ }\textbf {\bibinfo {volume}
  {448}}~(\bibinfo {number} {7152}),\ \bibinfo {pages} {452}}\BibitemShut
  {NoStop}%
\bibitem [{\citenamefont {Bednorz}\ and\ \citenamefont
  {M\"uller}(1986)}]{1986_Bednorz_Muller_PBCM}%
  \BibitemOpen
  \bibfield  {author} {\bibinfo {author} {\bibnamefont {Bednorz}, \bibfnamefont
  {J.}}, \ and\ \bibinfo {author} {\bibfnamefont {K.}~\bibnamefont {M\"uller}}}
  (\bibinfo {year} {1986}),\ \href {\doibase 10.1007/BF01303701} {\bibfield
  {journal} {\bibinfo  {journal} {Zeitschrift f\"ur Physik B Condensed Matter}\
  }\textbf {\bibinfo {volume} {64}}~(\bibinfo {number} {2}),\ \bibinfo {pages}
  {189}}\BibitemShut {NoStop}%
\bibitem [{\citenamefont {Belemuk}\ \emph {et~al.}(2014)\citenamefont
  {Belemuk}, \citenamefont {Chtchelkatchev},\ and\ \citenamefont
  {Mikheyenkov}}]{2014_Belemuk_Chtchelkatchev_arXiv}%
  \BibitemOpen
  \bibfield  {author} {\bibinfo {author} {\bibnamefont {Belemuk}, \bibfnamefont
  {A.~M.}}, \bibinfo {author} {\bibfnamefont {N.~M.}\ \bibnamefont
  {Chtchelkatchev}}, \ and\ \bibinfo {author} {\bibfnamefont {A.~V.}\
  \bibnamefont {Mikheyenkov}}} (\bibinfo {year} {2014}),\ \href {\doibase
  10.1103/PhysRevA.90.023625} {\bibfield  {journal} {\bibinfo  {journal} {Phys.
  Rev. A}\ }\textbf {\bibinfo {volume} {90}},\ \bibinfo {pages}
  {023625}}\BibitemShut {NoStop}%
\bibitem [{\citenamefont {Bloch}\ \emph {et~al.}(2008)\citenamefont {Bloch},
  \citenamefont {Dalibard},\ and\ \citenamefont
  {Zwerger}}]{2008_Bloch_Dalibard_RMP}%
  \BibitemOpen
  \bibfield  {author} {\bibinfo {author} {\bibnamefont {Bloch}, \bibfnamefont
  {I.}}, \bibinfo {author} {\bibfnamefont {J.}~\bibnamefont {Dalibard}}, \ and\
  \bibinfo {author} {\bibfnamefont {W.}~\bibnamefont {Zwerger}}} (\bibinfo
  {year} {2008}),\ \href {\doibase 10.1103/RevModPhys.80.885} {\bibfield
  {journal} {\bibinfo  {journal} {Rev. Mod. Phys.}\ }\textbf {\bibinfo {volume}
  {80}},\ \bibinfo {pages} {885}}\BibitemShut {NoStop}%
\bibitem [{\citenamefont {Browaeys}\ \emph {et~al.}(2005)\citenamefont
  {Browaeys}, \citenamefont {H\"affner}, \citenamefont {McKenzie},
  \citenamefont {Rolston}, \citenamefont {Helmerson},\ and\ \citenamefont
  {Phillips}}]{2005_Browaeys_Haffner_PRA}%
  \BibitemOpen
  \bibfield  {author} {\bibinfo {author} {\bibnamefont {Browaeys},
  \bibfnamefont {A.}}, \bibinfo {author} {\bibfnamefont {H.}~\bibnamefont
  {H\"affner}}, \bibinfo {author} {\bibfnamefont {C.}~\bibnamefont {McKenzie}},
  \bibinfo {author} {\bibfnamefont {S.~L.}\ \bibnamefont {Rolston}}, \bibinfo
  {author} {\bibfnamefont {K.}~\bibnamefont {Helmerson}}, \ and\ \bibinfo
  {author} {\bibfnamefont {W.~D.}\ \bibnamefont {Phillips}}} (\bibinfo {year}
  {2005}),\ \href {\doibase 10.1103/PhysRevA.72.053605} {\bibfield  {journal}
  {\bibinfo  {journal} {Phys. Rev. A}\ }\textbf {\bibinfo {volume} {72}},\
  \bibinfo {pages} {053605}}\BibitemShut {NoStop}%
\bibitem [{\citenamefont {Cai}\ \emph {et~al.}(2012{\natexlab{a}})\citenamefont
  {Cai}, \citenamefont {Duan},\ and\ \citenamefont {Wu}}]{2012_Cai_Duan_PRA}%
  \BibitemOpen
  \bibfield  {author} {\bibinfo {author} {\bibnamefont {Cai}, \bibfnamefont
  {Z.}}, \bibinfo {author} {\bibfnamefont {L.-M.}\ \bibnamefont {Duan}}, \ and\
  \bibinfo {author} {\bibfnamefont {C.}~\bibnamefont {Wu}}} (\bibinfo {year}
  {2012}{\natexlab{a}}),\ \href {\doibase 10.1103/PhysRevA.86.051601}
  {\bibfield  {journal} {\bibinfo  {journal} {Phys. Rev. A}\ }\textbf {\bibinfo
  {volume} {86}},\ \bibinfo {pages} {051601}}\BibitemShut {NoStop}%
\bibitem [{\citenamefont {Cai}\ \emph {et~al.}(2011)\citenamefont {Cai},
  \citenamefont {Wang},\ and\ \citenamefont {Wu}}]{2011_Cai_Wang_PRA}%
  \BibitemOpen
  \bibfield  {author} {\bibinfo {author} {\bibnamefont {Cai}, \bibfnamefont
  {Z.}}, \bibinfo {author} {\bibfnamefont {Y.}~\bibnamefont {Wang}}, \ and\
  \bibinfo {author} {\bibfnamefont {C.}~\bibnamefont {Wu}}} (\bibinfo {year}
  {2011}),\ \href {\doibase 10.1103/PhysRevA.83.063621} {\bibfield  {journal}
  {\bibinfo  {journal} {Phys. Rev. A}\ }\textbf {\bibinfo {volume} {83}},\
  \bibinfo {pages} {063621}}\BibitemShut {NoStop}%
\bibitem [{\citenamefont {Cai}\ \emph {et~al.}(2012{\natexlab{b}})\citenamefont
  {Cai}, \citenamefont {Wang},\ and\ \citenamefont {Wu}}]{2012_Cai_Wang_PRBR}%
  \BibitemOpen
  \bibfield  {author} {\bibinfo {author} {\bibnamefont {Cai}, \bibfnamefont
  {Z.}}, \bibinfo {author} {\bibfnamefont {Y.}~\bibnamefont {Wang}}, \ and\
  \bibinfo {author} {\bibfnamefont {C.}~\bibnamefont {Wu}}} (\bibinfo {year}
  {2012}{\natexlab{b}}),\ \href {\doibase 10.1103/PhysRevB.86.060517}
  {\bibfield  {journal} {\bibinfo  {journal} {Phys. Rev. B}\ }\textbf {\bibinfo
  {volume} {86}},\ \bibinfo {pages} {060517}}\BibitemShut {NoStop}%
\bibitem [{\citenamefont {Cai}\ and\ \citenamefont
  {Wu}(2011)}]{2011_Cai_Wu_PRA}%
  \BibitemOpen
  \bibfield  {author} {\bibinfo {author} {\bibnamefont {Cai}, \bibfnamefont
  {Z.}}, \ and\ \bibinfo {author} {\bibfnamefont {C.}~\bibnamefont {Wu}}}
  (\bibinfo {year} {2011}),\ \href {\doibase 10.1103/PhysRevA.84.033635}
  {\bibfield  {journal} {\bibinfo  {journal} {Phys. Rev. A}\ }\textbf {\bibinfo
  {volume} {84}},\ \bibinfo {pages} {033635}}\BibitemShut {NoStop}%
\bibitem [{\citenamefont {Casula}\ \emph {et~al.}(2008)\citenamefont {Casula},
  \citenamefont {Ceperley},\ and\ \citenamefont
  {Mueller}}]{2008_Casula_Ceperley_PRA}%
  \BibitemOpen
  \bibfield  {author} {\bibinfo {author} {\bibnamefont {Casula}, \bibfnamefont
  {M.}}, \bibinfo {author} {\bibfnamefont {D.~M.}\ \bibnamefont {Ceperley}}, \
  and\ \bibinfo {author} {\bibfnamefont {E.~J.}\ \bibnamefont {Mueller}}}
  (\bibinfo {year} {2008}),\ \href {\doibase 10.1103/PhysRevA.78.033607}
  {\bibfield  {journal} {\bibinfo  {journal} {Phys. Rev. A}\ }\textbf {\bibinfo
  {volume} {78}},\ \bibinfo {pages} {033607}}\BibitemShut {NoStop}%
\bibitem [{\citenamefont {Challis}\ \emph {et~al.}(2009)\citenamefont
  {Challis}, \citenamefont {Girvin},\ and\ \citenamefont
  {Glazman}}]{2009_Challis_Girvin_PRA}%
  \BibitemOpen
  \bibfield  {author} {\bibinfo {author} {\bibnamefont {Challis}, \bibfnamefont
  {J.~H.}}, \bibinfo {author} {\bibfnamefont {S.~M.}\ \bibnamefont {Girvin}}, \
  and\ \bibinfo {author} {\bibfnamefont {L.~I.}\ \bibnamefont {Glazman}}}
  (\bibinfo {year} {2009}),\ \href {\doibase 10.1103/PhysRevA.79.043609}
  {\bibfield  {journal} {\bibinfo  {journal} {Phys. Rev. A}\ }\textbf {\bibinfo
  {volume} {79}},\ \bibinfo {pages} {043609}}\BibitemShut {NoStop}%
\bibitem [{\citenamefont {Cheinet}\ \emph {et~al.}(2008)\citenamefont
  {Cheinet}, \citenamefont {Trotzky}, \citenamefont {Feld}, \citenamefont
  {Schnorrberger}, \citenamefont {Moreno-Cardoner}, \citenamefont {F\"olling},\
  and\ \citenamefont {Bloch}}]{2008_Cheinet_Trotzky_PRL}%
  \BibitemOpen
  \bibfield  {author} {\bibinfo {author} {\bibnamefont {Cheinet}, \bibfnamefont
  {P.}}, \bibinfo {author} {\bibfnamefont {S.}~\bibnamefont {Trotzky}},
  \bibinfo {author} {\bibfnamefont {M.}~\bibnamefont {Feld}}, \bibinfo {author}
  {\bibfnamefont {U.}~\bibnamefont {Schnorrberger}}, \bibinfo {author}
  {\bibfnamefont {M.}~\bibnamefont {Moreno-Cardoner}}, \bibinfo {author}
  {\bibfnamefont {S.}~\bibnamefont {F\"olling}}, \ and\ \bibinfo {author}
  {\bibfnamefont {I.}~\bibnamefont {Bloch}}} (\bibinfo {year} {2008}),\ \href
  {\doibase 10.1103/PhysRevLett.101.090404} {\bibfield  {journal} {\bibinfo
  {journal} {Phys. Rev. Lett.}\ }\textbf {\bibinfo {volume} {101}},\ \bibinfo
  {pages} {090404}}\BibitemShut {NoStop}%
\bibitem [{\citenamefont {Chen}\ and\ \citenamefont
  {Balents}(2013)}]{2013_Chen_Balents_PRL}%
  \BibitemOpen
  \bibfield  {author} {\bibinfo {author} {\bibnamefont {Chen}, \bibfnamefont
  {G.}}, \ and\ \bibinfo {author} {\bibfnamefont {L.}~\bibnamefont {Balents}}}
  (\bibinfo {year} {2013}),\ \href {\doibase 10.1103/PhysRevLett.110.206401}
  {\bibfield  {journal} {\bibinfo  {journal} {Phys. Rev. Lett.}\ }\textbf
  {\bibinfo {volume} {110}},\ \bibinfo {pages} {206401}}\BibitemShut {NoStop}%
\bibitem [{\citenamefont {Chern}\ and\ \citenamefont
  {Wu}(2011)}]{2011_Chern_Wu_PRE}%
  \BibitemOpen
  \bibfield  {author} {\bibinfo {author} {\bibnamefont {Chern}, \bibfnamefont
  {G.-W.}}, \ and\ \bibinfo {author} {\bibfnamefont {C.}~\bibnamefont {Wu}}}
  (\bibinfo {year} {2011}),\ \href {\doibase 10.1103/PhysRevE.84.061127}
  {\bibfield  {journal} {\bibinfo  {journal} {Phys. Rev. E}\ }\textbf {\bibinfo
  {volume} {84}},\ \bibinfo {pages} {061127}}\BibitemShut {NoStop}%
\bibitem [{\citenamefont {Chin}\ \emph {et~al.}(2010)\citenamefont {Chin},
  \citenamefont {Grimm}, \citenamefont {Julienne},\ and\ \citenamefont
  {Tiesinga}}]{2010_Chin_Grimm_RMP}%
  \BibitemOpen
  \bibfield  {author} {\bibinfo {author} {\bibnamefont {Chin}, \bibfnamefont
  {C.}}, \bibinfo {author} {\bibfnamefont {R.}~\bibnamefont {Grimm}}, \bibinfo
  {author} {\bibfnamefont {P.}~\bibnamefont {Julienne}}, \ and\ \bibinfo
  {author} {\bibfnamefont {E.}~\bibnamefont {Tiesinga}}} (\bibinfo {year}
  {2010}),\ \href {\doibase 10.1103/RevModPhys.82.1225} {\bibfield  {journal}
  {\bibinfo  {journal} {Rev. Mod. Phys.}\ }\textbf {\bibinfo {volume} {82}},\
  \bibinfo {pages} {1225}}\BibitemShut {NoStop}%
\bibitem [{\citenamefont {Collin}\ \emph {et~al.}(2010)\citenamefont {Collin},
  \citenamefont {Larson},\ and\ \citenamefont
  {Martikainen}}]{2010_Collin_Larson_PRA}%
  \BibitemOpen
  \bibfield  {author} {\bibinfo {author} {\bibnamefont {Collin}, \bibfnamefont
  {A.}}, \bibinfo {author} {\bibfnamefont {J.}~\bibnamefont {Larson}}, \ and\
  \bibinfo {author} {\bibfnamefont {J.~P.}\ \bibnamefont {Martikainen}}}
  (\bibinfo {year} {2010}),\ \href {\doibase 10.1103/PhysRevA.81.023605}
  {\bibfield  {journal} {\bibinfo  {journal} {Phys. Rev. A}\ }\textbf {\bibinfo
  {volume} {81}},\ \bibinfo {pages} {023605}}\BibitemShut {NoStop}%
\bibitem [{\citenamefont {Dalibard}\ \emph {et~al.}(2011)\citenamefont
  {Dalibard}, \citenamefont {Gerbier}, \citenamefont
  {Juzeli\ifmmode~\bar{u}\else \={u}\fi{}nas},\ and\ \citenamefont
  {\"Ohberg}}]{2011_Dalibard_Gerbier_RMP}%
  \BibitemOpen
  \bibfield  {author} {\bibinfo {author} {\bibnamefont {Dalibard},
  \bibfnamefont {J.}}, \bibinfo {author} {\bibfnamefont {F.}~\bibnamefont
  {Gerbier}}, \bibinfo {author} {\bibfnamefont {G.}~\bibnamefont
  {Juzeli\ifmmode~\bar{u}\else \={u}\fi{}nas}}, \ and\ \bibinfo {author}
  {\bibfnamefont {P.}~\bibnamefont {\"Ohberg}}} (\bibinfo {year} {2011}),\
  \href {\doibase 10.1103/RevModPhys.83.1523} {\bibfield  {journal} {\bibinfo
  {journal} {Rev. Mod. Phys.}\ }\textbf {\bibinfo {volume} {83}},\ \bibinfo
  {pages} {1523}}\BibitemShut {NoStop}%
\bibitem [{\citenamefont {Duan}(2006)}]{2006_Duan_PRL}%
  \BibitemOpen
  \bibfield  {author} {\bibinfo {author} {\bibnamefont {Duan}, \bibfnamefont
  {L.-M.}}} (\bibinfo {year} {2006}),\ \href {\doibase
  10.1103/PhysRevLett.96.103201} {\bibfield  {journal} {\bibinfo  {journal}
  {Phys. Rev. Lett.}\ }\textbf {\bibinfo {volume} {96}},\ \bibinfo {pages}
  {103201}}\BibitemShut {NoStop}%
\bibitem [{\citenamefont {Dutta}\ \emph {et~al.}(2011)\citenamefont {Dutta},
  \citenamefont {Eckardt}, \citenamefont {Hauke}, \citenamefont {Malomed},\
  and\ \citenamefont {Lewenstein}}]{2011_Dutta_Eckardt_NJP}%
  \BibitemOpen
  \bibfield  {author} {\bibinfo {author} {\bibnamefont {Dutta}, \bibfnamefont
  {O.}}, \bibinfo {author} {\bibfnamefont {A.}~\bibnamefont {Eckardt}},
  \bibinfo {author} {\bibfnamefont {P.}~\bibnamefont {Hauke}}, \bibinfo
  {author} {\bibfnamefont {B.}~\bibnamefont {Malomed}}, \ and\ \bibinfo
  {author} {\bibfnamefont {M.}~\bibnamefont {Lewenstein}}} (\bibinfo {year}
  {2011}),\ \href {http://stacks.iop.org/1367-2630/13/i=2/a=023019} {\bibfield
  {journal} {\bibinfo  {journal} {New Journal of Physics}\ }\textbf {\bibinfo
  {volume} {13}}~(\bibinfo {number} {2}),\ \bibinfo {pages}
  {023019}}\BibitemShut {NoStop}%
\bibitem [{\citenamefont {Dutta}\ \emph
  {et~al.}(2015{\natexlab{a}})\citenamefont {Dutta}, \citenamefont {Gajda},
  \citenamefont {Hauke}, \citenamefont {Lewenstein}, \citenamefont {Lühmann},
  \citenamefont {Malomed}, \citenamefont {Sowinski},\ and\ \citenamefont
  {Zakrzewski}}]{2015_Dutta_Gajda_RPP}%
  \BibitemOpen
  \bibfield  {author} {\bibinfo {author} {\bibnamefont {Dutta}, \bibfnamefont
  {O.}}, \bibinfo {author} {\bibfnamefont {M.}~\bibnamefont {Gajda}}, \bibinfo
  {author} {\bibfnamefont {P.}~\bibnamefont {Hauke}}, \bibinfo {author}
  {\bibfnamefont {M.}~\bibnamefont {Lewenstein}}, \bibinfo {author}
  {\bibfnamefont {D.-S.}\ \bibnamefont {Lühmann}}, \bibinfo {author}
  {\bibfnamefont {B.~A.}\ \bibnamefont {Malomed}}, \bibinfo {author}
  {\bibfnamefont {T.}~\bibnamefont {Sowinski}}, \ and\ \bibinfo {author}
  {\bibfnamefont {J.}~\bibnamefont {Zakrzewski}}} (\bibinfo {year}
  {2015}{\natexlab{a}}),\ \href
  {http://stacks.iop.org/0034-4885/78/i=6/a=066001} {\bibfield  {journal}
  {\bibinfo  {journal} {Rep. Prog. Phys.}\ }\textbf {\bibinfo {volume}
  {78}}~(\bibinfo {number} {6}),\ \bibinfo {pages} {066001}}\BibitemShut
  {NoStop}%
\bibitem [{\citenamefont {Dutta}\ \emph
  {et~al.}(2014{\natexlab{a}})\citenamefont {Dutta}, \citenamefont
  {Lewenstein},\ and\ \citenamefont {Zakrzewski}}]{2014_Dutta_Lewenstein_NJP}%
  \BibitemOpen
  \bibfield  {author} {\bibinfo {author} {\bibnamefont {Dutta}, \bibfnamefont
  {O.}}, \bibinfo {author} {\bibfnamefont {M.}~\bibnamefont {Lewenstein}}, \
  and\ \bibinfo {author} {\bibfnamefont {J.}~\bibnamefont {Zakrzewski}}}
  (\bibinfo {year} {2014}{\natexlab{a}}),\ \href
  {http://stacks.iop.org/1367-2630/16/i=5/a=052002} {\bibfield  {journal}
  {\bibinfo  {journal} {New Journal of Physics}\ }\textbf {\bibinfo {volume}
  {16}}~(\bibinfo {number} {5}),\ \bibinfo {pages} {052002}}\BibitemShut
  {NoStop}%
\bibitem [{\citenamefont {Dutta}\ \emph
  {et~al.}(2014{\natexlab{b}})\citenamefont {Dutta}, \citenamefont
  {Przysiezna},\ and\ \citenamefont {Lewenstein}}]{2014_Dutta_Przysiezna_PRA}%
  \BibitemOpen
  \bibfield  {author} {\bibinfo {author} {\bibnamefont {Dutta}, \bibfnamefont
  {O.}}, \bibinfo {author} {\bibfnamefont {A.}~\bibnamefont {Przysiezna}}, \
  and\ \bibinfo {author} {\bibfnamefont {M.}~\bibnamefont {Lewenstein}}}
  (\bibinfo {year} {2014}{\natexlab{b}}),\ \href {\doibase
  10.1103/PhysRevA.89.043602} {\bibfield  {journal} {\bibinfo  {journal} {Phys.
  Rev. A}\ }\textbf {\bibinfo {volume} {89}},\ \bibinfo {pages}
  {043602}}\BibitemShut {NoStop}%
\bibitem [{\citenamefont {Dutta}\ \emph
  {et~al.}(2015{\natexlab{b}})\citenamefont {Dutta}, \citenamefont
  {Przysiezna},\ and\ \citenamefont {Zakrzewski}}]{2015_Dutta_Przysiezna_SR}%
  \BibitemOpen
  \bibfield  {author} {\bibinfo {author} {\bibnamefont {Dutta}, \bibfnamefont
  {O.}}, \bibinfo {author} {\bibfnamefont {A.}~\bibnamefont {Przysiezna}}, \
  and\ \bibinfo {author} {\bibfnamefont {J.}~\bibnamefont {Zakrzewski}}}
  (\bibinfo {year} {2015}{\natexlab{b}}),\ \href
  {http://dx.doi.org/10.1038/srep11060} {\bibfield  {journal} {\bibinfo
  {journal} {Scientific Reports}\ }\textbf {\bibinfo {volume} {5}},\ \bibinfo
  {pages} {11060}}\BibitemShut {NoStop}%
\bibitem [{\citenamefont {Feiguin}\ and\ \citenamefont
  {Heidrich-Meisner}(2007)}]{2007_Feiguin_Heidrich_PRB}%
  \BibitemOpen
  \bibfield  {author} {\bibinfo {author} {\bibnamefont {Feiguin}, \bibfnamefont
  {A.~E.}}, \ and\ \bibinfo {author} {\bibfnamefont {F.}~\bibnamefont
  {Heidrich-Meisner}}} (\bibinfo {year} {2007}),\ \href {\doibase
  10.1103/PhysRevB.76.220508} {\bibfield  {journal} {\bibinfo  {journal} {Phys.
  Rev. B}\ }\textbf {\bibinfo {volume} {76}},\ \bibinfo {pages}
  {220508}}\BibitemShut {NoStop}%
\bibitem [{\citenamefont {Fort}\ \emph {et~al.}(2011)\citenamefont {Fort},
  \citenamefont {Fabbri}, \citenamefont {Fallani}, \citenamefont {Clément},\
  and\ \citenamefont {Inguscio}}]{2011_Fort_Fabbri_JPCS}%
  \BibitemOpen
  \bibfield  {author} {\bibinfo {author} {\bibnamefont {Fort}, \bibfnamefont
  {C.}}, \bibinfo {author} {\bibfnamefont {N.}~\bibnamefont {Fabbri}}, \bibinfo
  {author} {\bibfnamefont {L.}~\bibnamefont {Fallani}}, \bibinfo {author}
  {\bibfnamefont {D.}~\bibnamefont {Clément}}, \ and\ \bibinfo {author}
  {\bibfnamefont {M.}~\bibnamefont {Inguscio}}} (\bibinfo {year} {2011}),\
  \href {http://stacks.iop.org/1742-6596/264/i=1/a=012018} {\bibfield
  {journal} {\bibinfo  {journal} {Journal of Physics: Conference Series}\
  }\textbf {\bibinfo {volume} {264}}~(\bibinfo {number} {1}),\ \bibinfo {pages}
  {012018}}\BibitemShut {NoStop}%
\bibitem [{\citenamefont {Galitski}\ and\ \citenamefont
  {Spielman}(2013)}]{2013_Galitski_Nat}%
  \BibitemOpen
  \bibfield  {author} {\bibinfo {author} {\bibnamefont {Galitski},
  \bibfnamefont {V.}}, \ and\ \bibinfo {author} {\bibfnamefont {I.~B.}\
  \bibnamefont {Spielman}}} (\bibinfo {year} {2013}),\ \href
  {http://dx.doi.org/10.1038/nature11841} {\bibfield  {journal} {\bibinfo
  {journal} {Nature}\ }\textbf {\bibinfo {volume} {494}}~(\bibinfo {number}
  {7435}),\ \bibinfo {pages} {49}}\BibitemShut {NoStop}%
\bibitem [{\citenamefont {Ganczarek}\ \emph {et~al.}(2014)\citenamefont
  {Ganczarek}, \citenamefont {Modugno}, \citenamefont {Pettini},\ and\
  \citenamefont {Zakrzewski}}]{2014_Ganczarek_Modugno_PRA}%
  \BibitemOpen
  \bibfield  {author} {\bibinfo {author} {\bibnamefont {Ganczarek},
  \bibfnamefont {W.}}, \bibinfo {author} {\bibfnamefont {M.}~\bibnamefont
  {Modugno}}, \bibinfo {author} {\bibfnamefont {G.}~\bibnamefont {Pettini}}, \
  and\ \bibinfo {author} {\bibfnamefont {J.}~\bibnamefont {Zakrzewski}}}
  (\bibinfo {year} {2014}),\ \href {\doibase 10.1103/PhysRevA.90.033621}
  {\bibfield  {journal} {\bibinfo  {journal} {Phys. Rev. A}\ }\textbf {\bibinfo
  {volume} {90}},\ \bibinfo {pages} {033621}}\BibitemShut {NoStop}%
\bibitem [{\citenamefont {Garcia-March}\ \emph {et~al.}(2012)\citenamefont
  {Garcia-March}, \citenamefont {Dounas-Frazer},\ and\ \citenamefont
  {Carr}}]{2012_Garcia-March_Dounas-Frazer_FrontPhys}%
  \BibitemOpen
  \bibfield  {author} {\bibinfo {author} {\bibnamefont {Garcia-March},
  \bibfnamefont {M.}}, \bibinfo {author} {\bibfnamefont {D.}~\bibnamefont
  {Dounas-Frazer}}, \ and\ \bibinfo {author} {\bibfnamefont {L.}~\bibnamefont
  {Carr}}} (\bibinfo {year} {2012}),\ \href {\doibase
  10.1007/s11467-011-0236-6} {\bibfield  {journal} {\bibinfo  {journal}
  {Frontiers of Physics}\ }\textbf {\bibinfo {volume} {7}}~(\bibinfo {number}
  {1}),\ \bibinfo {pages} {131}}\BibitemShut {NoStop}%
\bibitem [{\citenamefont {Garcia-March}\ and\ \citenamefont
  {Carr}(2015)}]{2015_Garcia-March_Carr_PRA}%
  \BibitemOpen
  \bibfield  {author} {\bibinfo {author} {\bibnamefont {Garcia-March},
  \bibfnamefont {M.~A.}}, \ and\ \bibinfo {author} {\bibfnamefont {L.~D.}\
  \bibnamefont {Carr}}} (\bibinfo {year} {2015}),\ \href {\doibase
  10.1103/PhysRevA.91.033626} {\bibfield  {journal} {\bibinfo  {journal} {Phys.
  Rev. A}\ }\textbf {\bibinfo {volume} {91}},\ \bibinfo {pages}
  {033626}}\BibitemShut {NoStop}%
\bibitem [{\citenamefont {Garc\'{i}a-March}\ \emph {et~al.}(2011)\citenamefont
  {Garc\'{i}a-March}, \citenamefont {Dounas-Frazer},\ and\ \citenamefont
  {Carr}}]{2011_Garcia-March_Dounas-Frazer_PRA}%
  \BibitemOpen
  \bibfield  {author} {\bibinfo {author} {\bibnamefont {Garc\'{i}a-March},
  \bibfnamefont {M.~A.}}, \bibinfo {author} {\bibfnamefont {D.~R.}\
  \bibnamefont {Dounas-Frazer}}, \ and\ \bibinfo {author} {\bibfnamefont
  {L.~D.}\ \bibnamefont {Carr}}} (\bibinfo {year} {2011}),\ \href {\doibase
  10.1103/PhysRevA.83.043612} {\bibfield  {journal} {\bibinfo  {journal} {Phys.
  Rev. A}\ }\textbf {\bibinfo {volume} {83}},\ \bibinfo {pages}
  {043612}}\BibitemShut {NoStop}%
\bibitem [{\citenamefont {Gemelke}\ \emph {et~al.}(2010)\citenamefont
  {Gemelke}, \citenamefont {Sarajlic},\ and\ \citenamefont
  {Chu}}]{2010_Gemelke_Sarajlic_arXiv}%
  \BibitemOpen
  \bibfield  {author} {\bibinfo {author} {\bibnamefont {Gemelke}, \bibfnamefont
  {N.}}, \bibinfo {author} {\bibfnamefont {E.}~\bibnamefont {Sarajlic}}, \ and\
  \bibinfo {author} {\bibfnamefont {S.}~\bibnamefont {Chu}}} (\bibinfo {year}
  {2010}),\ \href {http://adsabs.harvard.edu/abs/2010arXiv1007.2677G} {\bibinfo
   {journal} {arXiv preprint arXiv:1007.2677}\ }\BibitemShut {NoStop}%
\bibitem [{\citenamefont {Gillet}\ \emph {et~al.}(2014)\citenamefont {Gillet},
  \citenamefont {Garcia-March}, \citenamefont {Busch},\ and\ \citenamefont
  {Sols}}]{2014_Gillet_Garcia-March_PRA}%
  \BibitemOpen
\bibfield  {journal} {  }\bibfield  {author} {\bibinfo {author} {\bibnamefont
  {Gillet}, \bibfnamefont {J.}}, \bibinfo {author} {\bibfnamefont {M.~A.}\
  \bibnamefont {Garcia-March}}, \bibinfo {author} {\bibfnamefont
  {T.}~\bibnamefont {Busch}}, \ and\ \bibinfo {author} {\bibfnamefont
  {F.}~\bibnamefont {Sols}}} (\bibinfo {year} {2014}),\ \href {\doibase
  10.1103/PhysRevA.89.023614} {\bibfield  {journal} {\bibinfo  {journal} {Phys.
  Rev. A}\ }\textbf {\bibinfo {volume} {89}},\ \bibinfo {pages}
  {023614}}\BibitemShut {NoStop}%
\bibitem [{\citenamefont {Greiner}\ \emph {et~al.}(2001)\citenamefont
  {Greiner}, \citenamefont {Bloch}, \citenamefont {Mandel}, \citenamefont
  {H\"ansch},\ and\ \citenamefont {Esslinger}}]{2001_Greiner_Bloch_PRL}%
  \BibitemOpen
  \bibfield  {author} {\bibinfo {author} {\bibnamefont {Greiner}, \bibfnamefont
  {M.}}, \bibinfo {author} {\bibfnamefont {I.}~\bibnamefont {Bloch}}, \bibinfo
  {author} {\bibfnamefont {O.}~\bibnamefont {Mandel}}, \bibinfo {author}
  {\bibfnamefont {T.~W.}\ \bibnamefont {H\"ansch}}, \ and\ \bibinfo {author}
  {\bibfnamefont {T.}~\bibnamefont {Esslinger}}} (\bibinfo {year} {2001}),\
  \href {\doibase 10.1103/PhysRevLett.87.160405} {\bibfield  {journal}
  {\bibinfo  {journal} {Phys. Rev. Lett.}\ }\textbf {\bibinfo {volume} {87}},\
  \bibinfo {pages} {160405}}\BibitemShut {NoStop}%
\bibitem [{\citenamefont {Guan}\ \emph {et~al.}(2007)\citenamefont {Guan},
  \citenamefont {Batchelor}, \citenamefont {Lee},\ and\ \citenamefont
  {Bortz}}]{2007_Guan_Batchelor_PRB}%
  \BibitemOpen
  \bibfield  {author} {\bibinfo {author} {\bibnamefont {Guan}, \bibfnamefont
  {X.~W.}}, \bibinfo {author} {\bibfnamefont {M.~T.}\ \bibnamefont
  {Batchelor}}, \bibinfo {author} {\bibfnamefont {C.}~\bibnamefont {Lee}}, \
  and\ \bibinfo {author} {\bibfnamefont {M.}~\bibnamefont {Bortz}}} (\bibinfo
  {year} {2007}),\ \href {\doibase 10.1103/PhysRevB.76.085120} {\bibfield
  {journal} {\bibinfo  {journal} {Phys. Rev. B}\ }\textbf {\bibinfo {volume}
  {76}},\ \bibinfo {pages} {085120}}\BibitemShut {NoStop}%
\bibitem [{\citenamefont {Guo}\ \emph {et~al.}(2007)\citenamefont {Guo},
  \citenamefont {Zhang},\ and\ \citenamefont {Chen}}]{2007_Guo_Zhang_PRA}%
  \BibitemOpen
  \bibfield  {author} {\bibinfo {author} {\bibnamefont {Guo}, \bibfnamefont
  {L.}}, \bibinfo {author} {\bibfnamefont {Y.}~\bibnamefont {Zhang}}, \ and\
  \bibinfo {author} {\bibfnamefont {S.}~\bibnamefont {Chen}}} (\bibinfo {year}
  {2007}),\ \href {\doibase 10.1103/PhysRevA.75.013622} {\bibfield  {journal}
  {\bibinfo  {journal} {Phys. Rev. A}\ }\textbf {\bibinfo {volume} {75}},\
  \bibinfo {pages} {013622}}\BibitemShut {NoStop}%
\bibitem [{\citenamefont {Hart}\ \emph {et~al.}(2015)\citenamefont {Hart},
  \citenamefont {Duarte}, \citenamefont {Yang}, \citenamefont {Liu},
  \citenamefont {Paiva}, \citenamefont {Khatami}, \citenamefont {Scalettar},
  \citenamefont {Trivedi}, \citenamefont {Huse},\ and\ \citenamefont
  {Hulet}}]{2015_Hart_Duarte_Nature}%
  \BibitemOpen
  \bibfield  {author} {\bibinfo {author} {\bibnamefont {Hart}, \bibfnamefont
  {R.~A.}}, \bibinfo {author} {\bibfnamefont {P.~M.}\ \bibnamefont {Duarte}},
  \bibinfo {author} {\bibfnamefont {T.-L.}\ \bibnamefont {Yang}}, \bibinfo
  {author} {\bibfnamefont {X.}~\bibnamefont {Liu}}, \bibinfo {author}
  {\bibfnamefont {T.}~\bibnamefont {Paiva}}, \bibinfo {author} {\bibfnamefont
  {E.}~\bibnamefont {Khatami}}, \bibinfo {author} {\bibfnamefont {R.~T.}\
  \bibnamefont {Scalettar}}, \bibinfo {author} {\bibfnamefont {N.}~\bibnamefont
  {Trivedi}}, \bibinfo {author} {\bibfnamefont {D.~A.}\ \bibnamefont {Huse}}, \
  and\ \bibinfo {author} {\bibfnamefont {R.~G.}\ \bibnamefont {Hulet}}}
  (\bibinfo {year} {2015}),\ \href {http://dx.doi.org/10.1038/nature14223}
  {\bibfield  {journal} {\bibinfo  {journal} {Nature}\ }\textbf {\bibinfo
  {volume} {519}}~(\bibinfo {number} {7542}),\ \bibinfo {pages}
  {211}}\BibitemShut {NoStop}%
\bibitem [{\citenamefont {Hauke}\ \emph {et~al.}(2011)\citenamefont {Hauke},
  \citenamefont {Zhao}, \citenamefont {Goyal}, \citenamefont {Deutsch},
  \citenamefont {Liu},\ and\ \citenamefont
  {Lewenstein}}]{2011_Hauke_Zhao_PRAR}%
  \BibitemOpen
  \bibfield  {author} {\bibinfo {author} {\bibnamefont {Hauke}, \bibfnamefont
  {P.}}, \bibinfo {author} {\bibfnamefont {E.}~\bibnamefont {Zhao}}, \bibinfo
  {author} {\bibfnamefont {K.}~\bibnamefont {Goyal}}, \bibinfo {author}
  {\bibfnamefont {I.~H.}\ \bibnamefont {Deutsch}}, \bibinfo {author}
  {\bibfnamefont {W.~V.}\ \bibnamefont {Liu}}, \ and\ \bibinfo {author}
  {\bibfnamefont {M.}~\bibnamefont {Lewenstein}}} (\bibinfo {year} {2011}),\
  \href {\doibase 10.1103/PhysRevA.84.051603} {\bibfield  {journal} {\bibinfo
  {journal} {Phys. Rev. A}\ }\textbf {\bibinfo {volume} {84}},\ \bibinfo
  {pages} {051603}}\BibitemShut {NoStop}%
\bibitem [{\citenamefont {H\'ebert}\ \emph {et~al.}(2013)\citenamefont
  {H\'ebert}, \citenamefont {Cai}, \citenamefont {Rousseau}, \citenamefont
  {Wu}, \citenamefont {Scalettar},\ and\ \citenamefont
  {Batrouni}}]{2013_Hebert_Cai_PRB}%
  \BibitemOpen
  \bibfield  {author} {\bibinfo {author} {\bibnamefont {H\'ebert},
  \bibfnamefont {F.}}, \bibinfo {author} {\bibfnamefont {Z.}~\bibnamefont
  {Cai}}, \bibinfo {author} {\bibfnamefont {V.~G.}\ \bibnamefont {Rousseau}},
  \bibinfo {author} {\bibfnamefont {C.}~\bibnamefont {Wu}}, \bibinfo {author}
  {\bibfnamefont {R.~T.}\ \bibnamefont {Scalettar}}, \ and\ \bibinfo {author}
  {\bibfnamefont {G.~G.}\ \bibnamefont {Batrouni}}} (\bibinfo {year} {2013}),\
  \href {\doibase 10.1103/PhysRevB.87.224505} {\bibfield  {journal} {\bibinfo
  {journal} {Phys. Rev. B}\ }\textbf {\bibinfo {volume} {87}},\ \bibinfo
  {pages} {224505}}\BibitemShut {NoStop}%
\bibitem [{\citenamefont {Hemmerich}(2014)}]{2015_Hemmerich_private}%
  \BibitemOpen
  \bibfield  {author} {\bibinfo {author} {\bibnamefont {Hemmerich},
  \bibfnamefont {A.}}} (\bibinfo {year} {2014}),\ \href@noop {} {}\bibinfo
  {howpublished} {private communication}\BibitemShut {NoStop}%
\bibitem [{\citenamefont {Hofer}\ \emph {et~al.}(2012)\citenamefont {Hofer},
  \citenamefont {Bruder},\ and\ \citenamefont {Stojanovi\ifmmode~\acute{c}\else
  \'{c}\fi{}}}]{2012_Hofer_Bruder_PRA}%
  \BibitemOpen
  \bibfield  {author} {\bibinfo {author} {\bibnamefont {Hofer}, \bibfnamefont
  {P.~P.}}, \bibinfo {author} {\bibfnamefont {C.}~\bibnamefont {Bruder}}, \
  and\ \bibinfo {author} {\bibfnamefont {V.~M.}\ \bibnamefont
  {Stojanovi\ifmmode~\acute{c}\else \'{c}\fi{}}}} (\bibinfo {year} {2012}),\
  \href {\doibase 10.1103/PhysRevA.86.033627} {\bibfield  {journal} {\bibinfo
  {journal} {Phys. Rev. A}\ }\textbf {\bibinfo {volume} {86}},\ \bibinfo
  {pages} {033627}}\BibitemShut {NoStop}%
\bibitem [{\citenamefont {{Hu}}\ \emph {et~al.}(2015)\citenamefont {{Hu}},
  \citenamefont {{Niu}}, \citenamefont {{Yang}}, \citenamefont {{Chen}},
  \citenamefont {{Wu}}, \citenamefont {{Xiong}},\ and\ \citenamefont
  {{Zhou}}}]{2015_Hu_Niu_arXiv}%
  \BibitemOpen
  \bibfield  {author} {\bibinfo {author} {\bibnamefont {{Hu}}, \bibfnamefont
  {D.}}, \bibinfo {author} {\bibfnamefont {L.}~\bibnamefont {{Niu}}}, \bibinfo
  {author} {\bibfnamefont {B.}~\bibnamefont {{Yang}}}, \bibinfo {author}
  {\bibfnamefont {X.}~\bibnamefont {{Chen}}}, \bibinfo {author} {\bibfnamefont
  {B.}~\bibnamefont {{Wu}}}, \bibinfo {author} {\bibfnamefont {H.}~\bibnamefont
  {{Xiong}}}, \ and\ \bibinfo {author} {\bibfnamefont {X.}~\bibnamefont
  {{Zhou}}}} (\bibinfo {year} {2015}),\ \href@noop {} {\bibfield  {journal}
  {\bibinfo  {journal} {ArXiv e-prints}\ }}\Eprint
  {http://arxiv.org/abs/1507.08014} {arXiv:1507.08014 [cond-mat.quant-gas]}
  \BibitemShut {NoStop}%
\bibitem [{\citenamefont {Hu}\ \emph {et~al.}(2007)\citenamefont {Hu},
  \citenamefont {Liu},\ and\ \citenamefont {Drummond}}]{2007_Hu_Liu_PRL}%
  \BibitemOpen
  \bibfield  {author} {\bibinfo {author} {\bibnamefont {Hu}, \bibfnamefont
  {H.}}, \bibinfo {author} {\bibfnamefont {X.-J.}\ \bibnamefont {Liu}}, \ and\
  \bibinfo {author} {\bibfnamefont {P.~D.}\ \bibnamefont {Drummond}}} (\bibinfo
  {year} {2007}),\ \href {\doibase 10.1103/PhysRevLett.98.070403} {\bibfield
  {journal} {\bibinfo  {journal} {Phys. Rev. Lett.}\ }\textbf {\bibinfo
  {volume} {98}},\ \bibinfo {pages} {070403}}\BibitemShut {NoStop}%
\bibitem [{\citenamefont {Hui}\ \emph {et~al.}(2012)\citenamefont {Hui},
  \citenamefont {Barnett}, \citenamefont {Porto},\ and\ \citenamefont
  {Das~Sarma}}]{2012_Hui_Barnett_PRA}%
  \BibitemOpen
  \bibfield  {author} {\bibinfo {author} {\bibnamefont {Hui}, \bibfnamefont
  {H.-Y.}}, \bibinfo {author} {\bibfnamefont {R.}~\bibnamefont {Barnett}},
  \bibinfo {author} {\bibfnamefont {J.~V.}\ \bibnamefont {Porto}}, \ and\
  \bibinfo {author} {\bibfnamefont {S.}~\bibnamefont {Das~Sarma}}} (\bibinfo
  {year} {2012}),\ \href {\doibase 10.1103/PhysRevA.86.063636} {\bibfield
  {journal} {\bibinfo  {journal} {Phys. Rev. A}\ }\textbf {\bibinfo {volume}
  {86}},\ \bibinfo {pages} {063636}}\BibitemShut {NoStop}%
\bibitem [{\citenamefont {Hung}\ \emph {et~al.}(2011)\citenamefont {Hung},
  \citenamefont {Lee},\ and\ \citenamefont {Wu}}]{2011_Hung_Lee_PRB}%
  \BibitemOpen
  \bibfield  {author} {\bibinfo {author} {\bibnamefont {Hung}, \bibfnamefont
  {H.-H.}}, \bibinfo {author} {\bibfnamefont {W.-C.}\ \bibnamefont {Lee}}, \
  and\ \bibinfo {author} {\bibfnamefont {C.}~\bibnamefont {Wu}}} (\bibinfo
  {year} {2011}),\ \href {\doibase 10.1103/PhysRevB.83.144506} {\bibfield
  {journal} {\bibinfo  {journal} {Phys. Rev. B}\ }\textbf {\bibinfo {volume}
  {83}},\ \bibinfo {pages} {144506}}\BibitemShut {NoStop}%
\bibitem [{\citenamefont {Isacsson}\ and\ \citenamefont
  {Girvin}(2005)}]{2005_Isacsson_Girvin_PRA}%
  \BibitemOpen
  \bibfield  {author} {\bibinfo {author} {\bibnamefont {Isacsson},
  \bibfnamefont {A.}}, \ and\ \bibinfo {author} {\bibfnamefont {S.~M.}\
  \bibnamefont {Girvin}}} (\bibinfo {year} {2005}),\ \href {\doibase
  10.1103/PhysRevA.72.053604} {\bibfield  {journal} {\bibinfo  {journal} {Phys.
  Rev. A}\ }\textbf {\bibinfo {volume} {72}},\ \bibinfo {pages}
  {053604}}\BibitemShut {NoStop}%
\bibitem [{\citenamefont {{Jahn}}\ and\ \citenamefont
  {{Teller}}(1937)}]{1937_Jahn_Teller_RSLPSA}%
  \BibitemOpen
  \bibfield  {author} {\bibinfo {author} {\bibnamefont {{Jahn}}, \bibfnamefont
  {H.~A.}}, \ and\ \bibinfo {author} {\bibfnamefont {E.}~\bibnamefont
  {{Teller}}}} (\bibinfo {year} {1937}),\ \href {\doibase
  10.1098/rspa.1937.0142} {\bibfield  {journal} {\bibinfo  {journal} {Royal
  Society of London Proceedings Series A}\ }\textbf {\bibinfo {volume} {161}},\
  \bibinfo {pages} {220}}\BibitemShut {NoStop}%
\bibitem [{\citenamefont {Jaksch}\ \emph {et~al.}(1998)\citenamefont {Jaksch},
  \citenamefont {Bruder}, \citenamefont {Cirac}, \citenamefont {Gardiner},\
  and\ \citenamefont {Zoller}}]{1998_Jaksch_Bruder_PRL}%
  \BibitemOpen
  \bibfield  {author} {\bibinfo {author} {\bibnamefont {Jaksch}, \bibfnamefont
  {D.}}, \bibinfo {author} {\bibfnamefont {C.}~\bibnamefont {Bruder}}, \bibinfo
  {author} {\bibfnamefont {J.~I.}\ \bibnamefont {Cirac}}, \bibinfo {author}
  {\bibfnamefont {C.~W.}\ \bibnamefont {Gardiner}}, \ and\ \bibinfo {author}
  {\bibfnamefont {P.}~\bibnamefont {Zoller}}} (\bibinfo {year} {1998}),\ \href
  {\doibase 10.1103/PhysRevLett.81.3108} {\bibfield  {journal} {\bibinfo
  {journal} {Phys. Rev. Lett.}\ }\textbf {\bibinfo {volume} {81}},\ \bibinfo
  {pages} {3108}}\BibitemShut {NoStop}%
\bibitem [{\citenamefont {Johnson}\ \emph {et~al.}(2009)\citenamefont
  {Johnson}, \citenamefont {Tiesinga}, \citenamefont {Porto},\ and\
  \citenamefont {Williams}}]{2009_Johnson_Tiesinga_NJP}%
  \BibitemOpen
  \bibfield  {author} {\bibinfo {author} {\bibnamefont {Johnson}, \bibfnamefont
  {P.~R.}}, \bibinfo {author} {\bibfnamefont {E.}~\bibnamefont {Tiesinga}},
  \bibinfo {author} {\bibfnamefont {J.~V.}\ \bibnamefont {Porto}}, \ and\
  \bibinfo {author} {\bibfnamefont {C.~J.}\ \bibnamefont {Williams}}} (\bibinfo
  {year} {2009}),\ \href {http://stacks.iop.org/1367-2630/11/i=9/a=093022}
  {\bibfield  {journal} {\bibinfo  {journal} {New Journal of Physics}\ }\textbf
  {\bibinfo {volume} {11}}~(\bibinfo {number} {9}),\ \bibinfo {pages}
  {093022}}\BibitemShut {NoStop}%
\bibitem [{\citenamefont {Jona-Lasinio}\ \emph {et~al.}(2003)\citenamefont
  {Jona-Lasinio}, \citenamefont {Morsch}, \citenamefont {Cristiani},
  \citenamefont {Malossi}, \citenamefont {M\"uller}, \citenamefont {Courtade},
  \citenamefont {Anderlini},\ and\ \citenamefont
  {Arimondo}}]{2003_Lasinio_Morsch_PRL}%
  \BibitemOpen
  \bibfield  {author} {\bibinfo {author} {\bibnamefont {Jona-Lasinio},
  \bibfnamefont {M.}}, \bibinfo {author} {\bibfnamefont {O.}~\bibnamefont
  {Morsch}}, \bibinfo {author} {\bibfnamefont {M.}~\bibnamefont {Cristiani}},
  \bibinfo {author} {\bibfnamefont {N.}~\bibnamefont {Malossi}}, \bibinfo
  {author} {\bibfnamefont {J.~H.}\ \bibnamefont {M\"uller}}, \bibinfo {author}
  {\bibfnamefont {E.}~\bibnamefont {Courtade}}, \bibinfo {author}
  {\bibfnamefont {M.}~\bibnamefont {Anderlini}}, \ and\ \bibinfo {author}
  {\bibfnamefont {E.}~\bibnamefont {Arimondo}}} (\bibinfo {year} {2003}),\
  \href {\doibase 10.1103/PhysRevLett.91.230406} {\bibfield  {journal}
  {\bibinfo  {journal} {Phys. Rev. Lett.}\ }\textbf {\bibinfo {volume} {91}},\
  \bibinfo {pages} {230406}}\BibitemShut {NoStop}%
\bibitem [{\citenamefont {Jotzu}\ \emph {et~al.}(2014)\citenamefont {Jotzu},
  \citenamefont {Messer}, \citenamefont {Desbuquois}, \citenamefont {Lebrat},
  \citenamefont {Uehlinger}, \citenamefont {Greif},\ and\ \citenamefont
  {Esslinger}}]{2014_Jotzu_Messer_Nature}%
  \BibitemOpen
  \bibfield  {author} {\bibinfo {author} {\bibnamefont {Jotzu}, \bibfnamefont
  {G.}}, \bibinfo {author} {\bibfnamefont {M.}~\bibnamefont {Messer}}, \bibinfo
  {author} {\bibfnamefont {R.}~\bibnamefont {Desbuquois}}, \bibinfo {author}
  {\bibfnamefont {M.}~\bibnamefont {Lebrat}}, \bibinfo {author} {\bibfnamefont
  {T.}~\bibnamefont {Uehlinger}}, \bibinfo {author} {\bibfnamefont
  {D.}~\bibnamefont {Greif}}, \ and\ \bibinfo {author} {\bibfnamefont
  {T.}~\bibnamefont {Esslinger}}} (\bibinfo {year} {2014}),\ \href
  {http://dx.doi.org/10.1038/nature13915} {\bibfield  {journal} {\bibinfo
  {journal} {Nature}\ }\textbf {\bibinfo {volume} {515}}~(\bibinfo {number}
  {7526}),\ \bibinfo {pages} {237}}\BibitemShut {NoStop}%
\bibitem [{\citenamefont {Kakashvili}\ and\ \citenamefont
  {Bolech}(2009)}]{2009_Kakashvili_Bolech_PRA}%
  \BibitemOpen
  \bibfield  {author} {\bibinfo {author} {\bibnamefont {Kakashvili},
  \bibfnamefont {P.}}, \ and\ \bibinfo {author} {\bibfnamefont {C.~J.}\
  \bibnamefont {Bolech}}} (\bibinfo {year} {2009}),\ \href {\doibase
  10.1103/PhysRevA.79.041603} {\bibfield  {journal} {\bibinfo  {journal} {Phys.
  Rev. A}\ }\textbf {\bibinfo {volume} {79}},\ \bibinfo {pages}
  {041603}}\BibitemShut {NoStop}%
\bibitem [{\citenamefont {Kamihara}\ \emph {et~al.}(2006)\citenamefont
  {Kamihara}, \citenamefont {Hiramatsu}, \citenamefont {Hirano}, \citenamefont
  {Kawamura}, \citenamefont {Yanagi}, \citenamefont {Kamiya},\ and\
  \citenamefont {Hosono}}]{2006_Kamihara_Hiramatsu_JACS}%
  \BibitemOpen
  \bibfield  {author} {\bibinfo {author} {\bibnamefont {Kamihara},
  \bibfnamefont {Y.}}, \bibinfo {author} {\bibfnamefont {H.}~\bibnamefont
  {Hiramatsu}}, \bibinfo {author} {\bibfnamefont {M.}~\bibnamefont {Hirano}},
  \bibinfo {author} {\bibfnamefont {R.}~\bibnamefont {Kawamura}}, \bibinfo
  {author} {\bibfnamefont {H.}~\bibnamefont {Yanagi}}, \bibinfo {author}
  {\bibfnamefont {T.}~\bibnamefont {Kamiya}}, \ and\ \bibinfo {author}
  {\bibfnamefont {H.}~\bibnamefont {Hosono}}} (\bibinfo {year} {2006}),\ \href
  {\doibase 10.1021/ja063355c} {\bibfield  {journal} {\bibinfo  {journal}
  {Journal of the American Chemical Society}\ }\textbf {\bibinfo {volume}
  {128}}~(\bibinfo {number} {31}),\ \bibinfo {pages} {10012}},\ \bibinfo {note}
  {pMID: 16881620}\BibitemShut {NoStop}%
\bibitem [{\citenamefont {Kantian}\ \emph {et~al.}(2007)\citenamefont
  {Kantian}, \citenamefont {Daley},\ and\ \citenamefont
  {Zoller}}]{2007_Kantian_Daley_NJPHYS}%
  \BibitemOpen
  \bibfield  {author} {\bibinfo {author} {\bibnamefont {Kantian}, \bibfnamefont
  {A.}}, \bibinfo {author} {\bibfnamefont {A.}~\bibnamefont {Daley}}, \ and\
  \bibinfo {author} {\bibfnamefont {P.}~\bibnamefont {Zoller}}} (\bibinfo
  {year} {2007}),\ \href {\doibase 10.1088/1367-2630/9/11/407} {\bibfield
  {journal} {\bibinfo  {journal} {New Journal of Physics}\ }\textbf {\bibinfo
  {volume} {9}},\ \bibinfo {pages} {407}}\BibitemShut {NoStop}%
\bibitem [{\citenamefont {{Khamehchi}}\ \emph {et~al.}(2015)\citenamefont
  {{Khamehchi}}, \citenamefont {{Qu}}, \citenamefont {{Mossman}}, \citenamefont
  {{Zhang}},\ and\ \citenamefont {{Engels}}}]{2015_Khamehchi_Qu_arXiv}%
  \BibitemOpen
  \bibfield  {author} {\bibinfo {author} {\bibnamefont {{Khamehchi}},
  \bibfnamefont {M.~A.}}, \bibinfo {author} {\bibfnamefont {C.}~\bibnamefont
  {{Qu}}}, \bibinfo {author} {\bibfnamefont {M.~E.}\ \bibnamefont {{Mossman}}},
  \bibinfo {author} {\bibfnamefont {C.}~\bibnamefont {{Zhang}}}, \ and\
  \bibinfo {author} {\bibfnamefont {P.}~\bibnamefont {{Engels}}}} (\bibinfo
  {year} {2015}),\ \href@noop {} {\bibinfo  {journal} {Nat Commun}\ ,\ \bibinfo
  {pages} {10867}}\BibitemShut {NoStop}%
\bibitem [{\citenamefont {Kivelson}(1982)}]{1982_Kivelson_PRB}%
  \BibitemOpen
\bibfield  {journal} {  }\bibfield  {author} {\bibinfo {author} {\bibnamefont
  {Kivelson}, \bibfnamefont {S.}}} (\bibinfo {year} {1982}),\ \href {\doibase
  10.1103/PhysRevB.26.4269} {\bibfield  {journal} {\bibinfo  {journal} {Phys.
  Rev. B}\ }\textbf {\bibinfo {volume} {26}},\ \bibinfo {pages}
  {4269}}\BibitemShut {NoStop}%
\bibitem [{\citenamefont {Kock}\ \emph {et~al.}(2015)\citenamefont {Kock},
  \citenamefont {\"Olschl\"ager}, \citenamefont {Ewerbeck}, \citenamefont
  {Huang}, \citenamefont {Mathey},\ and\ \citenamefont
  {Hemmerich}}]{2015_Kock_Olschlager_PRL}%
  \BibitemOpen
  \bibfield  {author} {\bibinfo {author} {\bibnamefont {Kock}, \bibfnamefont
  {T.}}, \bibinfo {author} {\bibfnamefont {M.}~\bibnamefont {\"Olschl\"ager}},
  \bibinfo {author} {\bibfnamefont {A.}~\bibnamefont {Ewerbeck}}, \bibinfo
  {author} {\bibfnamefont {W.-M.}\ \bibnamefont {Huang}}, \bibinfo {author}
  {\bibfnamefont {L.}~\bibnamefont {Mathey}}, \ and\ \bibinfo {author}
  {\bibfnamefont {A.}~\bibnamefont {Hemmerich}}} (\bibinfo {year} {2015}),\
  \href {\doibase 10.1103/PhysRevLett.114.115301} {\bibfield  {journal}
  {\bibinfo  {journal} {Phys. Rev. Lett.}\ }\textbf {\bibinfo {volume} {114}},\
  \bibinfo {pages} {115301}}\BibitemShut {NoStop}%
\bibitem [{\citenamefont {Kuklov}(2006)}]{2006_Kuklov_PRL}%
  \BibitemOpen
  \bibfield  {author} {\bibinfo {author} {\bibnamefont {Kuklov}, \bibfnamefont
  {A.~B.}}} (\bibinfo {year} {2006}),\ \href {\doibase
  10.1103/PhysRevLett.97.110405} {\bibfield  {journal} {\bibinfo  {journal}
  {Phys. Rev. Lett.}\ }\textbf {\bibinfo {volume} {97}},\ \bibinfo {pages}
  {110405}}\BibitemShut {NoStop}%
\bibitem [{\citenamefont {Lacki}\ \emph {et~al.}(2013)\citenamefont {Lacki},
  \citenamefont {Delande},\ and\ \citenamefont
  {Zakrzewski}}]{2013_Lacki_Delande_NJP}%
  \BibitemOpen
  \bibfield  {author} {\bibinfo {author} {\bibnamefont {Lacki}, \bibfnamefont
  {M.}}, \bibinfo {author} {\bibfnamefont {D.}~\bibnamefont {Delande}}, \ and\
  \bibinfo {author} {\bibfnamefont {J.}~\bibnamefont {Zakrzewski}}} (\bibinfo
  {year} {2013}),\ \href {http://stacks.iop.org/1367-2630/15/i=1/a=013062}
  {\bibfield  {journal} {\bibinfo  {journal} {New Journal of Physics}\ }\textbf
  {\bibinfo {volume} {15}}~(\bibinfo {number} {1}),\ \bibinfo {pages}
  {013062}}\BibitemShut {NoStop}%
\bibitem [{\citenamefont {Lacki}\ and\ \citenamefont
  {Zakrzewski}(2013)}]{2013_Lachi_Zakrzewski_PRL}%
  \BibitemOpen
  \bibfield  {author} {\bibinfo {author} {\bibnamefont {Lacki}, \bibfnamefont
  {M.}}, \ and\ \bibinfo {author} {\bibfnamefont {J.}~\bibnamefont
  {Zakrzewski}}} (\bibinfo {year} {2013}),\ \href {\doibase
  10.1103/PhysRevLett.110.065301} {\bibfield  {journal} {\bibinfo  {journal}
  {Phys. Rev. Lett.}\ }\textbf {\bibinfo {volume} {110}},\ \bibinfo {pages}
  {065301}}\BibitemShut {NoStop}%
\bibitem [{\citenamefont {Larson}\ \emph {et~al.}(2009)\citenamefont {Larson},
  \citenamefont {Collin},\ and\ \citenamefont
  {Martikainen}}]{2009_Larson_Collin_PRA}%
  \BibitemOpen
  \bibfield  {author} {\bibinfo {author} {\bibnamefont {Larson}, \bibfnamefont
  {J.}}, \bibinfo {author} {\bibfnamefont {A.}~\bibnamefont {Collin}}, \ and\
  \bibinfo {author} {\bibfnamefont {J.-P.}\ \bibnamefont {Martikainen}}}
  (\bibinfo {year} {2009}),\ \href {\doibase 10.1103/PhysRevA.79.033603}
  {\bibfield  {journal} {\bibinfo  {journal} {Phys. Rev. A}\ }\textbf {\bibinfo
  {volume} {79}},\ \bibinfo {pages} {033603}}\BibitemShut {NoStop}%
\bibitem [{\citenamefont {Larson}\ and\ \citenamefont
  {Martikainen}(2011)}]{2011_Larson_Martikainen_PRA}%
  \BibitemOpen
  \bibfield  {author} {\bibinfo {author} {\bibnamefont {Larson}, \bibfnamefont
  {J.}}, \ and\ \bibinfo {author} {\bibfnamefont {J.-P.}\ \bibnamefont
  {Martikainen}}} (\bibinfo {year} {2011}),\ \href {\doibase
  10.1103/PhysRevA.84.023621} {\bibfield  {journal} {\bibinfo  {journal} {Phys.
  Rev. A}\ }\textbf {\bibinfo {volume} {84}},\ \bibinfo {pages}
  {023621}}\BibitemShut {NoStop}%
\bibitem [{\citenamefont {Lee}\ \emph {et~al.}(2010)\citenamefont {Lee},
  \citenamefont {Wu},\ and\ \citenamefont {Das~Sarma}}]{2010_Lee_Wu_PRA}%
  \BibitemOpen
  \bibfield  {author} {\bibinfo {author} {\bibnamefont {Lee}, \bibfnamefont
  {W.-C.}}, \bibinfo {author} {\bibfnamefont {C.}~\bibnamefont {Wu}}, \ and\
  \bibinfo {author} {\bibfnamefont {S.}~\bibnamefont {Das~Sarma}}} (\bibinfo
  {year} {2010}),\ \href {\doibase 10.1103/PhysRevA.82.053611} {\bibfield
  {journal} {\bibinfo  {journal} {Phys. Rev. A}\ }\textbf {\bibinfo {volume}
  {82}},\ \bibinfo {pages} {053611}}\BibitemShut {NoStop}%
\bibitem [{\citenamefont {Lewenstein}\ and\ \citenamefont
  {Liu}(2011)}]{2011_Lewenstein_Liu_NatPhys}%
  \BibitemOpen
  \bibfield  {author} {\bibinfo {author} {\bibnamefont {Lewenstein},
  \bibfnamefont {M.}}, \ and\ \bibinfo {author} {\bibfnamefont {W.~V.}\
  \bibnamefont {Liu}}} (\bibinfo {year} {2011}),\ \href {\doibase
  10.1038/nphys1894} {\bibfield  {journal} {\bibinfo  {journal} {Nature
  Physics}\ }\textbf {\bibinfo {volume} {7}},\ \bibinfo {pages}
  {101}}\BibitemShut {NoStop}%
\bibitem [{\citenamefont {Lewenstein}\ \emph {et~al.}(2012)\citenamefont
  {Lewenstein}, \citenamefont {Sanpera},\ and\ \citenamefont
  {Ahufinger}}]{2012_Lewenstein_Sanpera_book}%
  \BibitemOpen
  \bibfield  {author} {\bibinfo {author} {\bibnamefont {Lewenstein},
  \bibfnamefont {M.}}, \bibinfo {author} {\bibfnamefont {A.}~\bibnamefont
  {Sanpera}}, \ and\ \bibinfo {author} {\bibfnamefont {V.}~\bibnamefont
  {Ahufinger}}} (\bibinfo {year} {2012}),\ \href@noop {} {\emph {\bibinfo
  {title} {Ultracold atoms in optical lattices: simulating Quantum Many-Body
  systems}}}\ (\bibinfo  {publisher} {Oxford University Press})\BibitemShut
  {NoStop}%
\bibitem [{\citenamefont {Li}\ \emph {et~al.}(2011{\natexlab{a}})\citenamefont
  {Li}, \citenamefont {Liu},\ and\ \citenamefont {Lin}}]{2011_Li_Liu_PRAR}%
  \BibitemOpen
  \bibfield  {author} {\bibinfo {author} {\bibnamefont {Li}, \bibfnamefont
  {X.}}, \bibinfo {author} {\bibfnamefont {W.~V.}\ \bibnamefont {Liu}}, \ and\
  \bibinfo {author} {\bibfnamefont {C.}~\bibnamefont {Lin}}} (\bibinfo {year}
  {2011}{\natexlab{a}}),\ \href {\doibase 10.1103/PhysRevA.83.021602}
  {\bibfield  {journal} {\bibinfo  {journal} {Phys. Rev. A}\ }\textbf {\bibinfo
  {volume} {83}},\ \bibinfo {pages} {021602}}\BibitemShut {NoStop}%
\bibitem [{\citenamefont {Li}\ \emph {et~al.}(2014{\natexlab{a}})\citenamefont
  {Li}, \citenamefont {Paramekanti}, \citenamefont {Hemmerich},\ and\
  \citenamefont {Liu}}]{2013_Li_Paramekanti_NatComm}%
  \BibitemOpen
  \bibfield  {author} {\bibinfo {author} {\bibnamefont {Li}, \bibfnamefont
  {X.}}, \bibinfo {author} {\bibfnamefont {A.}~\bibnamefont {Paramekanti}},
  \bibinfo {author} {\bibfnamefont {A.}~\bibnamefont {Hemmerich}}, \ and\
  \bibinfo {author} {\bibfnamefont {W.~V.}\ \bibnamefont {Liu}}} (\bibinfo
  {year} {2014}{\natexlab{a}}),\ \href {http://dx.doi.org/10.1038/ncomms4205}
  {\bibfield  {journal} {\bibinfo  {journal} {Nat Commun}\ }\textbf {\bibinfo
  {volume} {5}},\ \bibinfo {pages} {3205}}\BibitemShut {NoStop}%
\bibitem [{\citenamefont {Li}\ \emph {et~al.}(2012)\citenamefont {Li},
  \citenamefont {Zhang},\ and\ \citenamefont {Liu}}]{2012_Li_Zhang_PRL}%
  \BibitemOpen
  \bibfield  {author} {\bibinfo {author} {\bibnamefont {Li}, \bibfnamefont
  {X.}}, \bibinfo {author} {\bibfnamefont {Z.}~\bibnamefont {Zhang}}, \ and\
  \bibinfo {author} {\bibfnamefont {W.~V.}\ \bibnamefont {Liu}}} (\bibinfo
  {year} {2012}),\ \href {\doibase 10.1103/PhysRevLett.108.175302} {\bibfield
  {journal} {\bibinfo  {journal} {Phys. Rev. Lett.}\ }\textbf {\bibinfo
  {volume} {108}},\ \bibinfo {pages} {175302}}\BibitemShut {NoStop}%
\bibitem [{\citenamefont {Li}\ \emph {et~al.}(2011{\natexlab{b}})\citenamefont
  {Li}, \citenamefont {Zhao},\ and\ \citenamefont {Liu}}]{2011_Li_Zhao_PRA}%
  \BibitemOpen
  \bibfield  {author} {\bibinfo {author} {\bibnamefont {Li}, \bibfnamefont
  {X.}}, \bibinfo {author} {\bibfnamefont {E.}~\bibnamefont {Zhao}}, \ and\
  \bibinfo {author} {\bibfnamefont {W.~V.}\ \bibnamefont {Liu}}} (\bibinfo
  {year} {2011}{\natexlab{b}}),\ \href {\doibase 10.1103/PhysRevA.83.063626}
  {\bibfield  {journal} {\bibinfo  {journal} {Phys. Rev. A}\ }\textbf {\bibinfo
  {volume} {83}},\ \bibinfo {pages} {063626}}\BibitemShut {NoStop}%
\bibitem [{\citenamefont {Li}\ \emph {et~al.}(2013)\citenamefont {Li},
  \citenamefont {Zhao},\ and\ \citenamefont {Liu}}]{2013_Li_Zhao_NatComm}%
  \BibitemOpen
  \bibfield  {author} {\bibinfo {author} {\bibnamefont {Li}, \bibfnamefont
  {X.}}, \bibinfo {author} {\bibfnamefont {E.}~\bibnamefont {Zhao}}, \ and\
  \bibinfo {author} {\bibfnamefont {W.~V.}\ \bibnamefont {Liu}}} (\bibinfo
  {year} {2013}),\ \href {http://dx.doi.org/10.1038/ncomms2523} {\bibfield
  {journal} {\bibinfo  {journal} {Nat Commun}\ }\textbf {\bibinfo {volume}
  {4}},\ \bibinfo {pages} {1523}}\BibitemShut {NoStop}%
\bibitem [{\citenamefont {Li}\ \emph {et~al.}(2014{\natexlab{b}})\citenamefont
  {Li}, \citenamefont {Lieb},\ and\ \citenamefont {Wu}}]{2013_Li_Lieb_PRL}%
  \BibitemOpen
  \bibfield  {author} {\bibinfo {author} {\bibnamefont {Li}, \bibfnamefont
  {Y.}}, \bibinfo {author} {\bibfnamefont {E.~H.}\ \bibnamefont {Lieb}}, \ and\
  \bibinfo {author} {\bibfnamefont {C.}~\bibnamefont {Wu}}} (\bibinfo {year}
  {2014}{\natexlab{b}}),\ \href {\doibase 10.1103/PhysRevLett.112.217201}
  {\bibfield  {journal} {\bibinfo  {journal} {Phys. Rev. Lett.}\ }\textbf
  {\bibinfo {volume} {112}},\ \bibinfo {pages} {217201}}\BibitemShut {NoStop}%
\bibitem [{\citenamefont {Liao}\ \emph {et~al.}(2010)\citenamefont {Liao},
  \citenamefont {Rittner}, \citenamefont {Paprotta}, \citenamefont {Li},
  \citenamefont {Partridge}, \citenamefont {Hulet}, \citenamefont {Baur},\ and\
  \citenamefont {Mueller}}]{2010_Liao_Rittner_Nature}%
  \BibitemOpen
  \bibfield  {author} {\bibinfo {author} {\bibnamefont {Liao}, \bibfnamefont
  {Y.-a.}}, \bibinfo {author} {\bibfnamefont {A.~S.~C.}\ \bibnamefont
  {Rittner}}, \bibinfo {author} {\bibfnamefont {T.}~\bibnamefont {Paprotta}},
  \bibinfo {author} {\bibfnamefont {W.}~\bibnamefont {Li}}, \bibinfo {author}
  {\bibfnamefont {G.~B.}\ \bibnamefont {Partridge}}, \bibinfo {author}
  {\bibfnamefont {R.~G.}\ \bibnamefont {Hulet}}, \bibinfo {author}
  {\bibfnamefont {S.~K.}\ \bibnamefont {Baur}}, \ and\ \bibinfo {author}
  {\bibfnamefont {E.~J.}\ \bibnamefont {Mueller}}} (\bibinfo {year} {2010}),\
  \href {http://dx.doi.org/10.1038/nature09393} {\bibfield  {journal} {\bibinfo
   {journal} {Nature}\ }\textbf {\bibinfo {volume} {467}}~(\bibinfo {number}
  {7315}),\ \bibinfo {pages} {567}}\BibitemShut {NoStop}%
\bibitem [{\citenamefont {Lim}\ \emph {et~al.}(2008)\citenamefont {Lim},
  \citenamefont {Smith},\ and\ \citenamefont {Hemmerich}}]{2008_Lim_Smith_PRL}%
  \BibitemOpen
  \bibfield  {author} {\bibinfo {author} {\bibnamefont {Lim}, \bibfnamefont
  {L.-K.}}, \bibinfo {author} {\bibfnamefont {C.~M.}\ \bibnamefont {Smith}}, \
  and\ \bibinfo {author} {\bibfnamefont {A.}~\bibnamefont {Hemmerich}}}
  (\bibinfo {year} {2008}),\ \href {\doibase 10.1103/PhysRevLett.100.130402}
  {\bibfield  {journal} {\bibinfo  {journal} {Phys. Rev. Lett.}\ }\textbf
  {\bibinfo {volume} {100}},\ \bibinfo {pages} {130402}}\BibitemShut {NoStop}%
\bibitem [{\citenamefont {Lin}\ \emph {et~al.}(2011)\citenamefont {Lin},
  \citenamefont {Jimenez-Garcia},\ and\ \citenamefont
  {Spielman}}]{2011_Lin_Spielman_Nature}%
  \BibitemOpen
  \bibfield  {author} {\bibinfo {author} {\bibnamefont {Lin}, \bibfnamefont
  {Y.-J.}}, \bibinfo {author} {\bibfnamefont {K.}~\bibnamefont
  {Jimenez-Garcia}}, \ and\ \bibinfo {author} {\bibfnamefont {I.~B.}\
  \bibnamefont {Spielman}}} (\bibinfo {year} {2011}),\ \href
  {http://dx.doi.org/10.1038/nature09887} {\bibfield  {journal} {\bibinfo
  {journal} {Nature}\ }\textbf {\bibinfo {volume} {471}}~(\bibinfo {number}
  {7336}),\ \bibinfo {pages} {83}}\BibitemShut {NoStop}%
\bibitem [{\citenamefont {{Liu}}\ \emph {et~al.}(2015)\citenamefont {{Liu}},
  \citenamefont {{Li}}, \citenamefont {{Hulet}},\ and\ \citenamefont
  {{Liu}}}]{2015_Liu_Li_arXiv}%
  \BibitemOpen
  \bibfield  {author} {\bibinfo {author} {\bibnamefont {{Liu}}, \bibfnamefont
  {B.}}, \bibinfo {author} {\bibfnamefont {X.}~\bibnamefont {{Li}}}, \bibinfo
  {author} {\bibfnamefont {R.~G.}\ \bibnamefont {{Hulet}}}, \ and\ \bibinfo
  {author} {\bibfnamefont {W.~V.}\ \bibnamefont {{Liu}}}} (\bibinfo {year}
  {2015}),\ \href@noop {} {\bibfield  {journal} {\bibinfo  {journal} {ArXiv
  e-prints}\ }}\Eprint {http://arxiv.org/abs/1505.08164} {arXiv:1505.08164
  [cond-mat.quant-gas]} \BibitemShut {NoStop}%
\bibitem [{\citenamefont {Liu}\ \emph {et~al.}(2016)\citenamefont {Liu},
  \citenamefont {Li},\ and\ \citenamefont {Liu}}]{2015_Liu_Li_arXiv2}%
  \BibitemOpen
  \bibfield  {author} {\bibinfo {author} {\bibnamefont {Liu}, \bibfnamefont
  {B.}}, \bibinfo {author} {\bibfnamefont {X.}~\bibnamefont {Li}}, \ and\
  \bibinfo {author} {\bibfnamefont {W.~V.}\ \bibnamefont {Liu}}} (\bibinfo
  {year} {2016}),\ \href {\doibase 10.1103/PhysRevA.93.033643} {\bibfield
  {journal} {\bibinfo  {journal} {Phys. Rev. A}\ }\textbf {\bibinfo {volume}
  {93}},\ \bibinfo {pages} {033643}}\BibitemShut {NoStop}%
\bibitem [{\citenamefont {Liu}\ \emph {et~al.}(2014)\citenamefont {Liu},
  \citenamefont {Li}, \citenamefont {Wu},\ and\ \citenamefont
  {Liu}}]{2014_Liu_Li_NatComm}%
  \BibitemOpen
  \bibfield  {author} {\bibinfo {author} {\bibnamefont {Liu}, \bibfnamefont
  {B.}}, \bibinfo {author} {\bibfnamefont {X.}~\bibnamefont {Li}}, \bibinfo
  {author} {\bibfnamefont {B.}~\bibnamefont {Wu}}, \ and\ \bibinfo {author}
  {\bibfnamefont {W.~V.}\ \bibnamefont {Liu}}} (\bibinfo {year} {2014}),\ \href
  {http://dx.doi.org/10.1038/ncomms6064} {\bibfield  {journal} {\bibinfo
  {journal} {Nat Commun}\ }\textbf {\bibinfo {volume} {5}},\ \bibinfo {pages}
  {5064}}\BibitemShut {NoStop}%
\bibitem [{\citenamefont {Liu}\ \emph {et~al.}(2013)\citenamefont {Liu},
  \citenamefont {Yu},\ and\ \citenamefont {Liu}}]{2013_Liu_Yu_PRA}%
  \BibitemOpen
  \bibfield  {author} {\bibinfo {author} {\bibnamefont {Liu}, \bibfnamefont
  {B.}}, \bibinfo {author} {\bibfnamefont {X.-L.}\ \bibnamefont {Yu}}, \ and\
  \bibinfo {author} {\bibfnamefont {W.-M.}\ \bibnamefont {Liu}}} (\bibinfo
  {year} {2013}),\ \href {\doibase 10.1103/PhysRevA.88.063605} {\bibfield
  {journal} {\bibinfo  {journal} {Phys. Rev. A}\ }\textbf {\bibinfo {volume}
  {88}},\ \bibinfo {pages} {063605}}\BibitemShut {NoStop}%
\bibitem [{\citenamefont {Liu}\ and\ \citenamefont
  {Wu}(2006)}]{2006_Liu_Wu_PRA}%
  \BibitemOpen
  \bibfield  {author} {\bibinfo {author} {\bibnamefont {Liu}, \bibfnamefont
  {W.~V.}}, \ and\ \bibinfo {author} {\bibfnamefont {C.}~\bibnamefont {Wu}}}
  (\bibinfo {year} {2006}),\ \href {\doibase 10.1103/PhysRevA.74.013607}
  {\bibfield  {journal} {\bibinfo  {journal} {Phys. Rev. A}\ }\textbf {\bibinfo
  {volume} {74}},\ \bibinfo {pages} {013607}}\BibitemShut {NoStop}%
\bibitem [{\citenamefont {Liu}\ \emph {et~al.}(2010)\citenamefont {Liu},
  \citenamefont {Liu}, \citenamefont {Wu},\ and\ \citenamefont
  {Sinova}}]{2010_Liu_Liu_PRA}%
  \BibitemOpen
  \bibfield  {author} {\bibinfo {author} {\bibnamefont {Liu}, \bibfnamefont
  {X.-J.}}, \bibinfo {author} {\bibfnamefont {X.}~\bibnamefont {Liu}}, \bibinfo
  {author} {\bibfnamefont {C.}~\bibnamefont {Wu}}, \ and\ \bibinfo {author}
  {\bibfnamefont {J.}~\bibnamefont {Sinova}}} (\bibinfo {year} {2010}),\ \href
  {\doibase 10.1103/PhysRevA.81.033622} {\bibfield  {journal} {\bibinfo
  {journal} {Phys. Rev. A}\ }\textbf {\bibinfo {volume} {81}},\ \bibinfo
  {pages} {033622}}\BibitemShut {NoStop}%
\bibitem [{\citenamefont {Loh}\ and\ \citenamefont
  {Trivedi}(2010)}]{2010_Loh_Trivedi_PRL}%
  \BibitemOpen
  \bibfield  {author} {\bibinfo {author} {\bibnamefont {Loh}, \bibfnamefont
  {Y.~L.}}, \ and\ \bibinfo {author} {\bibfnamefont {N.}~\bibnamefont
  {Trivedi}}} (\bibinfo {year} {2010}),\ \href {\doibase
  10.1103/PhysRevLett.104.165302} {\bibfield  {journal} {\bibinfo  {journal}
  {Phys. Rev. Lett.}\ }\textbf {\bibinfo {volume} {104}},\ \bibinfo {pages}
  {165302}}\BibitemShut {NoStop}%
\bibitem [{\citenamefont {Lu}\ and\ \citenamefont
  {Arrigoni}(2009)}]{2009_Lu_Arrigoni_PRB}%
  \BibitemOpen
  \bibfield  {author} {\bibinfo {author} {\bibnamefont {Lu}, \bibfnamefont
  {X.}}, \ and\ \bibinfo {author} {\bibfnamefont {E.}~\bibnamefont {Arrigoni}}}
  (\bibinfo {year} {2009}),\ \href {\doibase 10.1103/PhysRevB.79.245109}
  {\bibfield  {journal} {\bibinfo  {journal} {Phys. Rev. B}\ }\textbf {\bibinfo
  {volume} {79}},\ \bibinfo {pages} {245109}}\BibitemShut {NoStop}%
\bibitem [{\citenamefont {Luke}\ \emph {et~al.}(1998)\citenamefont {Luke},
  \citenamefont {Fudamoto}, \citenamefont {Kojima}, \citenamefont {Larkin},
  \citenamefont {Merrin}, \citenamefont {Nachumi}, \citenamefont {Uemura},
  \citenamefont {Maeno}, \citenamefont {Mao}, \citenamefont {Mori},
  \citenamefont {Nakamura},\ and\ \citenamefont
  {Sigrist}}]{1998_Luke_Fudamoto_Nature}%
  \BibitemOpen
  \bibfield  {author} {\bibinfo {author} {\bibnamefont {Luke}, \bibfnamefont
  {G.~M.}}, \bibinfo {author} {\bibfnamefont {Y.}~\bibnamefont {Fudamoto}},
  \bibinfo {author} {\bibfnamefont {K.~M.}\ \bibnamefont {Kojima}}, \bibinfo
  {author} {\bibfnamefont {M.~I.}\ \bibnamefont {Larkin}}, \bibinfo {author}
  {\bibfnamefont {J.}~\bibnamefont {Merrin}}, \bibinfo {author} {\bibfnamefont
  {B.}~\bibnamefont {Nachumi}}, \bibinfo {author} {\bibfnamefont {Y.~J.}\
  \bibnamefont {Uemura}}, \bibinfo {author} {\bibfnamefont {Y.}~\bibnamefont
  {Maeno}}, \bibinfo {author} {\bibfnamefont {Z.~Q.}\ \bibnamefont {Mao}},
  \bibinfo {author} {\bibfnamefont {Y.}~\bibnamefont {Mori}}, \bibinfo {author}
  {\bibfnamefont {H.}~\bibnamefont {Nakamura}}, \ and\ \bibinfo {author}
  {\bibfnamefont {M.}~\bibnamefont {Sigrist}}} (\bibinfo {year} {1998}),\ \href
  {http://dx.doi.org/10.1038/29038} {\bibfield  {journal} {\bibinfo  {journal}
  {Nature}\ }\textbf {\bibinfo {volume} {394}}~(\bibinfo {number} {6693}),\
  \bibinfo {pages} {558}}\BibitemShut {NoStop}%
\bibitem [{\citenamefont {Lundblad}\ \emph {et~al.}(2008)\citenamefont
  {Lundblad}, \citenamefont {Lee}, \citenamefont {Spielman}, \citenamefont
  {Brown}, \citenamefont {Phillips},\ and\ \citenamefont
  {Porto}}]{2008_Lundblad_Lee_PRL}%
  \BibitemOpen
  \bibfield  {author} {\bibinfo {author} {\bibnamefont {Lundblad},
  \bibfnamefont {N.}}, \bibinfo {author} {\bibfnamefont {P.~J.}\ \bibnamefont
  {Lee}}, \bibinfo {author} {\bibfnamefont {I.~B.}\ \bibnamefont {Spielman}},
  \bibinfo {author} {\bibfnamefont {B.~L.}\ \bibnamefont {Brown}}, \bibinfo
  {author} {\bibfnamefont {W.~D.}\ \bibnamefont {Phillips}}, \ and\ \bibinfo
  {author} {\bibfnamefont {J.~V.}\ \bibnamefont {Porto}}} (\bibinfo {year}
  {2008}),\ \href {\doibase 10.1103/PhysRevLett.100.150401} {\bibfield
  {journal} {\bibinfo  {journal} {Phys. Rev. Lett.}\ }\textbf {\bibinfo
  {volume} {100}},\ \bibinfo {pages} {150401}}\BibitemShut {NoStop}%
\bibitem [{\citenamefont {Lühmann}\ \emph {et~al.}(2012)\citenamefont
  {Lühmann}, \citenamefont {Jürgensen},\ and\ \citenamefont
  {Sengstock}}]{2012_Luhmann_Jurgensen_NJP}%
  \BibitemOpen
  \bibfield  {author} {\bibinfo {author} {\bibnamefont {Lühmann},
  \bibfnamefont {D.-S.}}, \bibinfo {author} {\bibfnamefont {O.}~\bibnamefont
  {Jürgensen}}, \ and\ \bibinfo {author} {\bibfnamefont {K.}~\bibnamefont
  {Sengstock}}} (\bibinfo {year} {2012}),\ \href
  {http://stacks.iop.org/1367-2630/14/i=3/a=033021} {\bibfield  {journal}
  {\bibinfo  {journal} {New Journal of Physics}\ }\textbf {\bibinfo {volume}
  {14}}~(\bibinfo {number} {3}),\ \bibinfo {pages} {033021}}\BibitemShut
  {NoStop}%
\bibitem [{\citenamefont {Martikainen}(2011)}]{2011_Martikainen_PRA}%
  \BibitemOpen
  \bibfield  {author} {\bibinfo {author} {\bibnamefont {Martikainen},
  \bibfnamefont {J.-P.}}} (\bibinfo {year} {2011}),\ \href {\doibase
  10.1103/PhysRevA.83.013610} {\bibfield  {journal} {\bibinfo  {journal} {Phys.
  Rev. A}\ }\textbf {\bibinfo {volume} {83}},\ \bibinfo {pages}
  {013610}}\BibitemShut {NoStop}%
\bibitem [{\citenamefont {Martikainen}\ and\ \citenamefont
  {Larson}(2012)}]{2012_Martikainen_Larson_PRA}%
  \BibitemOpen
  \bibfield  {author} {\bibinfo {author} {\bibnamefont {Martikainen},
  \bibfnamefont {J.-P.}}, \ and\ \bibinfo {author} {\bibfnamefont
  {J.}~\bibnamefont {Larson}}} (\bibinfo {year} {2012}),\ \href {\doibase
  10.1103/PhysRevA.86.023611} {\bibfield  {journal} {\bibinfo  {journal} {Phys.
  Rev. A}\ }\textbf {\bibinfo {volume} {86}},\ \bibinfo {pages}
  {023611}}\BibitemShut {NoStop}%
\bibitem [{\citenamefont {Marzari}\ \emph {et~al.}(2012)\citenamefont
  {Marzari}, \citenamefont {Mostofi}, \citenamefont {Yates}, \citenamefont
  {Souza},\ and\ \citenamefont {Vanderbilt}}]{2012_Marzari_Mostofi_RMP}%
  \BibitemOpen
  \bibfield  {author} {\bibinfo {author} {\bibnamefont {Marzari}, \bibfnamefont
  {N.}}, \bibinfo {author} {\bibfnamefont {A.~A.}\ \bibnamefont {Mostofi}},
  \bibinfo {author} {\bibfnamefont {J.~R.}\ \bibnamefont {Yates}}, \bibinfo
  {author} {\bibfnamefont {I.}~\bibnamefont {Souza}}, \ and\ \bibinfo {author}
  {\bibfnamefont {D.}~\bibnamefont {Vanderbilt}}} (\bibinfo {year} {2012}),\
  \href {\doibase 10.1103/RevModPhys.84.1419} {\bibfield  {journal} {\bibinfo
  {journal} {Rev. Mod. Phys.}\ }\textbf {\bibinfo {volume} {84}},\ \bibinfo
  {pages} {1419}}\BibitemShut {NoStop}%
\bibitem [{\citenamefont {Mering}\ and\ \citenamefont
  {Fleischhauer}(2011)}]{2011_Mering_Fleischhauer_PRA}%
  \BibitemOpen
  \bibfield  {author} {\bibinfo {author} {\bibnamefont {Mering}, \bibfnamefont
  {A.}}, \ and\ \bibinfo {author} {\bibfnamefont {M.}~\bibnamefont
  {Fleischhauer}}} (\bibinfo {year} {2011}),\ \href {\doibase
  10.1103/PhysRevA.83.063630} {\bibfield  {journal} {\bibinfo  {journal} {Phys.
  Rev. A}\ }\textbf {\bibinfo {volume} {83}},\ \bibinfo {pages}
  {063630}}\BibitemShut {NoStop}%
\bibitem [{\citenamefont {Miyake}\ \emph {et~al.}(2013)\citenamefont {Miyake},
  \citenamefont {Siviloglou}, \citenamefont {Kennedy}, \citenamefont {Burton},\
  and\ \citenamefont {Ketterle}}]{2013_Miyake_Siviloglou_PRL}%
  \BibitemOpen
  \bibfield  {author} {\bibinfo {author} {\bibnamefont {Miyake}, \bibfnamefont
  {H.}}, \bibinfo {author} {\bibfnamefont {G.~A.}\ \bibnamefont {Siviloglou}},
  \bibinfo {author} {\bibfnamefont {C.~J.}\ \bibnamefont {Kennedy}}, \bibinfo
  {author} {\bibfnamefont {W.~C.}\ \bibnamefont {Burton}}, \ and\ \bibinfo
  {author} {\bibfnamefont {W.}~\bibnamefont {Ketterle}}} (\bibinfo {year}
  {2013}),\ \href {\doibase 10.1103/PhysRevLett.111.185302} {\bibfield
  {journal} {\bibinfo  {journal} {Phys. Rev. Lett.}\ }\textbf {\bibinfo
  {volume} {111}},\ \bibinfo {pages} {185302}}\BibitemShut {NoStop}%
\bibitem [{\citenamefont {M\"uller}\ \emph {et~al.}(2007)\citenamefont
  {M\"uller}, \citenamefont {F\"olling}, \citenamefont {Widera},\ and\
  \citenamefont {Bloch}}]{2007_Muller_Folling_PRL}%
  \BibitemOpen
  \bibfield  {author} {\bibinfo {author} {\bibnamefont {M\"uller},
  \bibfnamefont {T.}}, \bibinfo {author} {\bibfnamefont {S.}~\bibnamefont
  {F\"olling}}, \bibinfo {author} {\bibfnamefont {A.}~\bibnamefont {Widera}}, \
  and\ \bibinfo {author} {\bibfnamefont {I.}~\bibnamefont {Bloch}}} (\bibinfo
  {year} {2007}),\ \href {\doibase 10.1103/PhysRevLett.99.200405} {\bibfield
  {journal} {\bibinfo  {journal} {Phys. Rev. Lett.}\ }\textbf {\bibinfo
  {volume} {99}},\ \bibinfo {pages} {200405}}\BibitemShut {NoStop}%
\bibitem [{\citenamefont {Niu}\ \emph {et~al.}(2015)\citenamefont {Niu},
  \citenamefont {Hu}, \citenamefont {Jin}, \citenamefont {Dong}, \citenamefont
  {Chen},\ and\ \citenamefont {Zhou}}]{2015_Niu_Hu_OptExp}%
  \BibitemOpen
  \bibfield  {author} {\bibinfo {author} {\bibnamefont {Niu}, \bibfnamefont
  {L.}}, \bibinfo {author} {\bibfnamefont {D.}~\bibnamefont {Hu}}, \bibinfo
  {author} {\bibfnamefont {S.}~\bibnamefont {Jin}}, \bibinfo {author}
  {\bibfnamefont {X.}~\bibnamefont {Dong}}, \bibinfo {author} {\bibfnamefont
  {X.}~\bibnamefont {Chen}}, \ and\ \bibinfo {author} {\bibfnamefont
  {X.}~\bibnamefont {Zhou}}} (\bibinfo {year} {2015}),\ \href {\doibase
  10.1364/OE.23.010064} {\bibfield  {journal} {\bibinfo  {journal} {Opt.
  Express}\ }\textbf {\bibinfo {volume} {23}}~(\bibinfo {number} {8}),\
  \bibinfo {pages} {10064}}\BibitemShut {NoStop}%
\bibitem [{\citenamefont {Ohtomo}\ and\ \citenamefont
  {Hwang}(2004)}]{2004_Ohtomo_Hwang_Nature}%
  \BibitemOpen
  \bibfield  {author} {\bibinfo {author} {\bibnamefont {Ohtomo}, \bibfnamefont
  {A.}}, \ and\ \bibinfo {author} {\bibfnamefont {H.~Y.}\ \bibnamefont
  {Hwang}}} (\bibinfo {year} {2004}),\ \href
  {http://dx.doi.org/10.1038/nature02308} {\bibfield  {journal} {\bibinfo
  {journal} {Nature}\ }\textbf {\bibinfo {volume} {427}}~(\bibinfo {number}
  {6973}),\ \bibinfo {pages} {423}}\BibitemShut {NoStop}%
\bibitem [{\citenamefont {\"Olschl\"ager}\ \emph {et~al.}(2013)\citenamefont
  {\"Olschl\"ager}, \citenamefont {Kock}, \citenamefont {Wirth}, \citenamefont
  {Ewerbeck}, \citenamefont {Smith},\ and\ \citenamefont
  {Hemmerich}}]{2013_Olschlager_Kock_NJPHYS}%
  \BibitemOpen
  \bibfield  {author} {\bibinfo {author} {\bibnamefont {\"Olschl\"ager},
  \bibfnamefont {M.}}, \bibinfo {author} {\bibfnamefont {T.}~\bibnamefont
  {Kock}}, \bibinfo {author} {\bibfnamefont {G.}~\bibnamefont {Wirth}},
  \bibinfo {author} {\bibfnamefont {A.}~\bibnamefont {Ewerbeck}}, \bibinfo
  {author} {\bibfnamefont {C.~M.}\ \bibnamefont {Smith}}, \ and\ \bibinfo
  {author} {\bibfnamefont {A.}~\bibnamefont {Hemmerich}}} (\bibinfo {year}
  {2013}),\ \href {http://stacks.iop.org/1367-2630/15/i=8/a=083041} {\bibfield
  {journal} {\bibinfo  {journal} {New Journal of Physics}\ }\textbf {\bibinfo
  {volume} {15}}~(\bibinfo {number} {8}),\ \bibinfo {pages}
  {083041}}\BibitemShut {NoStop}%
\bibitem [{\citenamefont {\"Olschl\"ager}\ \emph {et~al.}(2011)\citenamefont
  {\"Olschl\"ager}, \citenamefont {Wirth},\ and\ \citenamefont
  {Hemmerich}}]{2011_Olschlager_Wirth_PRL}%
  \BibitemOpen
  \bibfield  {author} {\bibinfo {author} {\bibnamefont {\"Olschl\"ager},
  \bibfnamefont {M.}}, \bibinfo {author} {\bibfnamefont {G.}~\bibnamefont
  {Wirth}}, \ and\ \bibinfo {author} {\bibfnamefont {A.}~\bibnamefont
  {Hemmerich}}} (\bibinfo {year} {2011}),\ \href {\doibase
  10.1103/PhysRevLett.106.015302} {\bibfield  {journal} {\bibinfo  {journal}
  {Phys. Rev. Lett.}\ }\textbf {\bibinfo {volume} {106}},\ \bibinfo {pages}
  {015302}}\BibitemShut {NoStop}%
\bibitem [{\citenamefont {\"Olschl\"ager}\ \emph {et~al.}(2012)\citenamefont
  {\"Olschl\"ager}, \citenamefont {Wirth}, \citenamefont {Kock},\ and\
  \citenamefont {Hemmerich}}]{2012_Olschlager_Wirth_PRL}%
  \BibitemOpen
  \bibfield  {author} {\bibinfo {author} {\bibnamefont {\"Olschl\"ager},
  \bibfnamefont {M.}}, \bibinfo {author} {\bibfnamefont {G.}~\bibnamefont
  {Wirth}}, \bibinfo {author} {\bibfnamefont {T.}~\bibnamefont {Kock}}, \ and\
  \bibinfo {author} {\bibfnamefont {A.}~\bibnamefont {Hemmerich}}} (\bibinfo
  {year} {2012}),\ \href {\doibase 10.1103/PhysRevLett.108.075302} {\bibfield
  {journal} {\bibinfo  {journal} {Phys. Rev. Lett.}\ }\textbf {\bibinfo
  {volume} {108}},\ \bibinfo {pages} {075302}}\BibitemShut {NoStop}%
\bibitem [{\citenamefont {Onsager}(1944)}]{1944_Onsager_PR}%
  \BibitemOpen
  \bibfield  {author} {\bibinfo {author} {\bibnamefont {Onsager}, \bibfnamefont
  {L.}}} (\bibinfo {year} {1944}),\ \href {\doibase 10.1103/PhysRev.65.117}
  {\bibfield  {journal} {\bibinfo  {journal} {Phys. Rev.}\ }\textbf {\bibinfo
  {volume} {65}},\ \bibinfo {pages} {117}}\BibitemShut {NoStop}%
\bibitem [{\citenamefont {van Oosten}\ \emph {et~al.}(2003)\citenamefont {van
  Oosten}, \citenamefont {van~der Straten},\ and\ \citenamefont
  {Stoof}}]{2003_Oosten_Straten_PRA}%
  \BibitemOpen
  \bibfield  {author} {\bibinfo {author} {\bibnamefont {van Oosten},
  \bibfnamefont {D.}}, \bibinfo {author} {\bibfnamefont {P.}~\bibnamefont
  {van~der Straten}}, \ and\ \bibinfo {author} {\bibfnamefont {H.~T.~C.}\
  \bibnamefont {Stoof}}} (\bibinfo {year} {2003}),\ \href {\doibase
  10.1103/PhysRevA.67.033606} {\bibfield  {journal} {\bibinfo  {journal} {Phys.
  Rev. A}\ }\textbf {\bibinfo {volume} {67}},\ \bibinfo {pages}
  {033606}}\BibitemShut {NoStop}%
\bibitem [{\citenamefont {Orso}(2007)}]{2007_Orso_PRL}%
  \BibitemOpen
  \bibfield  {author} {\bibinfo {author} {\bibnamefont {Orso}, \bibfnamefont
  {G.}}} (\bibinfo {year} {2007}),\ \href {\doibase
  10.1103/PhysRevLett.98.070402} {\bibfield  {journal} {\bibinfo  {journal}
  {Phys. Rev. Lett.}\ }\textbf {\bibinfo {volume} {98}},\ \bibinfo {pages}
  {070402}}\BibitemShut {NoStop}%
\bibitem [{\citenamefont {Parish}\ \emph {et~al.}(2007)\citenamefont {Parish},
  \citenamefont {Baur}, \citenamefont {Mueller},\ and\ \citenamefont
  {Huse}}]{2007_Parish_Baur_PRL}%
  \BibitemOpen
  \bibfield  {author} {\bibinfo {author} {\bibnamefont {Parish}, \bibfnamefont
  {M.~M.}}, \bibinfo {author} {\bibfnamefont {S.~K.}\ \bibnamefont {Baur}},
  \bibinfo {author} {\bibfnamefont {E.~J.}\ \bibnamefont {Mueller}}, \ and\
  \bibinfo {author} {\bibfnamefont {D.~A.}\ \bibnamefont {Huse}}} (\bibinfo
  {year} {2007}),\ \href {\doibase 10.1103/PhysRevLett.99.250403} {\bibfield
  {journal} {\bibinfo  {journal} {Phys. Rev. Lett.}\ }\textbf {\bibinfo
  {volume} {99}},\ \bibinfo {pages} {250403}}\BibitemShut {NoStop}%
\bibitem [{\citenamefont {Parker}\ \emph {et~al.}(2013)\citenamefont {Parker},
  \citenamefont {Ha},\ and\ \citenamefont {Chin}}]{2013_Parker_Ha_NatPhys}%
  \BibitemOpen
  \bibfield  {author} {\bibinfo {author} {\bibnamefont {Parker}, \bibfnamefont
  {C.~V.}}, \bibinfo {author} {\bibfnamefont {L.-C.}\ \bibnamefont {Ha}}, \
  and\ \bibinfo {author} {\bibfnamefont {C.}~\bibnamefont {Chin}}} (\bibinfo
  {year} {2013}),\ \href {http://dx.doi.org/10.1038/nphys2789} {\bibfield
  {journal} {\bibinfo  {journal} {Nat Phys}\ }\textbf {\bibinfo {volume}
  {9}}~(\bibinfo {number} {12}),\ \bibinfo {pages} {769}}\BibitemShut {NoStop}%
\bibitem [{\citenamefont {Paul}\ and\ \citenamefont
  {Tiesinga}(2013)}]{2013_Paul_Tiesinga_PRA}%
  \BibitemOpen
  \bibfield  {author} {\bibinfo {author} {\bibnamefont {Paul}, \bibfnamefont
  {S.}}, \ and\ \bibinfo {author} {\bibfnamefont {E.}~\bibnamefont {Tiesinga}}}
  (\bibinfo {year} {2013}),\ \href {\doibase 10.1103/PhysRevA.88.033615}
  {\bibfield  {journal} {\bibinfo  {journal} {Phys. Rev. A}\ }\textbf {\bibinfo
  {volume} {88}},\ \bibinfo {pages} {033615}}\BibitemShut {NoStop}%
\bibitem [{\citenamefont {Pethick}\ and\ \citenamefont
  {Smith}(2008)}]{2008_Pethick_Smith_book}%
  \BibitemOpen
  \bibfield  {author} {\bibinfo {author} {\bibnamefont {Pethick}, \bibfnamefont
  {C.}}, \ and\ \bibinfo {author} {\bibfnamefont {H.}~\bibnamefont {Smith}}}
  (\bibinfo {year} {2008}),\ \href@noop {} {\emph {\bibinfo {title}
  {Bose-Einstein Condensation in dilute gases}}}\ (\bibinfo  {publisher}
  {cambridge university press})\BibitemShut {NoStop}%
\bibitem [{\citenamefont {Pietraszewicz}\ \emph {et~al.}(2013)\citenamefont
  {Pietraszewicz}, \citenamefont {Sowi\ifmmode~\acute{n}\else \'{n}\fi{}ski},
  \citenamefont {Brewczyk}, \citenamefont {Lewenstein},\ and\ \citenamefont
  {Gajda}}]{2013_Pietraszewicz_Sowinski_PRA}%
  \BibitemOpen
  \bibfield  {author} {\bibinfo {author} {\bibnamefont {Pietraszewicz},
  \bibfnamefont {J.}}, \bibinfo {author} {\bibfnamefont {T.}~\bibnamefont
  {Sowi\ifmmode~\acute{n}\else \'{n}\fi{}ski}}, \bibinfo {author}
  {\bibfnamefont {M.}~\bibnamefont {Brewczyk}}, \bibinfo {author}
  {\bibfnamefont {M.}~\bibnamefont {Lewenstein}}, \ and\ \bibinfo {author}
  {\bibfnamefont {M.}~\bibnamefont {Gajda}}} (\bibinfo {year} {2013}),\ \href
  {\doibase 10.1103/PhysRevA.88.013608} {\bibfield  {journal} {\bibinfo
  {journal} {Phys. Rev. A}\ }\textbf {\bibinfo {volume} {88}},\ \bibinfo
  {pages} {013608}}\BibitemShut {NoStop}%
\bibitem [{\citenamefont {Pietraszewicz}\ \emph {et~al.}(2012)\citenamefont
  {Pietraszewicz}, \citenamefont {Sowi\ifmmode~\acute{n}\else \'{n}\fi{}ski},
  \citenamefont {Brewczyk}, \citenamefont {Zakrzewski}, \citenamefont
  {Lewenstein},\ and\ \citenamefont {Gajda}}]{2012_Pietraszewicz_Sowinski_PRA}%
  \BibitemOpen
  \bibfield  {author} {\bibinfo {author} {\bibnamefont {Pietraszewicz},
  \bibfnamefont {J.}}, \bibinfo {author} {\bibfnamefont {T.}~\bibnamefont
  {Sowi\ifmmode~\acute{n}\else \'{n}\fi{}ski}}, \bibinfo {author}
  {\bibfnamefont {M.}~\bibnamefont {Brewczyk}}, \bibinfo {author}
  {\bibfnamefont {J.}~\bibnamefont {Zakrzewski}}, \bibinfo {author}
  {\bibfnamefont {M.}~\bibnamefont {Lewenstein}}, \ and\ \bibinfo {author}
  {\bibfnamefont {M.}~\bibnamefont {Gajda}}} (\bibinfo {year} {2012}),\ \href
  {\doibase 10.1103/PhysRevA.85.053638} {\bibfield  {journal} {\bibinfo
  {journal} {Phys. Rev. A}\ }\textbf {\bibinfo {volume} {85}},\ \bibinfo
  {pages} {053638}}\BibitemShut {NoStop}%
\bibitem [{\citenamefont {Pinheiro}\ \emph {et~al.}(2013)\citenamefont
  {Pinheiro}, \citenamefont {Bruun}, \citenamefont {Martikainen},\ and\
  \citenamefont {Larson}}]{2013_Pinheiro_Bruun_PRL}%
  \BibitemOpen
  \bibfield  {author} {\bibinfo {author} {\bibnamefont {Pinheiro},
  \bibfnamefont {F.}}, \bibinfo {author} {\bibfnamefont {G.~M.}\ \bibnamefont
  {Bruun}}, \bibinfo {author} {\bibfnamefont {J.-P.}\ \bibnamefont
  {Martikainen}}, \ and\ \bibinfo {author} {\bibfnamefont {J.}~\bibnamefont
  {Larson}}} (\bibinfo {year} {2013}),\ \href {\doibase
  10.1103/PhysRevLett.111.205302} {\bibfield  {journal} {\bibinfo  {journal}
  {Phys. Rev. Lett.}\ }\textbf {\bibinfo {volume} {111}},\ \bibinfo {pages}
  {205302}}\BibitemShut {NoStop}%
\bibitem [{\citenamefont {Pinheiro}\ \emph {et~al.}(2012)\citenamefont
  {Pinheiro}, \citenamefont {Martikainen},\ and\ \citenamefont
  {Larson}}]{2012_Pinheiro_Martikainen_PRA}%
  \BibitemOpen
  \bibfield  {author} {\bibinfo {author} {\bibnamefont {Pinheiro},
  \bibfnamefont {F.}}, \bibinfo {author} {\bibfnamefont {J.-P.}\ \bibnamefont
  {Martikainen}}, \ and\ \bibinfo {author} {\bibfnamefont {J.}~\bibnamefont
  {Larson}}} (\bibinfo {year} {2012}),\ \href {\doibase
  10.1103/PhysRevA.85.033638} {\bibfield  {journal} {\bibinfo  {journal} {Phys.
  Rev. A}\ }\textbf {\bibinfo {volume} {85}},\ \bibinfo {pages}
  {033638}}\BibitemShut {NoStop}%
\bibitem [{\citenamefont {Pinheiro}\ \emph {et~al.}(2015)\citenamefont
  {Pinheiro}, \citenamefont {Matrikainen},\ and\ \citenamefont
  {Larson}}]{2015_Pinheiro_Matrikainen_NJP}%
  \BibitemOpen
  \bibfield  {author} {\bibinfo {author} {\bibnamefont {Pinheiro},
  \bibfnamefont {F.}}, \bibinfo {author} {\bibfnamefont {J.-P.}\ \bibnamefont
  {Matrikainen}}, \ and\ \bibinfo {author} {\bibfnamefont {J.}~\bibnamefont
  {Larson}}} (\bibinfo {year} {2015}),\ \href
  {http://stacks.iop.org/1367-2630/17/i=5/a=053004} {\bibfield  {journal}
  {\bibinfo  {journal} {New Journal of Physics}\ }\textbf {\bibinfo {volume}
  {17}}~(\bibinfo {number} {5}),\ \bibinfo {pages} {053004}}\BibitemShut
  {NoStop}%
\bibitem [{\citenamefont {Polkovnikov}\ \emph {et~al.}(2002)\citenamefont
  {Polkovnikov}, \citenamefont {Sachdev},\ and\ \citenamefont
  {Girvin}}]{2002_Polkovnikov_Sachdev_PRA}%
  \BibitemOpen
  \bibfield  {author} {\bibinfo {author} {\bibnamefont {Polkovnikov},
  \bibfnamefont {A.}}, \bibinfo {author} {\bibfnamefont {S.}~\bibnamefont
  {Sachdev}}, \ and\ \bibinfo {author} {\bibfnamefont {S.~M.}\ \bibnamefont
  {Girvin}}} (\bibinfo {year} {2002}),\ \href {\doibase
  10.1103/PhysRevA.66.053607} {\bibfield  {journal} {\bibinfo  {journal} {Phys.
  Rev. A}\ }\textbf {\bibinfo {volume} {66}},\ \bibinfo {pages}
  {053607}}\BibitemShut {NoStop}%
\bibitem [{\citenamefont {Przysiezna}\ \emph {et~al.}(2015)\citenamefont
  {Przysiezna}, \citenamefont {Dutta},\ and\ \citenamefont
  {Zakrzewski}}]{2015_Przysiezna_Dutta_NJP}%
  \BibitemOpen
  \bibfield  {author} {\bibinfo {author} {\bibnamefont {Przysiezna},
  \bibfnamefont {A.}}, \bibinfo {author} {\bibfnamefont {O.}~\bibnamefont
  {Dutta}}, \ and\ \bibinfo {author} {\bibfnamefont {J.}~\bibnamefont
  {Zakrzewski}}} (\bibinfo {year} {2015}),\ \href
  {http://stacks.iop.org/1367-2630/17/i=1/a=013018} {\bibfield  {journal}
  {\bibinfo  {journal} {New Journal of Physics}\ }\textbf {\bibinfo {volume}
  {17}}~(\bibinfo {number} {1}),\ \bibinfo {pages} {013018}}\BibitemShut
  {NoStop}%
\bibitem [{\citenamefont {Radzihovsky}\ and\ \citenamefont
  {Sheehy}(2010)}]{2010_Radzihovsky_Sheehy_RPP}%
  \BibitemOpen
  \bibfield  {author} {\bibinfo {author} {\bibnamefont {Radzihovsky},
  \bibfnamefont {L.}}, \ and\ \bibinfo {author} {\bibfnamefont {D.~E.}\
  \bibnamefont {Sheehy}}} (\bibinfo {year} {2010}),\ \href
  {http://stacks.iop.org/0034-4885/73/i=7/a=076501} {\bibfield  {journal}
  {\bibinfo  {journal} {Rep. Prog. Phys.}\ }\textbf {\bibinfo {volume}
  {73}}~(\bibinfo {number} {7}),\ \bibinfo {pages} {076501}}\BibitemShut
  {NoStop}%
\bibitem [{\citenamefont {Sachdev}(2011)}]{2011_Sachdev_Book}%
  \BibitemOpen
  \bibfield  {author} {\bibinfo {author} {\bibnamefont {Sachdev}, \bibfnamefont
  {S.}}} (\bibinfo {year} {2011}),\ \href
  {http://dx.doi.org/10.1017/CBO9780511973765} {\emph {\bibinfo {title}
  {Quantum Phase Transitions}}},\ \bibinfo {edition} {2nd}\ ed.\ (\bibinfo
  {publisher} {Cambridge University Press})\ \bibinfo {note} {cambridge Books
  Online}\BibitemShut {NoStop}%
\bibitem [{\citenamefont {Sakmann}\ \emph {et~al.}(2011)\citenamefont
  {Sakmann}, \citenamefont {Streltsov}, \citenamefont {Alon},\ and\
  \citenamefont {Cederbaum}}]{2011_Kaspar_Alexej_NJP}%
  \BibitemOpen
  \bibfield  {author} {\bibinfo {author} {\bibnamefont {Sakmann}, \bibfnamefont
  {K.}}, \bibinfo {author} {\bibfnamefont {A.~I.}\ \bibnamefont {Streltsov}},
  \bibinfo {author} {\bibfnamefont {O.~E.}\ \bibnamefont {Alon}}, \ and\
  \bibinfo {author} {\bibfnamefont {L.~S.}\ \bibnamefont {Cederbaum}}}
  (\bibinfo {year} {2011}),\ \href
  {http://stacks.iop.org/1367-2630/13/i=4/a=043003} {\bibfield  {journal}
  {\bibinfo  {journal} {New Journal of Physics}\ }\textbf {\bibinfo {volume}
  {13}}~(\bibinfo {number} {4}),\ \bibinfo {pages} {043003}}\BibitemShut
  {NoStop}%
\bibitem [{\citenamefont {Sau}\ \emph {et~al.}(2012)\citenamefont {Sau},
  \citenamefont {Wang},\ and\ \citenamefont {Das~Sarma}}]{2012_Sau_Wang_PRA}%
  \BibitemOpen
  \bibfield  {author} {\bibinfo {author} {\bibnamefont {Sau}, \bibfnamefont
  {J.~D.}}, \bibinfo {author} {\bibfnamefont {B.}~\bibnamefont {Wang}}, \ and\
  \bibinfo {author} {\bibfnamefont {S.}~\bibnamefont {Das~Sarma}}} (\bibinfo
  {year} {2012}),\ \href {\doibase 10.1103/PhysRevA.85.013644} {\bibfield
  {journal} {\bibinfo  {journal} {Phys. Rev. A}\ }\textbf {\bibinfo {volume}
  {85}},\ \bibinfo {pages} {013644}}\BibitemShut {NoStop}%
\bibitem [{\citenamefont {Sebby-Strabley}\ \emph {et~al.}(2006)\citenamefont
  {Sebby-Strabley}, \citenamefont {Anderlini}, \citenamefont {Jessen},\ and\
  \citenamefont {Porto}}]{2006_Sebby-Strabley_Anderlini_PRA}%
  \BibitemOpen
  \bibfield  {author} {\bibinfo {author} {\bibnamefont {Sebby-Strabley},
  \bibfnamefont {J.}}, \bibinfo {author} {\bibfnamefont {M.}~\bibnamefont
  {Anderlini}}, \bibinfo {author} {\bibfnamefont {P.~S.}\ \bibnamefont
  {Jessen}}, \ and\ \bibinfo {author} {\bibfnamefont {J.~V.}\ \bibnamefont
  {Porto}}} (\bibinfo {year} {2006}),\ \href {\doibase
  10.1103/PhysRevA.73.033605} {\bibfield  {journal} {\bibinfo  {journal} {Phys.
  Rev. A}\ }\textbf {\bibinfo {volume} {73}},\ \bibinfo {pages}
  {033605}}\BibitemShut {NoStop}%
\bibitem [{\citenamefont {Shchesnovich}(2012)}]{2012_Shchesnovich_PRA}%
  \BibitemOpen
  \bibfield  {author} {\bibinfo {author} {\bibnamefont {Shchesnovich},
  \bibfnamefont {V.~S.}}} (\bibinfo {year} {2012}),\ \href {\doibase
  10.1103/PhysRevA.85.013614} {\bibfield  {journal} {\bibinfo  {journal} {Phys.
  Rev. A}\ }\textbf {\bibinfo {volume} {85}},\ \bibinfo {pages}
  {013614}}\BibitemShut {NoStop}%
\bibitem [{\citenamefont {Sheng}\ \emph {et~al.}(2011)\citenamefont {Sheng},
  \citenamefont {Gu}, \citenamefont {Sun},\ and\ \citenamefont
  {Sheng}}]{2011_Sheng_Gu_NatComm}%
  \BibitemOpen
  \bibfield  {author} {\bibinfo {author} {\bibnamefont {Sheng}, \bibfnamefont
  {D.}}, \bibinfo {author} {\bibfnamefont {Z.-C.}\ \bibnamefont {Gu}}, \bibinfo
  {author} {\bibfnamefont {K.}~\bibnamefont {Sun}}, \ and\ \bibinfo {author}
  {\bibfnamefont {L.}~\bibnamefont {Sheng}}} (\bibinfo {year} {2011}),\ \href
  {http://dx.doi.org/10.1038/ncomms1380} {\bibfield  {journal} {\bibinfo
  {journal} {Nature communications}\ }\textbf {\bibinfo {volume} {2}},\
  \bibinfo {pages} {389}}\BibitemShut {NoStop}%
\bibitem [{\citenamefont {Smerzi}\ \emph {et~al.}(1997)\citenamefont {Smerzi},
  \citenamefont {Fantoni}, \citenamefont {Giovanazzi},\ and\ \citenamefont
  {Shenoy}}]{1997_Smerzi_Fantoni_PRL}%
  \BibitemOpen
  \bibfield  {author} {\bibinfo {author} {\bibnamefont {Smerzi}, \bibfnamefont
  {A.}}, \bibinfo {author} {\bibfnamefont {S.}~\bibnamefont {Fantoni}},
  \bibinfo {author} {\bibfnamefont {S.}~\bibnamefont {Giovanazzi}}, \ and\
  \bibinfo {author} {\bibfnamefont {S.~R.}\ \bibnamefont {Shenoy}}} (\bibinfo
  {year} {1997}),\ \href {\doibase 10.1103/PhysRevLett.79.4950} {\bibfield
  {journal} {\bibinfo  {journal} {Phys. Rev. Lett.}\ }\textbf {\bibinfo
  {volume} {79}},\ \bibinfo {pages} {4950}}\BibitemShut {NoStop}%
\bibitem [{\citenamefont {Soltan-Panahi}\ \emph {et~al.}(2012)\citenamefont
  {Soltan-Panahi}, \citenamefont {L\"uhmann}, \citenamefont {Struck},
  \citenamefont {Windpassinger},\ and\ \citenamefont
  {Sengstock}}]{2012_Soltan_Luhmann_NatPhys}%
  \BibitemOpen
  \bibfield  {author} {\bibinfo {author} {\bibnamefont {Soltan-Panahi},
  \bibfnamefont {P.}}, \bibinfo {author} {\bibfnamefont {D.-S.}\ \bibnamefont
  {L\"uhmann}}, \bibinfo {author} {\bibfnamefont {J.}~\bibnamefont {Struck}},
  \bibinfo {author} {\bibfnamefont {P.}~\bibnamefont {Windpassinger}}, \ and\
  \bibinfo {author} {\bibfnamefont {K.}~\bibnamefont {Sengstock}}} (\bibinfo
  {year} {2012}),\ \href {\doibase 10.1038/nphys2128} {\bibfield  {journal}
  {\bibinfo  {journal} {Nature Phys.}\ }\textbf {\bibinfo {volume} {8}},\
  \bibinfo {pages} {71}}\BibitemShut {NoStop}%
\bibitem [{\citenamefont {Sowinski}(2012)}]{2012_Sowinski_PRL}%
  \BibitemOpen
  \bibfield  {author} {\bibinfo {author} {\bibnamefont {Sowinski},
  \bibfnamefont {T.}}} (\bibinfo {year} {2012}),\ \href {\doibase
  10.1103/PhysRevLett.108.165301} {\bibfield  {journal} {\bibinfo  {journal}
  {Phys. Rev. Lett.}\ }\textbf {\bibinfo {volume} {108}},\ \bibinfo {pages}
  {165301}}\BibitemShut {NoStop}%
\bibitem [{\citenamefont {Sowinski}\ \emph {et~al.}(2013)\citenamefont
  {Sowinski}, \citenamefont {Lacki}, \citenamefont {Dutta}, \citenamefont
  {Pietraszewicz}, \citenamefont {Sierant}, \citenamefont {Gajda},
  \citenamefont {Zakrzewski},\ and\ \citenamefont
  {Lewenstein}}]{2013_Sowinski_Lacki_PRL}%
  \BibitemOpen
  \bibfield  {author} {\bibinfo {author} {\bibnamefont {Sowinski},
  \bibfnamefont {T.}}, \bibinfo {author} {\bibfnamefont {M.}~\bibnamefont
  {Lacki}}, \bibinfo {author} {\bibfnamefont {O.}~\bibnamefont {Dutta}},
  \bibinfo {author} {\bibfnamefont {J.}~\bibnamefont {Pietraszewicz}}, \bibinfo
  {author} {\bibfnamefont {P.}~\bibnamefont {Sierant}}, \bibinfo {author}
  {\bibfnamefont {M.}~\bibnamefont {Gajda}}, \bibinfo {author} {\bibfnamefont
  {J.}~\bibnamefont {Zakrzewski}}, \ and\ \bibinfo {author} {\bibfnamefont
  {M.}~\bibnamefont {Lewenstein}}} (\bibinfo {year} {2013}),\ \href {\doibase
  10.1103/PhysRevLett.111.215302} {\bibfield  {journal} {\bibinfo  {journal}
  {Phys. Rev. Lett.}\ }\textbf {\bibinfo {volume} {111}},\ \bibinfo {pages}
  {215302}}\BibitemShut {NoStop}%
\bibitem [{\citenamefont {Stasyuk}\ and\ \citenamefont
  {Velychko}(2011)}]{2011_Stasyuk_Velychko_CMP}%
  \BibitemOpen
  \bibfield  {author} {\bibinfo {author} {\bibnamefont {Stasyuk}, \bibfnamefont
  {I.}}, \ and\ \bibinfo {author} {\bibfnamefont {O.}~\bibnamefont {Velychko}}}
  (\bibinfo {year} {2011}),\ \href {\doibase 10.5488/CMP.14.13004} {\bibfield
  {journal} {\bibinfo  {journal} {Condensed Matter Physics}\ }\textbf {\bibinfo
  {volume} {14}}~(\bibinfo {number} {1}),\ \bibinfo {pages}
  {13004}}\BibitemShut {NoStop}%
\bibitem [{\citenamefont {Stasyuk}\ and\ \citenamefont
  {Velychko}(2012)}]{2012_Stasyuk_Velychko_CMP}%
  \BibitemOpen
  \bibfield  {author} {\bibinfo {author} {\bibnamefont {Stasyuk}, \bibfnamefont
  {I.}}, \ and\ \bibinfo {author} {\bibfnamefont {O.}~\bibnamefont {Velychko}}}
  (\bibinfo {year} {2012}),\ \href {\doibase 10.5488/CMP.15.33002} {\bibfield
  {journal} {\bibinfo  {journal} {Condensed Matter Physics}\ }\textbf {\bibinfo
  {volume} {15}},\ \bibinfo {pages} {33002}}\BibitemShut {NoStop}%
\bibitem [{\citenamefont {Stojanovi\'{c}}\ \emph {et~al.}(2008)\citenamefont
  {Stojanovi\'{c}}, \citenamefont {Wu}, \citenamefont {Liu},\ and\
  \citenamefont {Das~Sarma}}]{2008_Stojanovic_Wu_PRL}%
  \BibitemOpen
  \bibfield  {author} {\bibinfo {author} {\bibnamefont {Stojanovi\'{c}},
  \bibfnamefont {V.~M.}}, \bibinfo {author} {\bibfnamefont {C.}~\bibnamefont
  {Wu}}, \bibinfo {author} {\bibfnamefont {W.~V.}\ \bibnamefont {Liu}}, \ and\
  \bibinfo {author} {\bibfnamefont {S.}~\bibnamefont {Das~Sarma}}} (\bibinfo
  {year} {2008}),\ \href {\doibase 10.1103/PhysRevLett.101.125301} {\bibfield
  {journal} {\bibinfo  {journal} {Phys. Rev. Lett.}\ }\textbf {\bibinfo
  {volume} {101}},\ \bibinfo {pages} {125301}}\BibitemShut {NoStop}%
\bibitem [{\citenamefont {Str\"ater}\ and\ \citenamefont
  {Eckardt}(2015)}]{2015_Strater_Eckardt_PRA}%
  \BibitemOpen
  \bibfield  {author} {\bibinfo {author} {\bibnamefont {Str\"ater},
  \bibfnamefont {C.}}, \ and\ \bibinfo {author} {\bibfnamefont
  {A.}~\bibnamefont {Eckardt}}} (\bibinfo {year} {2015}),\ \href {\doibase
  10.1103/PhysRevA.91.053602} {\bibfield  {journal} {\bibinfo  {journal} {Phys.
  Rev. A}\ }\textbf {\bibinfo {volume} {91}},\ \bibinfo {pages}
  {053602}}\BibitemShut {NoStop}%
\bibitem [{\citenamefont {Struck}\ \emph {et~al.}(2013)\citenamefont {Struck},
  \citenamefont {Weinberg}, \citenamefont {Olschlager}, \citenamefont
  {Windpassinger}, \citenamefont {Simonet}, \citenamefont {Sengstock},
  \citenamefont {Hoppner}, \citenamefont {Hauke}, \citenamefont {Eckardt},
  \citenamefont {Lewenstein},\ and\ \citenamefont
  {Mathey}}]{2013_Struck_Weinberg_NatPhys}%
  \BibitemOpen
  \bibfield  {author} {\bibinfo {author} {\bibnamefont {Struck}, \bibfnamefont
  {J.}}, \bibinfo {author} {\bibfnamefont {M.}~\bibnamefont {Weinberg}},
  \bibinfo {author} {\bibfnamefont {C.}~\bibnamefont {Olschlager}}, \bibinfo
  {author} {\bibfnamefont {P.}~\bibnamefont {Windpassinger}}, \bibinfo {author}
  {\bibfnamefont {J.}~\bibnamefont {Simonet}}, \bibinfo {author} {\bibfnamefont
  {K.}~\bibnamefont {Sengstock}}, \bibinfo {author} {\bibfnamefont
  {R.}~\bibnamefont {Hoppner}}, \bibinfo {author} {\bibfnamefont
  {P.}~\bibnamefont {Hauke}}, \bibinfo {author} {\bibfnamefont
  {A.}~\bibnamefont {Eckardt}}, \bibinfo {author} {\bibfnamefont
  {M.}~\bibnamefont {Lewenstein}}, \ and\ \bibinfo {author} {\bibfnamefont
  {L.}~\bibnamefont {Mathey}}} (\bibinfo {year} {2013}),\ \href
  {http://dx.doi.org/10.1038/nphys2750} {\bibfield  {journal} {\bibinfo
  {journal} {Nat Phys}\ }\textbf {\bibinfo {volume} {9}}~(\bibinfo {number}
  {11}),\ \bibinfo {pages} {738}}\BibitemShut {NoStop}%
\bibitem [{\citenamefont {Sun}\ \emph {et~al.}(2012{\natexlab{a}})\citenamefont
  {Sun}, \citenamefont {Jackeli}, \citenamefont {Santos},\ and\ \citenamefont
  {Vekua}}]{2012_Sun_Jackeli_PRB}%
  \BibitemOpen
  \bibfield  {author} {\bibinfo {author} {\bibnamefont {Sun}, \bibfnamefont
  {G.}}, \bibinfo {author} {\bibfnamefont {G.}~\bibnamefont {Jackeli}},
  \bibinfo {author} {\bibfnamefont {L.}~\bibnamefont {Santos}}, \ and\ \bibinfo
  {author} {\bibfnamefont {T.}~\bibnamefont {Vekua}}} (\bibinfo {year}
  {2012}{\natexlab{a}}),\ \href {\doibase 10.1103/PhysRevB.86.155159}
  {\bibfield  {journal} {\bibinfo  {journal} {Phys. Rev. B}\ }\textbf {\bibinfo
  {volume} {86}},\ \bibinfo {pages} {155159}}\BibitemShut {NoStop}%
\bibitem [{\citenamefont {Sun}\ \emph {et~al.}(2011)\citenamefont {Sun},
  \citenamefont {Gu}, \citenamefont {Katsura},\ and\ \citenamefont
  {Das~Sarma}}]{2011_Sun_Gu_PRL}%
  \BibitemOpen
  \bibfield  {author} {\bibinfo {author} {\bibnamefont {Sun}, \bibfnamefont
  {K.}}, \bibinfo {author} {\bibfnamefont {Z.}~\bibnamefont {Gu}}, \bibinfo
  {author} {\bibfnamefont {H.}~\bibnamefont {Katsura}}, \ and\ \bibinfo
  {author} {\bibfnamefont {S.}~\bibnamefont {Das~Sarma}}} (\bibinfo {year}
  {2011}),\ \href {\doibase 10.1103/PhysRevLett.106.236803} {\bibfield
  {journal} {\bibinfo  {journal} {Phys. Rev. Lett.}\ }\textbf {\bibinfo
  {volume} {106}},\ \bibinfo {pages} {236803}}\BibitemShut {NoStop}%
\bibitem [{\citenamefont {Sun}\ \emph {et~al.}(2012{\natexlab{b}})\citenamefont
  {Sun}, \citenamefont {Liu}, \citenamefont {Hemmerich},\ and\ \citenamefont
  {Das~Sarma}}]{2011_Sun_Liu_NatPhys}%
  \BibitemOpen
  \bibfield  {author} {\bibinfo {author} {\bibnamefont {Sun}, \bibfnamefont
  {K.}}, \bibinfo {author} {\bibfnamefont {W.~V.}\ \bibnamefont {Liu}},
  \bibinfo {author} {\bibfnamefont {A.}~\bibnamefont {Hemmerich}}, \ and\
  \bibinfo {author} {\bibfnamefont {S.}~\bibnamefont {Das~Sarma}}} (\bibinfo
  {year} {2012}{\natexlab{b}}),\ \href {\doibase 10.1038/nphys2134} {\bibfield
  {journal} {\bibinfo  {journal} {Nature Physics}\ }\textbf {\bibinfo {volume}
  {8}}~(\bibinfo {number} {1}),\ \bibinfo {pages} {67}}\BibitemShut {NoStop}%
\bibitem [{\citenamefont {Sun}\ \emph {et~al.}(2009)\citenamefont {Sun},
  \citenamefont {Yao}, \citenamefont {Fradkin},\ and\ \citenamefont
  {Kivelson}}]{2009_Sun_Yao_PRL}%
  \BibitemOpen
  \bibfield  {author} {\bibinfo {author} {\bibnamefont {Sun}, \bibfnamefont
  {K.}}, \bibinfo {author} {\bibfnamefont {H.}~\bibnamefont {Yao}}, \bibinfo
  {author} {\bibfnamefont {E.}~\bibnamefont {Fradkin}}, \ and\ \bibinfo
  {author} {\bibfnamefont {S.~A.}\ \bibnamefont {Kivelson}}} (\bibinfo {year}
  {2009}),\ \href {\doibase 10.1103/PhysRevLett.103.046811} {\bibfield
  {journal} {\bibinfo  {journal} {Phys. Rev. Lett.}\ }\textbf {\bibinfo
  {volume} {103}},\ \bibinfo {pages} {046811}}\BibitemShut {NoStop}%
\bibitem [{\citenamefont {Sun}\ \emph {et~al.}(2010)\citenamefont {Sun},
  \citenamefont {Zhao},\ and\ \citenamefont {Liu}}]{2010_Sun_Zhao_PRL}%
  \BibitemOpen
  \bibfield  {author} {\bibinfo {author} {\bibnamefont {Sun}, \bibfnamefont
  {K.}}, \bibinfo {author} {\bibfnamefont {E.}~\bibnamefont {Zhao}}, \ and\
  \bibinfo {author} {\bibfnamefont {W.~V.}\ \bibnamefont {Liu}}} (\bibinfo
  {year} {2010}),\ \href {\doibase 10.1103/PhysRevLett.104.165303} {\bibfield
  {journal} {\bibinfo  {journal} {Phys. Rev. Lett.}\ }\textbf {\bibinfo
  {volume} {104}},\ \bibinfo {pages} {165303}}\BibitemShut {NoStop}%
\bibitem [{\citenamefont {Tarruell}\ \emph {et~al.}(2012)\citenamefont
  {Tarruell}, \citenamefont {Greif}, \citenamefont {Uehlinger}, \citenamefont
  {Jotzu},\ and\ \citenamefont {Esslinger}}]{2012_Tarruell_Greif_Nature}%
  \BibitemOpen
  \bibfield  {author} {\bibinfo {author} {\bibnamefont {Tarruell},
  \bibfnamefont {L.}}, \bibinfo {author} {\bibfnamefont {D.}~\bibnamefont
  {Greif}}, \bibinfo {author} {\bibfnamefont {T.}~\bibnamefont {Uehlinger}},
  \bibinfo {author} {\bibfnamefont {G.}~\bibnamefont {Jotzu}}, \ and\ \bibinfo
  {author} {\bibfnamefont {T.}~\bibnamefont {Esslinger}}} (\bibinfo {year}
  {2012}),\ \href {http://dx.doi.org/10.1038/nature10871} {\bibfield  {journal}
  {\bibinfo  {journal} {Nature}\ }\textbf {\bibinfo {volume} {483}}~(\bibinfo
  {number} {7389}),\ \bibinfo {pages} {302}}\BibitemShut {NoStop}%
\bibitem [{\citenamefont {Tokura}\ and\ \citenamefont
  {Nagaosa}(2000)}]{2000_Tokura_Nagaosa_Science}%
  \BibitemOpen
  \bibfield  {author} {\bibinfo {author} {\bibnamefont {Tokura}, \bibfnamefont
  {Y.}}, \ and\ \bibinfo {author} {\bibfnamefont {N.}~\bibnamefont {Nagaosa}}}
  (\bibinfo {year} {2000}),\ \href {\doibase 10.1126/science.288.5465.462}
  {\bibfield  {journal} {\bibinfo  {journal} {Science}\ }\textbf {\bibinfo
  {volume} {288}}~(\bibinfo {number} {5465}),\ \bibinfo {pages}
  {462}}\BibitemShut {NoStop}%
\bibitem [{\citenamefont {Trotzky}\ \emph {et~al.}(2008)\citenamefont
  {Trotzky}, \citenamefont {Cheinet}, \citenamefont {F{\"o}lling},
  \citenamefont {Feld}, \citenamefont {Schnorrberger}, \citenamefont {Rey},
  \citenamefont {Polkovnikov}, \citenamefont {Demler}, \citenamefont {Lukin},\
  and\ \citenamefont {Bloch}}]{2008_Trotzky_Cheinet_Science}%
  \BibitemOpen
  \bibfield  {author} {\bibinfo {author} {\bibnamefont {Trotzky}, \bibfnamefont
  {S.}}, \bibinfo {author} {\bibfnamefont {P.}~\bibnamefont {Cheinet}},
  \bibinfo {author} {\bibfnamefont {S.}~\bibnamefont {F{\"o}lling}}, \bibinfo
  {author} {\bibfnamefont {M.}~\bibnamefont {Feld}}, \bibinfo {author}
  {\bibfnamefont {U.}~\bibnamefont {Schnorrberger}}, \bibinfo {author}
  {\bibfnamefont {A.~M.}\ \bibnamefont {Rey}}, \bibinfo {author} {\bibfnamefont
  {A.}~\bibnamefont {Polkovnikov}}, \bibinfo {author} {\bibfnamefont {E.~A.}\
  \bibnamefont {Demler}}, \bibinfo {author} {\bibfnamefont {M.~D.}\
  \bibnamefont {Lukin}}, \ and\ \bibinfo {author} {\bibfnamefont
  {I.}~\bibnamefont {Bloch}}} (\bibinfo {year} {2008}),\ \href {\doibase
  10.1126/science.1150841} {\bibfield  {journal} {\bibinfo  {journal}
  {Science}\ }\textbf {\bibinfo {volume} {319}}~(\bibinfo {number} {5861}),\
  \bibinfo {pages} {295}}\BibitemShut {NoStop}%
\bibitem [{\citenamefont {Uehlinger}\ \emph {et~al.}(2013)\citenamefont
  {Uehlinger}, \citenamefont {Jotzu}, \citenamefont {Messer}, \citenamefont
  {Greif}, \citenamefont {Hofstetter}, \citenamefont {Bissbort},\ and\
  \citenamefont {Esslinger}}]{2013_Uehlinger_Jotzu_PRL}%
  \BibitemOpen
  \bibfield  {author} {\bibinfo {author} {\bibnamefont {Uehlinger},
  \bibfnamefont {T.}}, \bibinfo {author} {\bibfnamefont {G.}~\bibnamefont
  {Jotzu}}, \bibinfo {author} {\bibfnamefont {M.}~\bibnamefont {Messer}},
  \bibinfo {author} {\bibfnamefont {D.}~\bibnamefont {Greif}}, \bibinfo
  {author} {\bibfnamefont {W.}~\bibnamefont {Hofstetter}}, \bibinfo {author}
  {\bibfnamefont {U.}~\bibnamefont {Bissbort}}, \ and\ \bibinfo {author}
  {\bibfnamefont {T.}~\bibnamefont {Esslinger}}} (\bibinfo {year} {2013}),\
  \href {\doibase 10.1103/PhysRevLett.111.185307} {\bibfield  {journal}
  {\bibinfo  {journal} {Phys. Rev. Lett.}\ }\textbf {\bibinfo {volume} {111}},\
  \bibinfo {pages} {185307}}\BibitemShut {NoStop}%
\bibitem [{\citenamefont {Umucal{\i}lar}\ and\ \citenamefont
  {Oktel}(2008)}]{2008_Umucalar_Oktel_PRA}%
  \BibitemOpen
  \bibfield  {author} {\bibinfo {author} {\bibnamefont {Umucal{\i}lar},
  \bibfnamefont {R.}}, \ and\ \bibinfo {author} {\bibfnamefont {M.~O.}\
  \bibnamefont {Oktel}}} (\bibinfo {year} {2008}),\ \href {\doibase
  10.1103/PhysRevA.78.033602} {\bibfield  {journal} {\bibinfo  {journal} {Phys.
  Rev. A}\ }\textbf {\bibinfo {volume} {78}},\ \bibinfo {pages}
  {033602}}\BibitemShut {NoStop}%
\bibitem [{\citenamefont {Wang}\ \emph {et~al.}(2008)\citenamefont {Wang},
  \citenamefont {Dai}, \citenamefont {Chen},\ and\ \citenamefont
  {Xie}}]{2008_Wang_Dai_PRA}%
  \BibitemOpen
  \bibfield  {author} {\bibinfo {author} {\bibnamefont {Wang}, \bibfnamefont
  {L.}}, \bibinfo {author} {\bibfnamefont {X.}~\bibnamefont {Dai}}, \bibinfo
  {author} {\bibfnamefont {S.}~\bibnamefont {Chen}}, \ and\ \bibinfo {author}
  {\bibfnamefont {X.~C.}\ \bibnamefont {Xie}}} (\bibinfo {year} {2008}),\ \href
  {\doibase 10.1103/PhysRevA.78.023603} {\bibfield  {journal} {\bibinfo
  {journal} {Phys. Rev. A}\ }\textbf {\bibinfo {volume} {78}},\ \bibinfo
  {pages} {023603}}\BibitemShut {NoStop}%
\bibitem [{\citenamefont {Wang}\ and\ \citenamefont
  {Gong}(2010)}]{2010_Wang_Gong_PRB}%
  \BibitemOpen
  \bibfield  {author} {\bibinfo {author} {\bibnamefont {Wang}, \bibfnamefont
  {Y.-F.}}, \ and\ \bibinfo {author} {\bibfnamefont {C.-D.}\ \bibnamefont
  {Gong}}} (\bibinfo {year} {2010}),\ \href {\doibase
  10.1103/PhysRevB.82.113304} {\bibfield  {journal} {\bibinfo  {journal} {Phys.
  Rev. B}\ }\textbf {\bibinfo {volume} {82}},\ \bibinfo {pages}
  {113304}}\BibitemShut {NoStop}%
\bibitem [{\citenamefont {Wang}\ \emph {et~al.}(2011)\citenamefont {Wang},
  \citenamefont {Gu}, \citenamefont {Gong},\ and\ \citenamefont
  {Sheng}}]{2011_Wang_Gu_PRL}%
  \BibitemOpen
  \bibfield  {author} {\bibinfo {author} {\bibnamefont {Wang}, \bibfnamefont
  {Y.-F.}}, \bibinfo {author} {\bibfnamefont {Z.-C.}\ \bibnamefont {Gu}},
  \bibinfo {author} {\bibfnamefont {C.-D.}\ \bibnamefont {Gong}}, \ and\
  \bibinfo {author} {\bibfnamefont {D.~N.}\ \bibnamefont {Sheng}}} (\bibinfo
  {year} {2011}),\ \href {\doibase 10.1103/PhysRevLett.107.146803} {\bibfield
  {journal} {\bibinfo  {journal} {Phys. Rev. Lett.}\ }\textbf {\bibinfo
  {volume} {107}},\ \bibinfo {pages} {146803}}\BibitemShut {NoStop}%
\bibitem [{\citenamefont {Weinberg}\ \emph {et~al.}(2015)\citenamefont
  {Weinberg}, \citenamefont {\"Olschl\"ager}, \citenamefont {Str\"ater},
  \citenamefont {Prelle}, \citenamefont {Eckardt}, \citenamefont {Sengstock},\
  and\ \citenamefont {Simonet}}]{2015_Weinberg_Olschlager_PRA}%
  \BibitemOpen
  \bibfield  {author} {\bibinfo {author} {\bibnamefont {Weinberg},
  \bibfnamefont {M.}}, \bibinfo {author} {\bibfnamefont {C.}~\bibnamefont
  {\"Olschl\"ager}}, \bibinfo {author} {\bibfnamefont {C.}~\bibnamefont
  {Str\"ater}}, \bibinfo {author} {\bibfnamefont {S.}~\bibnamefont {Prelle}},
  \bibinfo {author} {\bibfnamefont {A.}~\bibnamefont {Eckardt}}, \bibinfo
  {author} {\bibfnamefont {K.}~\bibnamefont {Sengstock}}, \ and\ \bibinfo
  {author} {\bibfnamefont {J.}~\bibnamefont {Simonet}}} (\bibinfo {year}
  {2015}),\ \href {\doibase 10.1103/PhysRevA.92.043621} {\bibfield  {journal}
  {\bibinfo  {journal} {Phys. Rev. A}\ }\textbf {\bibinfo {volume} {92}},\
  \bibinfo {pages} {043621}}\BibitemShut {NoStop}%
\bibitem [{\citenamefont {Wen}(2012)}]{2012_Wen_PRB}%
  \BibitemOpen
  \bibfield  {author} {\bibinfo {author} {\bibnamefont {Wen}, \bibfnamefont
  {X.-G.}}} (\bibinfo {year} {2012}),\ \href {\doibase
  10.1103/PhysRevB.85.085103} {\bibfield  {journal} {\bibinfo  {journal} {Phys.
  Rev. B}\ }\textbf {\bibinfo {volume} {85}},\ \bibinfo {pages}
  {085103}}\BibitemShut {NoStop}%
\bibitem [{\citenamefont {Wirth}\ \emph {et~al.}(2011)\citenamefont {Wirth},
  \citenamefont {\"Olschl\"ager},\ and\ \citenamefont
  {Hemmerich}}]{2011_Wirth_Olschlager_NatPhys}%
  \BibitemOpen
  \bibfield  {author} {\bibinfo {author} {\bibnamefont {Wirth}, \bibfnamefont
  {G.}}, \bibinfo {author} {\bibfnamefont {M.}~\bibnamefont {\"Olschl\"ager}},
  \ and\ \bibinfo {author} {\bibfnamefont {A.}~\bibnamefont {Hemmerich}}}
  (\bibinfo {year} {2011}),\ \href {\doibase 10.1038/nphys1857} {\bibfield
  {journal} {\bibinfo  {journal} {Nature Physics}\ }\textbf {\bibinfo {volume}
  {7}},\ \bibinfo {pages} {147}}\BibitemShut {NoStop}%
\bibitem [{\citenamefont {Wu}\ and\ \citenamefont
  {Niu}(2001)}]{2001_Wu_Niu_PRA}%
  \BibitemOpen
  \bibfield  {author} {\bibinfo {author} {\bibnamefont {Wu}, \bibfnamefont
  {B.}}, \ and\ \bibinfo {author} {\bibfnamefont {Q.}~\bibnamefont {Niu}}}
  (\bibinfo {year} {2001}),\ \href {\doibase 10.1103/PhysRevA.64.061603}
  {\bibfield  {journal} {\bibinfo  {journal} {Phys. Rev. A}\ }\textbf {\bibinfo
  {volume} {64}},\ \bibinfo {pages} {061603}}\BibitemShut {NoStop}%
\bibitem [{\citenamefont {Wu}(2008{\natexlab{a}})}]{2008_Wu_PRL}%
  \BibitemOpen
  \bibfield  {author} {\bibinfo {author} {\bibnamefont {Wu}, \bibfnamefont
  {C.}}} (\bibinfo {year} {2008}{\natexlab{a}}),\ \href {\doibase
  10.1103/PhysRevLett.101.186807} {\bibfield  {journal} {\bibinfo  {journal}
  {Phys. Rev. Lett.}\ }\textbf {\bibinfo {volume} {101}},\ \bibinfo {pages}
  {186807}}\BibitemShut {NoStop}%
\bibitem [{\citenamefont {Wu}(2008{\natexlab{b}})}]{2008_Wu_PRL2}%
  \BibitemOpen
  \bibfield  {author} {\bibinfo {author} {\bibnamefont {Wu}, \bibfnamefont
  {C.}}} (\bibinfo {year} {2008}{\natexlab{b}}),\ \href {\doibase
  10.1103/PhysRevLett.100.200406} {\bibfield  {journal} {\bibinfo  {journal}
  {Phys. Rev. Lett.}\ }\textbf {\bibinfo {volume} {100}},\ \bibinfo {pages}
  {200406}}\BibitemShut {NoStop}%
\bibitem [{\citenamefont {Wu}(2009)}]{2009_Wu_MPLB}%
  \BibitemOpen
  \bibfield  {author} {\bibinfo {author} {\bibnamefont {Wu}, \bibfnamefont
  {C.}}} (\bibinfo {year} {2009}),\ \href {\doibase 10.1142/S0217984909017777}
  {\bibfield  {journal} {\bibinfo  {journal} {Mod. Phys. Lett. B}\ }\textbf
  {\bibinfo {volume} {23}}~(\bibinfo {number} {01}),\ \bibinfo {pages}
  {1}}\BibitemShut {NoStop}%
\bibitem [{\citenamefont {Wu}\ \emph {et~al.}(2007)\citenamefont {Wu},
  \citenamefont {Bergman}, \citenamefont {Balents},\ and\ \citenamefont
  {Das~Sarma}}]{2007_Wu_Bergman_PRL}%
  \BibitemOpen
  \bibfield  {author} {\bibinfo {author} {\bibnamefont {Wu}, \bibfnamefont
  {C.}}, \bibinfo {author} {\bibfnamefont {D.}~\bibnamefont {Bergman}},
  \bibinfo {author} {\bibfnamefont {L.}~\bibnamefont {Balents}}, \ and\
  \bibinfo {author} {\bibfnamefont {S.}~\bibnamefont {Das~Sarma}}} (\bibinfo
  {year} {2007}),\ \href {\doibase 10.1103/PhysRevLett.99.070401} {\bibfield
  {journal} {\bibinfo  {journal} {Phys. Rev. Lett.}\ }\textbf {\bibinfo
  {volume} {99}},\ \bibinfo {pages} {070401}}\BibitemShut {NoStop}%
\bibitem [{\citenamefont {Wu}\ and\ \citenamefont
  {Das~Sarma}(2008)}]{2008_Wu_Sarma_PRB}%
  \BibitemOpen
  \bibfield  {author} {\bibinfo {author} {\bibnamefont {Wu}, \bibfnamefont
  {C.}}, \ and\ \bibinfo {author} {\bibfnamefont {S.}~\bibnamefont
  {Das~Sarma}}} (\bibinfo {year} {2008}),\ \href {\doibase
  10.1103/PhysRevB.77.235107} {\bibfield  {journal} {\bibinfo  {journal} {Phys.
  Rev. B}\ }\textbf {\bibinfo {volume} {77}},\ \bibinfo {pages}
  {235107}}\BibitemShut {NoStop}%
\bibitem [{\citenamefont {Wu}\ \emph {et~al.}(2006)\citenamefont {Wu},
  \citenamefont {Liu}, \citenamefont {Moore},\ and\ \citenamefont
  {Das~Sarma}}]{2006_Wu_Liu_PRL}%
  \BibitemOpen
  \bibfield  {author} {\bibinfo {author} {\bibnamefont {Wu}, \bibfnamefont
  {C.}}, \bibinfo {author} {\bibfnamefont {W.~V.}\ \bibnamefont {Liu}},
  \bibinfo {author} {\bibfnamefont {J.}~\bibnamefont {Moore}}, \ and\ \bibinfo
  {author} {\bibfnamefont {S.}~\bibnamefont {Das~Sarma}}} (\bibinfo {year}
  {2006}),\ \href {\doibase 10.1103/PhysRevLett.97.190406} {\bibfield
  {journal} {\bibinfo  {journal} {Phys. Rev. Lett.}\ }\textbf {\bibinfo
  {volume} {97}},\ \bibinfo {pages} {190406}}\BibitemShut {NoStop}%
\bibitem [{\citenamefont {Wu}\ and\ \citenamefont
  {Zhai}(2008)}]{2008_Wu_Zhai_PRB}%
  \BibitemOpen
  \bibfield  {author} {\bibinfo {author} {\bibnamefont {Wu}, \bibfnamefont
  {K.}}, \ and\ \bibinfo {author} {\bibfnamefont {H.}~\bibnamefont {Zhai}}}
  (\bibinfo {year} {2008}),\ \href {\doibase 10.1103/PhysRevB.77.174431}
  {\bibfield  {journal} {\bibinfo  {journal} {Phys. Rev. B}\ }\textbf {\bibinfo
  {volume} {77}},\ \bibinfo {pages} {174431}}\BibitemShut {NoStop}%
\bibitem [{\citenamefont {Wu}\ \emph {et~al.}(2012)\citenamefont {Wu},
  \citenamefont {He}, \citenamefont {Zang},\ and\ \citenamefont
  {Kou}}]{2012_Wu_He_PRB}%
  \BibitemOpen
  \bibfield  {author} {\bibinfo {author} {\bibnamefont {Wu}, \bibfnamefont
  {Y.-J.}}, \bibinfo {author} {\bibfnamefont {J.}~\bibnamefont {He}}, \bibinfo
  {author} {\bibfnamefont {C.-L.}\ \bibnamefont {Zang}}, \ and\ \bibinfo
  {author} {\bibfnamefont {S.-P.}\ \bibnamefont {Kou}}} (\bibinfo {year}
  {2012}),\ \href {\doibase 10.1103/PhysRevB.86.085128} {\bibfield  {journal}
  {\bibinfo  {journal} {Phys. Rev. B}\ }\textbf {\bibinfo {volume} {86}},\
  \bibinfo {pages} {085128}}\BibitemShut {NoStop}%
\bibitem [{\citenamefont {Xu}\ and\ \citenamefont
  {Fisher}(2007)}]{2007_Xu_Fisher_PRB}%
  \BibitemOpen
  \bibfield  {author} {\bibinfo {author} {\bibnamefont {Xu}, \bibfnamefont
  {C.}}, \ and\ \bibinfo {author} {\bibfnamefont {M.~P.~A.}\ \bibnamefont
  {Fisher}}} (\bibinfo {year} {2007}),\ \href {\doibase
  10.1103/PhysRevB.75.104428} {\bibfield  {journal} {\bibinfo  {journal} {Phys.
  Rev. B}\ }\textbf {\bibinfo {volume} {75}},\ \bibinfo {pages}
  {104428}}\BibitemShut {NoStop}%
\bibitem [{\citenamefont {Xu}\ \emph {et~al.}(2013)\citenamefont {Xu},
  \citenamefont {Chen}, \citenamefont {Xiong}, \citenamefont {Liu},\ and\
  \citenamefont {Wu}}]{2013_Xu_Chen_PRA}%
  \BibitemOpen
  \bibfield  {author} {\bibinfo {author} {\bibnamefont {Xu}, \bibfnamefont
  {Y.}}, \bibinfo {author} {\bibfnamefont {Z.}~\bibnamefont {Chen}}, \bibinfo
  {author} {\bibfnamefont {H.}~\bibnamefont {Xiong}}, \bibinfo {author}
  {\bibfnamefont {W.~V.}\ \bibnamefont {Liu}}, \ and\ \bibinfo {author}
  {\bibfnamefont {B.}~\bibnamefont {Wu}}} (\bibinfo {year} {2013}),\ \href
  {\doibase 10.1103/PhysRevA.87.013635} {\bibfield  {journal} {\bibinfo
  {journal} {Phys. Rev. A}\ }\textbf {\bibinfo {volume} {87}},\ \bibinfo
  {pages} {013635}}\BibitemShut {NoStop}%
\bibitem [{\citenamefont {Yang}(2001)}]{2001_Yang_PRB}%
  \BibitemOpen
  \bibfield  {author} {\bibinfo {author} {\bibnamefont {Yang}, \bibfnamefont
  {K.}}} (\bibinfo {year} {2001}),\ \href {\doibase 10.1103/PhysRevB.63.140511}
  {\bibfield  {journal} {\bibinfo  {journal} {Phys. Rev. B}\ }\textbf {\bibinfo
  {volume} {63}},\ \bibinfo {pages} {140511}}\BibitemShut {NoStop}%
\bibitem [{\citenamefont {Yin}\ \emph {et~al.}(2015)\citenamefont {Yin},
  \citenamefont {Baarsma}, \citenamefont {Heikkinen}, \citenamefont
  {Martikainen},\ and\ \citenamefont {T\"orm\"a}}]{2015_Yin_Baarsma_arXiv}%
  \BibitemOpen
  \bibfield  {author} {\bibinfo {author} {\bibnamefont {Yin}, \bibfnamefont
  {S.}}, \bibinfo {author} {\bibfnamefont {J.~E.}\ \bibnamefont {Baarsma}},
  \bibinfo {author} {\bibfnamefont {M.~O.~J.}\ \bibnamefont {Heikkinen}},
  \bibinfo {author} {\bibfnamefont {J.-P.}\ \bibnamefont {Martikainen}}, \ and\
  \bibinfo {author} {\bibfnamefont {P.}~\bibnamefont {T\"orm\"a}}} (\bibinfo
  {year} {2015}),\ \href {\doibase 10.1103/PhysRevA.92.053616} {\bibfield
  {journal} {\bibinfo  {journal} {Phys. Rev. A}\ }\textbf {\bibinfo {volume}
  {92}},\ \bibinfo {pages} {053616}}\BibitemShut {NoStop}%
\bibitem [{\citenamefont {Zapata}\ \emph {et~al.}(1998)\citenamefont {Zapata},
  \citenamefont {Sols},\ and\ \citenamefont {Leggett}}]{1998_Zapata_Sols_PRA}%
  \BibitemOpen
  \bibfield  {author} {\bibinfo {author} {\bibnamefont {Zapata}, \bibfnamefont
  {I.}}, \bibinfo {author} {\bibfnamefont {F.}~\bibnamefont {Sols}}, \ and\
  \bibinfo {author} {\bibfnamefont {A.~J.}\ \bibnamefont {Leggett}}} (\bibinfo
  {year} {1998}),\ \href {\doibase 10.1103/PhysRevA.57.R28} {\bibfield
  {journal} {\bibinfo  {journal} {Phys. Rev. A}\ }\textbf {\bibinfo {volume}
  {57}},\ \bibinfo {pages} {R28}}\BibitemShut {NoStop}%
\bibitem [{\citenamefont {Zhai}(2015)}]{2015_Zhai_RPP}%
  \BibitemOpen
  \bibfield  {author} {\bibinfo {author} {\bibnamefont {Zhai}, \bibfnamefont
  {H.}}} (\bibinfo {year} {2015}),\ \href
  {http://stacks.iop.org/0034-4885/78/i=2/a=026001} {\bibfield  {journal}
  {\bibinfo  {journal} {Rep. Prog. Phys.}\ }\textbf {\bibinfo {volume}
  {78}}~(\bibinfo {number} {2}),\ \bibinfo {pages} {026001}}\BibitemShut
  {NoStop}%
\bibitem [{\citenamefont {Zhai}\ \emph {et~al.}(2013)\citenamefont {Zhai},
  \citenamefont {Yue}, \citenamefont {Wu}, \citenamefont {Chen}, \citenamefont
  {Zhang},\ and\ \citenamefont {Zhou}}]{2013_Zhai_Yue_PRA}%
  \BibitemOpen
  \bibfield  {author} {\bibinfo {author} {\bibnamefont {Zhai}, \bibfnamefont
  {Y.}}, \bibinfo {author} {\bibfnamefont {X.}~\bibnamefont {Yue}}, \bibinfo
  {author} {\bibfnamefont {Y.}~\bibnamefont {Wu}}, \bibinfo {author}
  {\bibfnamefont {X.}~\bibnamefont {Chen}}, \bibinfo {author} {\bibfnamefont
  {P.}~\bibnamefont {Zhang}}, \ and\ \bibinfo {author} {\bibfnamefont
  {X.}~\bibnamefont {Zhou}}} (\bibinfo {year} {2013}),\ \href {\doibase
  10.1103/PhysRevA.87.063638} {\bibfield  {journal} {\bibinfo  {journal} {Phys.
  Rev. A}\ }\textbf {\bibinfo {volume} {87}},\ \bibinfo {pages}
  {063638}}\BibitemShut {NoStop}%
\bibitem [{\citenamefont {Zhang}\ \emph {et~al.}(2011)\citenamefont {Zhang},
  \citenamefont {Hung}, \citenamefont {Zhang},\ and\ \citenamefont
  {Wu}}]{2011_Zhang_Hung_PRA}%
  \BibitemOpen
  \bibfield  {author} {\bibinfo {author} {\bibnamefont {Zhang}, \bibfnamefont
  {M.}}, \bibinfo {author} {\bibfnamefont {H.-h.}\ \bibnamefont {Hung}},
  \bibinfo {author} {\bibfnamefont {C.}~\bibnamefont {Zhang}}, \ and\ \bibinfo
  {author} {\bibfnamefont {C.}~\bibnamefont {Wu}}} (\bibinfo {year} {2011}),\
  \href {\doibase 10.1103/PhysRevA.83.023615} {\bibfield  {journal} {\bibinfo
  {journal} {Phys. Rev. A}\ }\textbf {\bibinfo {volume} {83}},\ \bibinfo
  {pages} {023615}}\BibitemShut {NoStop}%
\bibitem [{\citenamefont {Zhang}\ \emph
  {et~al.}(2010{\natexlab{a}})\citenamefont {Zhang}, \citenamefont {Hung},\
  and\ \citenamefont {Wu}}]{2010_Zhang_Hung_PRA2}%
  \BibitemOpen
  \bibfield  {author} {\bibinfo {author} {\bibnamefont {Zhang}, \bibfnamefont
  {S.}}, \bibinfo {author} {\bibfnamefont {H.-h.}\ \bibnamefont {Hung}}, \ and\
  \bibinfo {author} {\bibfnamefont {C.}~\bibnamefont {Wu}}} (\bibinfo {year}
  {2010}{\natexlab{a}}),\ \href {\doibase 10.1103/PhysRevA.82.053618}
  {\bibfield  {journal} {\bibinfo  {journal} {Phys. Rev. A}\ }\textbf {\bibinfo
  {volume} {82}},\ \bibinfo {pages} {053618}}\BibitemShut {NoStop}%
\bibitem [{\citenamefont {Zhang}\ \emph {et~al.}(2015)\citenamefont {Zhang},
  \citenamefont {Lang},\ and\ \citenamefont {Zhou}}]{2015_Zhang_Lang_arXiv}%
  \BibitemOpen
  \bibfield  {author} {\bibinfo {author} {\bibnamefont {Zhang}, \bibfnamefont
  {S.-L.}}, \bibinfo {author} {\bibfnamefont {L.-J.}\ \bibnamefont {Lang}}, \
  and\ \bibinfo {author} {\bibfnamefont {Q.}~\bibnamefont {Zhou}}} (\bibinfo
  {year} {2015}),\ \href {\doibase 10.1103/PhysRevLett.115.225301} {\bibfield
  {journal} {\bibinfo  {journal} {Phys. Rev. Lett.}\ }\textbf {\bibinfo
  {volume} {115}},\ \bibinfo {pages} {225301}}\BibitemShut {NoStop}%
\bibitem [{\citenamefont {Zhang}\ and\ \citenamefont
  {Zhou}(2014)}]{2014_Zhang_Zhou_PRA}%
  \BibitemOpen
  \bibfield  {author} {\bibinfo {author} {\bibnamefont {Zhang}, \bibfnamefont
  {S.-L.}}, \ and\ \bibinfo {author} {\bibfnamefont {Q.}~\bibnamefont {Zhou}}}
  (\bibinfo {year} {2014}),\ \href {\doibase 10.1103/PhysRevA.90.051601}
  {\bibfield  {journal} {\bibinfo  {journal} {Phys. Rev. A}\ }\textbf {\bibinfo
  {volume} {90}},\ \bibinfo {pages} {051601}}\BibitemShut {NoStop}%
\bibitem [{\citenamefont {Zhang}\ \emph
  {et~al.}(2010{\natexlab{b}})\citenamefont {Zhang}, \citenamefont {Hung},
  \citenamefont {Ho}, \citenamefont {Zhao},\ and\ \citenamefont
  {Liu}}]{2010_Zhang_Hung_PRA}%
  \BibitemOpen
  \bibfield  {author} {\bibinfo {author} {\bibnamefont {Zhang}, \bibfnamefont
  {Z.}}, \bibinfo {author} {\bibfnamefont {H.-H.}\ \bibnamefont {Hung}},
  \bibinfo {author} {\bibfnamefont {C.~M.}\ \bibnamefont {Ho}}, \bibinfo
  {author} {\bibfnamefont {E.}~\bibnamefont {Zhao}}, \ and\ \bibinfo {author}
  {\bibfnamefont {W.~V.}\ \bibnamefont {Liu}}} (\bibinfo {year}
  {2010}{\natexlab{b}}),\ \href {\doibase 10.1103/PhysRevA.82.033610}
  {\bibfield  {journal} {\bibinfo  {journal} {Phys. Rev. A}\ }\textbf {\bibinfo
  {volume} {82}},\ \bibinfo {pages} {033610}}\BibitemShut {NoStop}%
\bibitem [{\citenamefont {Zhang}\ \emph {et~al.}(2012)\citenamefont {Zhang},
  \citenamefont {Li},\ and\ \citenamefont {Liu}}]{2012_Zhang_Li_PRA}%
  \BibitemOpen
  \bibfield  {author} {\bibinfo {author} {\bibnamefont {Zhang}, \bibfnamefont
  {Z.}}, \bibinfo {author} {\bibfnamefont {X.}~\bibnamefont {Li}}, \ and\
  \bibinfo {author} {\bibfnamefont {W.~V.}\ \bibnamefont {Liu}}} (\bibinfo
  {year} {2012}),\ \href {\doibase 10.1103/PhysRevA.85.053606} {\bibfield
  {journal} {\bibinfo  {journal} {Phys. Rev. A}\ }\textbf {\bibinfo {volume}
  {85}},\ \bibinfo {pages} {053606}}\BibitemShut {NoStop}%
\bibitem [{\citenamefont {Zhao}\ and\ \citenamefont
  {Liu}(2008)}]{2008_Zhao_Liu_PRL}%
  \BibitemOpen
  \bibfield  {author} {\bibinfo {author} {\bibnamefont {Zhao}, \bibfnamefont
  {E.}}, \ and\ \bibinfo {author} {\bibfnamefont {W.~V.}\ \bibnamefont {Liu}}}
  (\bibinfo {year} {2008}),\ \href {\doibase 10.1103/PhysRevLett.100.160403}
  {\bibfield  {journal} {\bibinfo  {journal} {Phys. Rev. Lett.}\ }\textbf
  {\bibinfo {volume} {100}},\ \bibinfo {pages} {160403}}\BibitemShut {NoStop}%
\bibitem [{\citenamefont {Zhou}\ \emph
  {et~al.}(2011{\natexlab{a}})\citenamefont {Zhou}, \citenamefont {Porto},\
  and\ \citenamefont {Das~Sarma}}]{2011_Zhou_Porto_PRB}%
  \BibitemOpen
  \bibfield  {author} {\bibinfo {author} {\bibnamefont {Zhou}, \bibfnamefont
  {Q.}}, \bibinfo {author} {\bibfnamefont {J.~V.}\ \bibnamefont {Porto}}, \
  and\ \bibinfo {author} {\bibfnamefont {S.}~\bibnamefont {Das~Sarma}}}
  (\bibinfo {year} {2011}{\natexlab{a}}),\ \href {\doibase
  10.1103/PhysRevB.83.195106} {\bibfield  {journal} {\bibinfo  {journal} {Phys.
  Rev. B}\ }\textbf {\bibinfo {volume} {83}},\ \bibinfo {pages}
  {195106}}\BibitemShut {NoStop}%
\bibitem [{\citenamefont {Zhou}\ \emph
  {et~al.}(2011{\natexlab{b}})\citenamefont {Zhou}, \citenamefont {Porto},\
  and\ \citenamefont {Das~Sarma}}]{2011_Zhou_Porto_PRAR}%
  \BibitemOpen
  \bibfield  {author} {\bibinfo {author} {\bibnamefont {Zhou}, \bibfnamefont
  {Q.}}, \bibinfo {author} {\bibfnamefont {J.~V.}\ \bibnamefont {Porto}}, \
  and\ \bibinfo {author} {\bibfnamefont {S.}~\bibnamefont {Das~Sarma}}}
  (\bibinfo {year} {2011}{\natexlab{b}}),\ \href {\doibase
  10.1103/PhysRevA.84.031607} {\bibfield  {journal} {\bibinfo  {journal} {Phys.
  Rev. A}\ }\textbf {\bibinfo {volume} {84}},\ \bibinfo {pages}
  {031607}}\BibitemShut {NoStop}%
\bibitem [{\citenamefont {Zhou}\ \emph {et~al.}(2015)\citenamefont {Zhou},
  \citenamefont {Zhao},\ and\ \citenamefont {Liu}}]{2015_Zhou_Zhao_PRL}%
  \BibitemOpen
  \bibfield  {author} {\bibinfo {author} {\bibnamefont {Zhou}, \bibfnamefont
  {Z.}}, \bibinfo {author} {\bibfnamefont {E.}~\bibnamefont {Zhao}}, \ and\
  \bibinfo {author} {\bibfnamefont {W.~V.}\ \bibnamefont {Liu}}} (\bibinfo
  {year} {2015}),\ \href {\doibase 10.1103/PhysRevLett.114.100406} {\bibfield
  {journal} {\bibinfo  {journal} {Phys. Rev. Lett.}\ }\textbf {\bibinfo
  {volume} {114}},\ \bibinfo {pages} {100406}}\BibitemShut {NoStop}%
\end{thebibliography}%
